\documentclass[aps,prx,twocolumn,footinbib,superscriptaddress,floatfix]{revtex4-2}
\usepackage{graphicx}
\usepackage{indentfirst}
\usepackage{physics}
\usepackage{braket}
\usepackage{float}
\usepackage{amsmath}
\usepackage{amssymb}
\usepackage{mathtools}
\usepackage{epstopdf}
\usepackage{footnote}
\usepackage{CJK}
\usepackage{esint}
\usepackage{comment}
\usepackage{color}
\usepackage[T1]{fontenc}
\usepackage{subfigure}
\usepackage{amsfonts}
\usepackage{footmisc}
\usepackage{scrextend}
\usepackage{multirow}
\usepackage[hyperfootnotes=false]{hyperref}
\usepackage[acronym]{glossaries}
\usepackage[english]{babel}
\usepackage{url}
\usepackage{bm}
\usepackage{hyperref}
\usepackage{algpseudocode}
\usepackage{soul}
\usepackage{xcolor}
\definecolor{darkblue}{rgb}{0,0,0.5}
\hypersetup{
colorlinks=true,
linkcolor=black,
filecolor=blue,
citecolor=darkblue,  
urlcolor=black,
}

\newcommand*\diff{\mathop{}\!\mathrm{d}}

\usepackage{stmaryrd}
\usepackage{trimclip}
\usepackage[normalem]{ulem}

\makeatletter
\DeclareRobustCommand{\shortto}{%
  \mathrel{\mathpalette\short@to\relax}%
}

\newcommand{\short@to}[2]{%
  \mkern2mu
  \clipbox{{.5\width} 0 0 0}{$\m@th#1\vphantom{+}{\shortrightarrow}$}%
  }
\makeatother

\newtheorem{theorem}{Theorem}

\newtheorem{result}[theorem]{Result}

\newtheorem{lemma}[theorem]{Lemma}

\newenvironment{proof}[1][Proof]{\noindent\textbf{#1.} }{\ \rule{0.5em}{0.5em}}

\newcommand\argmin{\mathop{\mathrm{argmin}}}
\newcommand{\calA}{{\cal A}}

\newcommand{\calE}{{\cal E}}
\newcommand{\calF}{{\cal F}}

\newcommand{\calL}{{\cal L}}

\newcommand{\calO}{{\cal O}}

\newcommand{\calU}{{\cal U}}

\newcommand{\1}{^{(1)}}

\newcommand{\BZ}[1]{{{\textcolor{black}{#1}}}}
\newcommand{\QZ}[1]{{{\textcolor{black}{#1}}}}

\def\be{\begin{equation}}
\def\ee{\end{equation}}
\def\ba{\begin{eqnarray}}
\def\ea{\end{eqnarray}}

\begin{document}
%TC:ignore
\title{Quantum-data-driven dynamical transition in quantum learning}

%JL: suggested title: an analytic theory of quantum data in quantum learning dynamics
%the role of quantum data in quantum learning dynamics

\author{Bingzhi Zhang}
\affiliation{Department of Physics and Astronomy, University of Southern California, Los Angeles, CA 90089, USA}
\affiliation{Ming Hsieh Department of Electrical and Computer Engineering, University of Southern California, Los Angeles, CA 90089, USA}

\author{Junyu Liu}
\affiliation{Pritzker School of Molecular Engineering, The University of Chicago, Chicago, IL 60637, USA}
\affiliation{Department of Computer Science, The University of Chicago, Chicago, IL 60637, USA}
\affiliation{Kadanoff Center for Theoretical Physics, The University of Chicago, Chicago, IL 60637, USA}
\affiliation{Department of Computer Science, University of Pittsburgh, Pittsburgh, PA 15260, USA}

\author{Liang Jiang}
\affiliation{Pritzker School of Molecular Engineering, The University of Chicago, Chicago, IL 60637, USA}

\author{Quntao Zhuang}
\email{qzhuang@usc.edu}
\affiliation{Ming Hsieh Department of Electrical and Computer Engineering,
University of Southern California, Los Angeles, CA 90089, USA}
\affiliation{Department of Physics and Astronomy, University of Southern California, Los Angeles, CA 90089, USA}

\begin{abstract}
%200 words limit
Quantum neural networks, parameterized quantum circuits optimized under a specific cost function, provide a paradigm for achieving near-term quantum advantage in quantum information processing.
Understanding QNN training dynamics is crucial for optimizing their performance, however, the role of quantum data in training for supervised learning such as classification and regression remains unclear.
We reveal a quantum-data-driven dynamical transition where the target values and data determine the convergence of the training. 
Through analytical classification over the fixed points of the dynamical equation, we reveal a comprehensive `phase diagram' featuring seven distinct dynamics originating from a bifurcation with multiple codimension. Perturbative analyses identify both exponential and polynomial convergence class. We provide a non-perturbative theory to explain the transition via generalized restricted Haar ensemble.
The analytical results are confirmed with numerical simulations and experimentation on IBM quantum devices. 
Our findings provide guidance on constructing the cost function to accelerate convergence in QNN training.

\end{abstract}

%TC:endignore

\maketitle

\section{Introduction}

Classical neural networks are the crucial paradigm of machine learning that \BZ{drive} the surge of artificial intelligence. Generalizing the classical notion \BZ{to} quantum, quantum neural networks (QNN) or variational quantum algorithms~\cite{peruzzo2014variational,farhi2014quantum,mcclean2016theory,mcclean2018barren,mcardle2020quantum,cerezo2021variational,killoran2019continuous,niu2022}, have shown promise in solving complex problems involving different types of data. 
In variational quantum eigensolver (VQE)~\cite{peruzzo2014variational,kandala2017hardware} and quantum optimization~\cite{farhi2014quantum,ebadi2022quantum}, the goal is to prepare a state that minimizes a cost function, without the need \BZ{for} training data. However, supervised quantum machine learning relies on sufficient training data---\BZ{labeled} quantum states encoding either quantum or classical information. Such learning tasks have been widely explored in 
identifying phases within many-body quantum systems~\cite{cong2019quantum}, and classification \BZ{of} quantum sensing data~\cite{chen2021universal,zhang2022fast,zhuang2019,xia2021} or classical data~\cite{farhi2018classification,li2022quantum,grant2018hierarchical, li2015experimental,havlivcek2019supervised}.

%} On the other hand, the multi-data case where the cost function is a composite of data with multiple labels is important for quantum state discrimination~\cite{chen2021universal,zhang2022fast}, classical image classification~\cite{farhi2018classification,li2015experimental,li2022quantum}, many-body phase classification~\cite{cong2019quantum}, kernel support vector machine data classification~\cite{li2015experimental,havlivcek2019supervised} and quantum sensor data classification~\cite{zhuang2019,xia2021}.

With the rise of QNN applications in supervised learning, the fundamental study of their convergence properties becomes an important task, especially in the overparametrization region~\cite{larocca2023theory} where QNNs are empowered by a large number of layers. Recent progress in the theory of the Quantum Neural Tangent Kernel (QNTK)~\cite{liu2022representation,liu2023analytic,liu2022laziness,wang2022symmetric,yu2023expressibility} adopted the classical notion of neural tangent kernel to provide insight into the convergence dynamics. Furthermore, for QNNs with a quadratic loss function, a dynamical transition originating from the transcritical bifurcation \BZ{has been} revealed in the training dynamics of optimization tasks~\cite{zhang2023dynamical}. However, the results do not apply to supervised quantum machine learning, where complex quantum data \BZ{are} involved.

%\BZ{its target is limited to prepare a single target state, i.e. ground state of a physical Hamiltonian, thus is unable to characterize the convergence dynamics given training dataset with multiple quantum data in the field of supervised machine learning. \sout{only a single data is considered there, limiting its applicability to the aforementioned multi-data scenarios, which is typical in the field of quantum machine learning.}}

\begin{figure}[t]
    \centering
    \includegraphics[width=0.45\textwidth]{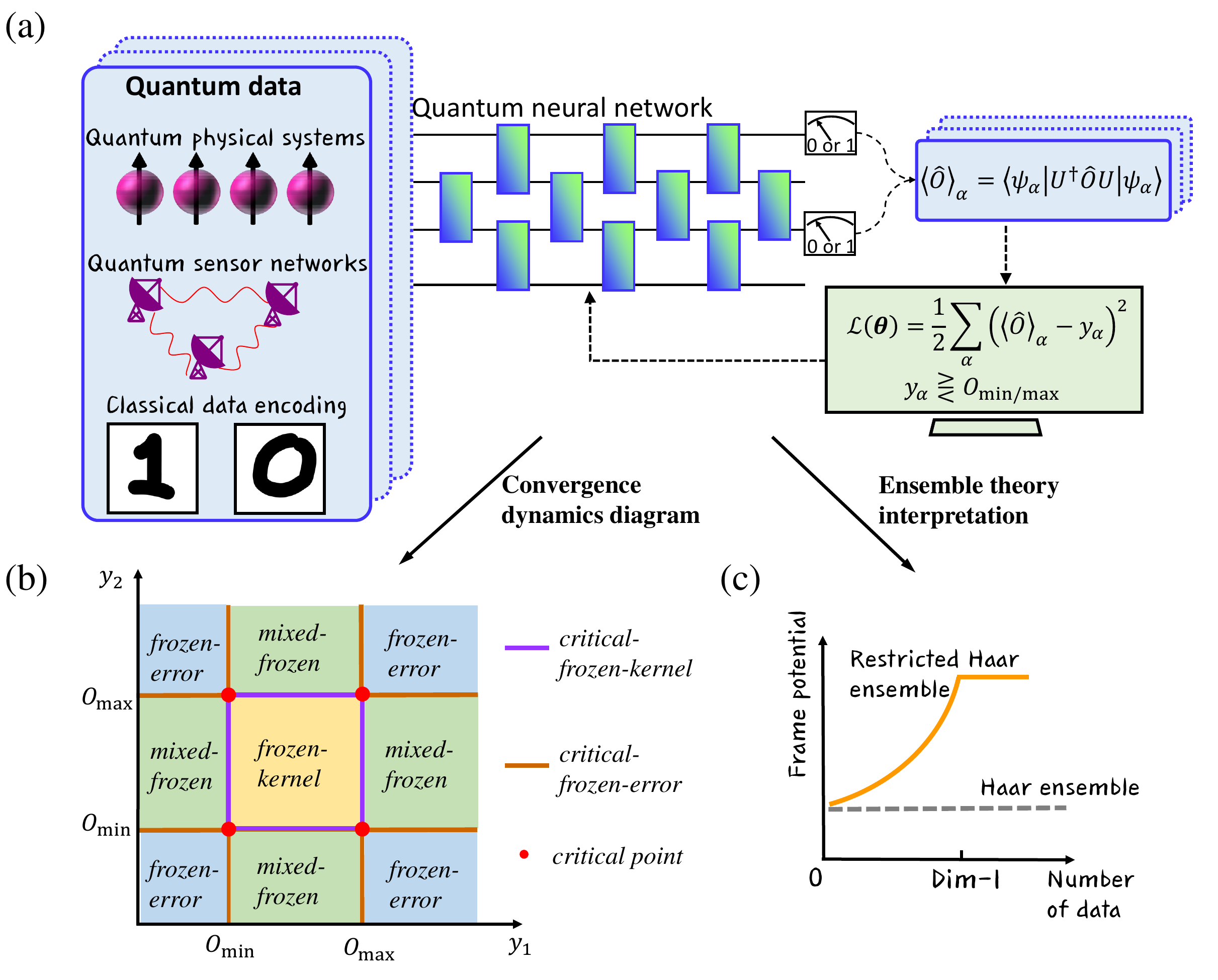}
    \caption{Illustration of the QNN for supervised learning and main results. \BZ{(a)} We study the training dynamics of errors and kernels in minimizing the MSE loss function $\calL = \frac{1}{2}\sum_\alpha (\braket{\hat{O}}_\alpha - y_\alpha)^2$, and develop a set of nonlinear dynamical equations (Eqs.~\eqref{eq:dyeqs_UT}). \BZ{(b)} We identify a dynamical transition among two convergence classes involving seven different dynamics in total (six types are shown here), and perturbatively solve its convergence dynamics. \BZ{(c)} We also provide a non-perturbative interpretation via restricted Haar ensemble theory to characterize the optimized circuits under constraints from data.}
    \label{fig:concept}
\end{figure}

In this work, we develop a quantum-data-driven theory of dynamical transition for supervised learning and reveal the complete multi-dimensional `phase diagram' in QNN training dynamics (see Fig.~\ref{fig:concept}\BZ{(b)}). Under the numerically supported assumption of the frozen relative \BZ{quantum meta-kernel (dQNTK)}, we obtain a group of nonlinear dynamical equations of the training error and kernels that \BZ{predict} seven different types of dynamics via the corresponding fixed points. Around each physical fixed point, we can define a fixed-point charge, determined by the choice of target value. When the target value crosses the boundary, \BZ{the} minimum/maximum eigenvalue of the observable, the fixed-point charge changes its sign and induces a stability transition on the fixed point, which can be identified as a bifurcation with multi-codimension.
Then, we perform a leading-order perturbative \BZ{analysis} and obtain the convergence speed of each of the seven dynamics, where an exponential convergence class and a polynomial convergence class are identified. All \BZ{the} analytical results are confirmed with numerical simulations of QNN training. Furthermore, we develop a non-perturbative unitary ensemble theory for the optimized quantum circuits to characterize the constrained randomness and to support the frozen relative dQNTK \BZ{assumption}. We also \BZ{verify} our results in examples of training dynamics with IBM quantum devices.
As the QNN training dynamics is determined by the target value choice, our results provide guidance on constructing the cost function to maximize the speed of convergence.

\section{Overview of results}

Given a QNN $\hat{U}(\bm \theta)$ with $L$ variational parameters $\bm \theta = (\theta_1,\dots,\theta_L)$, we consider a supervised learning task involving $N$ quantum data $\{\ket{\psi_\alpha}\}_{\alpha=1}^N$, each of which is associated with a real-valued target label $y_\alpha$. As shown in Fig.~\ref{fig:concept}\BZ{(a)}, the input data can be quantum states of a many-body systems~\cite{cong2019quantum}, states output from \BZ{\sout{a}} quantum sensor networks~\cite{zhuang2019} or quantum states encoding classical data~\cite{farhi2018classification}.

\BZ{For} input quantum data $\ket{\psi_\alpha}$, the QNN applies the unitary $\hat{U}(\bm \theta)$ to produce the output $\hat{U}(\bm \theta)\ket{\psi_\alpha}$ and then performs the measurement $\hat{O}$, whose result is adopted as the estimated label. Note that the target label $y_\alpha$ can be assigned arbitrarily according to different tasks, \BZ{although} the measurement $\hat{O}$ typically has bounded maximum and minimum values $O_{\rm min/max}$. For example, while Pauli measurements always provide expectation $\in[-1,1]$, in regression we may set the target values as $\pm 0.5$ and in binary classification we can also set the target values to be $\pm 2$. As indicated by the single data result in Ref.~\cite{zhang2023dynamical}, the choice of the target values has an important role in the training dynamics.

The error---the average deviation of the estimated label to the target label---associated with a data-target pair $(\ket{\psi_\alpha}, y_\alpha)$ is therefore
\begin{align}
    \epsilon_\alpha(\bm \theta) = \braket{\psi_\alpha|\hat{U}^\dagger(\bm \theta) \hat{O} \hat{U}(\bm \theta)|\psi_\alpha} - y_\alpha.
    \label{eq:epsilon_alpha_def}
\end{align}
To take into account the overall error over $N$ data, we define the \BZ{mean squared error} (MSE) loss as
\begin{align}
    \calL(\bm \theta) = \frac{1}{2N}\sum_{\alpha=1}^N \epsilon_\alpha(\bm \theta)^2.
    \label{L_cost}
\end{align}
%The QNN is optimized via gradient-descent over the above loss function depending on the gradient of each data's error $\nabla \epsilon_\alpha$ w.r.t. variational parameters $\bm \theta$, described by the $N\times N$ kernel matrix $K_{\alpha \beta}(\bm \theta) =\left\langle\nabla \epsilon_\alpha, \nabla \epsilon_\beta\right\rangle$, a inner product of gradients over parameter space. The kernel matrix can be regarded a generalization of the kernel scalar in VQE-like algorithms~\cite{zhang2023dynamical}. 
The training of QNN relies on gradient-descent update of the parameters $\bm \theta$, where each data's gradient of the error $\nabla \epsilon_\alpha(\bm \theta)$ (with respect to the parameters $\bm \theta$) plays an important role. Generalizing the kernel scalar in quantum optimization~\cite{zhang2023dynamical}, we introduce the kernel matrix $K_{\alpha \beta}(\bm \theta) =
    \left\langle\nabla \epsilon_\alpha, \nabla \epsilon_\beta\right\rangle$, an inner product of gradients over parameter space.

Our main result is that the target values $\{y_\alpha\}_{\alpha=1}^N$ determine the QNN training dynamics. The overall training can exhibit exponential converge when none of the target values are chosen as the boundary values $O_{\rm min/max}$; on the other hand, any coincidence of the target value and the boundary values of observable will lead to polynomial convergence. More specifically, depending on the interplay of the target values, seven different types of training dynamics can be identified. As shown in Fig.~\ref{fig:concept}\BZ{(b)} in a two data case, the target values $y_1$ and $y_2$ \BZ{divide} the parameter space into nine regions, with the lines $y_1=O_{\rm min/max}$ and $y_2=O_{\rm min/max}$. The four crossing points (red dots) are the {\em critical point} with polynomial convergence; the same polynomial convergence extends to the four lines, where {\em critical-frozen-error} (brown) and where {\em critical-frozen-kernel} (purple) dynamics are identified. The bulk regions enable exponential convergence and therefore are preferred. Furthermore, they are divided into three \BZ{different} dynamics, {\em frozen-kernel} (yellow), {\em mixed-frozen} (green) and {\em frozen-error} (blue). Besides the six dynamics depicted in Fig.~\ref{fig:concept}\BZ{(b)}, an additional type of training dynamics, {critical-mixed-frozen} dynamics, uniquely appears when the number of data $N>2$. 

We provide analytical theory to derive and explain behaviors of the above seven types of dynamics. Our analyses combine the solution of fixed point, the perturbative analyses around the fixed points to derive the convergence speed. In particular, we interpret the transition among different dynamics via the stability transition of fixed points, corresponding to a bifurcation transtion with multiple codimensions.

The dynamical transition is beyond the usual Haar random assumption of QNNs that only holds at initialization, as QNNs are under constraints from the convergence at late time. We develop the restricted Haar ensemble in a block-diagonal form
\begin{align}
    \calU_{\rm RH} = \left\{U\left\lvert U = \begin{pmatrix}
        Q & \bm 0\\
        \bm 0 & V
    \end{pmatrix}\right.\right\},
    \label{eq:rh_main_overview}
\end{align}
where $Q$ is a diagonal matrix with complex phases uniformly distributed to capture the convergence and $V$ is a Haar random unitary. \BZ{For any unitary ensemble, we can quantify its complexity via the frame potential~\cite{roberts2017chaos} (see detailed definition in Eq.~\eqref{eq:F2_def_main}), which is lower bounded by the value of the Haar measure.} As sketched in Fig.~\ref{fig:concept}\BZ{(c)} the ensemble has frame potential above the Haar value and increasing in a power-law with the number of data till saturation at close to the Hilbert space dimension. The frame potential is numerically verified in the QNN training.

\BZ{At the end of this section, we provide the intuition on the different choices of target values. Although it seems uncommon to choose a target value $y_\alpha > O_{\rm max}$ ($y_\alpha < O_{\rm min}$) to be nonphysical at the first glance, the minimization of loss function in Eq.~\eqref{L_cost} will force the QNN to output states with expectations of the bounded observable to be $O_{\rm max}$ ($O_{\rm min}$), which is as close as possible to the targeted nonphysical value. Thus, indeed we will obtain an optimized QNN identical to the one when setting the target values to be $O_{\rm max}$ ($O_{\rm min}$. Moreover, inspired by our previous work in optimization tasks~\cite{zhang2023dynamical}, we find that setting nonphysical target values can also further provide speedup in the supervised learning task.}

\section{Fundamental dynamical equations for training a QNN}
\label{sec:eqs}
In this section, we aim to develop the fundamental dynamical equations to simultaneously characterize the training dynamics of errors and kernels from \BZ{the first principle}.
During QNN training, we evaluate the cost function in Eq.~\eqref{L_cost} and minimize it using gradient descent to update each parameter,
\begin{align}
    \delta \theta_\ell(t) &\equiv \theta_\ell(t+1) - \theta_\ell(t) = -\eta \frac{\partial \calL(\bm \theta)}{\partial \theta_\ell} \nonumber\\
    &= -\frac{\eta}{N} \sum_{\alpha} \epsilon_\alpha(\bm \theta) \frac{\partial \epsilon_\alpha (\bm \theta)}{\partial \theta_\ell}\BZ{,}
    \label{eq:gradient_decent}
\end{align}
\BZ{where $\eta \ll 1$ is the learning rate in gradient descent.}
Accordingly, quantities depending on $\bm \theta$ also acquire new values in each training step, thus we only denote the time dependence explicitly for simplicity, e.g. $\epsilon_\alpha(t)\equiv \epsilon_\alpha (\bm \theta(t))$.
% Since $y_\alpha \in \mathbb{R}$ can be chosen freely which may lead to a nonzero converged constant of $\epsilon_\alpha$ with $t\to \infty$, we further expand the total error as 
% \begin{align}
%     \varepsilon_\alpha(\bm \theta) = \epsilon_\alpha(\bm \theta) - R_\alpha,
%     \label{eq:vareps_def}
% \end{align}
% where $R_\alpha = \lim_{t\to \infty} \epsilon_\alpha$, and the residual error $\varepsilon$ vanishes in asymptotic time limit.
From the first-order Taylor expansion, the total error $\epsilon_\alpha(t)$ is updated \BZ{using} Eq.~\eqref{eq:gradient_decent}
\begin{align}
    \delta \epsilon_\alpha(t) &= \sum_\ell \frac{\partial \epsilon_\alpha(\bm \theta)}{\partial \theta_\ell} \delta \theta_\ell + \calO(\eta^2)\\
    % &= \sum_\ell \frac{\partial \epsilon_\alpha(\bm \theta)}{\partial \theta_\ell} \left(-\frac{\eta}{N} \sum_{\beta} \epsilon_\beta(\bm \theta) \frac{\partial \epsilon_\beta(\bm \theta)}{\partial \theta_\ell}\right) + \calO(\eta^2)\\
    &= -\frac{\eta}{N} \sum_\beta K_{\alpha \beta} (\bm \theta) \epsilon_\beta(\bm \theta)  + \calO(\eta^2).
    \label{eq:eps_dyeq}
\end{align}

\BZ{Here}, we have defined the QNTK matrix as
\begin{align}
    K_{\alpha \beta}(\bm \theta) \equiv \sum_\ell \frac{\partial \epsilon_\alpha(\bm \theta)}{\partial \theta_\ell} \frac{\partial \epsilon_\beta(\bm \theta)}{\partial \theta_\ell} =
    \left\langle\nabla \epsilon_\alpha, \nabla \epsilon_\beta\right\rangle, 
    %\left\langle\frac{\partial \epsilon_\alpha(\bm \theta)}{\partial \bm \theta}, \frac{\partial \epsilon_\beta(\bm \theta)}{\partial \bm \theta}\right\rangle,
    \label{eq:K_def}
\end{align}
where $\nabla \epsilon_\alpha \equiv (\frac{\partial \epsilon_\alpha}{\partial \theta_1}, \dots, \frac{\partial \epsilon_\alpha}{\partial \theta_L})^T$ is the gradient vector of $\epsilon_\alpha$, and $\langle\cdot, \cdot \rangle$ represents the inner product over parameter space. 
By definition, the QNTK is a positive \BZ{semidefinite} symmetric matrix. The diagonal term $K_{\alpha\alpha}= \left\langle \nabla \epsilon_\alpha,\nabla \epsilon_\alpha   \right\rangle\equiv \|\nabla \epsilon_\alpha\|^2$ is the square of the norm of the gradient vector, while the off-diagonal term $K_{\alpha\beta}$ provides information about the angle between different gradient vectors. Indeed, following the definition of angle between gradient vectors, $\cos \angle\left[\nabla \epsilon_\alpha, \nabla\epsilon_\beta\right] = \left\langle\nabla \epsilon_\alpha, \nabla \epsilon_\beta\right\rangle /  \|\nabla\epsilon_\alpha\| \|\nabla\epsilon_\beta\|$, we can retrieve the geometric angle from the above defined QNTK as
\begin{align}
    \angle_{\alpha \beta}(\bm \theta) &\equiv\cos \angle\left[\nabla \epsilon_\alpha, \nabla\epsilon_\beta\right]= \frac{K_{\alpha \beta}}{\sqrt{K_{\alpha \alpha} K_{\beta \beta}}} 
    \label{eq:B_def}
\end{align}
where the matrix $\angle_{\alpha\beta}(\bm \theta)$ is introduced to simplify the notation. 

% The above difference equation becomes
% \begin{align}
%     \delta \epsilon_\alpha &= -\frac{\eta}{N} \sum_{\beta}\left(\sum_{\ell} J_{\alpha \ell} J_{\beta \ell}\right)\epsilon_\beta = -\frac{\eta}{N} \sum_{\beta} K_{\alpha \beta}\epsilon_\beta
% \end{align}

Our study focuses on the training dynamics of both errors and kernels of the QNNs. To study the convergence, we often separate the error into two parts:  
$
\epsilon_\alpha (t)\equiv \varepsilon_\alpha(t) + \epsilon_\alpha(\infty)
$
consists of a constant remaining term $\epsilon_\alpha(\infty)$
% $
% R_\alpha=\lim_{t\to \infty}\epsilon_\alpha(t)
% $
and a vanishing residual error
$
\varepsilon_\alpha(t).
$

With similar techniques in obtaining Eq.~\eqref{eq:eps_dyeq}, in Method %Appendix~\ref{app:QNTK} 
we derive the dynamical equation of QNTK. Combining with Eq.~\eqref{eq:eps_dyeq}, we have a set of coupled nonlinear dynamical equations for total error and QNTK
\begin{align}
    \left\{ \begin{array}{ll}
    \delta \epsilon_\alpha(t) = -\frac{\eta}{N} \sum_\beta K_{\alpha \beta}(t) \epsilon_\beta(t) ;\\
    \delta K_{\alpha \beta}(t) = -\frac{\eta}{N} \sum_\gamma \epsilon_\gamma(t) \left[\mu_{\gamma \beta \alpha}\left(t\right) + \mu_{\gamma \alpha \beta}\left(t\right) \right]
    .\end{array} \right.
    \label{eq:dyeqs_full}
\end{align}
where the dQNTK $\mu_{\gamma \alpha \beta}$ is defined as
\begin{align}
    \mu_{\gamma \alpha \beta}(\bm \theta) = \sum_{\ell^\prime, \ell} \frac{\partial \epsilon_\gamma (\bm \theta)}{\partial \theta_\ell}  \frac{\partial^2 \epsilon_\alpha (\bm \theta)}{\partial \theta_{\ell} \partial \theta_\ell^\prime}  \frac{\partial \epsilon_\beta (\bm \theta)}{\partial \theta_{\ell^\prime}},
    \label{eq:mu_def}
\end{align}
which is a bilinear form of total error's gradient and \BZ{Hessian}. Since we utilize a quadratic loss function Eq.~\eqref{L_cost}, there exists a gauge invariance under the orthogonal group $O(N)$ on the data space for loss function, thus on the gradient descent update in Eq.~\eqref{eq:gradient_decent} and dynamical equations in Eqs.~\eqref{eq:dyeqs_full}, \QZ{as we show in Supplementary Note 3}. 
%(See details in Appendix~\ref{app:gauge_invariance}).
However, quantities of inner products over parameter space, e.g. QNTK and dQNTK, are not gauge invariant.

\QZ{Before moving on, we emphasize that the dynamical equations in this section actually apply to the gradient-descent training of any quadrature loss function in Eq.~\eqref{L_cost}, regardless of whether it regards a QNN or classical systems.}

\section{Assumption of fixed relative dQNTK} 
\begin{figure}[t]
    \centering
    \includegraphics[width=0.45\textwidth]{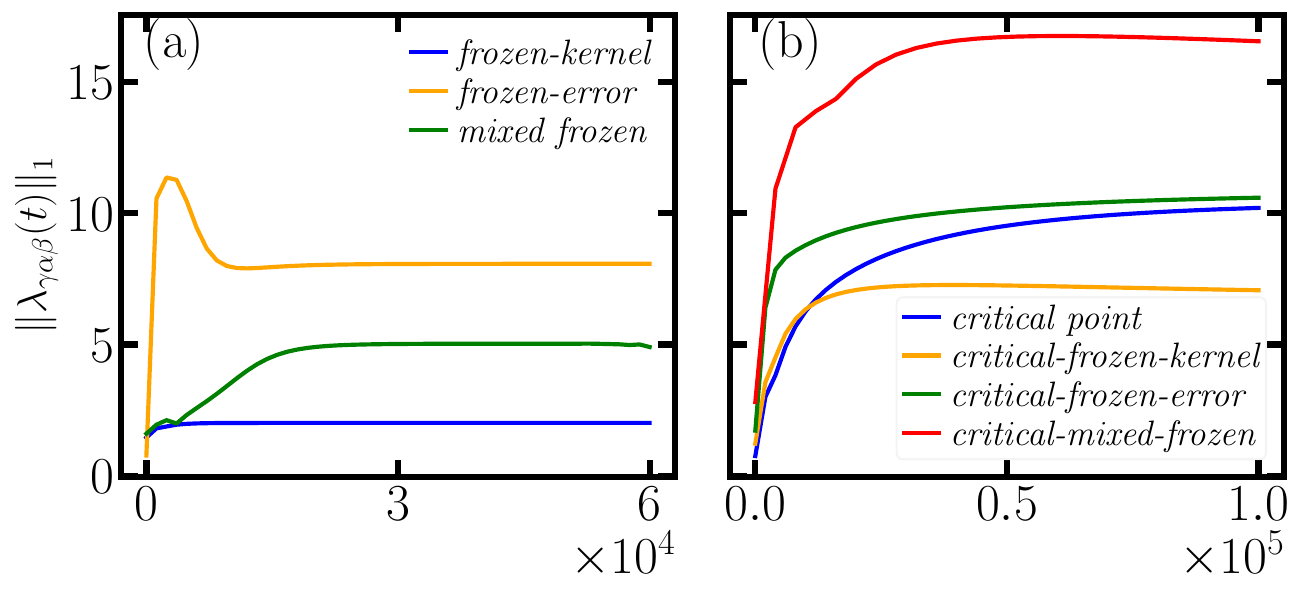}
    \caption{Convergence of relative dQNTK. We show the norm $\|\lambda_{\gamma \alpha \beta}(t)\|_1 \equiv \sum_{\gamma \alpha \beta}|\lambda_{\gamma \alpha \beta}(t)|$ for (a) exponential convergence class and (b) polynomial convergence class (detailed in `Classifying the dynamics' section %Sec.~\ref{sec:dy_classify}
    ). The targets for orthogonal data states are $y_1 = 0.3, y_2 = -0.5$ (blue), $y_1 = 5, y_2 = -6$ (orange) and $y_1 = 0.4, -5$ (green) in (a); $y_1 = 1, y_2 = -1$ (blue), $y_1 = 0.4, y_2 = -1$ (orange), $y_1 = 1, y_2 = -5$ (green) and $y_1 = 0.4, y_2 = 1, y_3 = -5$ (red) in (b). The corresponding dynamics are identified in Fig.~\ref{fig:venn} and Table.~\ref{tab:phases}.
    Here random Pauli ansatz (RPA) consists of $L= 48$ variational parameters on $n=4$ qubits with $\hat{O} = \hat{\sigma}_1^z$, Pauli-Z operator on the first qubit.
    }
    \label{fig:lda}
\end{figure}

In this section, we propose the key assumption (supported in `Ensemble average results' section) in order to analytically study the training dynamics through reduction on the number of independent variables in Eqs.~\eqref{eq:dyeqs_full}.
In a typical training process \BZ{toward} reaching a local minimum, the \BZ{Hessian} $\frac{\partial^2 \epsilon_\alpha}{\partial \theta_{\ell} \partial \theta_{\ell^\prime}}$ converges to a constant \BZ{during} late-time training. Therefore, according to the definition of dQNTK in Eq.~\eqref{eq:mu_def}, we can expect that $\mu_{\gamma \alpha \beta} \sim K_{\gamma \beta}$ has the same scaling. 
This intuition motivates us to define the relative dQNTK $\lambda_{\gamma \alpha \beta}(t)$ as
\begin{align}
    \lambda_{\gamma \alpha \beta}(t) = \frac{\mu_{\gamma \alpha \beta} (t)}{\sqrt{K_{\gamma \gamma}(t) K_{\beta \beta}(t)}},
    \label{eq:lambda_def_main}
\end{align}
which reduces to the scalar version in Ref.~\cite{zhang2023dynamical} for optimization when $N=1$. 
Our major assumption in this work is that the relative dQNTK converges to a constant $\lambda_{\gamma \alpha \beta}(t) \to \lambda_{\gamma \alpha \beta}$ in the late time. 
We numerically verify the assumption in various cases, as we detail in \QZ{Supplementary Note 6}.
%\ref{app:numeric_detail}
In Fig.~\ref{fig:lda}, we plot the sum of the absolute values, $\|\lambda_{\gamma \alpha \beta}\|_1 \equiv \sum_{\gamma\alpha \beta}|\lambda_{\gamma \alpha \beta}|$, to show the convergence. 
This assumption is not only motivated by previous results of Ref.~\cite{zhang2023dynamical}, but also supported by the unitary ensemble theory in `Ensemble average results' section.

Under the constant relative dQNTK assumption, the dynamical equations of Eq.~\eqref{eq:dyeqs_full} then \BZ{become}
% \begin{align}
%     \left\{ \begin{array}{ll}
%     \delta \epsilon_\alpha = -\frac{\eta}{N} \sum_\beta K_{\alpha \beta} \epsilon_\beta ;\\
%     \delta K_{\alpha \beta} = -\frac{\eta}{N} \sum_\gamma \epsilon_\gamma \sqrt{K_{\gamma \gamma}} \left(\lambda_{\gamma \beta \alpha}\sqrt{K_{\alpha \alpha}} + \lambda_{\gamma \alpha \beta} \sqrt{K_{\beta \beta}}\right)
%     .\end{array} \right.
%     \label{eK_dynamics}
% \end{align}
\begin{align}
    \left\{ \begin{array}{ll}
    \partial_t \epsilon_\alpha(t) = -\frac{\eta}{N} \sum_\beta K_{\alpha \beta}(t) \epsilon_\beta(t) ;\\
    \partial_t K_{\alpha \beta}(t) = -\frac{\eta}{N}  \left(f_{\beta \alpha}(t)\sqrt{K_{\alpha \alpha}(t)} + f_{\alpha\beta}(t) \sqrt{K_{\beta \beta}(t)}\right)
    .\end{array} \right.
    \label{eK_dynamics}
\end{align}
where we have defined the functions
\be 
f_{\alpha\beta}(t)=\sum_\gamma \sqrt{K_{\gamma\gamma}(t)} \epsilon_\gamma(t) \lambda_{\gamma \alpha \beta}
\label{f_def}
\ee 
for convenience and taken the continuous-time limit.

Our major result is the classification of the training dynamics of QNN in supervised learning based on Eq.~\eqref{eK_dynamics}. In the next section, we obtain the fixed points representing each dynamics under similar assumptions as in Ref.~\cite{zhang2023dynamical}. In `Convergence towards fixed points', we further provide perturbative analyses on the late-time training dynamics to obtain the convergence speed towards the fixed points. In `Ensemble average results' section, we develop the unitary ensemble theory to support the assumption proposed above. In `Experiment' section, we present experimental results on IBM quantum devices.

\QZ{We point out that our main conclusions hold generally for gradient-descent training of bounded observables under quadratic loss function, assuming the the fixed relative dQNTK assumption, regardless of the detailed dynamics---quantum or classical.}

\section{Solving the fixed points}
From Eqs.~\eqref{eK_dynamics}, we can obtain the fixed points below. 
\begin{result}
\label{th:main}
{\it (Frozen gradient angle and error-kernel duality)}
\BZ{There exists} a family of fixed points of the training dynamics of Eq.~\eqref{eK_dynamics} \BZ{satisfying}
\ba 
& \epsilon_\alpha K_{\alpha \alpha} =0, \forall \alpha,
\\
&\angle_{\alpha\beta}={\rm const}.
\ea 
\end{result}
In other words, in late-time training, (1) the error $\epsilon_\alpha$ and kernel $K_{\alpha\alpha}$ \BZ{satisfy} a duality---either one of the two is zero or both are zero; (2) the relative orientation among gradient vectors associated with each data is fixed. We \BZ{claim} the above conclusion as a result instead of a theorem, as there is a weak assumption behind it: the functions $f_{\alpha\beta}(t)$ have the same scaling \BZ{versus} $t$ despite different $\alpha$ and $\beta$.

To show Result~\ref{th:main}, we begin with the following lemma
\begin{lemma}
\label{lemma_B}
When the ratio
\be 
\calA_{\alpha\beta}= \lim_{t\to\infty} \frac{\left(\frac{f_{\beta\alpha}(t)}{\sqrt{K_{\beta\beta}(t)}} + \frac{f_{\alpha\beta}(t)}{\sqrt{K_{\alpha\alpha}(t)}} \right)}{\left(\frac{f_{\beta\beta}(t)}{\sqrt{K_{\beta\beta}(t)}} + \frac{f_{\alpha \alpha}(t)}{\sqrt{K_{\alpha\alpha}(t)}}\right)}={\rm const},
\ee 
is a finite constant \BZ{in the interval} $[-1,1]$.
Then $\angle_{\alpha\beta}(\infty)=\calA_{\alpha\beta}$ is a fixed point \QZ{of Eq.~\eqref{eK_dynamics}}.
\end{lemma}
We provide the proof in \QZ{Supplementary Note 1} %\QZ{\ref{app:proof_lemma_B}}.
We expect the conditions in Lemma~\ref{lemma_B} to hold, as the functions $f_{\alpha\beta}(t)$ defined in Eq.~\eqref{f_def} have the same scaling with time $t$ for different indices $\alpha,\beta$ at late time. Indeed, this is true unless the constants $\lambda_{\gamma\alpha\beta}$'s are particularly chosen such that certain terms can exactly cancel out in the summation of Eq.~\eqref{f_def}.
Under the assumption that the functions $f_{\alpha\beta}(t)$ have the same scaling, we find that $\calA_{\alpha\beta}$'s are indeed constants by symmetry of the expression. Furthermore, our numerical results \QZ{(see Supplementary Note 6)} %\QZ{(see~\ref{app:numeric_detail})} 
indeed support that the constant is between $[-1,1]$. 

% An interesting phenomena to note is that under the time-independent $N$-dimensional orthogonal transform on data space which can be regarded as a global gauge, both geometric angle $\angle_{\alpha \beta}$ and relative dQTNK $\lambda_{\gamma \alpha \beta}$ remain time-invariant in late time detailed in Appendix~\ref{app:gauge_invariance}, which may suggest the universality of convergence  $\angle_{\alpha \beta}$ and $\lambda_{\gamma \alpha \beta}$ due to its independence on the choice of data space. 

%Since $\angle_{\alpha \beta}\in [-1,1]$ is bounded, it is reasonable to assume it goes to the fixed point whenever the fixed point is within the region of $[-1,1]$.

From \BZ{the} definition in Eq.~\eqref{eq:B_def}, with $\angle_{\alpha \beta}(t) = \angle_{\alpha \beta}$ being a constant, $K_{\alpha\beta}(t)=\angle_{\alpha\beta}\sqrt{K_{\alpha\alpha}(t) K_{\beta\beta}(t)}$ is entirely determined by the diagonal kernels. Therefore, in the kernel-error dynamical equation \eqref{eK_dynamics},  the only independent variables are $\{\epsilon_\alpha(t), K_{\alpha \alpha}(t)\}_{\alpha = 1}^N$ and the relevant dynamical equations among Eq.~\eqref{eK_dynamics} can be simplified to
\begin{align}
    \left\{ \begin{array}{ll}
    \partial_t \epsilon_\alpha(t) = -\frac{\eta}{N} \sum_\beta \angle_{\alpha \beta} \sqrt{K_{\alpha \alpha}(t)} \sqrt{K_{\beta \beta}(t)} \epsilon_\beta(t) ;\\
    \partial_t \sqrt{K_{\alpha \alpha}(t)} = -\frac{\eta}{N} \sum_\beta   \lambda_{\alpha \alpha \beta} \sqrt{K_{\beta \beta} (t)}\epsilon_\beta (t)
    .\end{array} \right.
    \label{eq:dyeqs_UT}
\end{align}
% where we define another matrix $A_{\alpha \gamma} \equiv \lambda_{\gamma \alpha \alpha}$ for convenience.  
%In continuous limit, we can write it in ordinary differential equations as
% \begin{align}
%     \left\{ \begin{array}{ll}
%     \diff_t \epsilon_\alpha = -\frac{\eta}{N} \sum_\beta \angle_{\alpha \beta} \sqrt{K_{\alpha \alpha}} \sqrt{K_{\beta \beta}} \epsilon_\beta ;\\
%     \diff_t K_{\alpha \alpha} = -\frac{2\eta}{N} \sum_\beta A_{\alpha \beta}  \sqrt{K_{\alpha \alpha}} \sqrt{K_{\beta \beta}}\epsilon_\beta
%     .\end{array} \right.
% \end{align}
From here, we can conclude that $\{K_{\alpha\alpha}\epsilon_\alpha=0, \forall \alpha\}$ forms a family of fixed points, which arrives at Result~\ref{th:main}. 

\section{Classification of the dynamics}
\label{sec:dy_classify}
\begin{figure}[t]
    \centering
    \includegraphics[width=0.45\textwidth]{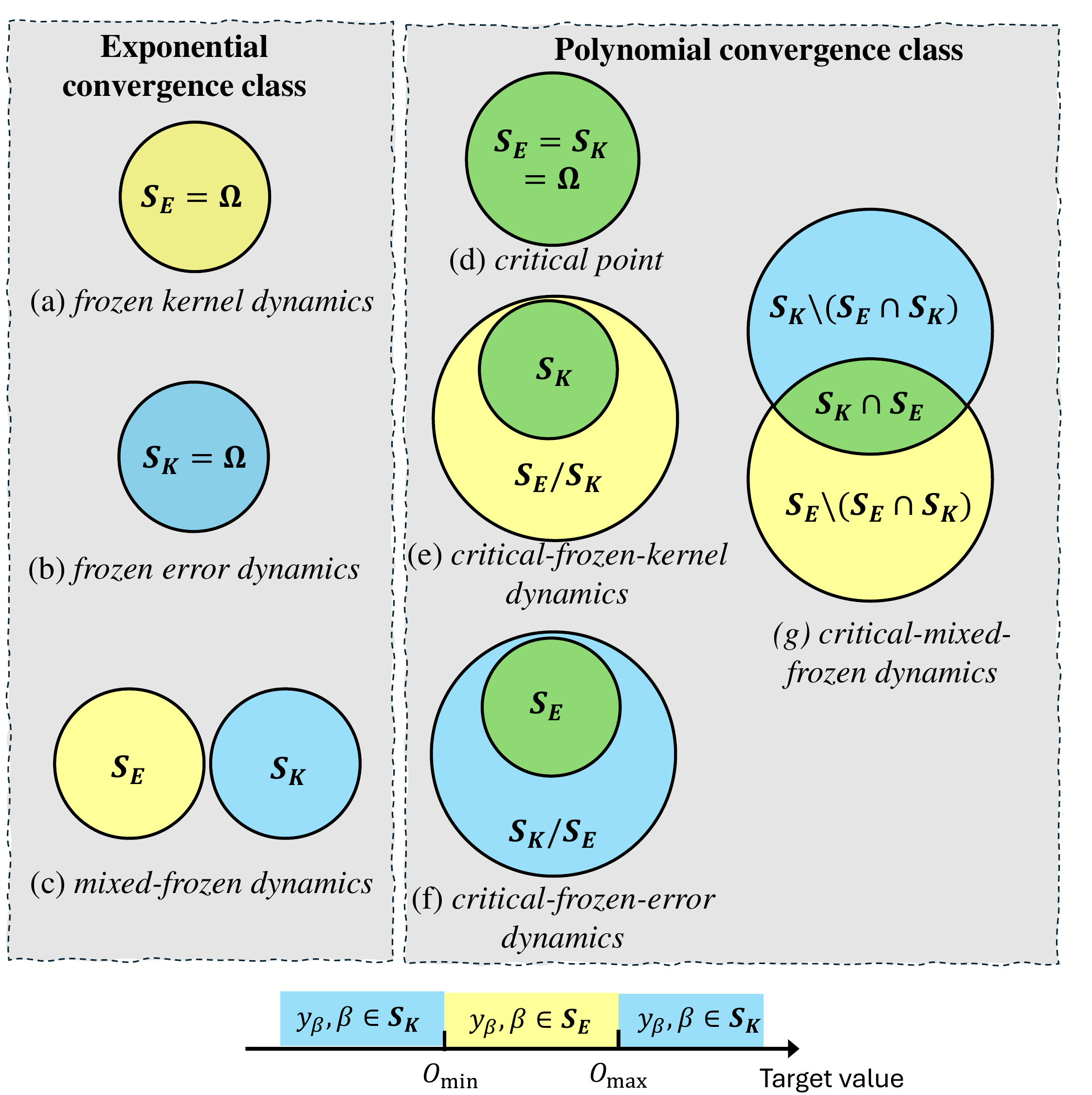}
    \caption{Venn diagram of classes of dynamics. In all cases, we have $S_E \cup S_K = \Omega$. The corresponding dynamics are explained in `Convergence towards fixed points' section. The bottom legend shows the the connection of the set $S_E$ and $S_k$ to the target value configuration. }
    \label{fig:venn}
\end{figure}

As indicated in Result~\ref{th:main}, $\{K_{\alpha \alpha} \epsilon_\alpha = 0, \ \forall \alpha\}$ defines a family of fixed points. Since $K_{\alpha \alpha} \epsilon_\alpha = 0$ can be achieved by either $K_{\alpha \alpha}=0$ or $\epsilon_\alpha = 0$ or both of them are zeros, we can have various different fixed points. Below we systematically classify the QNN dynamics based on the fixed points. Denote $\Omega = \{\beta\}_{\beta=1}^N$ to be the whole set of data indices, we can define two sets of indices $S_E, S_K$ conditioned on the convergence of errors and kernels as
\begin{align}
    \left\{\begin{array}{ll}
    S_E \equiv \{\beta|\lim_{t\to \infty} \epsilon_\beta(t) = 0\}; \\
    S_K \equiv \{\beta|\lim_{t\to \infty} K_{\beta \beta}(t) = 0\},
    \end{array}\right.
    \label{eq:Se_Sk_def}
\end{align}
where $S_E \cup S_K = \Omega$ always holds.
% and the set of labels $\beta\in S_E$ and $\beta \in S_K$ for those with $\epsilon_\beta (t) \to 0$, and $K_\beta (t) \to 0$ correspondingly. With $S_E \cup S_K = \Omega$ always, 
The fixed points can thus be classified in terms of the relation between the zero-error indices $S_E$ and the zero-kernel indices $S_K$, as we list in the table below
\begin{table}[H]
    \centering
    \begin{tabular}{c c c}
    \hline
        \multicolumn{2}{c}{$S_E\cap S_K= \emptyset$}  & Exponential convergence class  \\
        \hline
         \ \ \ \  \ \ \ \ & $S_K = \emptyset$ & {\it frozen-kernel dynamics}\\
         \ \ \ \  \ \ \ \ &$S_E = \emptyset$& {\it frozen-error dynamics}\\
         &$S_E, S_K \neq \emptyset$& {\it mixed-frozen dynamics}\\
         \hline 
         \hline 
         \multicolumn{2}{c}{$S_E \cap S_K \neq \emptyset$}  & Polynomial convergence class  \\
         \hline 
         &$S_E = S_K =\Omega$& {\it critical point} \\
         & $S_K \subsetneq S_E = \Omega$& {\it critical-frozen-kernel dynamics}\\
         & $S_E \subsetneq S_K = \Omega$& {\it critical-frozen-error dynamics}\\
         % & $S_E \subsetneq S_K, S_K \subsetneq S_E$
         &$S_E \not\subset S_K, S_K \not\subset S_E$ & {\it critical-mixed-frozen dynamics}\\
         \hline 
    \end{tabular}
    \caption{Summary of the relation between zero error and kernel index sets $S_E, S_K$ and the corresponding different types of QNN training dynamics. All types of dynamics are explained in `Convergence towards fixed points' section.}
    \label{tab:phases}
\end{table}
We also depict the Venn diagram \BZ{of} each \BZ{type} of dynamics to visually represent the table above in Fig.~\ref{fig:venn}. All the names of the dynamics and the overall classification of exponential versus polynomial convergence (in the residual error) will be explained in `Convergence towards fixed points' section. Compared with the case of optimization algorithms considered in Ref.~\cite{zhang2023dynamical}, QNNs for supervised learning have four extra types of dynamics, {\it mixed-frozen, critical-frozen-kernel, critical frozen-error} and {\it critical-mixed-frozen dynamics} due to the interaction between data through convergence. 

%As a nonlinear dynamical equation, typically the fixed point that a dynamics will converge towards is determined by the initial point. While in the single data case, conserved quantity involving kernel and error is explicitly solvable and therefore the phase diagram for each fixed point is obtained~\cite{zhang2023dynamical}, in this multi-data case we are not able to find explicit conserved quantity for phase diagram. Instead, we can adopt intuitive approach to determine the phase diagram based on empirical results in the single-data case. 

To determine which set a data state belongs to in Eq.~\eqref{eq:Se_Sk_def}, we need to identify for a particular data index $\beta$ whether the kernel $K_{\beta\beta}(t)$ or the error $\epsilon_\beta(t)$  will decay to zero at late time. While the exact determination will require training the QNN to late time, we can obtain intuition from the relation between target value $y_\beta$ and achievable values for the observable $\hat{O}$.
When a target value $y_\beta$ lies within the achievable region $(O_{\rm min}, O_{\rm max})$, the error $\epsilon_\beta(t)$ is expected to converge to zero when the circuit is deep, implying $\beta \in S_E$; When a target value is not in the achievable region, then we expect $\epsilon_\beta (t)$ to converge to nonzero constants. Thus, the fixed point condition in Result~\ref{th:main} requires $K_{\beta\beta} (t)$ vanishing to zero, and thus $\beta\in S_K$; when the target value is at the boundary $y_\beta = O_{\rm min/max}$, then we expect the special case of critical phenomena with both error and kernel vanishing at late time thus $\beta \in S_E \cap S_K$. The above intuition about target value and `phase diagram' can be summarized as the following
\begin{align}
\left\{ \begin{array}{ll}
         \beta \in S_E, & \mbox{if $y_\beta \in [O_{\rm min}, O_{\rm max}]$};\\
        \beta \in S_K, & \mbox{if $y_\beta \in (-\infty, O_{\rm min}] \cup [O_{\rm max}, +\infty)$}.\end{array} \right.
\end{align}
When $y_\beta = O_{\rm min}$ or $O_{\rm max}$, we have $\beta \in S_E \cap S_K$.
The Venn diagrams summarize the classification of fixed points and connection to target value configuration for each case, as shown in Fig.~\ref{fig:venn}.

Numerical analysis confirms that this classification holds for the orthogonal data case, where $\braket{\psi_\alpha|\psi_\beta} = \delta_{\alpha \beta}$, as detailed in the following section. Although the orthogonality property does not hold always in machine learning tasks, we take the orthogonal data as a typical case to unveil the fruitful physical phenomena within the training dynamics. In practice, typical random states in high-dimensional space are expected to be exponentially close to orthogonal states. Important quantum machine learning tasks involving state discrimination and classification also benefit from orthogonal data encoding due to the  Helstrom limit~\cite{helstrom1967minimum, helstrom1969quantum}. 

Since the dynamical equations in Eq.~\eqref{eq:dyeqs_full} are gauge invariant, the fixed point identified in Result~\ref{th:main} is also gauge invariant. However, the classification of the dynamics will be dependent on the choice of gauge---different ways of defining the error as combinations of the natural basis in Eq.~\eqref{eq:epsilon_alpha_def}. This is intuitive, as the dynamical transitions are driven by the data and the target values are naturally tuned according to each observable.

\begin{figure}[t]
    \centering
    \includegraphics[width=0.5\textwidth]{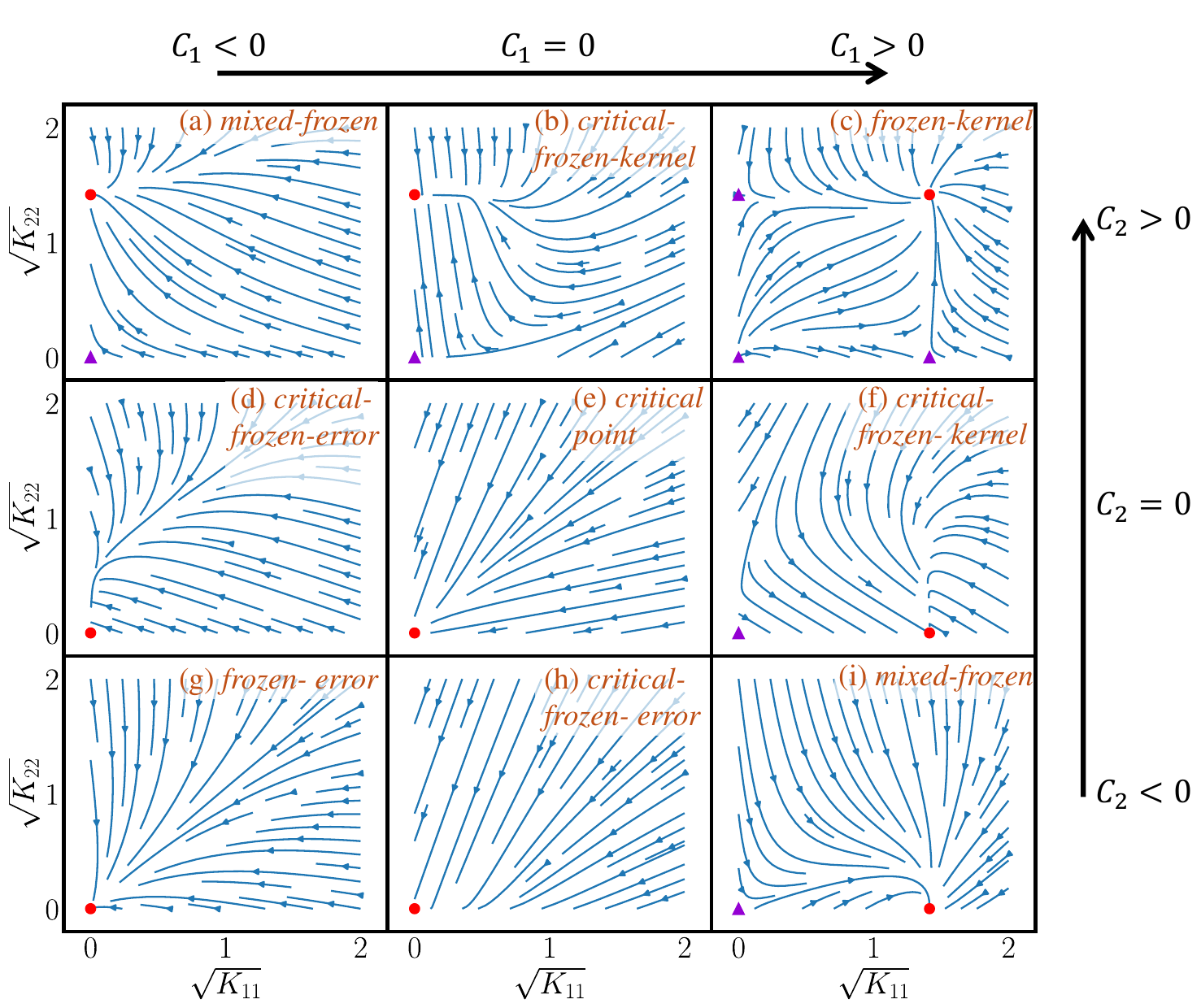}
    \caption{Flow diagram for convergence toward fixed points. The flow diagram is described by Eq.~\eqref{eq:dyeq_g_main}. Red dots in each subplot represent the only \BZ{physically} accessible stable fixed point, while purple triangles represent unstable fixed points. Here we choose $C_1, C_2$ to be $\pm 2, 0$. }
    \label{fig:flow}
\end{figure}

\section{Stability transition of fixed points: bifurcation}
\label{sec:stability}
We have identified the family of fixed points for the dynamical equations (Eq.~\eqref{eq:dyeqs_UT}) in Result~\ref{th:main}, and seen the classification of dynamics in `Classifying the dynamics' section. %Section~\ref{sec:dy_classify}. 
In this part, we aim to study the stability of every possible fixed point, which provides theoretical support on the convergence of each dynamics discussed above, and reveals the nature of the transition among different dynamics. 

Around any fixed point $(\epsilon_\alpha^*, K_{\alpha \alpha}^*)$ of the dynamical equations in Eq.~\eqref{eq:dyeqs_UT}, we can define a group of constant fixed-point charges as
\begin{align}
    C_\alpha = K_{\alpha \alpha}^* - 2\lambda_{\alpha \alpha \alpha} \epsilon_\alpha^*, \forall \alpha.
    \label{eq:C_def}
\end{align}
\BZ{Note that the above fixed-point charges \BZ{are} only well-defined around the fixed point. We introduce them to analyze the stability of fixed point as we will detail below. It is \BZ{\sout{the}} different from the conserved quantity identified in the optimization learning task~\cite{zhang2023dynamical} which holds for the entire late-time training supported by the corresponding dynamical equation.}
Thanks to the constants $C_\alpha$, we can decouple the dynamical equation near the fixed point, and reduce it to a set of equations dependent only on $K_{\alpha \alpha}(t)$,
\begin{align}
    \partial_t \sqrt{K_{\alpha \alpha}(t)} &= -\frac{\eta}{2N} \sum_\beta \frac{\lambda_{\alpha \alpha \beta}}{\lambda_{\beta \beta \beta}}  \sqrt{K_{\beta \beta}(t)} \left(K_{\beta \beta}(t) - C_\beta\right) \label{eq:dyeq_g_main_1} \\
    &\equiv \frac{\eta}{2N} G_\alpha (\{K_{\beta \beta}\}, \{C_\beta \}),
        \label{eq:dyeq_g_main}
\end{align}
where we introduce the function $G_\alpha (\{K_{\beta \beta}\}, \{C_\beta \})$ for convenience. Note that Eq.~\eqref{eq:dyeq_g_main} only holds near the fixed point. Through the linearization at fixed point $\{K_{\alpha \alpha}^*\}$ (see details in Method %Appendix~\ref{app:fix_point_stability}
), we have
\begin{align}
    &\partial_t \sqrt{K_{\alpha \alpha}(t)}\nonumber\\
    &= \frac{\eta}{2N} \sum_{\beta} M_{\alpha \beta}(\{K_{\beta \beta}^*\}, \{C_\beta \}) \left(\sqrt{K_{\beta \beta}(t)} - \sqrt{K_{\beta \beta}^*}\right),
    \label{eq:K_eq_linear}
\end{align}
where the matrix $M_{\alpha \beta}(\{K_{\beta \beta}^*\}, \{C_\beta \})$ is the Jacobian of $G_{\alpha}$ w.r.t. each kernel element $\sqrt{K_{\beta \beta}}$ at the fixed point $\{K_{\beta \beta}^*\}$
\begin{align}
    M_{\alpha \beta}(\{K_{\beta \beta}\}, \{C_\alpha \}) \equiv \left.\frac{\partial G_\alpha (\{K_{\beta \beta}\}, \{C_\beta \})}{\partial \sqrt{K_{\beta \beta}}}\right\rvert_{\{K_{\beta \beta}^*\}}.
    \label{eq:M_def_main}
\end{align}
The stability of the fixed point  $\{K_{\beta \beta}^*\}$ can thus be determined from the spectrum of the matrix $M_{\alpha \beta}(\{K_{\beta \beta}^*\}, \{C_\beta \})$. Once an eigenvalue with a positive real part appears, the fixed point becomes unstable. Combining the stable fixed point and $\{C_\alpha\}$, we can directly derive the classification in Fig.~\ref{fig:venn}, and therefore connect the each fixed point to the corresponding class of training dynamics.

We take the two-data case as an example to reveal the stability transition of the fixed points under the change of $\{C_\beta\}$. In this case, the eigenvalue of the 2-by-2 matrix $M$ is a function of $\tr(M)$ and $\det(M)$ only.
One can easily find the trace and determinant as
\begin{align}
    \left\{ \begin{array}{ll}
    \tr(M) = C_1 + C_2 - 3(K_{11}^* + K_{22}^*),\\
    \det(M) \propto \left(C_1 - 3 K_{11}^*\right)\left(C_2 - 3 K_{22}^*\right).
    \end{array}\right.
    \label{eq:tr_det_M}
\end{align}
Recall that $K_{\alpha \alpha}$ is defined to be the $2$-norm of total error's gradient w.r.t. variational parameters, the \BZ{physically} accessible fixed point can only be $(K_{11}^*, K_{22}^*) = (C_1, C_2), (C_1, 0), (0, C_2)$ and $(0, 0)$. 
Via tuning $(C_1, C_2)$, the stability of each fixed point would undergo a transition, illustrated by the flow diagrams in Fig.~\ref{fig:flow}.
When $C_1, C_2 >0$, all the four fixed points are physically accessible (Fig.~\ref{fig:flow}(c)). However, only $(K_{11}^*, K_{22}^*)=(C_1, C_2)$ (red dot) is a stable fixed point with $\tr(M)<0, \det(M)>0$ where every flow points toward it, while the others (purple triangles) are all unstable to be either a saddle point or a source. As $C_1, C_2>0$ are both positive, its convergence toward $(C_1, C_2)$ corresponds to the {\it frozen-kernel dynamics}.
When we hold one of the charge to be positive while tuning the other one, for instance, decreasing $C_2$ from positive to negative with $C_1>0$ ((c)-(f)-(i)), due to the requirement that $K_{\alpha \alpha}>0$, only the fixed points $(C_1, 0)$ and $(0, 0)$ are physically accessible, then we find that $(C_1, 0)$ becomes a stable fixed point (red dots in (f), (i)), while $(0, 0)$ (purple triangles in (f), (i)) is still unstable, corresponding to the {\it critical-frozen-kernel dynamics} and {\it mixed-frozen dynamics} separately. Similar analysis holds for tuning $C_1$ while holding $C_2>0$ ((c)-(b)-(a)), resulting in the same dynamical transition. When we have $C_2 <0$ while decreasing $C_1$ from positive to negative, we see the only \BZ{physically} accessible and stable fixed point is $(0, 0)$ (red dots in (g)(h)), leading to the {\it critical-frozen-error dynamics} and {\it frozen-error dynamics} separately. Specifically, when we have both $C_1 = C_2 = 0$, all fixed points collide and leads to {\it critical point}. 
Therefore, we can identify the stability transition of the fixed point as a bifurcation transition with multiple codimensions. Although the linearized dynamics in Eq.~\eqref{eq:K_eq_linear} only hold close to the fixed point, the bifurcation transition in supervised learning we uncover holds generally. While the fixed point location changes under gauge transform $O(N)$, its stability property persists since the spectrum of $M_{\alpha \beta}$ is gauge invariant.

\section{Convergence towards fixed points: Exponential convergence class}
\label{sec:convergence}
Now we assume the dynamical quantities---the errors and QNTKs---converge towards the fixed point given in Result~\ref{th:main} and study the convergence speed for different dynamics identified above in Table~\ref{tab:phases}.  To unveil the scaling of convergence for each dynamics, we solve the dynamical equations in Eqs.~\eqref{eq:dyeqs_UT} close to the known stable fixed point identified above in `Stability transition of fixed points: bifurcation' section, and present the corresponding solution in leading order, verify our theoretical predictions with numerical simulations. 
% To unveil the scaling of convergence for each phase, we present the reduced dynamical equation for each possible fixed point, present the corresponding solution in leading order, and verify our theoretical predictions with numerical simulations.

In the numerical simulations to verify our solutions, without loss of generality, we consider the random Pauli ansatz (RPA)~\cite{liu2023analytic, zhang2023dynamical} constructed as
$
\hat{U}(\bm \theta) = \prod_{\ell = 1}^{D} \hat{W}_\ell \hat{V}_{\ell}(\theta_\ell),
$
where $\bm \theta = (\theta_1,\dots,\theta_{L})$ are the variational parameters.
Here $\{\hat{W}_\ell\}_{\ell=1}^L \in \mathcal{U}_{\rm Haar}(d)$ is a set of unitaries with dimension $d = 2^n$ sampled from Haar ensemble, and $\hat{V}_\ell$ is a global $n$-qubit rotation gate defined to be
$
    \hat{V}_\ell(\theta_\ell) = e^{-i\theta_\ell \hat{X}_\ell/2},
$
where $\hat{X}_\ell\in\{\hat{\sigma}^x,\hat{\sigma}^y,\hat{\sigma}^z\}^{\otimes n}$ is a randomly-sampled $n$-qubit Pauli operator nontrivially supported on every qubit. Note that $\{\hat{X}_\ell,\hat{W}_\ell\}_{\ell=1}^L$ remain unchanged through the training. The observable is chosen as Pauli-Z, which has the minimum and maximum achievable values $O_{\rm min/max}=\pm1$. Without \BZ{losing} generality, the $N$ orthogonal data states in the simulation are generated by applying a unitary sampled from Haar ensemble onto $N$ different computational bases. The loss function of RPA in numerical simulations is minimized with learning rate $\eta = 10^{-3}$, and all numerical simulations are implemented with \texttt{TensorCircuit}~\cite{zhang2023tensorcircuit}.

% We will begin with the exponential convergence class and then continue to the polynomial convergence class.

% \subsubsection{Exponential convergence class}
We begin with the exponential convergence class of dynamics, which corresponds to the cases where each data can only have either zero error or zero kernel, $S_E \cap S_K = \emptyset$, as we indicate in Fig.~\ref{fig:venn} and Table~\ref{tab:phases}. 

\subsection{Frozen-kernel dynamics}
For {\it frozen-kernel dynamics} (Fig.~\ref{fig:venn}a), we have an empty set of zero-kernel indices, $S_K = \emptyset$, and a full set of zero-error indices, $S_E=\Omega$, leading to the fixed point as $\{(\epsilon_\beta (\infty) = 0, K_{\beta \beta}(\infty) >0)\}_{\beta \in \Omega}$. Around the fixed point, we can perform \BZ{the} leading-order perturbative \BZ{analysis} from Eq.~\eqref{eq:dyeqs_UT} and obtain 
\begin{align}
    \partial_t \epsilon_\alpha(t)
    = -\frac{\eta}{N} \sum_{\beta \in \Omega} K_{\alpha \beta}(\infty) \epsilon_\beta (t),
    \label{eq:frozen_k_eq}
\end{align}
for all indices $\alpha$, where $K_{\alpha \beta}(\infty)  \equiv \angle_{\alpha \beta} \sqrt{K_{\alpha \alpha}(\infty)} \sqrt{K_{\beta \beta}(\infty)}$ is the late-time QNTK matrix. As the QNTK matrix is symmetric and positive definite, the linearized equation leads to the exponential convergence of all errors $\{\epsilon_\alpha (t)\}$ at the same rate and subsequently the exponential convergence of the kernels $\{K_{\alpha \alpha} (t)\}$ towards the constant non-zero values as
\begin{align}
    \epsilon_\alpha(t), K_{\alpha \alpha}(t) - K_{\alpha \alpha}(\infty) \propto e^{-\eta w^*t}, \forall \alpha \in \Omega,
\end{align}
where $w^*$ is the minimum eigenvalue of QNTK matrix $K_{\alpha \beta}(\infty)$. Since all errors vanish exponentially and $S_K = \emptyset$, this is a generalization of the {\it frozen-kernel dynamics} in QNN-based optimization algorithms found in Ref.~\cite{zhang2023dynamical}

\begin{figure*}[t]
    \centering
    \includegraphics[width=0.75\textwidth]{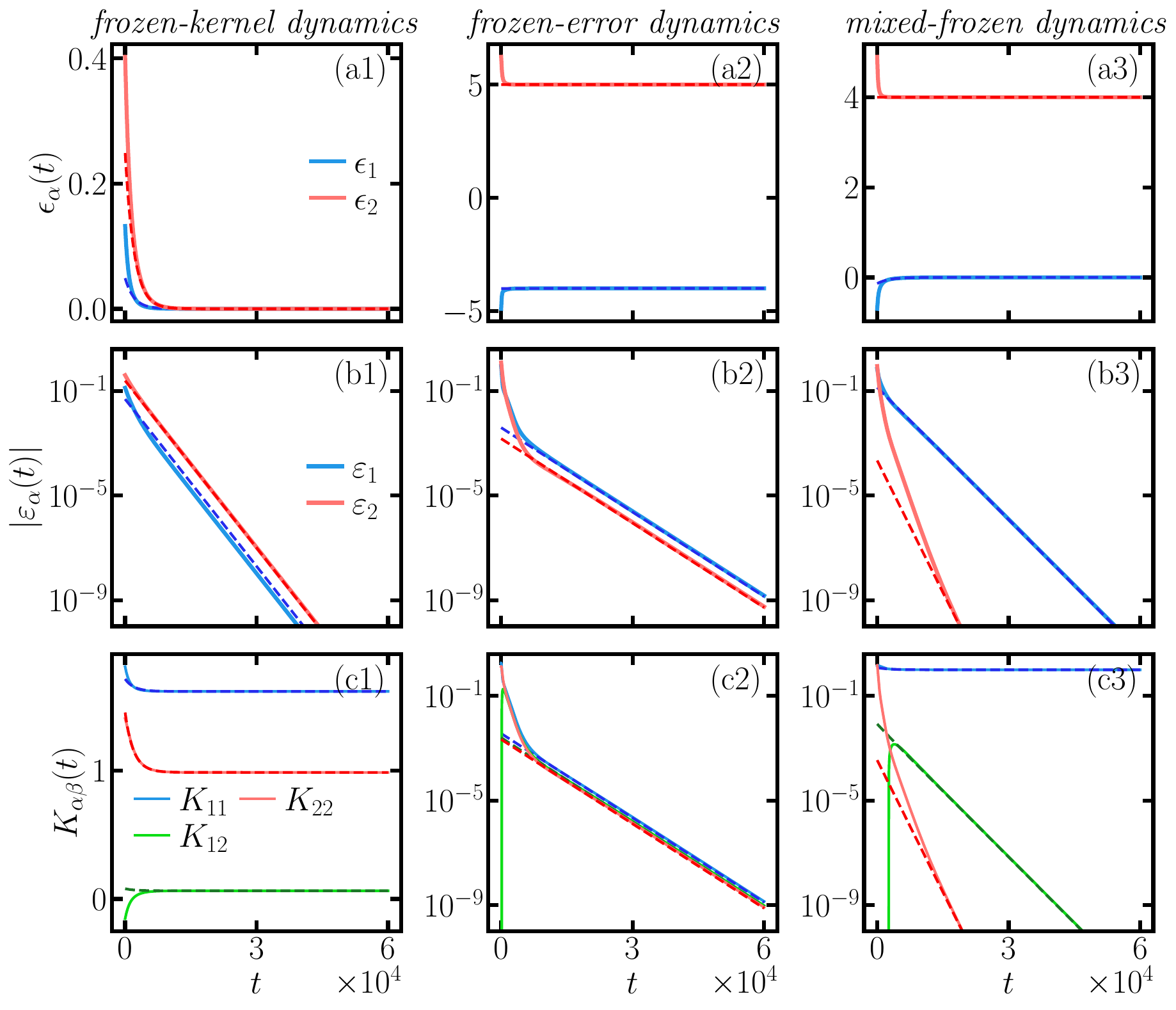}
    \caption{Exponential convergence class dynamics in QNN with orthogonal data. From left to right we show the error and QNTK dynamics of {\it frozen-kernel dynamics}, {\it frozen-error dynamics} and {\it mixed-frozen dynamics}. From top to bottom we plot total error $\epsilon_\alpha(t)$, residual error $\varepsilon_\alpha(t) = \epsilon_\alpha(t) - \epsilon_\alpha(\infty)$, and QNTK $K_{\alpha \beta}(t)$. Subplots in each row share the same legend. Light solid and dark dashed curves with same color represent numerical simulations and corresponding theoretical predictions %Eqs.~\eqref{eq:frozen_k_sol},~\eqref{eq:frozen_err_sol}, ~\eqref{eq:mix_frozen_sol1} and~\eqref{eq:mix_frozen_sol2} 
    for each data \QZ{(see Supplementary Note 4)}. %\ref{app:details_convergence}. 
    Subplots in each row share the same legend. Here random Pauli ansatz (RPA) consists of $L = 48$ variational parameters on $n = 4$ qubits with $\hat{O} = \hat{\sigma}^z_1$, Pauli-Z operator on the first qubit. There are $N=2$ orthogonal data states targeted at $y_1 = 0.3, y_2 = -0.5$ (left), $y_1 = 5, y_2 = -6$ (middle) and $y_1 = 0.4, y_2 = -5$ (right).}
    \label{fig:exp_class_N2}
\end{figure*}

Now we compare the above theory results with the numerical simulations of QNN training.
In Fig.~\ref{fig:exp_class_N2} left panels (a1), (b1), and (c1), we provide the numerical results (solid curves) of $N=2$ data states with $y_1 = 0.3, y_2 = -0.5$, and see alignment with our theoretical predictions (dashed curves), where the error exponentially vanishes (b1) while the kernels converge to a nonzero constant (c1). Note that in {\it frozen-kernel dynamics} the residual error equals the total error, $\epsilon_\alpha(t) = \varepsilon_\alpha(t)$, as the errors all converge to  $\epsilon_\alpha(\infty) = 0$ at late time.

% Let us denote $g_\alpha \equiv \sqrt{K_{\alpha \alpha}}$, then we have
% \begin{align}
%     \left\{ \begin{array}{ll}
%     \diff_t \epsilon_\alpha(t) = -\frac{\eta}{N} \sum_\beta \angle_{\alpha \beta} g_\alpha(t)  g_\beta(t) \epsilon_\beta(t) ;\\
%     \diff_t g_\alpha(t) = -\frac{\eta}{N} \sum_\beta A_{\alpha \beta} g_\beta (t) \epsilon_\beta (t)
%     .\end{array} \right.
%     \label{eq:dyeqs_UT_cont}
% \end{align}

\subsection{Frozen-error dynamics} 
Similar to the {\it frozen-kernel dynamics}, in the {\it frozen-error dynamics} (Fig.~\ref{fig:venn}b), we have $S_E = \emptyset$ with the fixed point $\{(\epsilon_\beta(\infty) \neq 0, K_{\beta \beta}(\infty) = 0)\}_{\beta \in \Omega}$. Around the fixed point, leading-order perturbative analyses of Eq.~\eqref{eq:dyeqs_UT} leads to
\begin{align}
    \partial_t \sqrt{K_{\alpha \alpha}(t)} = -\frac{\eta}{N} \sum_{\beta \in \Omega} F_{\alpha \beta} \sqrt{K_{\beta \beta} (t)},
    \label{eq:frozen_err_eq}
\end{align}
where $F_{\alpha \beta} \equiv \lambda_{\alpha \alpha \beta} \epsilon_\beta(\infty)$ is a constant matrix with positive eigenvalues at late time. Therefore, the convergence towards the fixed point is again exponential and all quantities have the same convergence rate as
\begin{align}
    \epsilon_\alpha(t) - \epsilon_\alpha(\infty), K_{\alpha \alpha}(t) \propto e^{-\eta w^*t}, \forall \alpha \in \Omega,
\end{align}
where $w^*$ is the minimum eigenvalue of $F_{\alpha \beta}$.
% Although $F_{\alpha \beta}$ is not symmetric in general, we can still solve its dynamics through diagonalization as
% \begin{align}
%     \left\{ \begin{array}{ll}
%     \epsilon_\alpha(t) = b_\alpha^E e^{-2\eta u^* t/N} + \epsilon_\alpha(\infty); \\
%     \sqrt{K_{\alpha \alpha}(t)} = b_\alpha^G e^{-\eta u^* t/N},
%     \end{array}\right.
%     \label{eq:frozen_err_sol}
% \end{align}
% where $u^* = \min_{\beta} u_\beta$ is the minimum eigenvalue of $F_{\alpha \beta}$, and $b_\alpha^G, b_\alpha^E$ are also free fitting parameters. 
As all kernels vanish exponentially while all errors converge to constant, this is a generalization of the {\it frozen-error dynamics} in QNN-based optimization algorithms in Ref.~\cite{zhang2023dynamical}. 

The numerical results are compared with the above theory in Fig.~\ref{fig:exp_class_N2} middle panels (a2), (b2) and (c2). The total error $\epsilon_\alpha(t)$ converges to a nonzero constant (a2) since the target $y_1 = 5, y_2 = -6$ is out of reach from measurement; meanwhile, the residual error $\varepsilon_\alpha(t)$ and QNTK $K_{\alpha \beta}(t)$ vanishes exponentially (b2-c2), as predicted by the theory.

\subsection{Mixed-frozen dynamics}
When both the zero-error indices $S_E$ and zero-kernel indices $S_K$ are not empty (and have no overlap), the fixed point has only the error going to zero or only the kernel going to zero---$\{(\epsilon_\beta(\infty) = 0, K_{\beta \beta}(\infty) > 0)\}_{\beta \in S_E} \cup \{(\epsilon_\beta(\infty) \neq 0, K_{\beta \beta}(\infty) = 0)\}_{\beta \in S_K}$. This is a combination of fixed points of the {\it frozen-kernel dynamics} and {\it frozen-error dynamics}, leading to a {\it mixed-frozen dynamics} (Fig.~\ref{fig:venn}c).  Similar to the previous two types of dynamics, we can perform perturbative analyses from Eq.~\eqref{eq:dyeqs_UT}, and obtain the leading-order solution
\begin{align}
\epsilon_\alpha(t), K_{\alpha \alpha}(t)-K_{\alpha \alpha}(\infty) \propto e^{-\eta w^* t/N}, \forall \alpha \in S_E
    \label{eq:mix_frozen_sol1}
\end{align}
and
\begin{align}
    \epsilon_\beta (t) -\epsilon_\beta(\infty), K_{\beta \beta}(t) \propto e^{-2\eta w^* t/N}, \forall \beta \in S_K
    \label{eq:mix_frozen_sol2}
\end{align}
where $w^*$ is a positive constant determined by a matrix in terms of frozen error and kernels, and the corresponding relative dQNTK and geometric angles.

From Fig.~\ref{fig:exp_class_N2} right panels (a3), (b3) and (c3), since our measurement is $\hat{O} = \hat{\sigma}^z_1$, for $\alpha \in S_E$ with $y_\alpha = 0.4 \in (O_{\rm min}, O_{\rm max})$, we see the error decreases exponentially toward zero (blue in (a3)-(b3)) and its corresponding QNTK $K_{\alpha \alpha}(t)$ converges to a positive constant (blue in (c3)). For $\beta \in S_K$ with $y_\beta = -5<O_{\rm min}$, the total error ends at a positive constant, while the residual error $\varepsilon_{\beta}(t)$ and QNTK $K_{\beta \beta}(t)$ decay exponentially (red in (b3)-(c3)). For off-diagonal kernels $K_{\alpha \beta}$ with $\alpha \neq \beta$ that can be inferred from Eq.~\eqref{eq:B_def}, it converges to a positive constant $\forall \alpha, \beta \in S_E$, or vanishes exponentially otherwise.
An interesting phenomena induced by the interaction between data targeted within different types of dynamics is that the decay exponent of $\varepsilon_\beta(t), K_{\beta \beta}(t), \forall \beta\in S_K$ is about two times as large as the one from $\varepsilon_\alpha(t), \forall \alpha \in S_E$ and $K_{\alpha \beta}(t), \forall \alpha \in S_E, \beta \in S_K$.

\section{Convergence toward fixed points: Polynomial convergence class}
In this part, we address the cases of overlapping zero-error indices and zero-kernel indices, $S_E \cap S_K \neq \emptyset$, leading to the polynomial convergence class of dynamics, as we indicate in Fig.~\ref{fig:venn}. 

\subsection{Critical point}
The simplest case is the {\it critical point} with both \BZ{sets} of indices full, $S_E = S_K = \Omega$, as shown in Fig.~\ref{fig:venn}d. In this case, the fixed point has all errors and kernels vanishing, $\{(\epsilon_\alpha (\infty)=0, K_{\alpha \alpha}(\infty) = 0)\}_{\alpha \in \Omega}$. From Eqs.~\eqref{eq:dyeqs_UT}, we can obtain the leading-order decay of all quantities as
\begin{align}
    \epsilon_\alpha (t), K_{\alpha \alpha}(t) \propto 1/t, \forall \alpha \in \Omega.
\end{align}
% \begin{align}
%     \left\{ \begin{array}{ll}
%     \epsilon_\alpha(t) \propto 1/t,\\
%     \sqrt{K_{\alpha \alpha}(t)} = c_\alpha^G/\sqrt{c_0 + \eta t/N},
%     \end{array} \right.
%     \label{eq:crit_sol}
% \end{align}
% where $c_0, a_\alpha^E, c_\alpha^G$ are free fitting parameters. 
In Fig.~\ref{fig:poly_class_N2} left panels (a1), (b1) and (c1), indeed we see that both error and QNTK  decay polynomially as $\epsilon_\alpha(t), K_{\alpha \beta}(t) \sim 1/t$, which can be regarded as a generalization of \BZ{the} {\it critical point} identified in QNN-based optimization algorithms from Ref.~\cite{zhang2023dynamical}.

\subsection{Critical-frozen-kernel dynamics}
When the zero-kernel indices form a strict subset of zero-error indices, $S_K \subsetneq S_E = \Omega$, we have the {\it critical-frozen-kernel dynamics} (Fig.~\ref{fig:venn}e), where the fixed point is a mixture of both quantities vanishing and only the error vanishing---$\{(\epsilon_\beta(\infty)=0, K_{\beta \beta}(\infty) = 0)\}_{\beta \in S_K} \cup \{(\epsilon_\beta(\infty) = 0, K_{\beta \beta}(\infty)>0)\}_{\beta \in S_E \setminus S_K}$. This is a combination of corresponding fixed points from {\it critical point} and {\it frozen-kernel dynamics}. Initially without noticeable interactions between data from $S_K$ and $S_E\setminus S_K$, we expect that error and QNTK from each set should vary with time nearly independently following the dynamics from {\it critical point} and {\it frozen-kernel dynamics} studied above, leading to the fact that $\sqrt{K_{\beta \beta}(t)}\epsilon_\beta(t), \forall \beta \in S_K$ decays much slower than that with indices $\forall \beta \in S_E\setminus S_K$. Therefore, in late time, we approximate the dynamics of $\epsilon_\alpha(t), K_{\alpha \alpha}(t), \forall \alpha \in S_K$ to be self-governed as a ``free-field'', and maintains $1/t$ decay as in the {\it critical point}.
% described by the reduced dynamical equations
% \begin{align}
%     \left\{ \begin{array}{ll}
%     &\partial_t \epsilon_\alpha(t) = -\frac{\eta}{N} \sum_{\alpha' \in S_K} \angle_{\alpha \alpha'}\sqrt{K_{\alpha \alpha}(t)} \sqrt{K_{\alpha' \alpha'}(t)} \epsilon_{\alpha'}(t) ;\\
%     &\partial_t \sqrt{K_{\alpha \alpha}(t)} = -\frac{2\eta}{N} \sum_{\alpha' \in S_K}   \lambda_{\alpha \alpha \alpha'} \sqrt{K_{\alpha' \alpha'} (t)}\epsilon_{\alpha'} (t), 
%     \end{array} \right.
%     \label{eq:crit_frozen_kernel_eq1}
% \end{align}
% for all $\alpha \in S_K$
% and we can take the same ansatz solution as Eqs.~\eqref{eq:crit_sol}.

With the solution $\forall \beta \in S_K$ in hand, we can then perturbatively solve the rest and obtain the overall solution,
\be 
\epsilon_\alpha(t), K_{\alpha \alpha}(t) \propto  1/t, \forall \alpha \in S_K,
\ee 
and
% the PDE for the total error $\epsilon_\beta(t), \forall \beta \in S_E \setminus S_K$ becomes
% \begin{align}
%     \partial_t \epsilon_\beta(t) &= -\frac{\eta}{N} \sqrt{K_{\beta \beta}(\infty)} \left[\sum_{\beta' \in S_E \setminus S_K} \angle_{\beta \beta'} \sqrt{K_{\beta' \beta'} (\infty)} \epsilon_{\beta'}(t)\right. \nonumber \\  
%     &\qquad \qquad \left. + \sum_{\alpha' \in S_K} \frac{\angle_{\beta \alpha'}c_{\alpha'}^E c_{\alpha'}^G}{\left(c_0 + \eta t/N\right)^{3/2}}\right], \forall \beta \in S_E \setminus S_K.
%     \label{eq:crit_frozen_kernel_eq2}
% \end{align}
% We can find out a solution via balancing the contribution of r.h.s. To summarize, the dynamics is described by 
\begin{align}
        \epsilon_\beta(t) \propto 1/t^{3/2},
        K_{\beta \beta}(t)-K_{\beta \beta}(\infty) \propto 1/t, \forall \beta \in S_E \setminus S_K.
    \label{eq:crit_frozen_kernel_sol}
\end{align}
Here $S_E \setminus S_K = \{\beta| \beta\in S_E, \beta\notin S_K\}$ is the set difference between sets $S_E, S_K$ and $K_{\beta \beta}(\infty)$'s are the corresponding converged kernel values. The off-diagonal kernels $K_{\alpha\beta}$ for $\alpha \neq \beta$ can be determined from Eq.~\eqref{eq:B_def}, and have the same scaling as corresponding diagonal counterparts if both indices $\alpha, \beta$ belongs to the same set, $S_E\setminus S_K$ or $S_K$, while $\sim 1/\sqrt{t}$ for $\alpha \in S_E \setminus S_K, \beta \in S_K$.

\begin{figure*}[t]
    \centering
    \includegraphics[width=0.75\textwidth]{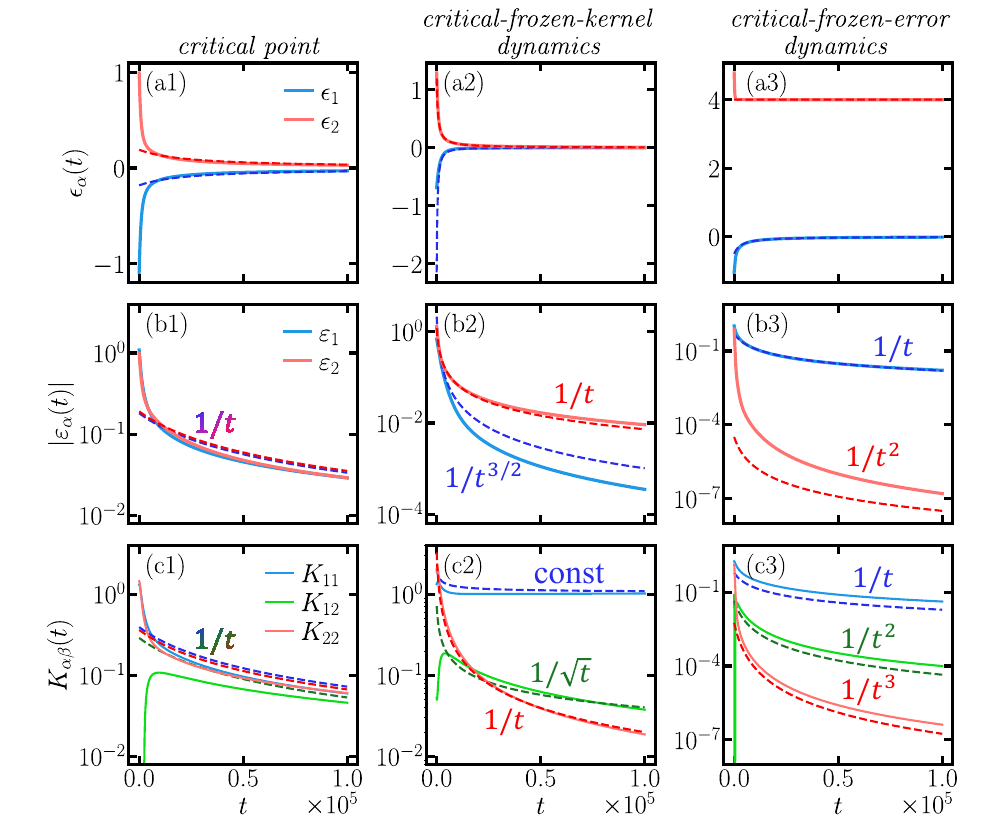}
    \caption{Polynomial convergence class dynamics in QNN with orthogonal data. From left to right we show the error and QNTK dynamics of {\it critical point}, {\it critical-frozen-kernel dynamics} and {\it critical-frozen-error dynamics}. From top to bottom we plot total error $\epsilon_\alpha(t)$, residual error $\varepsilon_\alpha(t) = \epsilon_\alpha(t) - \epsilon_\alpha(\infty)$, and QNTK $K_{\alpha \beta}(t)$. Light solid and dark dashed curves with same color represent numerical simulations and corresponding theoretical predictions for each data (see Supplementary Note 4).
    %\QZ{(see \ref{app:details_convergence})}. 
    Subplots in each row share the same legend. Here random Pauli ansatz (RPA) consists of $L = 48$ variational parameters on $n = 4$ qubits with $\hat{O} = \hat{\sigma}^z_1$, the Pauli-Z operator on first qubit. There are $N=2$ orthogonal data states targeted at $y_1 = 1, y_2 = -1$ (left), $y_1 = 0.4, y_2 = -1$ (middle) and $y_1 = 1, y_2 = -5$ (right).}
    \label{fig:poly_class_N2}
\end{figure*}

We verify our above theoretical predictions with numerical simulations in Fig.~\ref{fig:poly_class_N2} middle panels (a2), (b2) and (c2). The ``free-field theory'' approach utilized above is valid as the corresponding error and QNTK decays $\sim 1/t$ (see red curves (a2)-(c2)), just as the {\it critical point}. The interaction \BZ{between data} dynamics induces the higher-order polynomial decay of error $\sim t^{-3/2}$ (blue in (b2)) on data $\alpha \in S_E \setminus S_K$ at late time. Compared with the {\it frozen-kernel dynamics} dynamics, here the corresponding kernel $K_{\beta \beta}(t)$ for indices $\beta \in S_E \setminus S_K$ also converges to a positive constant though at a much slower speed $\sim 1/\sqrt{t}$ affected by the slowest decay from data targeted at the boundary.

\subsection{Critical-frozen-error dynamics}
Similarly, when the zero-error indices form a strict subset of the zero-kernel indices, $S_E \subsetneq S_K = \Omega$, we have the {\it critical-frozen-error dynamics} (Fig.~\ref{fig:venn}f) with the fixed point described by $\{(\epsilon_\beta (\infty) =0, K_{\beta \beta}(\infty) = 0\}_{\beta \in S_E} \cup \{(\epsilon_\beta(\infty)\neq 0, K_{\beta \beta}(\infty) = 0)\}_{\beta \in S_K \setminus S_E}$, just a combination of {\it critical point} and {\it frozen-error dynamics}. Due to the same reason as in {\it critical-frozen-kernel dynamics} discussed above, the late-time dynamics of $\epsilon_\alpha(t), K_{\alpha \alpha}(t), \forall \alpha \in S_E$ \BZ{are} also self-governed as the ``free field'' and can be satisfied by the polynomial solution $\propto 1/t$. 

Then the rest of the variables can then be solved asymptotically and lead to the {\it critical-frozen-error dynamics} dynamics: 
\be 
\epsilon_\alpha(t), K_{\alpha \alpha}(t) \propto  1/t, \forall \alpha \in S_E,
\ee 
and
\begin{align}
        \epsilon_\beta(t)-\epsilon_\beta(\infty) \propto 1/t^2, 
        K_{\beta \beta}(t) \propto 1/t^{3}, \forall \beta \in S_K \setminus S_E.
    \label{eq:crit_frozen_error_sol}    
\end{align}
The nontrivial off-diagonal terms of $K_{\alpha\beta}$ for $\alpha\in S_E, \beta \in S_K \setminus S_E$ are given by Eq.~\eqref{eq:B_def} and can have scaling of $1/t^2$ at late time.

As shown in Fig.~\ref{fig:poly_class_N2} right panels (a3), (b3) and (c3), the error and kernel of data targeted at boundary decays polynomially as $\sim 1/t$ (blue in (a3)-(c3)), on the other hand, the total error of data targeted beyond accessible values still converges to a nonzero constants (red in (a3)), but the residual error $\varepsilon_\beta(t), \forall \beta \in S_K \setminus S_E$ vanishes only at a higher-order polynomial speed of $\sim 1/t^2$ (red in (b3)), which is induced by the interaction with data targeted at the boundary, thus much slower compared to the {\it mixed-frozen dynamics}.

\begin{figure}[t]
    \centering
    \includegraphics[width=0.45\textwidth]{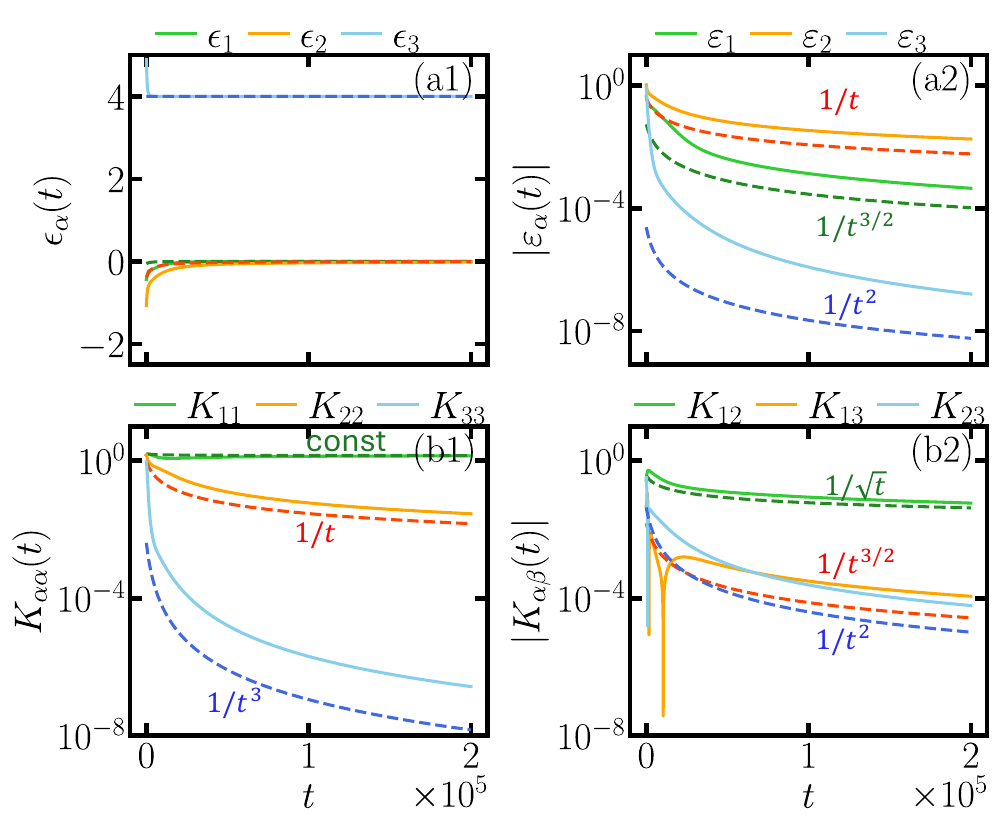}
    \caption{Convergence of {\it critical-mixed-frozen dynamics} in QNN with orthogonal data.  We plot total error $\epsilon_\alpha(t)$, residual error $\varepsilon_\alpha(t) = \epsilon_\alpha(t) - \epsilon_\alpha(\infty)$ in top panel, and diagonal $K_{\alpha \alpha}(t)$ and off-diagonal QNTK $K_{\alpha \beta}(t)$. Light solid and dark dashed curves with same color represent numerical simulations and corresponding theoretical predictions for each data. Here random Pauli ansatz (RPA) consists of $L = 48$ variational parameters ($D=L$ for RPA) on $n = 4$ qubits with $\hat{O} = \hat{\sigma}^z_1$, the Pauli-Z operator on first qubit. There are $N=3$ orthogonal data states targeted at $y_1 = 0.4, y_2 = 1, y_3 = -5$.}
    \label{fig:crit_mix_fro}
\end{figure}

\subsection{Critical-mixed-frozen dynamics}
Finally, we consider the most complex case where none of the sets contains the other, $S_E \not\subset S_K$ and $S_K \not\subset S_E$, and two sets have nonempty overlap $S_E \cap S_K \neq \emptyset$, which corresponds to the {\it critical-mixed-frozen dynamics} (Fig.~\ref{fig:venn}g). This dynamics only takes place for supervised learning with at least $N\ge 3$ input quantum data. The fixed point is described by $\{(\epsilon_\beta(\infty)=0,K_{\beta \beta}(\infty) = 0)\}_{\beta \in S_E \cap S_K} \cup \{(\epsilon_\beta(\infty)=0, K_{\beta \beta}(\infty)> 0)\}_{\beta \in S_E \setminus (S_E\cap S_K)} \cup \{(\epsilon_\beta(\infty)\neq 0, K_{\beta \beta}(\infty)= 0)\}_{\beta \in S_K \setminus (S_E\cap S_K)}$. Due to the existence of data targeted at the boundary for $\beta \in S_E \cap S_K$, we can still solve its corresponding dynamics via the ``free-field'' approach which brings us the $1/t$ decay. Then, we can reduce the dynamical equations for the rest of quantities and obtain the leading-order result:
\be 
\epsilon_\alpha(t), K_{\alpha \alpha}(t) \propto  1/t, 
\ee
for all data $\forall \alpha \in S_E \cap S_K$,
\be 
\epsilon_\alpha(t) \propto 1/t^{3/2},  K_{\alpha \alpha}(t)-K_{\alpha \alpha}(\infty) \propto 1/t,  
\ee 
for all data $\forall \alpha \in S_E \setminus (S_E \cap S_K)$,
and
\be 
 \epsilon_\alpha(t)-\epsilon_\alpha(\infty) \propto 1/t^2, K_{\alpha \alpha}(t) \propto 1/t^3,
\ee
for the rest data $\forall \alpha \in S_K \setminus (S_E \cap S_K)$. The off-diagonal terms of $K_{\alpha \beta}$ for $\alpha \neq \beta$ can still be determined from Eq.~\eqref{eq:B_def} and for these with index crossing dynamics, it can have scaling of $\sim 1/\sqrt{t}$ for all indices $\alpha \in S_E \setminus(S_E \cap S_K), \beta \in S_E \cap S_K$, $\sim 1/t^{3/2}$ for all indices $\alpha \in S_E \setminus(S_E \cap S_K), \beta \in S_K \setminus(S_E \cap S_K)$ and $\sim 1/t^2$ for all indices $\alpha \in S_E \cap S_K, \beta \in S_K \setminus(S_E \cap S_K)$.
% \begin{align}
%     \left \{ \begin{array}{ll}
%     \epsilon_\alpha(t) \propto 1/t, \forall \alpha \in S_E \cap S_K;\\
%     \epsilon_\alpha(t) \propto 1/t^{3/2}, \forall \alpha \in S_E \setminus (S_E \cap S_K);\\
%     \epsilon_\alpha(t)-\epsilon_\alpha(\infty) \propto 1/t^2 , \forall \alpha \in S_K \setminus (S_E \cap S_K) ;\\
%     K_{\alpha \alpha}(t) \propto 1/t, \forall \alpha \in S_E \cap S_K;\\
%     K_{\alpha \alpha}(t)-K_{\alpha \alpha}(\infty) \propto 1/\sqrt{t} , \forall \alpha \in S_E \setminus (S_E \cap S_K);\\
%     K_{\alpha \alpha}(t) \propto 1/t^3, \forall \alpha \in S_K \setminus (S_E \cap S_K).
%     \end{array} \right.
%     \label{eq:crit_mix_fro_sol}
% \end{align}

In Fig.~\ref{fig:crit_mix_fro}, we verify our above theory predictions with numerical simulations. The error and kernel of data targeted at the boundary $y_\alpha=\pm 1$ decays polynomially as $\sim 1/t$ (orange in (a1), (a2), (b1)), well captured by the ``free-field'' approach. Meanwhile, for data targeted within the accessible region, the error decays polynomially \BZ{with} a faster speed at $\sim 1/t^{3/2}$ (green in (a1), (a2)) with kernel \BZ{approaching} a constant (green in (b1)). On the other hand, for data targeted outside the accessible region, the total error can only converge to a nonzero constant (blue in (a1)), however, the residual error $\varepsilon_\alpha(t)$ vanishes quadratically $\sim 1/t^2$ (blue in (a2)), and the kernel decays cubically $\sim 1/t^3$ (blue in (b1)). In addition, the cross-dynamics off-diagonal terms of $K_{\alpha\beta}$ also agree with the theory predictions---polynomial decay with $1/\sqrt{t}, 1/t^{3/2}$ and $1/t^2$ scalings, as shown in (b2).

From the convergence of polynomial convergence class discussed above, we see that as long as there exists a data state targeted at the boundary, either $O_{\rm min}$ or $O_{\rm max}$, the convergence dynamics for all data will be suppressed to polynomial decay though with potential different orders, in contrast to the exponential convergence class. Therefore, our results imply that in quantum machine learning, a proper design of loss function is important to enable fast convergence towards the same QNN configuration.

\section{Ensemble average results}
\label{sec:ensemble}
In this section, we provide physical insight and analytical results to resolve the only assumption for deriving the dynamical equations Eq.~\eqref{eq:dyeqs_UT} that the relative dQNTK $\lambda_{\alpha \alpha \beta}$ approaches a constant at late time. Our results rely on large depth $D \gg 1$ (equivalently $L \gg 1$), where the converged circuit unitaries optimized from random initialization can be modeled as a specific \BZ{unitary ensemble}, the restricted Haar ensemble. 

Under random initialization, the circuit unitary can be represented as a typical sample from Haar random ensemble, as long as the circuit ansatz is universal~\cite{mcclean2018barren, liu2023analytic, cerezo2021cost}. However, as the training starts, the circuit unitary quickly deviates from the Haar random unitary to map each of the input data state $\ket{\psi_\alpha}$ to the corresponding target state $\ket{\Phi_\alpha}$ due to the constraint \BZ{imposed by the} target value $y_\alpha$; therefore, we model the converged circuit unitaries as the restricted Haar ensemble in a block-diagonal form
\begin{align}
    \calU_{\rm RH} = \left\{U\left\lvert U = \begin{pmatrix}
        Q & \bm 0\\
        \bm 0 & V
    \end{pmatrix}\right.\right\},
    \label{eq:rh_main}
\end{align}
where $Q = \oplus_{\alpha=1}^N e^{i\phi_{\alpha}}$ is a diagonal matrix with complex phases uniformly distributed $\phi_\alpha \sim \mathbb{U}[0, 2\pi)$ (also known as random diagonal-unitary matrix in Ref.~\cite{nakata2013diagonal}) and $V$ is a Haar random unitary of dimension $d-N$. The rows and columns are represented in basis of input and target states. Specifically, for $N \ge d-1$, the unitary in \BZ{the} restricted Haar ensemble becomes a diagonal matrix with complex phases only; while for $N=1$, the ensemble reduces to the restricted Haar unitary considered in QNN-based optimization algorithms~\cite{zhang2023dynamical}.

We consider the multi-state preparation task as there are less degrees of freedom in the targets to provide insights into the ensemble-average results. As we discussed above, the input data states are orthogonal, $\braket{\psi_\alpha|\psi_\beta} = \delta_{\alpha \beta}$, which can be generated from a random unitary applied on the computational basis. The observable for each data state is a state projector to its corresponding target state $\hat{O}_\alpha = \ketbra{\Phi_\alpha}{\Phi_\alpha}$ with orthogonality $\braket{\Phi_\alpha|\Phi_\beta} = \delta_{\alpha \beta}$. To quantify the evolution of the QNN unitary ensemble, we study the frame potential, a widely utilized tool in quantum information science and quantum chaos~\cite{roberts2017chaos}. Here, we choose the second-order frame potential 
\be
\calF^{(2)}_{\calU} = \int_\calU \diff U \diff U' |\tr(U^\dagger U^\prime)|^{4},
\label{eq:F2_def_main}
\ee
as a typical nontrivial measure on the unitary ensemble $\calU$, and results for higher-order frame potential are presented in \QZ{Supplementary Note 5}.
%\QZ{\ref{app:rh}}. 
A smaller value of the frame potential indicates a higher level of randomness for an unitary ensemble---the minimum value of the $k$-th-order frame potential, $\min_{\calU}\calF^{(k)}_{\calU}=k!$ , is achieved by the Haar random ensemble (more generally the $k$-design~\cite{roberts2017chaos}).

\begin{figure}[t]
    \centering
    \includegraphics[width=0.45\textwidth]{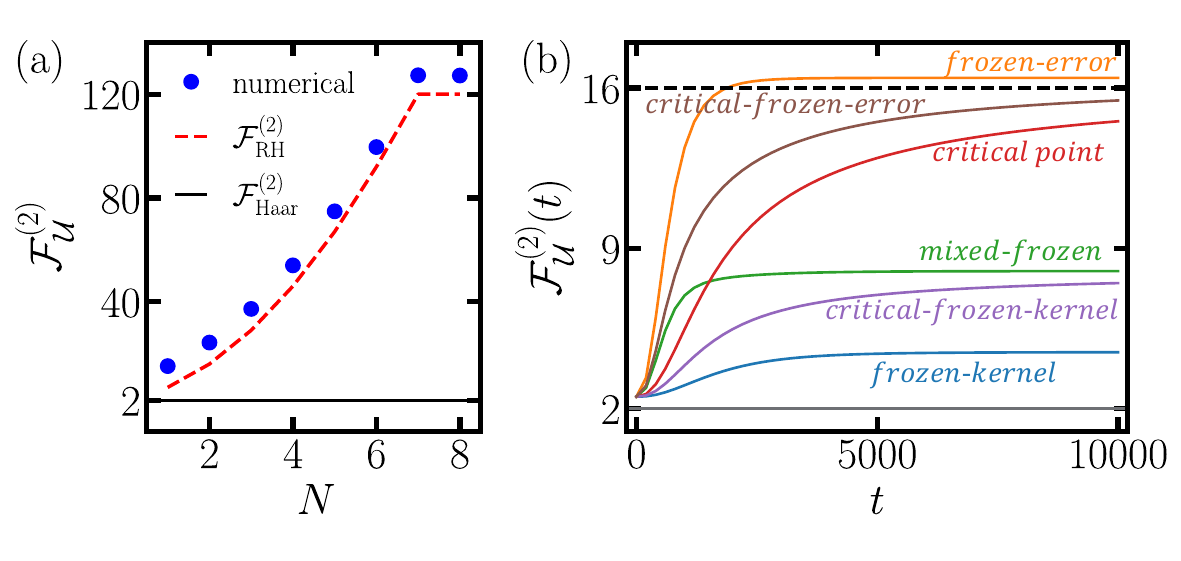}
    \caption{Second-order frame potential of circuit unitaries of QNNs for multi-state preparation. In (a) we plot the frame potential of circuit unitaries of QNNs versus number of data states. Red dashed curve and grey solid line show the frame potential of restricted Haar ensemble Eq.~\eqref{eq:framePot_main} and Haar unitary ensemble $\calF^{(2)}_{\rm Haar}=2$. In (b) we plot the dynamics of $\calF^{(2)}(t)$ in training with targets set in various types of dynamics represented by different colors. The black dashed line represents $\calF^{(2)}_{\rm RH} = 16$. Here in (a) random Pauli ansatz (RPA) consists of $L = 128$ parameters on $n=3$ qubits, and the targets for $N$ orthogonal data states are set within {\it frozen-error dynamics} $y_1, y_2 > 1$. In (b) the RPA consists of $L=64$ parameters on $n=2$ qubits with $N=2$ input orthogonal data states. In both cases, the target states are chosen to be computational basis. }
    \label{fig:framePot}
\end{figure}

For restricted Haar ensemble, we analytically obtain its frame potential as
\begin{align}
    \calF^{(2)}_{\rm RH} = \begin{cases}
        2 N^2 + 3N + 2, & N\le d-2,\\
        2d^2 - d, & N \ge d-1.
    \end{cases}
    \label{eq:framePot_main}
\end{align}
We see $\calF_{\rm RH}^{(2)}$ grows quadratically with number of data until \BZ{saturates at} the squared Hilbert space dimension when $N \ge d-1$, which is in sharp contrast to the Haar random ensemble result $\calF_{\rm Haar}^{(2)} = 2$ independent of \BZ{both} system dimension \BZ{and} number of data (additional calculations can be found in \QZ{Supplementary Note 5}). 
%\QZ{\ref{app:rh}}
As a sanity check, the $N=0$ no data case agrees with the Haar random case. At large $N$, the frame potential saturates to $2d^2-d$, limited by the Hilbert space dimension due to orthogonal condition on input data. Such a phenomena can be understood from the reduction in the degree of freedom driven by the increasing number of data. The analytical formula is plot in Fig.~\ref{fig:framePot}(a) as the red dashed curve. 

\begin{figure}[t]
    \centering
    \includegraphics[width=0.45\textwidth]{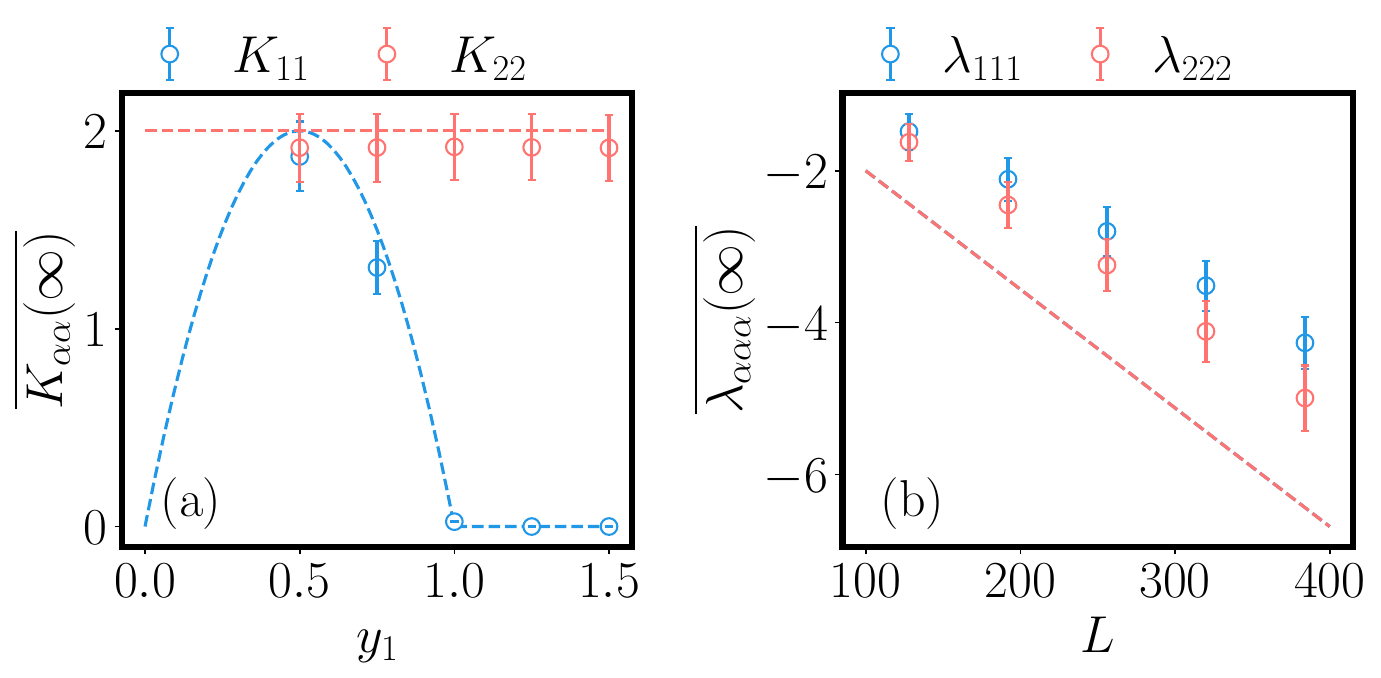}
    \caption{Average results under restricted Haar ensemble. We plot (a) $K_{\alpha \alpha}(\infty)$ versus $y_1$ with $y_2 = 0.5$ and $L = 256$ fixed, (b) $\lambda_{\alpha \alpha \alpha}(\infty)$ versus $L$ with $y_1 = 5, y_2 = 6$ fixed. Blue and red dashed lines in (a) represent Eq.~\eqref{eq:K1_rh_main}. Blue and red dashed lines (overlapped) in (b) represent Eq.~\eqref{eq:lambda1_rh_main}. Here random Pauli ansatz (RPA) consists of $L$ variational parameters on $n=4$ qubits. There are $N=2$ orthogonal data states and the corresponding target states are computational basis $\ket{0000}, \ket{0001}$.}
    \label{fig:rh_result}
\end{figure}

We expect when the converged state is unique, for example in the {\em frozen-error dynamics}, the frame potential will converge to the restricted Haar ensemble's prediction. To provide a quantitative understanding,  we show the frame potential from numerical simulation at late-time (blue dots) with various data states and see a good agreement with theory from restricted Haar ensemble (red dashed line) in Fig.~\ref{fig:framePot}(a). 
Overall, similar convergence of frame potential can also be found in {\it frozen-error, critical-point} and {\it critical-frozen-error}, as we show in Fig.~\ref{fig:framePot}(b). Their deviations from the exact theoretical result (black dashed) are due to finite samples in the ensemble, and slow convergence of unitary in dynamics belonging to polynomial convergence class. 
For non-unique converged states of dynamics with at least one target value chosen within accessible region $y_\alpha \in (O_{\rm min}, O_{\rm max})$, the frame potential of unitary ensemble $\calU$ can lie between the values of Haar and restricted Haar ensembles, $\calF_{\rm Haar}^{(2)} < \calF_{\calU}^{(2)} < \calF_{\rm RH}^{(2)}$, due to extra randomness allowed in the unitary, as shown by the green, purple and blue lines in Fig.~\ref{fig:framePot}(b).

Given the sub-block unitary $V$ forms a $4$-design, we have the following results.
\begin{theorem}
\label{theorem:RH}
    For multi-state preparation task with observable $\hat{O}_\alpha = \ketbra{\Phi_\alpha}{\Phi_\alpha}$ satisfying $\braket{\Phi_\alpha|\Phi_\beta} = \delta_{\alpha \beta}$ with $N < d-1$, when the circuit satisfies restricted Haar ensemble and the input data states are orthogonal, the ensemble average of QNTK and relative dQNTK for each data (unified indices) are
    \begin{align}
         \overline{K_{\alpha \alpha}(\infty)} &= \frac{L}{2d}o_\alpha (1-o_\alpha), \label{eq:K1_rh_main} \\
          % \overline{K_{\alpha \beta}(\infty)} &\simeq -\frac{L}{2d^2} \sqrt{o_\alpha o_\beta (1-o_\alpha) (1-o_\beta)},
          % \label{eq:K2_rh_main} \\
          % \overline{\angle_{\alpha \beta}(\infty)} &\simeq -\frac{1}{d},
          % \label{eq:angle_rh_main}\\
        \overline{\lambda_{\alpha \alpha \alpha}(\infty)} &= -\frac{1}{4d} \left[2(d o_\alpha - 2) + L(2o_\alpha - 1)\right], 
         \label{eq:lambda1_rh_main}
         % \overline{\lambda_{\alpha \alpha \beta}(\infty)} &\simeq \frac{1}{4d^2} \left[2(d o_\alpha - 2) + L(2o_\alpha - 1)\right], 
         % \label{eq:lambda2_rh_main}
    \end{align}
    at the $L\gg1, d\gg1$ limit,
    where $o_\alpha = \epsilon_\alpha(\infty) + y_\alpha$.
\end{theorem}
Note that the average relative dQNTK are taken to be the ratio of corresponding average quantities, and we expect the change of order of average does not affect the result significantly due to self-averaging. In Fig.~\ref{fig:rh_result}(a), we see a clear dependence of the converged QNTK $\overline{K_{11}(\infty)}$ on different target values $y_1$ while $\overline{K_{22}(\infty)}$ remains the same as $y_2$ is fixed, and both are captured by the restricted Haar ensemble average result in Eq.~\eqref{eq:K1_rh_main}. In Fig.~\ref{fig:rh_result}(b), the converged relative dQNTK $\overline{\lambda_{\alpha \alpha \alpha}(\infty)}$ scales linearly with the number of variational parameters in the ansatz, as predicted from Eq.~\eqref{eq:lambda1_rh_main}. The accurate prediction on other components of interest $\overline{K_{\alpha \beta}(\infty)}, \overline{\lambda_{\alpha \alpha \beta}(\infty)}$ requires more information such as the infidelity between output state and other target states, which we defer to future works.

\section{Experiment} 
\label{sec:exp}

\begin{figure}[t]
    \centering
    \includegraphics[width=0.45\textwidth]{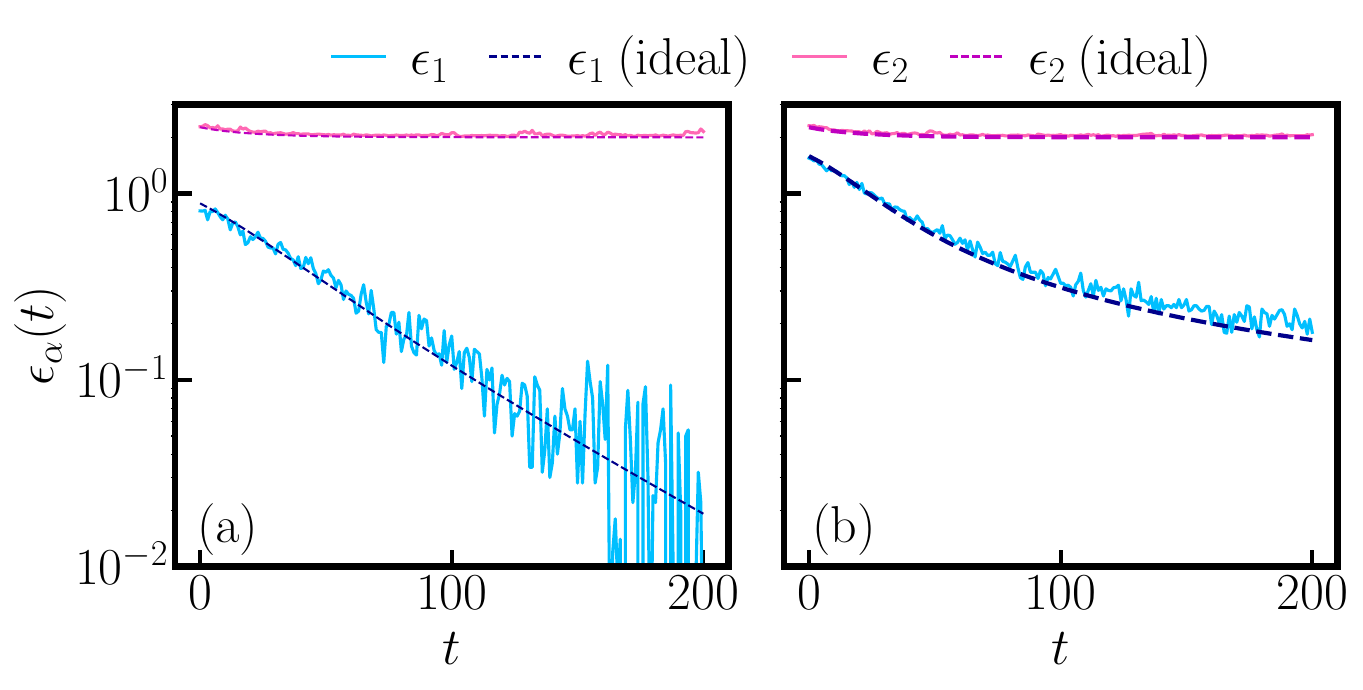}
    \caption{Training dynamics of total error $\epsilon_\alpha (t)$ on IBM quantum devices, Kyiv. 
    In (a) and (b), the target values are chosen to be $y_1=-0.3, y_2=-3$ and $y_1=-1, y_2=-3$ separately, corresponding to the {\it mixed-frozen dynamics} and {\it critical-frozen-error dynamics}. Solid light blue and purple curves represent experimental results for $\epsilon_1(t)$ and $\epsilon_2(t)$, dashed dark blue and pink curves represent corresponding ideal simulation results. An $n=2$ qubit $D=6$-layer hardware efficient ansatz (with $L=24$ parameters) is utilized to minimize loss function with input states $\ket{\psi_1}=\ket{01}$, $\ket{\psi_2}=\ket{10}$, and the observable is $\hat{O} = \hat{\sigma}^z_1$, Pauli-Z operator on the first qubit.}
    \label{fig:experiment}
\end{figure}

In this section, we validate some of the unique training dynamics in the multi-data scenario on IBM quantum devices. Our experiments are implemented on the hardware \texttt{IBM Kyiv}, an IBM \texttt{Eagle r3} hardware with 127 qubits, via \texttt{Pennylane}~\cite{bergholm2018pennylane} and \texttt{IBM Qiskit}~\cite{Qiskit}. The device has median $T_1\sim 
251.87 \text{us}$, median $T_2\sim 114.09 \text{us}$, median ECR error $\sim 1.117\times 10^{-2}$, median SX error $\sim 3.097 \times 10^{-4}$, and median readout error $\sim 9.000\times 10^{-3}$. We adopt the QNN with the \BZ{experimentally friendly} hardware-efficient ansatz (HEA), where each layer consists of single-qubit rotations along Y and Z directions, followed by CNOT gates on  nearest neighbors in a brickwall \BZ{pattern}~\cite{kandala2017hardware}. As an example, we choose two different computational bases as the input data states, $\ket{\psi_1}=\ket{01}, \ket{\psi_2}=\ket{10}$. Through complete \BZ{state} tomography (see Methods) %Appendix~\ref{app:experiment_detail})
, the initial states are prepared with high fidelity at $\braket{01|\rho_1|01} =0.996 \pm 0.0018$ and $\braket{10|\rho_2|10} = 0.994 \pm 0.0020$ for prepared states $\rho_1, \rho_2$ (mixed state in general due to hardware noise) averaged over 12 rounds. The high fidelity guarantees the condition of orthogonal data underlying our analyses. We randomly assign initial angles uniformly sampled from $[0, 2\pi)$ to the parameterized gates in HEA, and maintain consistency across all experiments. For the observable, we consider the Pauli-Z operator of the first qubit, as a simple but sufficient \BZ{demonstration} of our theory.

In Fig.~\ref{fig:experiment}, we choose the target values to be (a) $y_1=-0.3, y_2=-3$ and (b) $y_1=-1, y_2=-3$, corresponding to the {\it mixed-frozen dynamics} and {\it critical-frozen-error dynamics}, both of which are unique for supervised learning  compared to optimization algorithms studied in Ref.~\cite{zhang2023dynamical}. In both cases, the experimental data (solid) agree well with the ideal simulation results (dashed), indicating the constant error within both dynamics for data targeted at $y_\alpha < O_{\rm min}$ (pink), the exponential convergence for data with target $O_{\rm min} <y_\alpha< O_{\rm max}$ (blue in (a)) and polynomial convergence for data with target at $y_\alpha = O_{\rm min}$ (blue in (b)) up to some fluctuations due to shot and hardware noise. To suppress error, we repeat experiments two times for each case.

\section{Discussion}
Our results go beyond the data-induced barren plateau \BZ{phenomenon} from random initializations in the paradigm of quantum machine learning~\cite{thanasilp2023subtleties, ragone2024lie}, and identify two distinct convergence classes including seven different dynamics in total via analytically solving the convergence of error and kernel of each data. The dynamical transition originating from bifurcation with multi codimensions is driven by the data in supervised learning, suggesting fruitful physics and a new source for dynamical transition in the framework of quantum machine learning. The effect of data is also revealed in the restricted Haar ensemble via its constrained randomness controlled by the number of data. In practical applications, our findings guide the design of \BZ{the} loss function to speedup the training of QNNs.

Our findings also connect to the observation in Ref.~\cite{you2023analyzing}. When the target value is chosen to be $\pm 1$ in Pauli measurements, only a polynomial convergence is observed; while a rescaling of the observable, equivalent to shifting the target values within $(-1, 1)$ leads to an exponential convergence though reaching to different solutions, which are fully explained by the {\it critical point} and {\it frozen-kernel dynamics} in our work. Ref.~\cite{liu2022representation} considered supervise learning only in the frozen-kernel dynamics, while the dynamical transition is not uncovered there.

%\BZ{***We may need a few sentences here to discuss the difference between our results and ~\cite{liu2022representation} where both multi-data are considered.}

The two convergence classes with seven different dynamics we identified are focused on the orthogonal input data states. For \BZ{a} more general case where input data are allowed to be non-orthogonal, one can expect that the accessible region of the measurement observable and thus the dynamical ``phase'' diagram will be changed induced by the overlaps among input data states, therefore we leave it as an open question for future study to understand the training dynamics with data correlations. In addition, it is an open problem whether a time-dependent tuning of target values can enhance the overall training of QNNs, given the different convergence dynamics in the time-independent cases considered in this work.

% Our findings also explains the observation in Ref.~\cite{you2023analyzing}: 
% for the critical case of target values of $\pm 1$ in Pauli measurements, polynomial convergence is observed; while a rescaling of the observable will force the QNN into the frozen kernel dynamics with target values in $(-1,1)$, leading to exponential convergence.

While comparison between linear loss functions and quadratic loss functions is considered in previous work for optimization tasks~\cite{zhang2023dynamical}, a linear loss function does not work for classification of more than two classes of data, since linear loss functions push the observable only to boundaries.

%TC:ignore
\section{Methods}

\subsection{Experimental details}
\label{app:experiment_detail}
In this section, we provide additional details on our experiment on the \texttt{IBM Quantum} devices. In the experiment, we take $500$ shots to estimate the expectation value of the measurement operator, and the learning rate in the experiment is chosen to be $\eta = 0.01$. Compared with the theory simulation choice of $\eta=0.001$,  we choose a \BZ{relatively} larger learning rate in \BZ{the} experiment to speed up the convergence and to mitigate the effect of noise from experimental imperfections.

\begin{figure}[h]
    \centering
    \centering
    \includegraphics[width=0.45\textwidth]{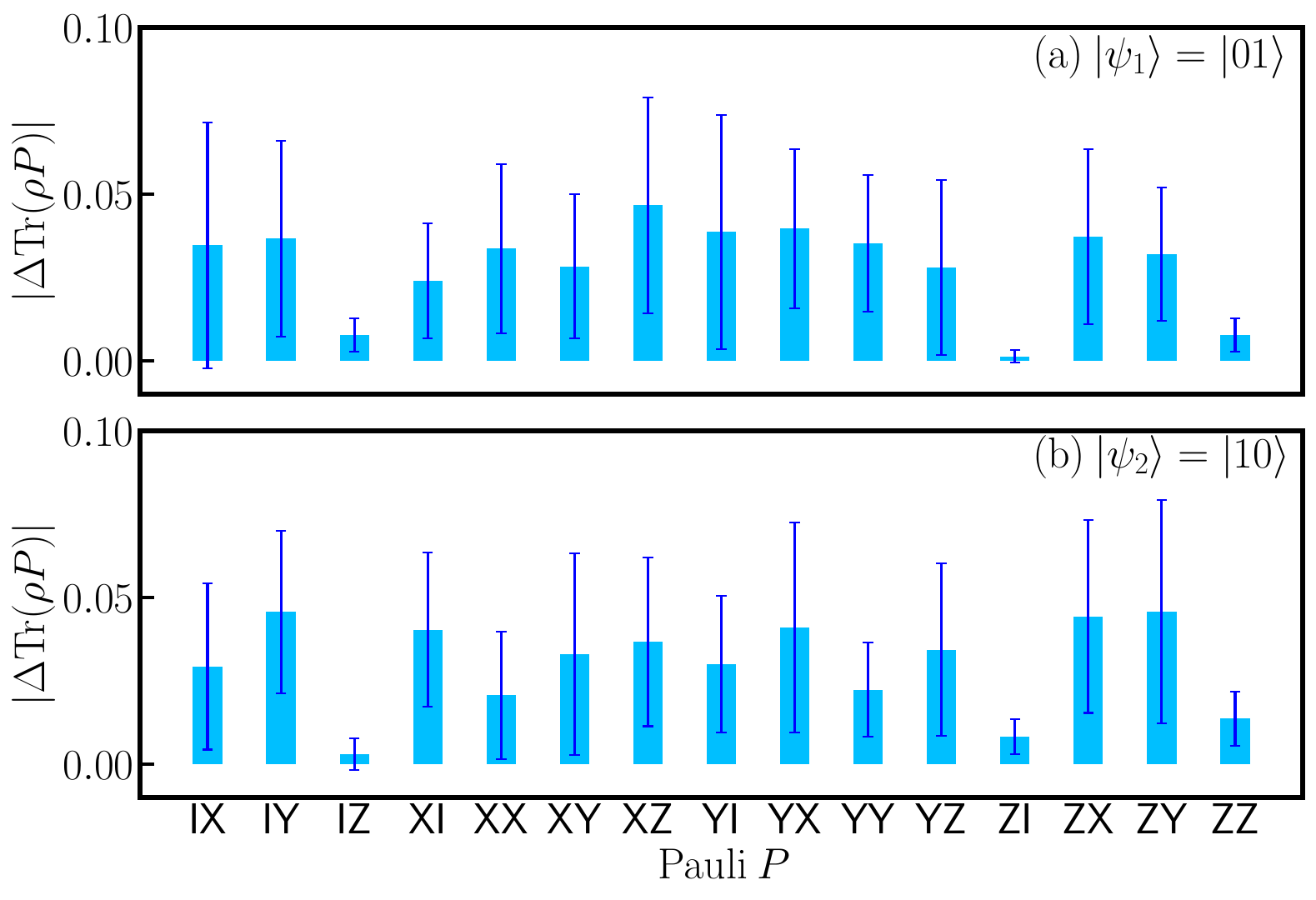}
    \caption{Deviation of prepared states $\rho$ from corresponding ideal state $\ket{\psi}$ in state tomography. The deviation is defined as $|\Delta \tr(\rho P)| = |\tr(\rho P) - \braket{\psi|P|\psi}|$. The top and bottom \BZ{panels show} deviation for $\ket{01}$ and $\ket{10}$ separately. Blue \BZ{bars show} the average deviation over $12$ rounds and error bars represent the standard deviation.    \label{fig:tomography}}
\end{figure}

We provide the detailed tomography results on the actual states prepared on the quantum devices, and compare it to ideal results. 
In Fig.~\ref{fig:tomography}, we show the deviations of tomography results $|\Delta \tr(\rho P)| = |\tr(\rho P) - \braket{\psi|P|\psi}|$ over all nontrivial Pauli operators $P$, with $\rho$ being the actual state prepared on the device and $\ket{\psi}$ the ideal state. Each of the Pauli expectation values is measured repeatedly for $12$ times. For all Pauli operators, the averaged deviation are less than $0.05$ (blue bars) with fluctuations due to hardware drift noise. Overall, the input data states are prepared with high fidelity, thus the overlap between prepared states violating the orthogonal condition can be neglected.

\subsection{Dynamics of QNTK}
\label{app:QNTK}

In this section, we derive the dynamical equation for QNTK matrix.
The dynamics of $K_{\alpha \beta}(t)$ can be further evaluated as
\begin{align}
    \delta K_{\alpha \beta}(t) &= \sum_\ell \delta\left( \frac{\partial \epsilon_\alpha (t)}{\partial \theta_\ell} \frac{\partial \epsilon_\beta (t)}{\partial \theta_\ell} \right)\\
    &= \sum_\ell \left(\frac{\partial \epsilon_\alpha (t)}{\partial \theta_\ell} \delta \left(\frac{\partial \epsilon_\beta (t)}{\partial \theta_\ell}\right) + \delta\left(\frac{\partial \epsilon_\alpha (t)}{\partial \theta_\ell} \right) \frac{\partial \epsilon_\beta (t)}{\partial \theta_\ell} \right.\nonumber\\ &\quad \quad \quad \left.+ \delta\left(\frac{\partial \epsilon_\alpha (t)}{\partial \theta_\ell}\right) \delta \left(\frac{\partial \epsilon_\beta (t)}{\partial \theta_\ell}\right)\right).
    \label{eq:K_tdiff1}
\end{align}
The last term is higher order in $\eta \ll 1$, and we neglect it.

We can evaluate time difference of total error's gradient via the first-order Taylor expansion
\begin{align}
    \delta\left(\frac{\partial \epsilon_\alpha (t)}{\partial \theta_\ell}\right) &= \sum_{\ell^\prime} \frac{\partial^2 \epsilon_\alpha (t)}{\partial \theta_{\ell^\prime} \partial \theta_\ell} \delta \theta_{\ell^\prime}(t)\\
    &= -\frac{\eta}{N}  \sum_{\beta} \epsilon_{\beta}(t) \sum_{\ell^\prime} \frac{\partial \epsilon_{\beta} (t)}{\partial \theta_{\ell^\prime}} \frac{\partial^2 \epsilon_\alpha (t)}{\partial \theta_{\ell^\prime} \partial \theta_\ell}\\
    &= -\frac{\eta}{N} \sum_\beta \sum_{\ell^\prime} H_{\alpha \ell \ell^\prime}(t) J_{\beta \ell^\prime}(t) \epsilon_\beta(t),
\end{align}
where we apply gradient descent rule Eq.~\eqref{eq:gradient_decent} in the second line, and we introduce the Hessian of total error $H_{\alpha \ell \ell^\prime}(t) = \frac{\partial^2 \epsilon_\alpha (t)}{\partial \theta_{\ell} \partial \theta_\ell^\prime}$. $J_{\alpha \ell}(t) = \partial \epsilon_\alpha / \partial \theta_\ell$ is the gradient of total error as we introduced in the main text.
Thus the time difference of $K_{\alpha\beta}(t)$ in Eq.~\eqref{eq:K_tdiff1} becomes
\begin{align}
     \delta K_{\alpha \beta}(t) &=  \sum_\ell \left[\frac{\partial \epsilon_\alpha }{\partial \theta_\ell} \delta \left(\frac{\partial \epsilon_\beta }{\partial \theta_\ell}\right) + \delta\left(\frac{\partial \epsilon_\alpha }{\partial \theta_\ell} \right) \frac{\partial \epsilon_\beta }{\partial \theta_\ell}\right] + \calO(\eta^2)\\
     &= -\frac{\eta}{N}\sum_\gamma \sum_{\ell^\prime, \ell} \left[J_{\alpha \ell} H_{\beta \ell \ell^\prime} J_{\gamma \ell^\prime} \epsilon_{\gamma}   +  \epsilon_{\gamma} J_{\gamma \ell^\prime}  H_{\alpha \ell^\prime \ell}  J_{\beta \ell} \right] \\
     &= -\frac{\eta}{N} \sum_\gamma \epsilon_\gamma (t) \left(\mu_{\gamma \beta \alpha}(t) + \mu_{\gamma \alpha \beta}(t)\right),
\end{align}
where $\mu_{\gamma \alpha \beta} \equiv \sum_{\ell, \ell'} J_{\gamma \ell'} H_{\alpha \ell' \ell}  J_{\beta \ell}$ is the dQNTK we defined in Eq.~\eqref{eq:mu_def}. Therefore, the above equation is the exact dynamical equation presented in Eq.~\eqref{eq:dyeqs_full}.

\begin{figure}[t]
    \centering
    \includegraphics[width=0.45\textwidth]{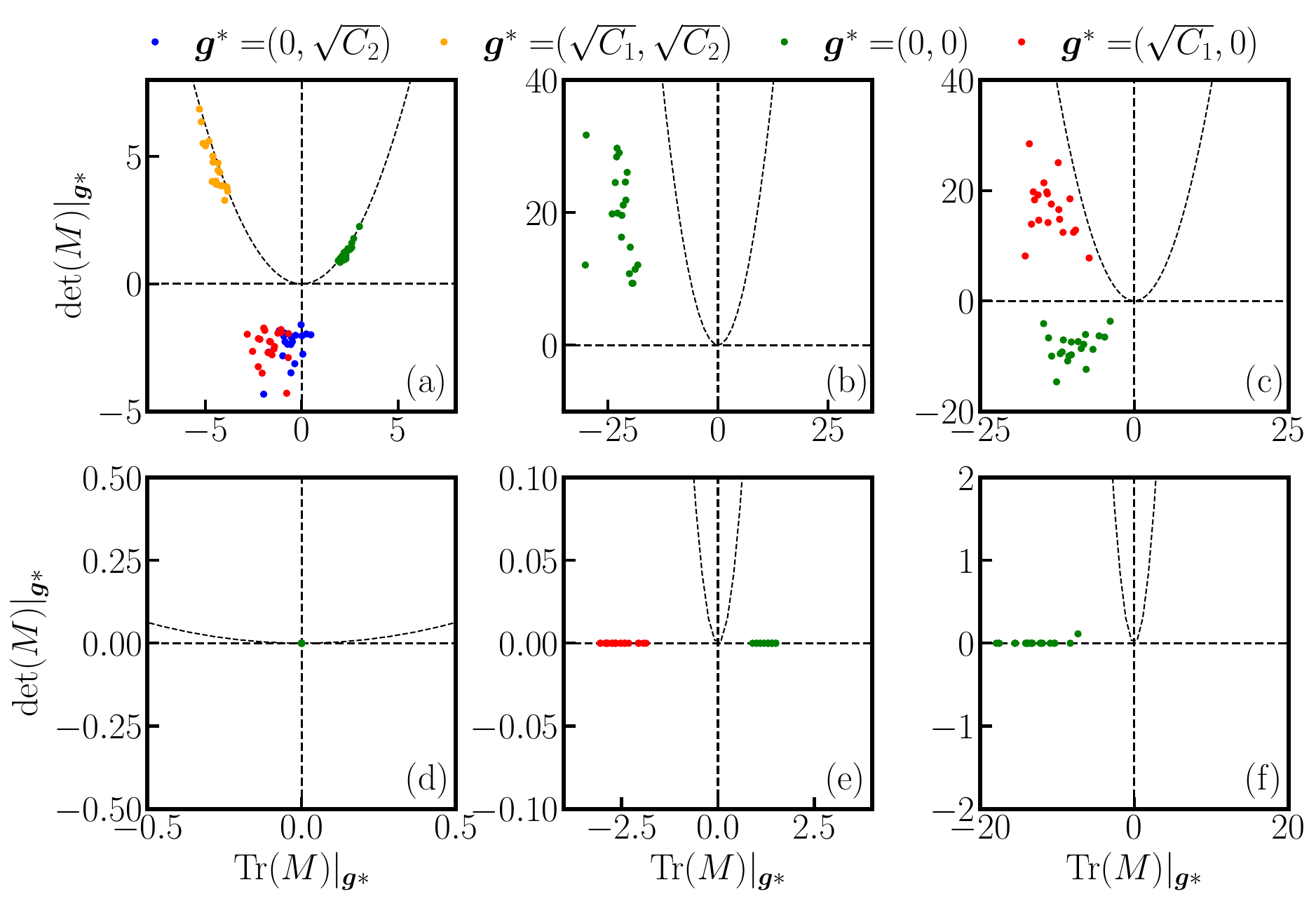}
    \caption{Poincar\'e diagram of fixed points for QNN dynamics with two data. The top and bottom panels show exponential and polynomial convergence classes with {\it frozen-kernel, frozen-error, mixed-frozen} (a-c) and {\it critical point, critical-frozen-kernel, critical-frozen-error} (d-f). Colored dots represent different \BZ{physically} accessible fixed points with different initialization of training parameters. Black horizontal and vertical dashed lines indicate $\det(M) = 0$ and $\tr(M)=0$ for reference. Grey dashed curve shows $\tr(M)^2=4\det(M)$, a criteria to determine whether there exists a spiral surrounding the fixed point. All settings are the same as in Fig.~\ref{fig:lda}.    \label{fig:poincare_diagram}}
\end{figure}

\subsection{Stability transition of fixed points}
\label{app:fix_point_stability}

\begin{figure}[h]
    \centering
    \includegraphics[width=0.45\textwidth]{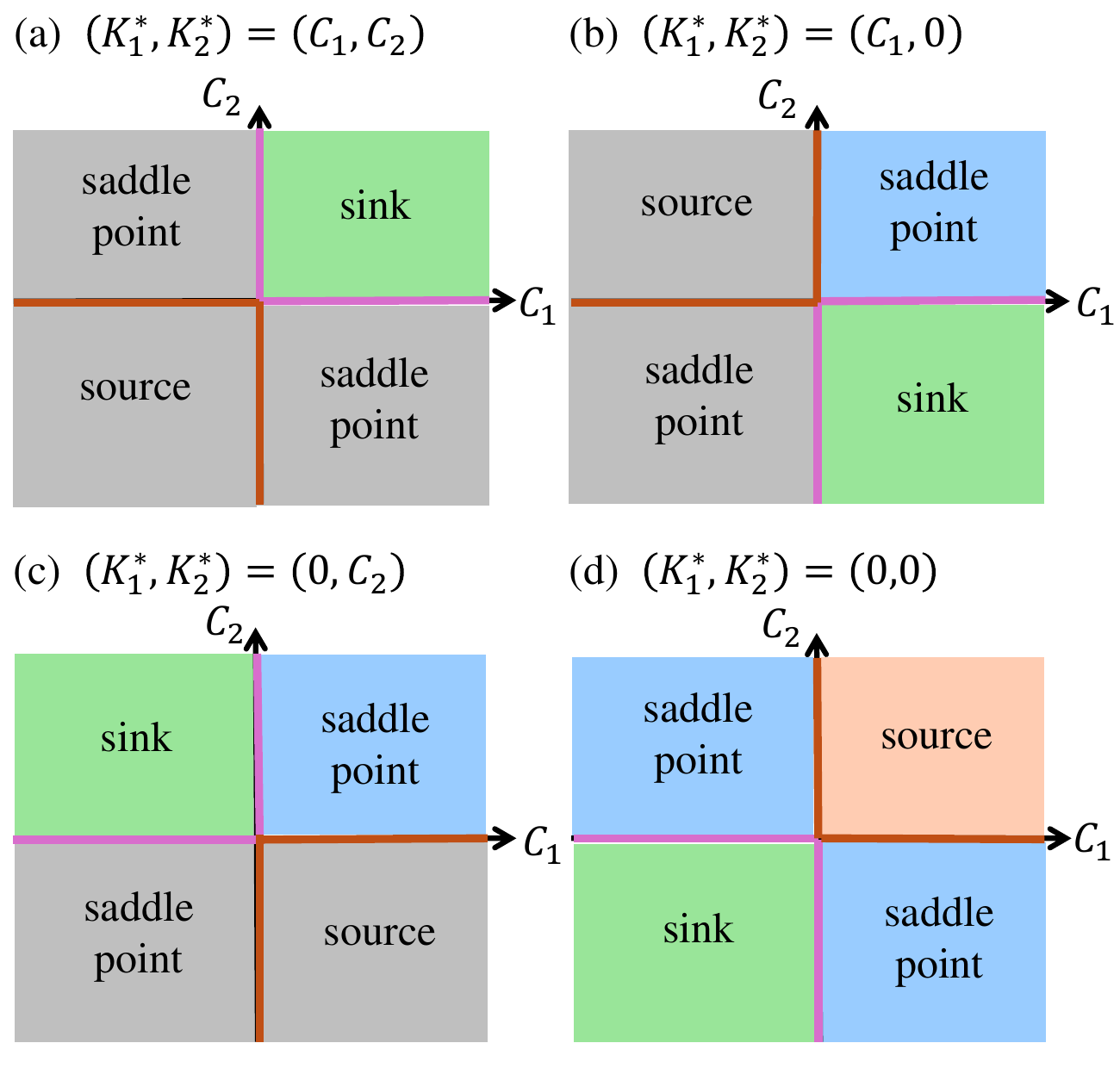}
    \caption{Stability of each fixed point. The fixed point can be classified as a sink (green), a saddle point (blue) or a source (red) depending on the values of $C_1, C_2$. The brown and pink colored axis represent the fixed point to be a line of unstable/stable fixed point. The grey-shaded regions indicate that the fixed point cannot be physically accessed under the current choice of $C_1$ and $C_2$.    \label{fig:fix_point_stability}}
\end{figure}

In this section, we present additional details on the stability transition of fixed points by tuning the fixed-point charges $\{C_\beta\}_\beta $ defined in Eq.~\eqref{eq:C_def}. Starting from the linearized equation Eq.~\eqref{eq:K_eq_linear} in the main text, the matrix Eq.~\eqref{eq:M_def_main} can be explicitly written out for the two data case as
\begin{align}
    M(\bm g, \bm C) = \begin{pmatrix}
        C_1 - 3g_1^2 & z_{12} \left(C_2-3 g_2^2\right)\\
        z_{21} \left(C_1-3 g_1^2\right) & C_2 - 3g_2^2
    \end{pmatrix},
\end{align}
where for simplicity we define 
\begin{align}
    g_\alpha(t) &\equiv \sqrt{K_{\alpha \alpha}(t)}, \label{eq:g_def}\\
    z_{\alpha \beta} &\equiv \frac{\lambda_{\alpha \alpha \beta}}{\lambda_{\beta \beta \beta}}, 
\end{align}
Its eigenvalue can be solved as
\begin{align}
    \nu_{\pm} = \frac{\tr(M) \pm \sqrt{\tr(M)^2 - 4\det(M)}}{2}.
    \label{eq:M_eigenvalue}
\end{align}
Therefore, the stability of any fixed point can be fully characterized by the trace and determinant of $M$ as $(\tr(M), \det(M))$. Both terms are functions of the fixed-point charges $C_1, C_2$ as
\begin{align}
    \left\{ \begin{array}{ll}
    \tr(M) = C_1 + C_2 - 3(g_1^2 + g_2^2),\\
    \det(M) = \left(C_1 - 3g_1^2\right)\left(C_2 - 3g_2^2\right)\left(1 - z_{12} z_{21}\right),
    \end{array}\right.
\end{align}
which is exactly what we see in Eq.~\eqref{eq:tr_det_M} in the main text with typical $z_{12}z_{21}<1$. One can thus determine whether a fixed point is a stable one (`sink'), unstable one (`source') or a saddle point from the signs of the $\tr(M)$ and $\det(M)$:
\begin{enumerate}
\item 
When $\det(M) <0$, we always have $\nu_{-} <0$ and $\nu_{+} >0$, indicating the fixed point to be a saddle point;
\item 
If $\det(M) = 0$ and $\tr(M) < 0$, the eigenvalues become $\nu_{-} = \tr(M)<0$ and $\nu_{+} = 0$, we have a line of stable fixed point as one of the degree of freedoms vanishes; 
\item 
When $\det(M) > 0$ and $\tr(M) < 0$, the real part of $\nu_{\pm}$ is negative and leads to the stable fixed point, identified as `sink'. Precisely speaking, for $\tr(M)^2 \gtreqless 0$ inducing either two different real eigenvalues, a single identical real eigenvalue, or two complex conjugate eigenvalues, the sink can be classified to be a regular sink, degenerate sink and spiral sink;
\item 
For $\det(M) \ge 0$ and $\tr(M) >0$, the fixed point can be classified in a similar way, leading to the `source' and line of unstable fixed point. 
\end{enumerate}
Therefore, for any fixed point $\bm g^*$, we can identify its stability given arbitrary values of fixed-point charges $C_1, C_2$, as shown in Fig.~\ref{fig:fix_point_stability}. On the other hand, the shift of charges would induce a stability transition for every fixed point.

At the end of this section, we connect the above stability analyses on the fixed point to QNN training. For a data with index $\alpha \in S_E \setminus (S_E \cap S_K)$, we can directly see that $C_\alpha >0$, on the other hand for $\alpha \in S_K \setminus (S_E \cap S_K)$, the quantity becomes $C_\alpha <0$. Specifically when $\alpha \in S_E \cap S_K$, $C_\alpha = 0$. In Fig.~\ref{fig:poincare_diagram}, we plot the Poincar\'e diagram for different \BZ{physically} accessible fixed points within different dynamics. The only stable fixed points are those with $\tr(M)\le 0$ and $\det(M)\ge 0$ living in the second quadrant. The dashed curve in each figure represents the equation $\tr(M)^2 - 4\det(M) = 0$ which determines the imaginary part of eigenvalues from Eq.~\eqref{eq:M_eigenvalue} leading to the property of degeneracy and spiral. Here we see that from different initializations, the fine dynamical property of fixed points within each dynamics could be different, which leaves us an interesting open question beyond the scope of our work. Overall, the only stable fixed point within each dynamics aligns with our classification via $S_E, S_K$ in the main text.

%\BZ{When we propose $L$ as ``system size'', should we still think QNN as $(0+1)$ or $(1+1)$-dimension system?}

%\bibliographystyle{apsrev4-2}
%\bibliography{myref}

\begin{acknowledgements}
BZ and QZ acknowledges ONR Grant No. N00014-23-1-2296, NSF (CAREER CCF-2240641, 2330310, 2350153 and OMA-2326746), and DARPA HR00112490453. JL is supported in part by the University of Pittsburgh, International Business Machines (IBM) Quantum through the Chicago Quantum Exchange, and the Pritzker School of Molecular Engineering at the University of Chicago through AFOSR MURI (FA9550-21-1-0209). LJ acknowledges support from the ARO (W911NF-23-1-0077), ARO MURI (W911NF-21-1-0325), AFOSR MURI (FA9550-19-1-0399, FA9550-21-1-0209, FA9550-23-1-0338), NSF (OMA-1936118, ERC-1941583, OMA-2137642, OSI-2326767, CCF-2312755), NTT Research, Packard Foundation (2020-71479), and the Marshall and Arlene Bennett Family Research Program. This material is based upon work supported by the U.S. Department of Energy, Office of Science, National Quantum Information Science Research Centers. The experimental part of the research was conducted using IBM Quantum Systems provided through USC’s IBM Quantum Innovation Center. 
\end{acknowledgements}

%TC:endignore

\newpage

\begin{widetext}

\appendix

\section{Proof of Lemma~\ref{lemma_B}}
\label{app:proof_lemma_B}
In this section, we provide the proof of Lemma~\ref{lemma_B}.
\begin{proof}
Recall $f_{\alpha\beta}(t)$ defined in Eq.~\eqref{f_def}, for convenience we also define 
\be 
g_\gamma(t)=\sqrt{K_{\gamma\gamma}(t)},
\ee 
such that 
$
f_{\alpha\beta}(t)=\sum_\gamma g_\gamma(t) \epsilon_\gamma(t) \lambda_{\gamma \alpha \beta}.
$

We can derive the time-derivative of $\angle_{\alpha \beta}$ as \BZ{follows}.
    \begin{align}
        &\BZ{\partial_t} \angle_{\alpha \beta} = \frac{(\BZ{\partial_t} K_{\alpha \beta}) \sqrt{K_{\alpha \alpha} K_{\beta \beta}} - K_{\alpha \beta} \BZ{\partial_t}\sqrt{K_{\alpha \alpha} K_{\beta \beta}} }{K_{\alpha \alpha} K_{\beta \beta}}\\
        &= -\frac{\eta}{N} \frac{\sum_\gamma \epsilon_\gamma \sqrt{K_{\gamma \gamma}}\left(\lambda_{\gamma \beta \alpha} \sqrt{K_{\alpha \alpha}} + \lambda_{\gamma \alpha \beta} \sqrt{K_{\beta \beta}}\right)}{\sqrt{K_{\alpha \alpha} K_{\beta \beta}}}\nonumber\\
        &\quad - \frac{K_{\alpha \beta}}{K_{\alpha \alpha} K_{\beta \beta}} \frac{(\BZ{\partial_t} K_{\alpha \alpha}) K_{\beta \beta} + K_{\alpha \alpha} \BZ{\partial_t} K_{\beta \beta}}{2\sqrt{K_{\alpha \alpha} K_{\beta \beta}}}\\
        &= -\frac{\eta}{N} \frac{\sum_\gamma \epsilon_\gamma \sqrt{K_{\gamma \gamma}}\left(\lambda_{\gamma \beta \alpha} \sqrt{K_{\alpha \alpha}} + \lambda_{\gamma \alpha \beta} \sqrt{K_{\beta \beta}}\right)}{\sqrt{K_{\alpha \alpha} K_{\beta \beta}}}\nonumber\\
        &\quad + \frac{2\eta}{N}\frac{K_{\alpha \beta}}{K_{\alpha \alpha} K_{\beta \beta}} \left(\frac{\sum_\gamma \epsilon_\gamma \sqrt{K_{\gamma \gamma}} \lambda_{\gamma \alpha \alpha} \sqrt{K_{\alpha \alpha}} K_{\beta \beta}}{2\sqrt{K_{\alpha \alpha} K_{\beta \beta}}}\right.\nonumber  \\
        &\qquad \qquad\qquad \qquad \left. + \frac{K_{\alpha \alpha}\sum_\gamma \epsilon_\gamma \sqrt{K_{\gamma \gamma}} \lambda_{\gamma \beta \beta} \sqrt{K_{\beta \beta}} }{2\sqrt{K_{\alpha \alpha} K_{\beta \beta}}}\right)\\
        &= -\frac{\eta}{N} \sum_{\gamma} \frac{\epsilon_\gamma \sqrt{K_{\gamma \gamma}}}{\sqrt{K_{\alpha \alpha} K_{\beta \beta}}} \left[\lambda_{\gamma \beta \alpha} \sqrt{K_{\alpha \alpha}} + \lambda_{\gamma \alpha \beta} \sqrt{K_{\beta \beta}}\right.\nonumber\\
        & \qquad \qquad\qquad \qquad \qquad \left. -K_{\alpha \beta} \left(\frac{\lambda_{\gamma \alpha \alpha}}{\sqrt{K_{\alpha \alpha}}} + \frac{\lambda_{\gamma \beta \beta}} {\sqrt{K_{\beta \beta}}}\right) \right]
        \\
        &=-\frac{\eta}{N} \sum_{\gamma} \epsilon_\gamma \sqrt{K_{\gamma \gamma}} \left[\frac{\lambda_{\gamma\beta \alpha}-\angle_{\alpha\beta}\lambda_{\gamma \beta \beta }}{\sqrt{K_{\beta\beta}}}+ \frac{\lambda_{\gamma \alpha\beta}-\angle_{\alpha\beta}\lambda_{\gamma\alpha\alpha}}{\sqrt{K_{\alpha\alpha}}}\right]\\
        &=-\frac{\eta}{N} \sum_\gamma \epsilon_\gamma g_\gamma \left[\left(\frac{\lambda_{\gamma \beta \alpha}}{g_\beta} + \frac{\lambda_{\gamma \alpha \beta}}{g_\alpha} \right) - \left(\frac{\lambda_{\gamma \beta \beta}}{g_\beta} + \frac{\lambda_{\gamma \alpha \alpha}}{g_\alpha }\right)\angle_{\alpha \beta}\right].
        \label{eq:B_dyeq}
    \end{align}

Then Eq.~\eqref{eq:B_dyeq} can be simplified as
\begin{align}
    \diff_t \angle_{\alpha \beta}(t) &= -\frac{\eta}{N}  \left[\left(\frac{f_{\beta\alpha}(t)}{g_\beta(t)} + \frac{f_{\alpha\beta}(t)}{g_\alpha(t)} \right)\right.\nonumber\\
    &\left. \qquad \qquad \qquad \quad - \left(\frac{f_{\beta\beta}(t)}{g_\beta(t)} + \frac{f_{\alpha \alpha}(t)}{g_\alpha(t)}\right)\angle_{\alpha \beta}(t)\right].
    \label{B_dynamics_simp}
\end{align}

Suppose 
\be 
\calA_{\alpha\beta}\equiv \lim_{t\to\infty} \frac{\left(\frac{f_{\beta\alpha}(t)}{g_\beta(t)} + \frac{f_{\alpha\beta}(t)}{g_\alpha(t)} \right)}{\left(\frac{f_{\beta\beta}(t)}{g_\beta(t)} + \frac{f_{\alpha \alpha}(t)}{g_\alpha(t)}\right)}={\rm const},
\ee 
is a non-zero constant in $[-1,1]$, at late time Eq.~\eqref{B_dynamics_simp} can be simplified as
\begin{align}
    \diff_t \angle_{\alpha \beta}(t) &= -\frac{\eta}{N}{\left(\frac{f_{\beta\beta}(t)}{g_\beta(t)} + \frac{f_{\alpha \alpha}(t)}{g_\alpha(t)}\right)}  \left[\calA_{\alpha\beta} - \angle_{\alpha \beta}(t)\right].
    \label{B_dynamics_simp2}
\end{align}
Therefore we obtain the fixed point
\be 
\angle_{\alpha \beta}(t)=\calA_{\alpha\beta}.
\ee

\end{proof}

\begin{figure}
    \centering
    \includegraphics[width=0.45\textwidth]{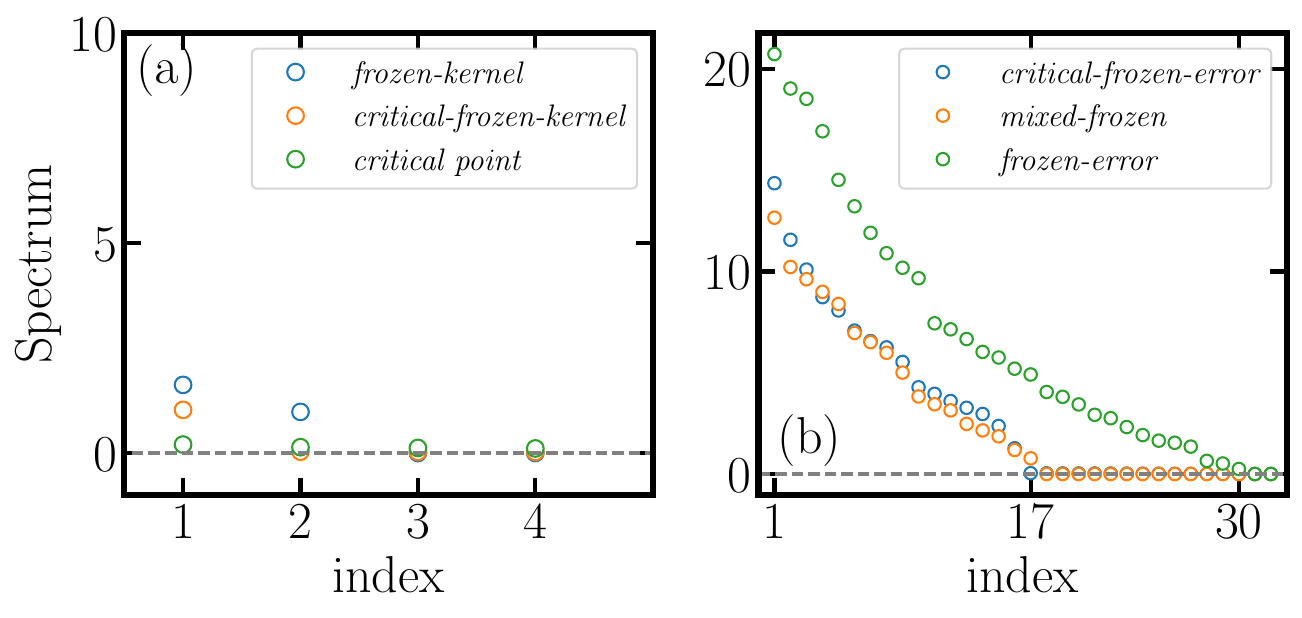}
    \caption{Spectrum of Hessian of loss function for different QNN training dynamics with two data. We plot the $4$ and $32$ largest eigenvalues in (a) and (b) separately. The setting is the same as in Fig.~\ref{fig:lda}.}
    \label{fig:spectrum}
\end{figure}

\section{Hessian spectrum interpretation}
In this section, we interpret the dynamical transition via the spectrum of Hessian of loss function in Eq.~\eqref{L_cost}. To see this, \BZ{we} begin with the dynamical equation of variational parameters at the stable fixed point $\bm \theta^*$ as
\be
    \delta \bm \theta \simeq -\eta {\bf H}(\bm \theta^*) \left(\bm \theta - \bm \theta^*\right),
    \label{H_dynamics}
\ee
where ${\bf H}(\bm \theta^*)$ is the Hessian matrix of loss function with dimension $L \times L$ defined as
\begin{align}
    &{\bf H}_{\ell_1 \ell_2}(\bm \theta) = \frac{\partial^2 \calL}{\partial \theta_{\ell_1} \partial \theta_{\ell_2}} = \sum_{\beta \in \Omega} \left(\frac{\partial \epsilon_\beta}{\partial \theta_{\ell_1}} \frac{\partial \epsilon_\beta}{\partial \theta_{\ell_2}} + \epsilon_\beta \frac{\partial^2 \epsilon_\beta}{\partial \theta_{\ell_1} \partial \theta_{\ell_2}}\right)\\
    &= \sum_{\beta \in S_E \setminus (S_E \cap S_K)} \frac{\partial \epsilon_\beta}{\partial \theta_{\ell_1}} \frac{\partial \epsilon_\beta}{\partial \theta_{\ell_2}} + \sum_{\beta \in S_K \setminus (S_E \cap S_K)} \epsilon_\beta \frac{\partial^2 \epsilon_\beta}{\partial \theta_{\ell_1} \partial \theta_{\ell_2}}.
\end{align}
In the above, the first equation comes from definition and the second equation adopts the definition of $S_E$ and $S_K$. We can regard Eq.~\eqref{H_dynamics} as an imaginary-time Schr\"odinger equation with ${\bf H}(\bm \theta^*)$ as an effective Hamiltonian. Therefore, it is natural to study the spectrum of ${\bf H}(\bm \theta^*)$. Clearly, we see the matrices in the first summation are only rank-$1$, while the others in general have rank much larger than one.
For {\it frozen-kernel dynamics} with $S_E = \Omega$, the \BZ{Hessian} ${\bf H}_{\ell_1 \ell_2}= \sum_{\beta \in S_E} \frac{\partial \epsilon_\beta}{\partial \theta_{\ell_1}} \frac{\partial \epsilon_\beta}{\partial \theta_{\ell_2}}$ becomes sums of rank-1 matrices, resulting in a rank-$N$ matrix given \BZ{an} orthogonal input data \BZ{set}. Furthermore, one can see that the trace of \BZ{Hessian} is simply the trace of QNTK matrix $\tr({\bf H}_{\ell_1 \ell_2}) = \sum_\beta K_{\beta \beta}$. When part of the data are targeted at the boundary leading to the {\it critical-frozen-kernel dynamics} with $S_K \subsetneq S_E = \Omega$, the rank of the Hamiltonian directly decreases to $N - |S_K|$. Specifically, at {\it critical point} with all data targeted at the boundary, all eigenvalues in the spectrum vanish at the fixed point. The above results are verified in Fig.~\ref{fig:spectrum}(a). On the other hand, when there are data targeted beyond the accessible region, the \BZ{Hessian} of total error $\frac{\partial^2 \epsilon_\beta}{\partial \theta_{\ell_1} \partial \theta_{\ell_2}}$ would significantly \BZ{increase} the number of positive eigenvalues in the spectrum. In fact, through numerical simulation (see Fig.~\ref{fig:spectrum}(b)) we find that the number of positive eigenvalues in {\it mixed-frozen dynamics} is just $|S_E\setminus(S_E \cap S_K)|$ more than that for {\it critical-frozen-error dynamics}, and the {\it frozen-error dynamics} has many more positive eigenvalues compared to the others. Meanwhile, how the spectrum behaves with more data involved \BZ{\sout{still}} remains unexplored as the rank may saturate to the number of parameters $L$. We leave that as an open question in future research.

\section{Gauge invariance in training dynamics}
\label{app:gauge_invariance}

In this section, we study the training dynamics under basis transformation.
We begin with the MSE loss $\calL = \frac{1}{2N}\sum_\alpha \epsilon_\alpha^2$. The inner product enables us to introduce an orthogonal matrix $S\in O(N)$, independent of both $\bm \theta$ and $t$, to transform the total error vector to 
\begin{align}
    \epsilon_\alpha(\bm \theta) \to \sum_{\alpha'} S_{\alpha \alpha'}\epsilon_{\alpha'}(\bm \theta) \equiv \tilde{\epsilon}_\alpha(\bm \theta). 
\end{align}
A direct result is that the MSE loss function is gauge invariant as
\begin{align}
    \tilde{\calL}(\bm \theta) &= \frac{1}{2N}\sum_\alpha \tilde{\epsilon}_\alpha^2(\bm \theta)
     =\frac{1}{2N} \sum_\alpha \sum_{\alpha_1, \alpha_2} S_{\alpha \alpha_1} \epsilon_{\alpha_1}(\bm \theta) S_{\alpha \alpha_2} \epsilon_{\alpha_2}(\bm \theta) 
     \\
    &= \frac{1}{2N} \sum_{\alpha} \epsilon_{\alpha}^2(\bm \theta) = \calL(\bm \theta),
\end{align}
where in the second line we apply $\sum_{\alpha} S_{\alpha \alpha_1} S_{\alpha \alpha_2} = \delta_{\alpha_1 \alpha_2}$. Thus we can identify the orthogonal group as a global gauge \BZ{symmetry} since it is independent of $t$ and $\bm \theta$ as we state above. The gauge invariance can be concluded from its inner product structure.
Following the definitions of QNTK and dQNTK in Eqs.~\eqref{eq:K_def},~\eqref{eq:mu_def} of the main text, they are transformed as
\begin{align}
    K_{\alpha \beta}(\bm \theta) &\to \sum_{\alpha', \beta'} S_{\alpha \alpha'} K_{\alpha' \beta'}(\bm \theta) S_{\beta \beta'} \equiv \tilde{K}_{\alpha \beta}(\bm \theta),\\
    \mu_{\gamma \alpha \beta}(\bm \theta) &\to \sum_{\gamma', \alpha', \beta'} S_{\gamma \gamma'} S_{\alpha \alpha'} \mu_{\gamma' \alpha' \beta'}(\bm \theta) S_{\beta \beta'} \equiv \tilde{\mu}_{\gamma \alpha \beta}(\bm \theta).
\end{align}
One can directly see that the QNTK and dQNTK do not \BZ{exhibit} the gauge invariance due to their outer product structure. However, one can easily check that $\tr(K) = \sum_\alpha K_{\alpha \alpha}$ is gauge invariant under the transformation. 

For the dynamics, we begin with the gradient descent rule.
\begin{align}
    \delta \theta_\ell(t)
    &= -\frac{\eta}{N} \sum_\alpha \tilde{\epsilon}_\alpha (\bm \theta) \frac{\partial \tilde{\epsilon}_\alpha (\bm \theta)}{\partial \theta_\ell}
    \\
    &= -\frac{\eta}{N} \sum_\alpha \sum_{\alpha_1, \alpha_2} S_{\alpha \alpha_1} \epsilon_{\alpha_1}(\bm \theta) S_{\alpha \alpha_2}\frac{\partial \epsilon_{\alpha_2} (\bm \theta)}{\partial \theta_\ell} 
    \\
    &= -\frac{\eta}{N} \sum_\alpha \epsilon_\alpha(\bm \theta) \frac{\partial \epsilon_\alpha(\bm \theta)}{\partial \theta_\ell},
\end{align}
which also preserves the gauge invariance.

For the first dynamical equation in Eqs.~\eqref{eq:dyeqs_full} of the main text, we have
\begin{align}
    &\delta \tilde{\epsilon}_{\alpha}(t) + \frac{\eta}{N} \sum_{\beta} \tilde{K}_{\alpha \beta}(t) \tilde{\epsilon}_\beta (t)\\
    &= \sum_{\alpha'} S_{\alpha \alpha'} \delta \epsilon_{\alpha'}(t) + \frac{\eta}{N} \sum_{\beta, \alpha', \beta_1, \beta_2} S_{\alpha \alpha'} K_{\alpha' \beta_1}(t) S_{\beta \beta_1} S_{\beta \beta_2} \epsilon_{\beta_2}(t)\\
    &= \sum_{\alpha'} S_{\alpha \alpha'} \left(\delta \epsilon_{\alpha'}(t) + \frac{\eta}{N} \sum_{\beta_1} K_{\alpha' \beta_1}(t) \epsilon_{\beta_1}(t) \right) = 0.
\end{align}
Similarly, for the second one, we have
\begin{align}
    &\delta \tilde{K}_{\alpha \beta}(t) + \frac{\eta}{N} \sum_{\gamma } \tilde{\epsilon}_{\gamma}(t)\left[\tilde{\mu}_{\gamma \alpha \beta}(t) + \tilde{\mu}_{\gamma \beta \alpha }(t)\right]\\
    &= \sum_{\alpha', \beta'} S_{\alpha \alpha'} \delta K_{\alpha' \beta'}(t) S_{\beta \beta'} + 
    \frac{\eta}{N} \sum_{\substack{\gamma, \gamma_1, \gamma_2,\\ \alpha', \beta'}} S_{\gamma \gamma_1} \epsilon_{\gamma_1} (t) \left[S_{\gamma \gamma_2} S_{\alpha \alpha'} \mu_{\gamma_2 \alpha' \beta'}(t) S_{\beta \beta'} + S_{\gamma \gamma_2}  S_{\beta \beta'} \mu_{\gamma_2 \beta' \alpha' }(t) S_{\alpha \alpha'} \right]\\
    &= \sum_{\alpha', \beta'} S_{\alpha \alpha'} \left[\delta K_{\alpha' \beta'}(t) + \frac{\eta}{N} \sum_{\gamma_1} \epsilon_{\gamma_1} \left(\mu_{\gamma_1 \alpha' \beta'}(t) + \mu_{\gamma_1 \beta' \alpha'}(t)\right)\right] S_{\beta \beta'}\\
    &= 0.
\end{align}
Therefore we can conclude that the dynamical equations in Eqs.~\eqref{eq:dyeqs_full} of the main text are gauge invariant under basis transformation from orthogonal group $O(N)$, which also suggests that $\tilde{\epsilon}_\alpha(t) \tilde{K}_{\alpha \alpha}(t)=0, \forall \alpha$ are fixed points.

\section{Detailed solutions for the convergence dynamics}
\label{app:details_convergence}

In this section, we present the details on deriving the convergence solution perturbatively around the stable fixed point. For convenience, we re-print the dynamical equations of \eqref{eq:dyeqs_UT} in the main text here
\begin{align}
    \left\{ \begin{array}{ll}
    \partial_t \epsilon_\alpha(t) = -\frac{\eta}{N} \sum_\beta \angle_{\alpha \beta} g_\alpha(t) g_{\beta}(t) \epsilon_\beta(t) ;\\
    \partial_t g_\alpha (t) = -\frac{\eta}{N} \sum_\beta   \lambda_{\alpha \alpha \beta} g_\beta (t) \epsilon_\beta (t),
    \end{array} \right.
    \label{eq:eqs_full}
\end{align}
where we define $g_\alpha(t) \equiv \sqrt{K_{\alpha \alpha}(t)}$ to simplify the notation.

\subsection{Exponential convergence class}

In this part, we study the exponential convergence class where $S_E \cap S_K = \emptyset$. The main idea \BZ{for} perturbatively \BZ{solving} the convergence dynamics towards a fixed point is to first focus on those quantities converging towards zero, and then apply the obtained solutions back to equations of the other equations.

\subsubsection{frozen-kernel dynamics}

For {\it frozen-kernel dynamics}, the fixed point is $\{(\epsilon_\alpha(\infty) = 0, K_{\alpha \alpha}(\infty) >0)\}_{\alpha \in \Omega}$. The leading order of the first PDE in Eqs.~\eqref{eq:eqs_full} becomes
\begin{align}
    \partial_t \epsilon_\alpha(t) &= -\frac{\eta}{N} \sum_{\beta \in \Omega} g_\alpha(\infty) \angle_{\alpha \beta}  g_\beta(\infty) \epsilon_\beta(t)\\
    &= -\frac{\eta}{N} \sum_{\beta \in \Omega} K_{\alpha \beta}(\infty) \epsilon_\beta(t).
\end{align}
As $K_{\alpha \beta}(\infty)$ is symmetric and positive definite, we can diagonalize it as $K_{\alpha \beta} = \sum_{\alpha', \beta'} P_{\alpha \alpha'} \Lambda_{\alpha' \beta'} P^T_{\beta' \beta}$, where $\Lambda_{\alpha' \beta'}$ is a diagonal matrix consisting of eigenvalues $\{w_\alpha\}_{\alpha=1}^N$ of $K_{\alpha \beta}$. Thus, we can solve $\epsilon_\alpha$ as
\begin{align}
    \epsilon_\alpha(t) = \sum_{\beta \in \Omega} b_\beta P_{\alpha \beta} e^{-\eta w_\beta t/N},
    \label{eq:frozen_kernel_sol1}
\end{align}
where $b_\beta$ are fitting parameters. 

% Since Eqs.~\eqref{eq:eqs_full} only holds for late time, one can consider a late time momennt $t'$ to solve the 
% solve the following linear equations
% \begin{align}
%     \sum_{\beta \in \Omega} b_{\beta}^E P_{\alpha \beta} e^{-\eta v_\beta t'/N} = \epsilon_\alpha(t').
%     \label{eq:frozen_kernel_coff}
% \end{align}

Plugging it into the second PDE in Eqs.~\eqref{eq:eqs_full}, we have
\begin{align}
    \partial_t g_\alpha(t) &= -\frac{\eta}{N} \sum_{\beta \in \Omega} \lambda_{\alpha \alpha \beta} g_\beta(\infty) \sum_{\gamma \in \Omega} b_\gamma P_{\beta \gamma} e^{-\eta w_\gamma t/N}\\
    &= -\frac{\eta}{N} \sum_{\gamma \in \Omega} \left(\sum_{\beta \in \Omega} \lambda_{\alpha \alpha \beta} g_\beta(\infty) P_{\beta \gamma} \right) b_\gamma e^{-\eta w_\gamma t/N},
\end{align}
which can be solved as
\begin{align}
    g_\alpha(t) &= g_\alpha(\infty) + \sum_{\gamma \in \Omega} \frac{b_\gamma}{w_\gamma}\left(\sum_{\beta \in \Omega} \lambda_{\alpha \alpha \beta} g_\beta(\infty) P_{\beta \gamma} \right) e^{-\eta v_\gamma t/N},
    \label{eq:frozen_kernel_sol2}
\end{align}
In the asymptotic limit of $t\gg 1$, we can only keep track on the exponent with the smallest eigenvalue $w^*=\min\{w_\beta\}$, which determines the leading-order behavior, resulting in simpler solutions as
\begin{align}
    \left\{ \begin{array}{ll}
     \epsilon_\alpha(t) =  b_{\gamma^*} P_{\alpha \gamma^*} e^{-\eta w^* t/N};\\
     g_\alpha (t) = g_\alpha(\infty) + \left(\sum_{\beta \in \Omega} \lambda_{\alpha \alpha \beta} g_\beta(\infty) P_{\beta \gamma^*} \right) \frac{b_{\gamma^*}}{w_{\gamma^*}}e^{-\eta w^* t/N},
    \end{array} \right.
    \label{eq:frozen_kernel_sol_simplify}
\end{align}
where $\gamma^* = \argmin_\beta w_\gamma$.

\subsubsection{frozen-error dynamics}

Inversely, for the {\it frozen-error dynamics}, the fixed point is $\{(\epsilon_\alpha(\infty) \neq 0, K_{\alpha \alpha}(\infty) =0)\}_{\alpha \in \Omega}$, the second PDE in Eqs.~\eqref{eq:eqs_full} is reduced to
\begin{align}
    \partial_t g_\alpha (t) &= -\frac{\eta}{N} \sum_{\beta \in \Omega} \lambda_{\alpha \alpha \beta} \epsilon_\beta(\infty) g_\beta(t) = -\frac{\eta}{N} \sum_\beta F_{\alpha \beta} g_\beta(t),
\end{align}
where we define $F_{\alpha \beta} \equiv A_{\alpha \beta} \epsilon_\beta(\infty)$. \BZ{Although} $F_{\alpha \beta}$ is not symmetric in general, we can still perform diagonalization to obtain $F_{\alpha \beta} = \sum_{\alpha', \beta'} = P_{\alpha \alpha'} \Lambda_{\alpha' \beta'} P^{-1}_{\beta' \beta}$, where $\Lambda_{\alpha' \beta'} = w_{\alpha' \alpha'} \delta_{\alpha' \beta'}$ is the diagonal matrix of eigenvalues. Then $g_\alpha(t)$ can be solved as
\begin{align}
    g_\alpha(t) = \sum_{\beta \in \Omega} b_\beta P_{\alpha \beta} e^{-\eta w_\beta t/N},
    \label{eq:frozen_error_sol1}
\end{align}
where $b_\beta$ are also free fitting parameters.
% \begin{align}
%     \sum_{\beta \in \Omega} b_\beta^G P_{\alpha \beta} e^{-\eta u_\beta t'/N} = g_\alpha(t').
%     \label{eq:frozen_error_coeff}
% \end{align}
One can then solve the dynamics of $\epsilon_\alpha(t)$ as
\begin{align}
    \partial_t \epsilon_\alpha(t) &= -\frac{\eta}{N} \sum_{\beta \in \Omega}  \angle_{\alpha \beta} \sum_{\gamma \in \Omega} b_\gamma P_{\alpha \gamma} e^{-\eta w_\gamma t/N} \sum_{\gamma' \in \Omega} b_{\gamma'} P_{\beta \gamma'} e^{-\eta w_{\gamma'} t/N} \epsilon_\beta(\infty)\\
    &= -\frac{\eta}{N} \sum_{\gamma, \gamma'}\left(\sum_\beta \angle_{\alpha \beta} \epsilon_\beta(\infty) P_{\beta \gamma'}\right) P_{\alpha \gamma} b_\gamma b_{\gamma'}  e^{-\eta \left(w_\gamma + w_{\gamma'} \right) t/N},
\end{align}   
which leads to the solution as
\begin{align}
    \epsilon_\alpha (t) &= \epsilon_\alpha (\infty) + \sum_{\gamma, \gamma' \in \Omega} \left(\sum_\beta \angle_{\alpha \beta} \epsilon_\beta(\infty) P_{\beta \gamma'}\right) \frac{P_{\alpha \gamma} b_\gamma b_{\gamma'}}{(w_\gamma + w_{\gamma'})} e^{-\eta (w_\gamma + w_{\gamma'}) t/N}.
    \label{eq:frozen_error_sol2}
\end{align}
In the asymptotic limit, the leading-order solution is
\begin{align}
    \left\{ \begin{array}{ll}
         \epsilon_\alpha(t) = \epsilon_\alpha(\infty) + \left(\sum_\beta \angle_{\alpha \beta} \epsilon_\beta(\infty) P_{\beta \gamma^*}\right) \frac{P_{\alpha \gamma^*} b_{\gamma^*}^2}{2 w^*} e^{-2\eta w^* t/N};\\
         g_\alpha(t) = b_{\gamma^*}P_{\alpha \gamma^*} e^{-\eta w^* t/N},
    \end{array}\right.
\end{align}
where $\gamma^* = \argmin_\gamma w_\gamma$ and $w^* = w_{\gamma^*}$.

\subsubsection{mixed-frozen dynamics}

For the {\it mixed-frozen dynamics}, the fixed point is $\{(\epsilon_\alpha(\infty) = 0, K_{\alpha \alpha}(\infty) >0)\}_{\alpha \in S_E}\cup \{(\epsilon_\alpha(\infty) \neq 0, K_{\alpha \alpha}(\infty) = 0)\}_{\alpha \in S_K}$. We first study the PDEs of $\{\epsilon_\alpha(t), \forall \alpha \in S_E\}$ and $\{g_\alpha(t), \forall \alpha \in S_K\}$, which can be reduced from Eqs.~\eqref{eq:eqs_full} as
\begin{align}
    \left\{ \begin{array}{ll}
    \partial_t \epsilon_\alpha(t) &= -\frac{\eta}{N} \left(\sum_{\beta\in S_E} g_\alpha(\infty) \angle_{\alpha \beta} g_\beta(\infty) \epsilon_\beta(t) + \sum_{\beta\in S_K} g_\alpha(\infty) \angle_{\alpha \beta}  \epsilon_\beta(\infty) g_\beta(t)\right), \forall \alpha \in S_E;\\
    \partial_t g_\alpha(t) &= -\frac{\eta}{N} \left(\sum_{\beta \in S_E} \lambda_{\alpha \alpha \beta} g_\beta (\infty) \epsilon_\beta (t) + \sum_{\beta \in S_K} \lambda_{\alpha \alpha \beta} \epsilon_\beta (\infty)g_\beta (t)\right)  
    , \forall \alpha \in S_K
    .\end{array} \right.
    \label{eq:mix_frozen_eq1}
\end{align}
Observing that the above linear PDEs can be reformed in a matrix form as
\begin{align}
    \partial_t \begin{pmatrix}
        [\epsilon_\alpha (t)]_{\alpha \in S_E}\\
        [g_\alpha(t)]_{\alpha \in S_K}
    \end{pmatrix} &= -\frac{\eta}{N} \begin{pmatrix}
         [K_{\alpha \beta}(\infty)]_{\alpha,\beta \in S_E} &  [g_\alpha(\infty) \angle_{\alpha \beta}  \epsilon_\beta(\infty)]_{\alpha \in S_E, \beta\in S_K}\\
         [\lambda_{\alpha \alpha \beta} g_\beta (\infty)]_{\alpha \in S_K, \beta \in S_E} & [\lambda_{\alpha \alpha \beta} \epsilon_\beta (\infty)]_{\alpha, \beta \in S_K}
    \end{pmatrix} \begin{pmatrix}
        [\epsilon_\beta (t)]_{\beta \in S_E}\\
        [g_\beta(t)]_{\beta \in S_K}
    \end{pmatrix},
\end{align}
where $`[\cdot ]_{\{\dots\}}{}'$ indicate the vector or matrix form with indices constraints.
Through the eigen-decomposition of the above matrix $P_{\alpha \alpha'} \Lambda_{\alpha' \beta'} P_{\beta' \beta}^{-1}$ with eigen-matrix $\Lambda_{\alpha' \beta'} = {\rm Diag}\{w_1, \cdots, w_N\}$, we obtain
\begin{align}
    \left\{ \begin{array}{ll}
    \epsilon_\alpha(t) = \sum_{\beta \in \Omega} b_\beta P_{\alpha \beta} e^{-\eta w_\beta t/N}, \forall \alpha \in S_E;\\
    g_\alpha(t) = \sum_{\beta \in \Omega} b_\beta P_{\alpha \beta} e^{-\eta w_\beta t/N}, \forall \alpha \in S_K,
    \end{array} \right.
\end{align}
where $\{b_\beta\}_{\beta \in \Omega}$ are free fitting parameters.
% where free parameters $b_\beta$ can be determined as
% \begin{align}
% \left\{ \begin{array}{ll}
%     \sum_{\beta \in \Omega} b_\beta P_{\alpha \beta} e^{-\eta w_\beta t'/N} = \epsilon_\alpha(t'), \forall \alpha \in S_E;\\
%     \sum_{\beta \in \Omega} b_\beta P_{\alpha \beta} e^{-\eta w_\beta t'/N} = g_\alpha(t'), \forall \alpha \in S_K.
% \end{array} \right.
% \end{align}
The PDE for $\{\epsilon_\alpha(t), \forall \alpha \in S_K\}$ becomes
\begin{align}
    \partial_t \epsilon_\alpha(t)
    &= -\frac{\eta}{N} \sum_{\alpha'\in \Omega} b_{\alpha'} P_{\alpha \alpha'} e^{-\eta w_{\alpha'} t/N} \left(\sum_{\beta \in S_E} \angle_{\alpha \beta} g_\beta(\infty)  \sum_{\beta' \in \Omega} b_{\beta'} P_{\beta \beta'} e^{-\eta w_{\beta'} t/N} + \sum_{\beta \in S_K} \angle_{\alpha \beta} \epsilon_\beta(\infty) \sum_{\beta'\in \Omega} b_{\beta'} P_{\beta \beta'} e^{-\eta w_{\beta'} t/N} \right)\\
    &= -\frac{\eta}{N} \sum_{\alpha', \beta'\in \Omega} \left( \sum_{\beta \in S_E} \angle_{\alpha \beta} g_\beta(\infty) P_{\beta \beta'} +  \sum_{\beta \in S_K} \angle_{\alpha \beta} \epsilon_\beta(\infty) P_{\beta \beta'} \right)  b_{\alpha'} b_{\beta'} P_{\alpha \alpha'} e^{-\eta \left(w_{\alpha'} + w_{\beta'}\right) t/N},
\end{align}
leading to the solution
\begin{align}   
    \epsilon_\alpha(t) &= \epsilon_\alpha(\infty) + \sum_{\alpha', \beta'\in \Omega} \left( \sum_{\beta \in S_E} \angle_{\alpha \beta} g_\beta(\infty) P_{\beta \beta'} +  \sum_{\beta \in S_K} \angle_{\alpha \beta} \epsilon_\beta(\infty) P_{\beta \beta'} \right) \frac{b_{\alpha'}
    b_{\beta'} P_{\alpha \alpha'}}{w_{\alpha'} + w_{\beta'}} e^{-\eta \left(w_{\alpha'} + w_{\beta'}\right) t/N}, \forall \alpha \in S_K.
\end{align}
Similarly, for $\{g_{\alpha \alpha}(t), \forall \alpha \in S_E\}$, we have
\begin{align}
    \partial_t g_\alpha(t) &= -\frac{\eta}{N} \left(\sum_{\beta \in S_E} \lambda_{\alpha \alpha \beta} g_\beta (\infty) \sum_{\beta' \in \Omega} b_{\beta'} P_{\beta \beta'} e^{-\eta w_{\beta'} t/N} + \sum_{\beta \in S_K} \lambda_{\alpha \alpha \beta} \epsilon_\beta (\infty) \sum_{\beta' \in \Omega} b_{\beta'} P_{\beta \beta'} e^{-\eta w_{\beta'} t/N} \right)\\
    &= -\frac{\eta}{N} \sum_{\beta' \in \Omega} \left(\sum_{\beta \in S_E} \lambda_{\alpha \alpha \beta} g_\beta (\infty) P_{\beta \beta'} + \sum_{\beta \in S_K} \lambda_{\alpha \alpha \beta} \epsilon_\beta (\infty) P_{\beta \beta'} \right)  b_{\beta'} e^{-\eta w_{\beta'} t/N},
\end{align}
resulting in the solution
\begin{align}
    g_\alpha(t) &= g_\alpha(\infty) + \sum_{\beta' \in \Omega} \left(\sum_{\beta \in S_E} \lambda_{\alpha \alpha \beta} g_\beta (\infty) P_{\beta \beta'} + \sum_{\beta \in S_K} \lambda_{\alpha \alpha \beta} \epsilon_\beta (\infty) P_{\beta \beta'} \right)  \frac{b_{\beta'}}{w_{\beta'}} e^{-\eta w_{\beta'} t/N}, \forall \alpha \in S_E.
\end{align}
In the asymptotic limit $t\gg 1$, we have the leading-order solution as
\begin{align}
    \left\{ \begin{array}{ll}
    \epsilon_\alpha(t) = b_{\gamma^*} P_{\alpha \gamma^*} e^{-\eta w^* t/N}, \forall \alpha \in S_E; \\
    \epsilon_\alpha(t) = \epsilon_\alpha (\infty) + \left( \sum_{\beta \in S_E} \angle_{\alpha \beta} g_\beta(\infty) P_{\beta \gamma^*} +  \sum_{\beta \in S_K} \angle_{\alpha \beta} \epsilon_\beta(\infty) P_{\beta \gamma^*} \right) \frac{b_{\gamma^*}^2
    P_{\alpha \gamma^*}}{2 w_{\gamma^*} } e^{-2 \eta w^* t/N}, \forall \alpha \in S_K;\\
    g_\alpha(t) = g_\alpha(\infty) + \left(\sum_{\beta \in S_E} \lambda_{\alpha \alpha \beta} g_\beta (\infty) P_{\beta \gamma^*} + \sum_{\beta \in S_K} \lambda_{\alpha \alpha \beta} \epsilon_\beta (\infty) P_{\beta \gamma^*} \right)  \frac{b_{\gamma^*}}{w_{\gamma^*}} e^{-\eta w^* t/N}, \forall \alpha \in S_E;\\
    g_\alpha(t) = b_{\gamma^*}P_{\alpha \gamma^*} e^{-\eta w^* t/N}, \forall \alpha \in S_K,
    \end{array}\right.
\end{align}
where $\gamma^* = \argmin_\gamma w_\gamma$ and $w^* = w_{\gamma^*}$.

\subsection{Polynomial convergence class}

In this section, we consider $S_E \cap S_K \neq \emptyset$, which corresponds to the polynomial convergence class.

\subsubsection{Critical point}

When $S_E = S_K = \Omega$, it corresponds to the {\it critical point} with the fixed point $\{(\epsilon_\alpha(\infty) = 0, K_{\alpha \alpha} (\infty) = 0)\}_{\alpha \in \Omega}$. The PDEs for error and kernel are the same as in Eqs.~\eqref{eq:eqs_full}, and to solve it, we take an ansatz solution
\begin{align}
    \left\{ \begin{array}{ll}
    \epsilon_\alpha(t) = c_\alpha^E/(c_0 + \eta t/N);\\
    g_\alpha(t) = c_\alpha^G/\sqrt{c_0 + \eta t/N}
    ,\end{array} \right.
    \label{eq:crit_sol_supp}
\end{align}
with fitting parameters $\{c_\alpha^E, c_\alpha^G\}$. 
% can be determined from
% \begin{align}
%     \left\{ \begin{array}{ll}
%         \sum_{\beta \in \Omega} \angle_{\alpha \beta} c_\alpha^G c_\beta^G c_\beta^E = c_\alpha^E;\\
%     \sum_{\beta \in \Omega} \lambda_{\alpha \beta} c_\beta^E c_\beta^G = c_\alpha^G/2.
%     \end{array} \right.
% \end{align}
% The proper choice of $c_\alpha^E, c_\alpha^G$, $c_0$ can also be found by matching the solution at late time $t'$. 

\subsubsection{Critical-frozen-kernel dynamics}

When $S_K \subsetneq S_E = \Omega$, we have the fixed points $\{(\epsilon_\alpha (\infty) =0, K_{\alpha \alpha}(\infty) = 0)\}_{\alpha \in S_K} \cup \{(\epsilon_\alpha(\infty) = 0, K_{\alpha \alpha} (\infty) >0)\}_{\alpha \in S_E \setminus S_K}$. Initially\BZ{,} the interaction between different data is negligible, and we can expect that data from $S_K$ follows the dynamics of {\it critical point} while the one from $S_E \setminus S_K$ follows the dynamics of {\it frozen-kernel dynamics}, which suggests that the convergence of $\epsilon_\beta(t) g_\beta (t)$ from $S_K$, governed by Eqs.~\eqref{eq:eqs_full}, is much faster \BZ{compared} to $S_E \setminus S_K$. Therefore, for the dynamics of error and kernel from $S_K$, we treat them as self-governed in a ``free-field'' theory as
\begin{align}
\left\{ \begin{array}{ll}
    \partial_t \epsilon_\alpha(t) &= -\frac{\eta}{N} \sum_{\beta \in  S_K}  g_\alpha(t)  \angle_{\alpha \beta}
    g_\beta(t) \epsilon_\beta(t), \forall \alpha \in S_K ;\\
    \partial_t g_\alpha(t) &= -\frac{\eta}{N} \sum_{\beta \in S_K} \lambda_{\alpha \alpha \beta} g_\beta (t) \epsilon_\beta (t) , \forall \alpha \in S_K .
    \end{array} \right.
\end{align}
The solution of these ``free-field'' part can be described by Eqs.~\eqref{eq:crit_sol_supp}.

Plugging in the polynomial solutions, the PDE for error from $\alpha \in S_E \setminus S_K$ is
\begin{align}
    \partial_t \epsilon_\alpha(t) &= -\frac{\eta}{N} g_\alpha(\infty)\left(\sum_{\beta \in S_E \setminus S_K}  \angle_{\alpha \beta} g_\beta(\infty) \epsilon_\beta(t) + \sum_{\beta \in S_K}  \frac{\angle_{\alpha \beta} c_\beta^E c_\beta^G}{\left(c_0 + \eta t/N \right)^{3/2}} \right).
\end{align}
At late time, when $\{g_\beta(\infty) \epsilon_\beta(t), \forall \beta \in S_E \setminus S_K\}$ is comparable to $\left(c_0 + \eta t/N \right)^{-3/2}$, the interactions cannot be neglected, and thus we take 
\begin{align}
    \epsilon_\alpha(t) = \frac{b_\alpha}{(c_0 + \eta t/N)^{3/2}}, \forall \alpha \in S_E \setminus S_K,
\end{align}
with fitting parameters $b_\alpha$.
% The fitting parameters $a_\alpha^E, a_0$ can be determined as the following. Note that in the late time limit of $t\to \infty$, $b/(c+qt)^{3/2} \simeq b/(qt)^{3/2} - 3 bc/(2(qt)^{5/2})$, both sides of the above PDE can be expanded in late time limit, and via matching the coefficient of $t^{-3/2}$, $a_\alpha^E, \forall \alpha \in S_E \setminus (S_E \cap S_K)$ becomes
% \begin{align}
%     \sum_{\beta \in S_E \setminus  S_K}  \angle_{\alpha \beta} g_\beta(\infty) a_\beta^E + \sum_{\beta \in S_K}  \angle_{\alpha \beta} c_\beta^E c_\beta^G = 0.
% \end{align}
% The matching of coefficient for $t^{-5/2}$, $a_0$ can be further found to be 
% \begin{align}
%     a_0 = -\frac{a_\alpha^E/g_\alpha(\infty) + c_0\sum_{\beta \in S_K} \angle_{\alpha \beta} c_\beta^E c_\beta^G}{\sum_{\beta \in S_E \setminus S_K} \angle_{\alpha \beta} g_\beta(\infty)a_\beta^E}.
% \end{align}
Then $g_\alpha(t), \alpha \in S_E \setminus S_K$ can be obtained from
\begin{align}
    \partial_t g_\alpha(t) &= -\frac{\eta}{N} \left(\sum_{\beta \in S_E \setminus S_K }  \frac{\lambda_{\alpha \alpha \beta} g_\beta(\infty) b_\alpha}{(c_0 + \eta t/N)^{3/2}} + \sum_{\beta \in S_K} \frac{\lambda_{\alpha \alpha \beta} c_\beta^E c_\beta^G}{\left(c_0 + \eta t/N \right)^{3/2}}\right),
\end{align}
leading to
\begin{align}
    g_\alpha(t) &= \left(\sum_{\beta \in S_E \setminus S_K} \lambda_{\alpha \alpha \beta} g_\beta(\infty) b_\alpha + \sum_{\beta \in S_K} \lambda_{\alpha \alpha \beta} c_\beta^E c_\beta^G \right)\frac{2}{\sqrt{c_0 + \eta t/N}} + g_\alpha(\infty), \forall \alpha \in S_E \setminus S_K.
\end{align}
To summarize, we have
\begin{align}
    \left\{\begin{array}{ll}
       \epsilon_\alpha(t) = c_\alpha^E/(c_0 + \eta t/N), \forall \alpha \in  S_K;  \\
       \epsilon_\alpha(t) = b_\alpha/(c_0 + \eta t/N)^{3/2}, \forall \alpha \in S_E \setminus S_K; \\
       g_\alpha(t) =  c_\alpha^G/\sqrt{c_0 + \eta t/N}, \forall \alpha \in  S_K;  \\
       g_\alpha(t) = 2\left(\sum_{\beta \in S_E \setminus S_K}\lambda_{\alpha \alpha \beta} g_\beta(\infty) b_\alpha + \sum_{\beta \in S_K} \lambda_{\alpha \alpha \beta} c_\beta^E c_\beta^G \right)/\sqrt{c_0 + \eta t/N} + g_\alpha(\infty), \forall \alpha \in S_E \setminus S_K.
    \end{array} \right.
\end{align}

\subsubsection{Critical-frozen-error dynamics}

When $S_E \subsetneq S_K = \Omega$, the fixed point is described by: $\{(\epsilon_\alpha (\infty) = 0, K_{\alpha \alpha}(\infty) = 0)\}_{\alpha \in S_E} \cup \{(\epsilon_\alpha(\infty) \neq 0, K_{\alpha \alpha} (\infty) =0)\}_{\alpha \in S_K \setminus S_E}$. Similar to the previous case, we apply the same method to solve the dynamics. For data from $S_E $, it is still  described by Eq.~\eqref{eq:crit_sol_supp}, and for $g_\alpha, \forall \alpha \in S_K \setminus S_E$, the PDE for $g_\alpha (t)$ becomes
\begin{align}
    \partial_t g_\alpha(t) &= -\frac{\eta}{N} \left(\sum_{\beta \in S_E }  \frac{\lambda_{\alpha \alpha \beta} c_\beta^E c_\beta^G}{(c_0 + \eta t/N)^{3/2}} + \sum_{\beta \in S_K \setminus S_E} \lambda_{\alpha \alpha \beta} \epsilon_\beta(\infty) g_\beta(t) \right).
\end{align}
From the balance of r.h.s., we have 
\begin{align}
    g_\alpha(t) = \frac{b_\alpha}{(c_0 + \eta t/N)^{3/2}}, \forall \alpha \in S_K \setminus S_E,
\end{align}
with free fitting parameters $b_\alpha$.
% Through the same expansion utilized in the above subsection, $a_\alpha^G$ can be obtained from $N_{\rm out}$ linear equations through the matching of coefficient for $t^{-3/2}$ as
% \begin{align}
%     \sum_{\beta \in S_K \setminus S_E} \lambda_{\alpha \alpha \beta} \epsilon_\beta(\infty) a_\beta^G + \sum_{\beta \in S_E} \lambda_{\alpha \alpha \beta} c_\beta^E c_\beta^G = 0,
% \end{align}
% and from matching coefficient for $t^{-5/2}$ $a_0$ can be found to be 
% \begin{align}
%     a_0 = -\frac{a_\alpha^G + c_0\sum_{\beta \in S_E}\lambda_{\alpha \alpha \beta} c_\beta^E c_\beta^G}{\sum_{\beta \in S_K \setminus S_E} \lambda_{\alpha \alpha \beta} \epsilon_\beta(\infty) a_\beta^G}.
% \end{align}
One can then integrate over $t$ to find the dynamics for $\epsilon_\alpha(t), \forall \alpha \in S_K \setminus (S_E \cap S_K)$. Overall, we have
\begin{align}
    \left\{ \begin{array}{ll}
    \epsilon_\alpha(t) = c_\alpha^E/(c_0 + \eta t/N), \forall \alpha \in S_E;  \\
    \epsilon_\alpha(t) = \frac{1}{2}\left[\sum_{\beta\in S_E} \angle_{\alpha \beta} b_\alpha c_\beta^E c_\beta^G + \sum_{\beta \in S_K \setminus S_E} \angle_{\alpha \beta}  b_\alpha b_\beta \epsilon_\beta(\infty) \right]/\left(c_0 + \eta t/N \right)^2  + \epsilon_\alpha(\infty), \forall \alpha \in S_K \setminus S_E; \\
    g_\alpha(t) = c_\alpha^G/\sqrt{c_0 + \eta t/N}, \forall \alpha \in S_E;  \\
    g_\alpha(t) = b_\alpha/(c_0 + \eta t/N)^{3/2}, \forall \alpha \in S_K \setminus S_E.
    \end{array}\right. 
\end{align}

\subsubsection{Critical-mixed-frozen dynamics}
Finally, we extend our analyses to the case where the target values lie in all possible regions $\mathbb{R}$. With the same ``free-field'' approach, the data from $S_E \cap S_K$ can be described by 
Eq.~\eqref{eq:crit_sol_supp}. Then the dynamical equations for $\{\epsilon_\alpha, \forall \alpha \in S_E \setminus (S_E \cap S_K)\}$ and $\{g_\alpha, \forall S_K \setminus (S_E \cap S_K)\}$ become
\begin{align}
    \left\{ \begin{array}{ll}
    \partial_t \epsilon_\alpha(t) &= -\frac{\eta}{N}g_\alpha(\infty)\left( \sum_{\beta \in S_E \setminus (S_E \cap S_K)}  \angle_{\alpha \beta} g_\beta(\infty) \epsilon_\beta(t) + \sum_{\beta \in S_E \cap S_K} \angle_{\alpha \beta} g_\beta(t) \epsilon_\beta(t)  + \sum_{\beta \in 
    S_K \setminus (S_E \cap S_K)}  \angle_{\alpha \beta} \epsilon_\beta(\infty) g_\beta(t) \right);\\
    \partial_t g_\alpha(t) &= -\frac{\eta}{N} \left(\sum_{\beta \in S_E \setminus (S_E \cap S_K)} \lambda_{\alpha \alpha \beta} g_\beta (\infty) \epsilon_\beta (t) + \sum_{\beta \in S_E \cap S_K} \lambda_{\alpha \alpha \beta} g_\beta (t) \epsilon_\beta (t) + \sum_{\beta \in S_K \setminus (S_E \cap S_K)} \lambda_{\alpha \alpha \beta} \epsilon_\beta (\infty) g_\beta (\infty) \right)
    .\end{array} \right.
\end{align}
As $\epsilon_\beta(t) g_\beta(t) = c_\beta^E c_\beta^G/(c_0 + \eta t/N)^{3/2}, \forall \beta \in S_E \cap S_K$, we here take
\begin{align}
    \epsilon_\alpha(t) &= \frac{b_\alpha^E}{\left(c_0 + \eta t/N\right)^{3/2}}, \forall \alpha \in S_E \setminus (S_E \cap S_K),\\
    g_\alpha(t) &= \frac{b_\alpha^G}{\left(c_0 + \eta t/N\right)^{3/2}}, \forall \alpha \in S_K \setminus (S_E \cap S_K),
\end{align}
with free fitting parameters $b_\alpha^E, b_\alpha^G$.
% where again the fitting parameters can be determined from matching coefficients of $t^{-3/2}$ and $t^{-5/2}$, which leads to the following linear equations.
% \begin{align}
% \left\{ \begin{array}{ll}
%     \sum_{\beta \in S_E \setminus (S_E \cap S_K)} \angle_{\alpha \beta} g_\beta(\infty) a_\beta^E + \sum_{\beta \in S_K \setminus (S_E \cap S_K)}  \angle_{\alpha \beta} \epsilon_\beta(\infty) a_\beta^G 
%     = -\sum_{\beta \in S_E \cap S_K} \angle_{\alpha \beta} c_\beta^E c_\beta^G;\\
%      \sum_{\beta \in S_E \setminus (S_E \cap S_K)} \lambda_{\alpha \alpha \beta} g_\beta (\infty) a_\beta^E + \sum_{\beta \in S_K \setminus (S_E \cap S_K)} \lambda_{\alpha \alpha \beta} \epsilon_\beta (\infty) a_\beta^G 
%     = - \sum_{\beta \in S_E \cap S_K} \lambda_{\alpha \alpha \beta} c_\beta^E c_\beta^G
%     .\end{array} \right.
% \end{align}
% and
% \begin{align}
% \left\{ \begin{array}{ll}
% a_0^E \sum_{\beta \in S_E \setminus (S_E \cap S_K)} \angle_{\alpha \beta} g_\beta(\infty) a_\beta^E + a_0^G \sum_{\beta \in S_K \setminus (S_E \cap S_K)} \angle_{\alpha \beta} \epsilon_\beta(\infty) a_\beta^G 
% = -a_\alpha^E/g_\alpha(\infty) -c_0  \sum_{\beta \in S_E \cap S_K} \angle_{\alpha \beta} c_\beta^E c_\beta^G;\\
% a_0^E \sum_{\beta \in S_E \setminus (S_E \cap S_K)} \lambda_{\alpha \alpha \beta} g_\beta (\infty) a_\beta^E + a_0^G \sum_{\beta \in S_K \setminus (S_E \cap S_K)} \lambda_{\alpha \alpha \beta} \epsilon_\beta (\infty) a_\beta^G 
% = -a_\alpha^G  - c_0  \sum_{\beta \in S_E \cap S_K} \lambda_{\alpha \alpha \beta} c_\beta^E c_\beta^G
% .\end{array} \right.
% \end{align}
\BZ{By taking one additional step}, one can find the solutions for the other errors and QNTKs. 
We summarize the solutions for errors and QNTKs as
\begin{align}
    \left\{ \begin{array}{ll}
    \epsilon_\alpha(t) = c_\alpha^E/(c_0 + \eta t/N), \forall \alpha \in S_E \cap S_K;\\
    \epsilon_\alpha(t) = b_\alpha^E/(c_0 + \eta t/N)^{3/2}, \forall \alpha \in S_E \setminus (S_E \cap S_K); \\
    \epsilon_\alpha(t) = \frac{1}{2}\left(\sum_{\beta\in S_E \setminus (S_E \cap S_K)}  \angle_{\alpha \beta} g_\beta(\infty) b_\beta^E + \sum_{\beta\in S_E \cap S_K} \angle_{\alpha \beta} c_\beta^E c_\beta^G  + \sum_{\beta\in S_K \setminus (S_E \cap S_K)} \angle_{\alpha \beta} \epsilon_\beta(\infty) b_\beta^G \right) b_\alpha^G/\left(c_0 + \eta t/N\right)^2\\
    \qquad \qquad + \epsilon_\alpha(\infty), \forall \alpha \in S_K \setminus (S_E \cap S_K); \\
    g_\alpha(t) = 2\left(\sum_{\beta\in S_E \setminus (S_E \cap S_K)} \lambda_{\alpha \alpha \beta} g_\beta(\infty) b_\beta^E + \sum_{\beta \in S_E \cap S_K} \lambda_{\alpha \alpha \beta} c_\beta^E c_\beta^G + \sum_{\beta \in S_K \setminus (S_E \cap S_K)} \lambda_{\alpha \alpha \beta} \epsilon_\beta(\infty) b_\beta^G\right)/\sqrt{c_0 + \eta t/N}\\
    \qquad \qquad + g_\alpha(\infty),\alpha \in S_E \setminus (S_E \cap S_K); \\
    g_\alpha(t) = c_\alpha^G/\sqrt{c_0 + \eta t/N}, \forall \alpha \in S_E \cap S_K; \\
    g_\alpha(t) = b_\alpha^G/(c_0 + \eta t/N)^{3/2}, \forall \alpha \in S_K \setminus (S_E \cap S_K).
    \end{array} \right. 
\end{align}

\section{Restricted Haar random ensemble}
\label{app:rh}

To provide an insight on the converged unitary in late time, we consider a multi-state preparation task \BZ{with both input and target states $\{\ket{\psi_\alpha}\}$, $\{\ket{\Phi_\alpha}\}$ forming orthonormal sets}, $\braket{\psi_\alpha|\psi_\beta} = \braket{\Phi_\alpha|\Phi_\beta} = \delta_{\alpha \beta}$. 
We can then formulate the ensemble of unitary (up to permutation) for the multi-state preparation task as
\begin{align}
    \calU_{\rm RH} = \left\{U \left \lvert U = \begin{pmatrix}
        Q_N &\bm 0 \\
        \bm 0 & V
    \end{pmatrix},\right.
    Q_N={\rm diag}\left(e^{i\phi_1}, \dots, e^{i\phi_N}\right), \{\phi_\alpha\}_{\alpha=1}^N \sim \mathbb{U}[0, 2\pi), V \in \calU_{\rm Haar}(d- N) \right\}.
    \label{eq:rh_def}
\end{align}
The unitary in the ensemble consists of two blocks, the first block $Q$ is a diagonal matrix of complex numbers with unity modulus and their corresponding angles are uniformly distributed within $[0, 2\pi)$ \BZ{as there is no preferred distribution for the complex phases}. The second block $V$ is sampled from Haar random unitaries with dimension $d-N$. Specifically, when $N \ge d-1$, $V$ \BZ{reduces} to a complex scalar $e^{i \phi}$ with $\phi$ uniformly distributed in $[0, 2\pi)$ as well. 
The uniform distribution of $\phi_\alpha$ is verified in Fig.~\ref{fig:angle_dist} (a) and (b) up to some fluctuations.
Note that the ensemble $\calU_{\rm RH}$ is a generalization of single-data restricted Haar ensemble discussed in Ref.~\cite{zhang2023dynamical}.

\begin{figure}[h]
    \centering
    \includegraphics[width=0.65\linewidth]{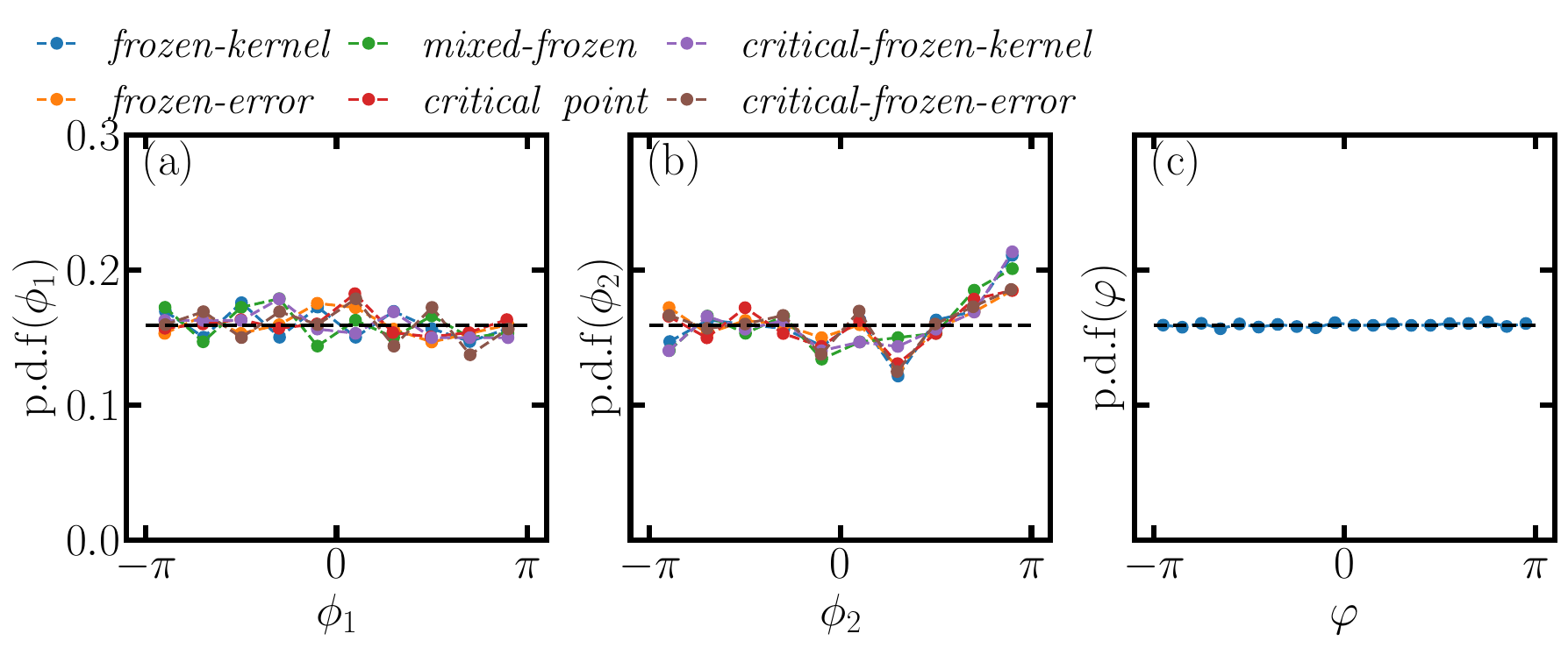}
    \caption{Distribution of complex angles. In (a), (b), we show the distribution of $\phi_1$ and $\phi_2$ from circuit unitaries at late time. We consider a $n=2$ qubit multi-state preparation task, and the RPA consists of $L=64$ parameters. In (c) we show the distribution of $\varphi$ generated from $d=8$ haar random unitaries. The black dashed lines represent the p.d.f of uniform distribution $\mathbb{U}[-\pi,\pi)$.}
    \label{fig:angle_dist}
\end{figure}

To unveil ensemble properties of the restricted Haar ensemble, we focus on its frame potential~\cite{roberts2017chaos}, a quantity to represent the randomness of unitaries within the ensemble. Ahead of presenting the calculation details, we summarize the calculation results here.
The $k$th frame potential of restricted Haar ensemble can be lower bounded by
\begin{align}
    \calF^{(k)}_{\rm RH} \ge \begin{cases}
        \sum_{k_1=\rm even}^k \sum_{k_2=0}^{k-k_1} \frac{k!}{((k_1/2)!)^2 k_2! (k-k_1-k_2)!} N^{k-k_1-k_2} \calF_{\rm Haar}^{(k_1/2 + k_2)}, & 1\le N < d-1\\
        \sum_{k_1=\rm even}^k \frac{k!}{((k_1/2)!)^2 (k- k_1)!} d^{k - k_1}, & d-1 \le N \le d
    \end{cases},
    \label{eq:rstFP_lb}
\end{align}
where the frame potential of Haar random unitaries is $\calF_{\rm Haar}^{(k)} = k!$~\cite{roberts2017chaos}.
Specifically for $k=2$, the frame potential can be exactly solved as
\begin{align}
    \calF^{(2)}_{\rm RH} = \begin{cases}
        2N^2 + 3N + 2, & 1\le N < d-1\\
        2d^2 -d, & d-1 \le N \le d
    \end{cases}.
    \label{eq:rstFP2}
\end{align}
In Fig.~\ref{fig:rstFP}(a)-(b), we see that our lower bound (Eq.~\eqref{eq:rstFP_lb}) can characterize the leading order scaling of the exact $k$th frame potential for restricted Haar ensemble. Specifically, for $k=2$, Eq.~\eqref{eq:rstFP2} (red line in Fig.~\ref{fig:rstFP} (b)) agrees with numerical results. In Fig.~\ref{fig:rstFP}(a), the gap between $\calF^{(k)}_{\rm RH}$ and $\calF^{(k)}_{\rm Haar}$ enlarges with increasing $k$ for a fixed number of data $N$. On the other hand, in Fig.~\ref{fig:rstFP}(b) for a specific order $k$ for example $k=2$, the $\calF^{(k)}_{\rm RH}$ increases with $N$ until convergence to a $d$-dependent constant, which is significantly different from the constant $\calF^{(k)}_{\rm Haar} = k!$ of Haar ensemble. We can interpret the phenomena by the increasing number of constraints thus less degree of randomness of unitaries from $\calU_{\rm RH}$ given more input data, leading to a larger frame potential.

The detailed calculations of QNTK matrix and relative dQNTK averaged over restricted Haar ensemble can be Appendix~\ref{app:ensemble_average_calculation}.

\begin{figure}[t]
    \centering
    \includegraphics[width=0.5\linewidth]{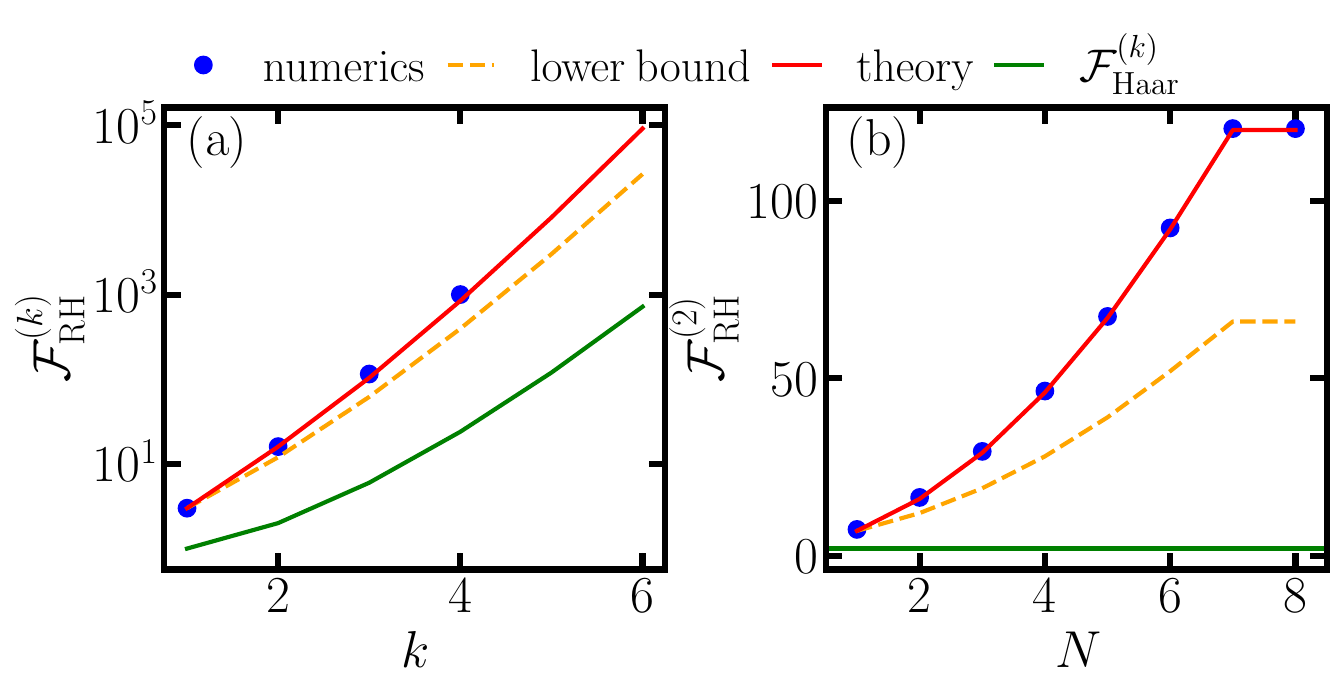}
    \caption{Frame potential of restricted Haar ensemble. In (a) the restricted Haar ensemble is in dimension of $d=4$ with $N=2$ data. In (b), the restricted Haar ensemble is in dimension $d=8$ with various $N$. Blue dots are numerical results of an ensemble of $10^4$ unitaries sampled from $\calU_{\rm RH}$. Red solid lines in (a) and (b) represent exact analytical results calculated from Eq.~\eqref{eq:framePot_rst_l2} and Eq.~\eqref{eq:rstFP2}. The orange dashed lines represent the lower bound from Eq.~\eqref{eq:rstFP_lb}. Green lines show the corresponding frame potential of haar random unitaries.}
    \label{fig:rstFP}
\end{figure}

\subsection{Calculation details of frame potential}

Following the definition, the $k$th frame potential of the restricted Haar ensemble unitaries becomes
\begin{align}
    \calF^{(k)}_{\rm RH} &= \frac{1}{|\calE|^2} \sum_{U,U^\prime \in \calU_{\rm RH}} |\tr(U^\dagger U^\prime)|^{2k}\\
    &= \frac{1}{|\calE|^2} \sum_{V,V^\prime \in \calU_{\rm Haar}} |\tr(Q_N^\dagger Q_N') + \tr(V^\dagger V^\prime)|^{2k}\\
    &= \int_{\mathbb{U}[0, 2\pi)} \prod_{\alpha=1}^N \diff \phi_{\alpha}  \diff \phi'_{\alpha} \int_{\calU_{\rm Haar}} \diff V  \diff V^\prime \left\lvert \sum_{\alpha =1}^N e^{i(\phi'_\alpha - \phi_\alpha)} +  \tr(V^\dagger V^\prime)\right \rvert^{2k}
    \label{eq:framePot_rst_l1}
\end{align}
For convenience, we denote $z\equiv \tr(V^\dagger V') = |z|e^{i\varphi}$, which is a complex scalar in general. As $V, V'\sim \calU_{\rm Haar}$ without other limitations, we expect $\varphi\sim \mathbb{U}[0, 2\pi)$ (see example in Fig.~\ref{fig:angle_dist} (c)), then we have
\begin{align}
    \calF^{(k)}_{\rm RH} &= \int_{\mathbb{U}[0, 2\pi)} \prod_{\alpha=1}^N \diff \phi_{\alpha} \diff \varphi \int_{\calU_{\rm Haar}} \diff |z|  \left[N + 2\sum_{\alpha < \beta}\cos(\phi'_\alpha - \phi_\alpha -\phi'_\beta + \phi_\beta ) + 2|z|\sum_\alpha \cos(\phi'_\alpha - \phi_\alpha - \varphi) + |z|^2\right]^k\\
    &= \int  \prod_{i=1}^{\binom{N}{2}} \diff x_i\prod_{j=1}^N \diff y_j p_X(x_i) p_Y(y_j)   \int_{\calU_{\rm Haar}} \diff |z| \left[N + 2\sum_{i=1}^{\binom{N}{2}}\cos(x_i) + 2|z|\sum_{j=1}^N \cos(y_j) + |z|^2\right]^k \label{eq:framePot_rst_l2}\\
    &\ge \int \diff y_1 p_Y(y_1) \int_{\calU_{\rm Haar}} \diff |z| \left[N + 2|z| \cos(y_1) + |z|^2\right]^k\\
    &= \sum_{\substack{k_1,k_2=0\\k_1 + k_2 \le k}} \int \diff y_1 p_Y(y_1) \int_{\calU_{\rm Haar}} \diff |z|  \binom{k}{k_1,k_2} 2^{k_1} N^{k-k_1-k_2}\cos^{k_1}(y_1) |z|^{k_1 + 2k_2}\\
    &= \sum_{\substack{k_1,k_2=0\\k_1 + k_2 \le k}} \binom{k}{k_1,k_2} 2^{k_1} N^{k-k_1-k_2} \int \diff y_1 p_Y(y_1) \cos^{k_1}(y_1) \int_{\calU_{\rm Haar}} \diff |z|  |z|^{k_1 + 2k_2} \\
    &= \sum_{\substack{k_1,k_2=0\\k_1 + k_2 \le k}} \binom{k}{k_1,k_2} 2^{k_1} N^{k-k_1-k_2} \mathbb{E}_{p_Y}\left[ \cos^{k_1}(y_1)\right] \calF_{\rm Haar}^{(k1/2+k_2)}, 
\end{align}
where in Eq.~\eqref{eq:framePot_rst_l2} we introduce the notation $x_i \equiv \phi_\alpha - \phi_\alpha - \phi'_\beta + \phi_\beta$ for $\alpha < \beta$ and $y_j \equiv \phi_\alpha - \phi_\alpha - \varphi$ for simplicity, and thus in total there are $\binom{N}{2}$ variables $x_i$ and $N$ variables $y_i$. As $\phi_\alpha, \varphi \sim \mathbb{U}[0, 2\pi)$, the distribution of $x_i$ and $y_j$ can be found to be
\begin{align}
    &p_X(x) = \begin{cases}
        \frac{(x+4 \pi )^3}{96 \pi ^4}, & -4\pi \le x \le -2\pi \\
        \frac{32 \pi^3 - 12 \pi x^2 - 3 x^3}{96 \pi^4}, & -2\pi \le x \le 0\\
        \frac{3 x^3-12 \pi x^2+32 \pi ^3}{96 \pi^4}, & 0 \le x \le 2\pi\\
        \frac{(4 \pi -x)^3}{96 \pi ^4}, & 2\pi \le x \le 4\pi
    \end{cases},
    &p_Y(y) = \begin{cases}
        \frac{(y+4 \pi )^2}{16 \pi ^3},  & -4\pi \le y \le -2\pi\\
        \frac{2\pi^2 - 2\pi y - y^2}{8 \pi ^3}, & -2\pi \le y \le 0\\
        \frac{(y-2 \pi )^2}{16 \pi ^3}, & 0 \le y \le 2\pi
    \end{cases}
\end{align}
The average $\mathbb{E}_{p_Y}[\cos^{k_1}(y_1)]$ can thus be evaluated as
\begin{align}
    &\mathbb{E}_{p_Y}[\cos^{k_1}(y_1)] = \int_{-4\pi}^{2\pi} \diff y_1 p_Y(y_1) \cos^{k_1}(y_1)\\
    &= \int_{-4\pi}^{2\pi} \diff y_1 \frac{(y+4 \pi )^2}{16 \pi ^3} \cos^{k_1}(y_1) + \int_{-2\pi}^0 \diff y_1 \frac{2\pi^2 - 2\pi y - y^2}{8 \pi ^3} \cos^{k_1}(y_1) + \int_{0}^{2\pi} \diff y_1 \frac{(y-2 \pi )^2}{16 \pi ^3} \cos^{k_1}(y_1)\\
    &= \int_0^{2\pi} \diff y_1 \frac{y^2}{16 \pi ^3} \cos^{k_1}(y_1-4\pi) + \int_0^{2\pi} \diff y_1 \frac{2\pi^2 - 2\pi (y-2\pi) - (y-2\pi)^2}{8 \pi ^3} \cos^{k_1}(y_1-2\pi) + \int_0^{2\pi} \diff y_1 \frac{(y-2 \pi )^2}{16 \pi ^3} \cos^{k_1}(y_1) \label{eq:framePot_l3}\\
    &= \int_0^{2\pi} \diff y_1 \frac{1}{2\pi}  \cos^{k_1}(y_1)\\
    &= \frac{\left((-1)^{k_1}+1\right)^2 \Gamma \left(\frac{k_1+1}{2}\right)}{4 \sqrt{\pi } \Gamma \left(\frac{k_1}{2}+1\right)},
\end{align}
where in Eq.~\eqref{eq:framePot_l3} we make the change of variables.
Therefore, the frame potential can be reduced to
\begin{align}
    \calF^{(k)}_{\rm RH} &\ge \sum_{\substack{k_1,k_2=0\\k_1 + k_2 \le k}} \binom{k}{k_1,k_2} 2^{k_1} N^{k-k_1-k_2} \mathbb{E}_{p_Y}\left[ \cos^{k_1}(y_1)\right] \calF_{\rm Haar}^{(k1/2+k_2)}\\
    &= \sum_{\substack{k_1,k_2=0\\k_1 + k_2 \le k}} \binom{k}{k_1,k_2} 2^{k_1} N^{k-k_1-k_2} \frac{\left((-1)^{k_1}+1\right)^2 \Gamma \left(\frac{k_1+1}{2}\right)}{4 \sqrt{\pi } \Gamma \left(\frac{k_1}{2}+1\right)} \calF_{\rm Haar}^{(k1/2+k_2)} \\
    &= \sum_{k_1=\rm even}^k \sum_{k_2=0}^{k-k_1}  \binom{k}{k_1, k_2} 2^{k_1} N^{k-k_1-k_2} \frac{\Gamma(k_1/2+1/2)}{\sqrt{\pi} \Gamma(k_1/2 + 1)} \calF_{\rm Haar}^{(k1/2+k_2)}\\
    &= \sum_{k_1=\rm even}^k \sum_{k_2=0}^{k-k_1} \frac{k!}{k_1! k_2! (k-k_1-k_2)!} 2^{k_1} N^{k-k_1-k_2} \frac{2^{-k_1} \sqrt{\pi} \Gamma(k_1 + 1)}{\sqrt{\pi} \Gamma(k_1/2 + 1)^2}  \calF_{\rm Haar}^{(k1/2+k_2)} \\
    &= \sum_{k_1=\rm even}^k \sum_{k_2=0}^{k-k_1} \frac{k!}{((k_1/2)!)^2 k_2! (k-k_1-k_2)!} N^{k-k_1-k_2} \calF_{\rm Haar}^{(k_1/2 + k_2)},
\end{align}
which holds for $N < d-1$. For a fixed $k$-th order, the leading order of the frame potential scales as $\calF_{\rm RH}^{(k)} \sim N^{k}$. Specifically, for $k=2$, we can find the exact result from Eq.~\eqref{eq:framePot_rst_l2} as
\begin{align}
    \calF_{\rm RH}^{(2)} &= \int  \prod_{i=1}^{\binom{N}{2}} \diff x_i\prod_{j=1}^N \diff y_j p_X(x_i) p_Y(y_j)   \int_{\calU_{\rm Haar}} \diff |z| \left[N + 2\sum_{i=1}^{\binom{N}{2}}\cos(x_i) + 2|z|\sum_{j=1}^N \cos(y_j) + |z|^2\right]^2\\
    &= \int  \prod_{i=1}^{\binom{N}{2}} \diff x_i\prod_{j=1}^N \diff y_j p_X(x_i) p_Y(y_j) \int_{\calU_{\rm Haar}} \diff |z| \left[N^2 + 4\sum_{i,i'=1}^{\binom{N}{2}} \cos(x_i)\cos(x_{i'}) + 4|z|^2 \sum_{j,j'=1}^N \cos(y_j)\cos(y_{j'}) + |z|^4 \right.\nonumber\\
    &\qquad \qquad \qquad \qquad \qquad \qquad \qquad \qquad \qquad \quad + 4N\sum_{i=1}^{\binom{N}{2}}\cos(x_i) + 4N|z|\sum_{j=1}^N \cos(y_j) + 2N|z|^2 + 8|z|\sum_{i,j}\cos(x_i)\cos(y_j)\nonumber\\
    &\qquad \qquad \qquad \qquad \qquad \qquad \qquad \qquad \qquad \quad \left. + 4|z|^2 \sum_i \cos(x_i) + 2|z|^3 \sum_j \cos(y_j)\right]\\
    &= N^2 + 2\binom{N}{2} + 2 N \calF_{\rm Haar}^{(1)} + \calF_{\rm Haar}^{(2)} + 2N \calF_{\rm Haar}^{(1)}\\
    &= 2N^2 + 3N + 2,
\end{align}
where we utilize $\mathbb{E}_{p_X}[\cos(x)] = \mathbb{E}_{p_Y}[\cos(y)] = 0$ and $\mathbb{E}_{p_X}[\cos(x_i)\cos(x_{i'})] = \mathbb{E}_{p_Y}[\cos(y_i)\cos(y_{i'})] = \delta_{i,i'}/2$. 

For $N \ge d-1$, the $k$th frame potential is reduced to
\begin{align}
    \calF^{(k)}_{\rm RH} &= \int_{\mathbb{U}[0, 2\pi)} \prod_{\alpha=1}^d \diff \phi_{\alpha}  \diff \phi'_{\alpha} \left\lvert \sum_{\alpha =1}^d e^{i(\phi'_\alpha - \phi_\alpha)} \right \rvert^{2k} \\
    &= \int_{\mathbb{U}[0, 2\pi)} \prod_{\alpha=1}^d \diff \phi_{\alpha}   \left[d + 2\sum_{\alpha < \beta}\cos(\phi'_\alpha - \phi_\alpha -\phi'_\beta + \phi_\beta )\right]^k  \\
    &= \int  \prod_{i=1}^{\binom{d}{2}} \diff x_i p_X(x_i) \left[d + 2\sum_{i=1}^{\binom{d}{2}}\cos(x_i)\right]^k \label{eq:framePot_l4}\\
    &\ge \sum_{k_1} \binom{k}{k_1} 2^{k_1} \int \diff x_1 p_X(x_1) \cos^{k_1} (x_1) d^{k-k_1}\\
    &= \sum_{k_1=\rm even}^k \binom{k}{k_1} 2^{k_1} d^{k - k_1} \frac{\Gamma(k_1/2+1/2)}{\sqrt{\pi} \Gamma(k_1/2 + 1)}\\
    &= \sum_{k_1=\rm even}^k \frac{k!}{((k_1/2)!)^2 (k- k_1)!} d^{k - k_1}. 
\end{align}
Here we see that the $k$-th order frame potential leads to a constant only depending on the system dimension $d=2^n$.
For $k=2$, we can also obtain the exact analytical result from Eq.~\ref{eq:framePot_l4} as
\begin{align}
    \calF^{(2)}_{\rm RH} &= \int  \prod_{i=1}^{\binom{d}{2}} \diff x_i p_X(x_i) \left[d + 2\sum_{i=1}^{\binom{d}{2}}\cos(x_i)\right]^2\\
    &= \int  \prod_{i=1}^{\binom{d}{2}} \diff x_i p_X(x_i) \left[d^2 + 4\sum_{i,i'=1}^{\binom{d}{2}} \cos(x_i)\cos(x_{i'}) + 4d \sum_{i=1}^{\binom{d}{2}}\cos(x_i)\right]\\
    &= 2d^2 -d.
\end{align}

\section{Additional numerical results}
\label{app:numeric_detail}

In the main text, \BZ{we} develop the coupled dynamical equations Eqs.~\eqref{eq:dyeqs_UT} \BZ{relying} on an assumption that the relative dQNTK $\lambda_{\gamma \alpha \beta}(t) = \mu_{\gamma \alpha \beta} (t)/\sqrt{K_{\gamma \gamma}(t) K_{\beta \beta}(t)}$ converges to a constant \BZ{at} late time, and provide numerical results based on a generalized norm. In the following, we show the additional numerical evidence to support it for each \BZ{type of} dynamics. From the definition of $\lambda_{\gamma \alpha \beta}$, we see that $\lambda_{\gamma \alpha \beta} = \lambda_{\beta \alpha \gamma}$, and thus in the following we only present the independent elements. In Fig.~\ref{fig:lda_frozen_kernel},~\ref{fig:lda_frozen_error} and~\ref{fig:lda_mix_frozen}, we show the convergence of $\lambda_{\gamma \alpha \beta}$ for {\it frozen-kernel dynamics}, {\it frozen-error dynamics} and {\it mixed-frozen dynamics} in the exponential convergence class. In Fig.~\ref{fig:lda_critical},~\ref{fig:lda_critical_frozen_kernel},~\ref{fig:lda_critical_frozen_error},~\ref{fig:lda_critical_mix_frozen}, we plot its convergence for {\it critical point}, {\it critical-frozen-kernel dynamics}, {\it critical-frozen-error dynamics} and {\it critical-mixed-frozen dynamics} in the polynomial convergence class. In both convergence classes of dynamics, we see that every element of the relative dQNTK $\lambda_{\gamma \alpha \beta}$ converges to a constant in late time of training.

\begin{figure}[h]
    \centering
    \includegraphics[width=0.6\textwidth]{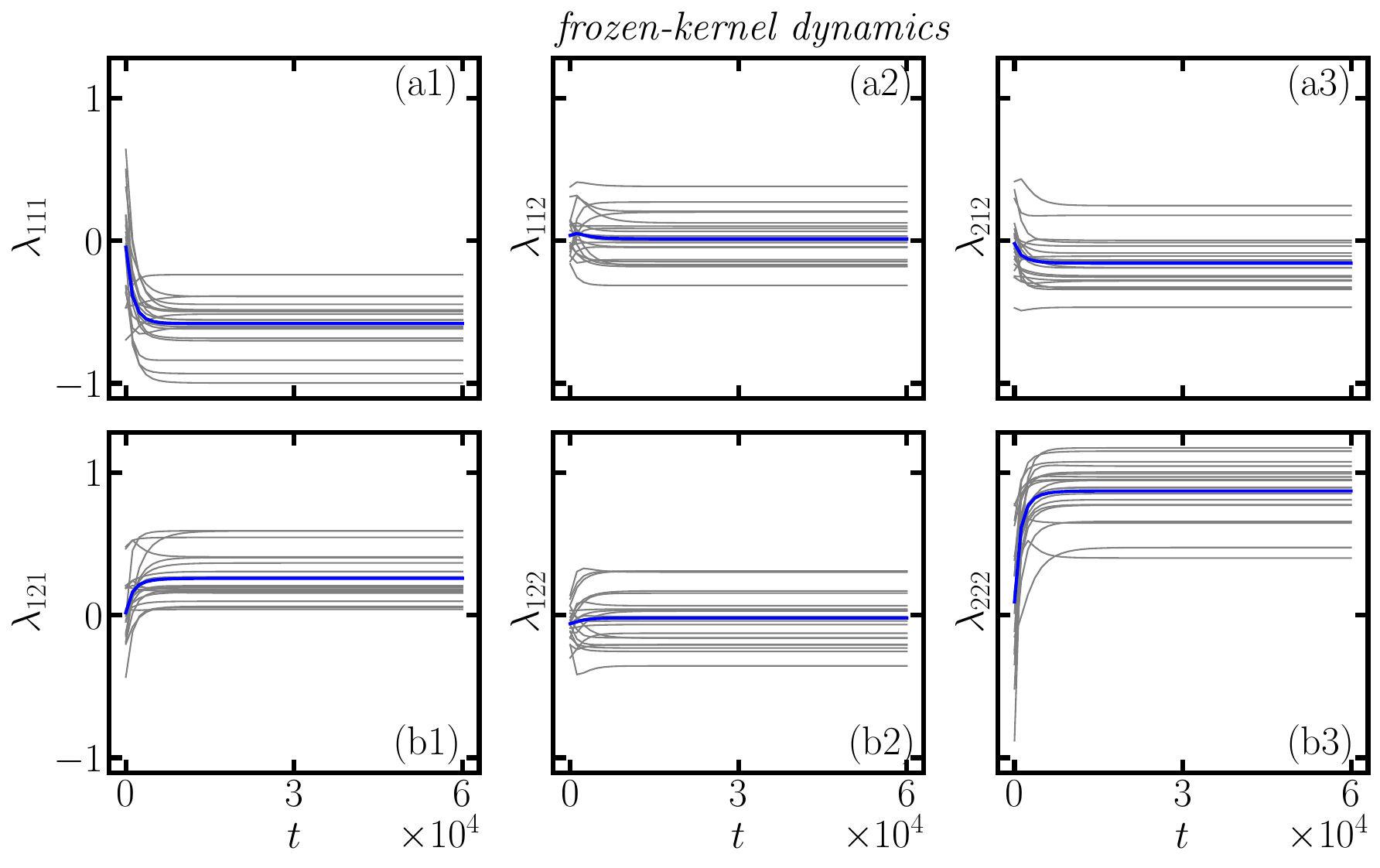}
    \caption{Dynamics of $\lambda_{\gamma \alpha \beta}$ for {\it frozen-kernel dynamics} in Fig.~\ref{fig:exp_class_N2} (a1)-(c1) of the main text. Grey lines represent $\lambda_{\gamma \alpha \beta}$ of each random sample, and the blue lines represent the corresponding average.}
    \label{fig:lda_frozen_kernel}
\end{figure}

\begin{figure}[h]
    \centering
    \includegraphics[width=0.6\textwidth]{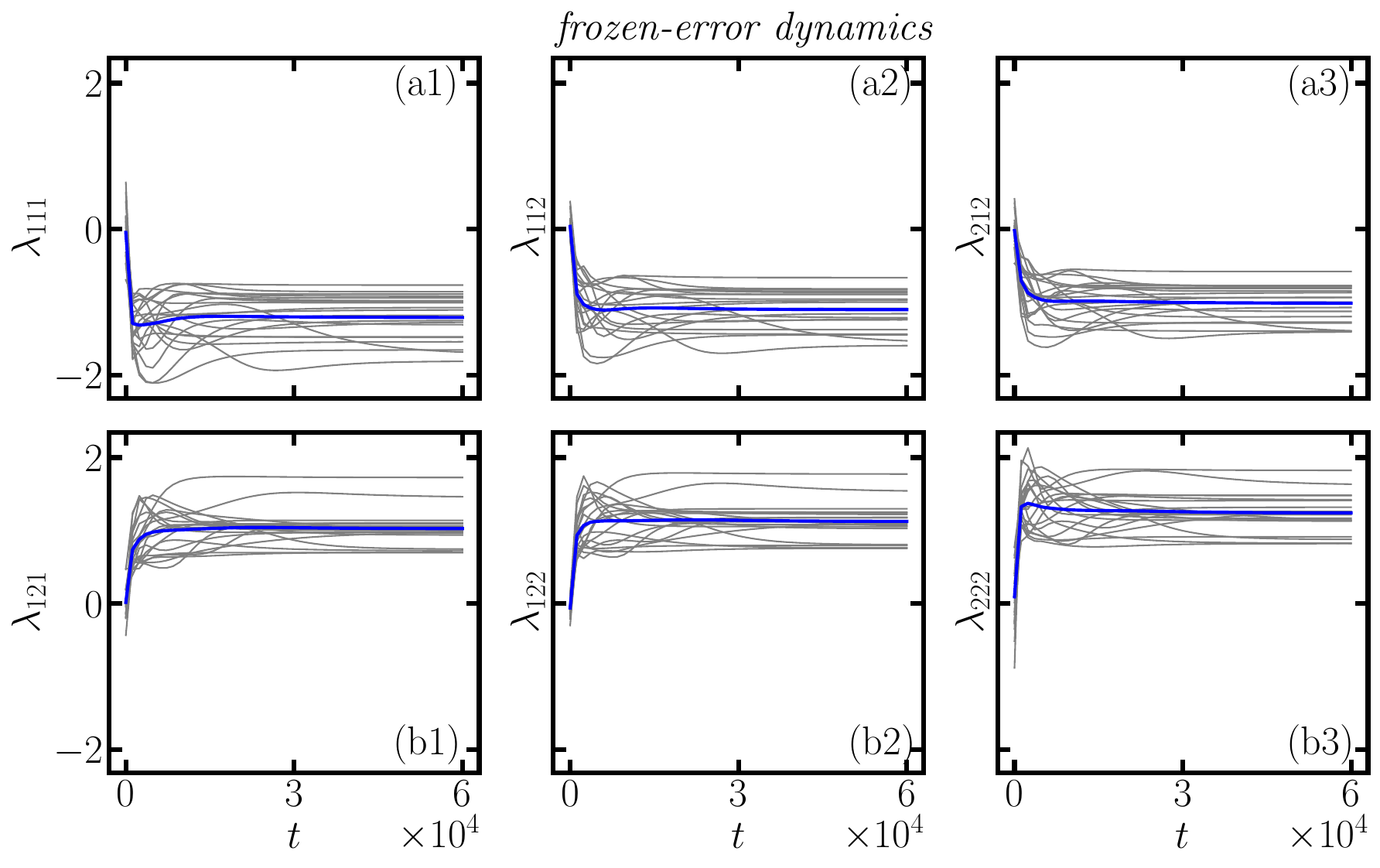}
    \caption{Dynamics of $\lambda_{\gamma \alpha \beta}$ for {\it frozen-error dynamics} in Fig.~\ref{fig:exp_class_N2} (a2)-(c2) of the main text. Grey lines represent $\lambda_{\gamma \alpha \beta}$ of each random sample, and the blue lines represent the corresponding average.}     \label{fig:lda_frozen_error}
\end{figure}

\begin{figure}[h]
    \centering
    \includegraphics[width=0.6\textwidth]{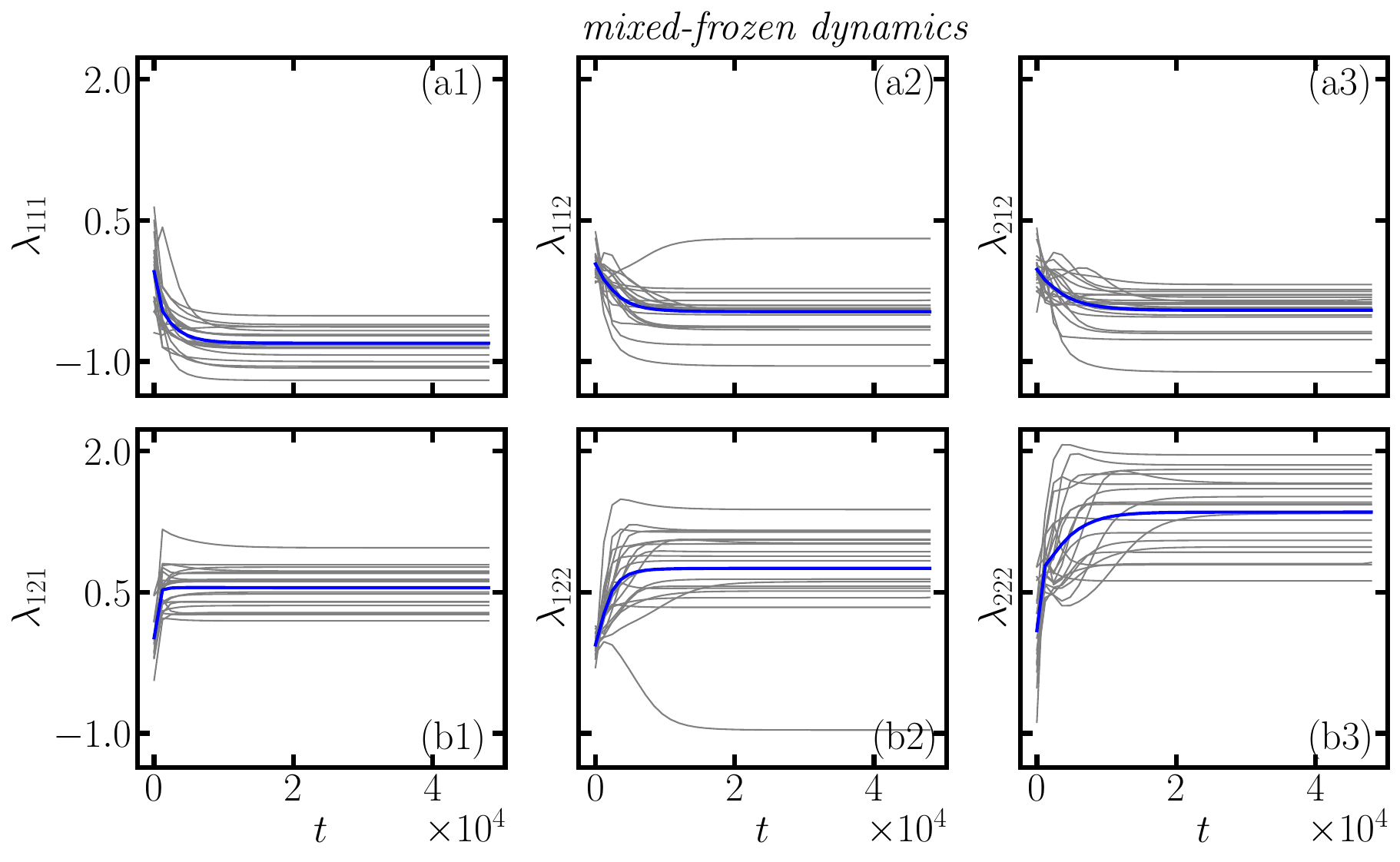}
    \caption{Dynamics of $\lambda_{\gamma \alpha \beta}$ for {\it mixed-frozen dynamics} in Fig.~\ref{fig:exp_class_N2} (a3)-(c3) of the main text. Grey lines represent $\lambda_{\gamma \alpha \beta}$ of each random sample, and the blue lines represent the corresponding average.}   \label{fig:lda_mix_frozen}
\end{figure}

\begin{figure}[h]
    \centering
    \includegraphics[width=0.6\textwidth]{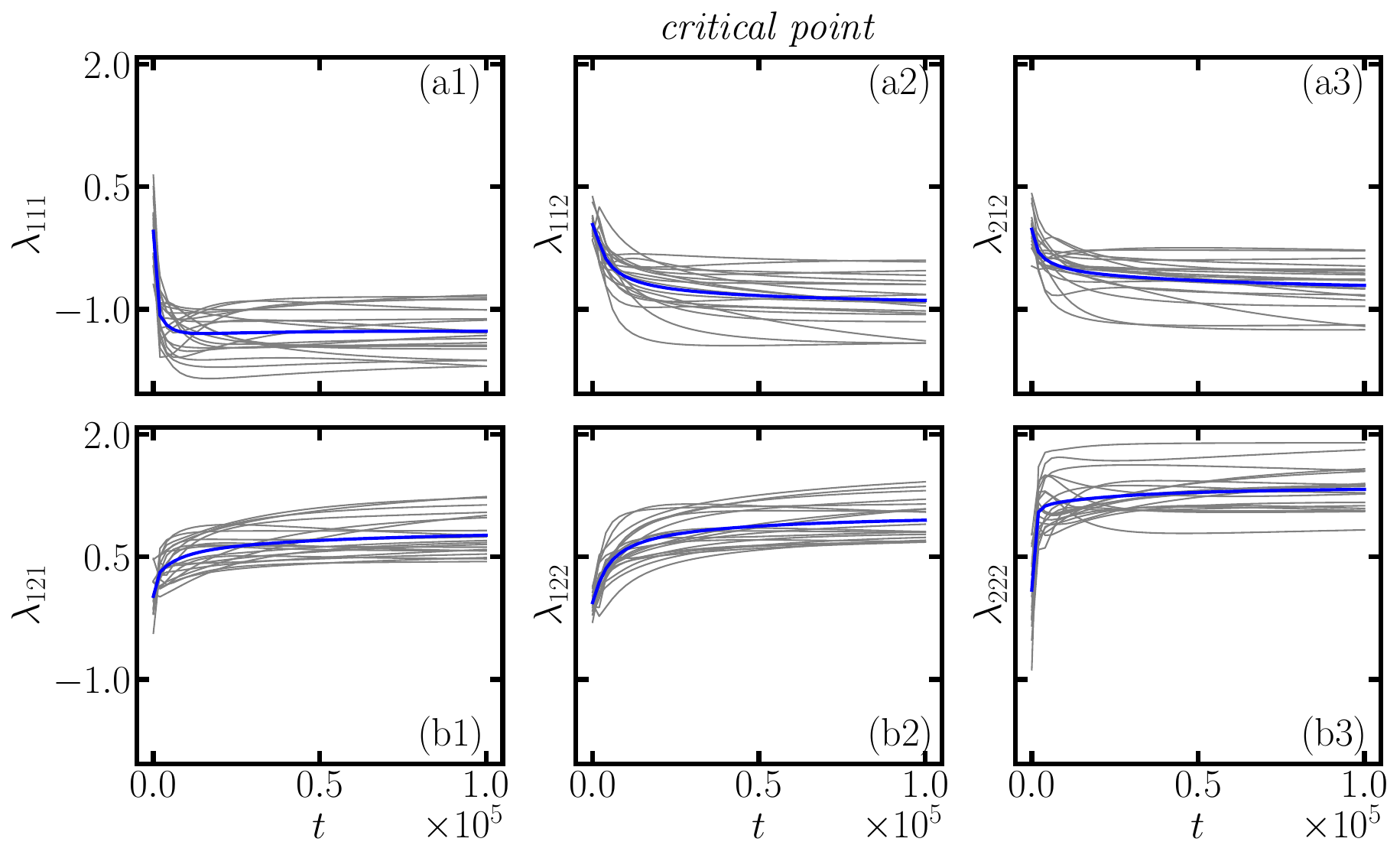}
    \caption{Dynamics of $\lambda_{\gamma \alpha \beta}$ for {\it critical point} in Fig.~\ref{fig:poly_class_N2} (a1)-(c1) of the main text. Grey lines represent $\lambda_{\gamma \alpha \beta}$ of each random sample, and the blue lines represent the corresponding average. }   \label{fig:lda_critical}
\end{figure}

\begin{figure}[h]
    \centering
    \includegraphics[width=0.6\textwidth]{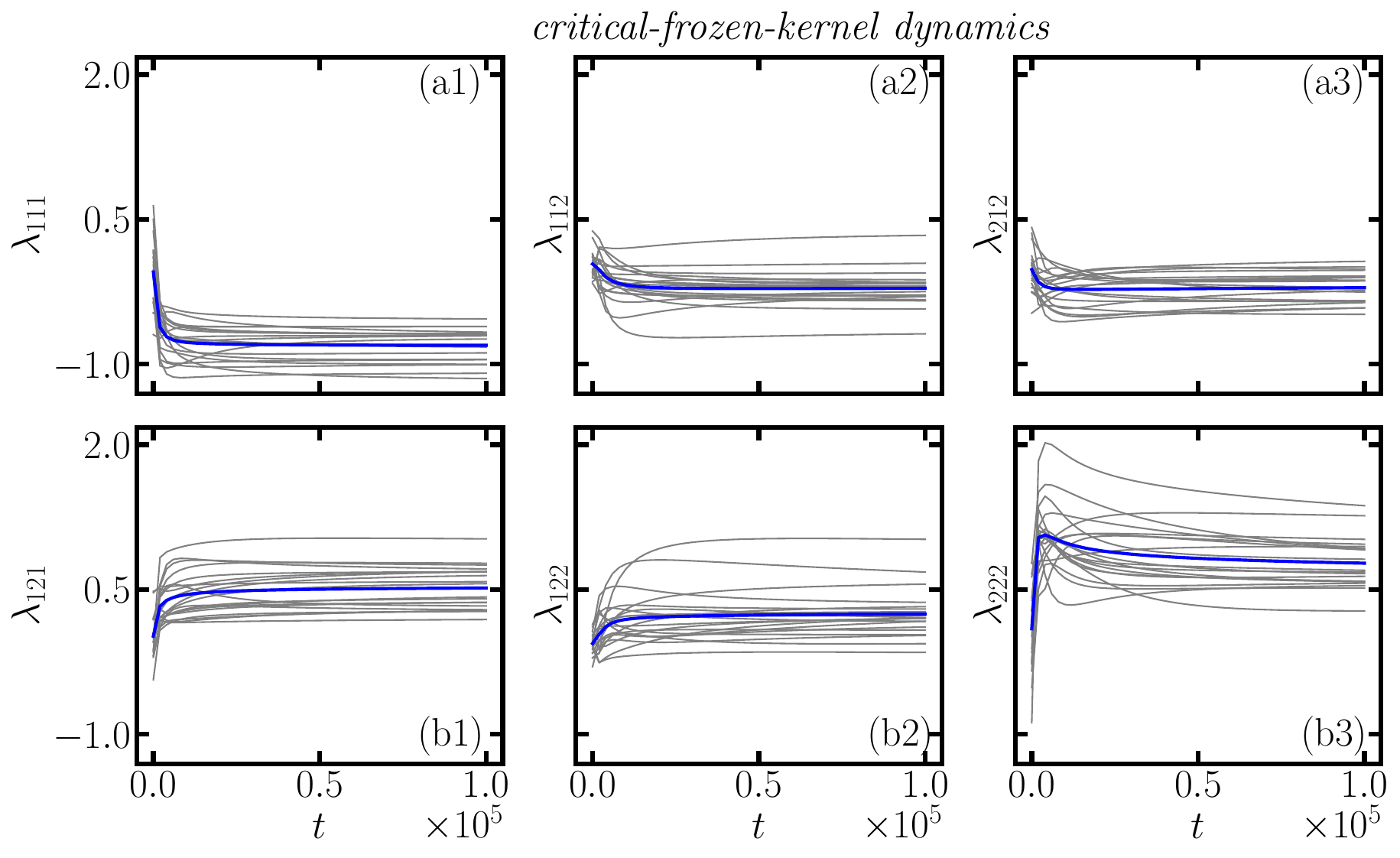}
    \caption{Dynamics of $\lambda_{\gamma \alpha \beta}$ for {\it critical-frozen-kernel dynamics} in Fig.~\ref{fig:poly_class_N2} (a2)-(c2) of the main text. Grey lines represent $\lambda_{\gamma \alpha \beta}$ of each random sample, and the blue lines represent the corresponding average.} \label{fig:lda_critical_frozen_kernel}
\end{figure}

\begin{figure}[h]
    \centering
    \includegraphics[width=0.6\textwidth]{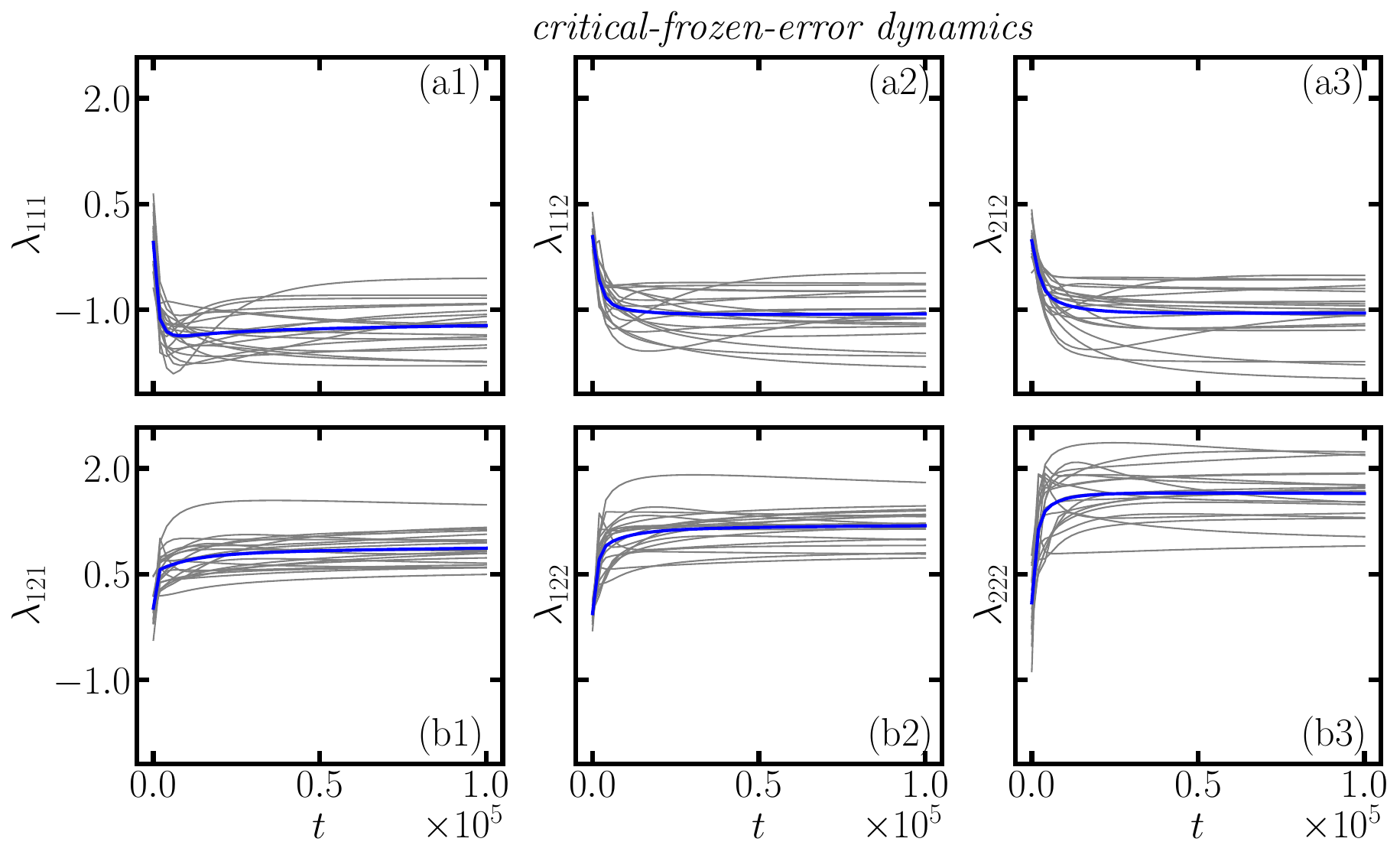}
    \caption{Dynamics of $\lambda_{\gamma \alpha \beta}$ for {\it critical-frozen-kernel dynamics} in Fig.~\ref{fig:poly_class_N2} (a3)-(c3) of the main text. Grey lines represent $\lambda_{\gamma \alpha \beta}$ of each random sample, and the blue lines represent the corresponding average. \label{fig:lda_critical_frozen_error}}
\end{figure}

\begin{figure}[h]
    \centering
    \includegraphics[width=0.9\textwidth]{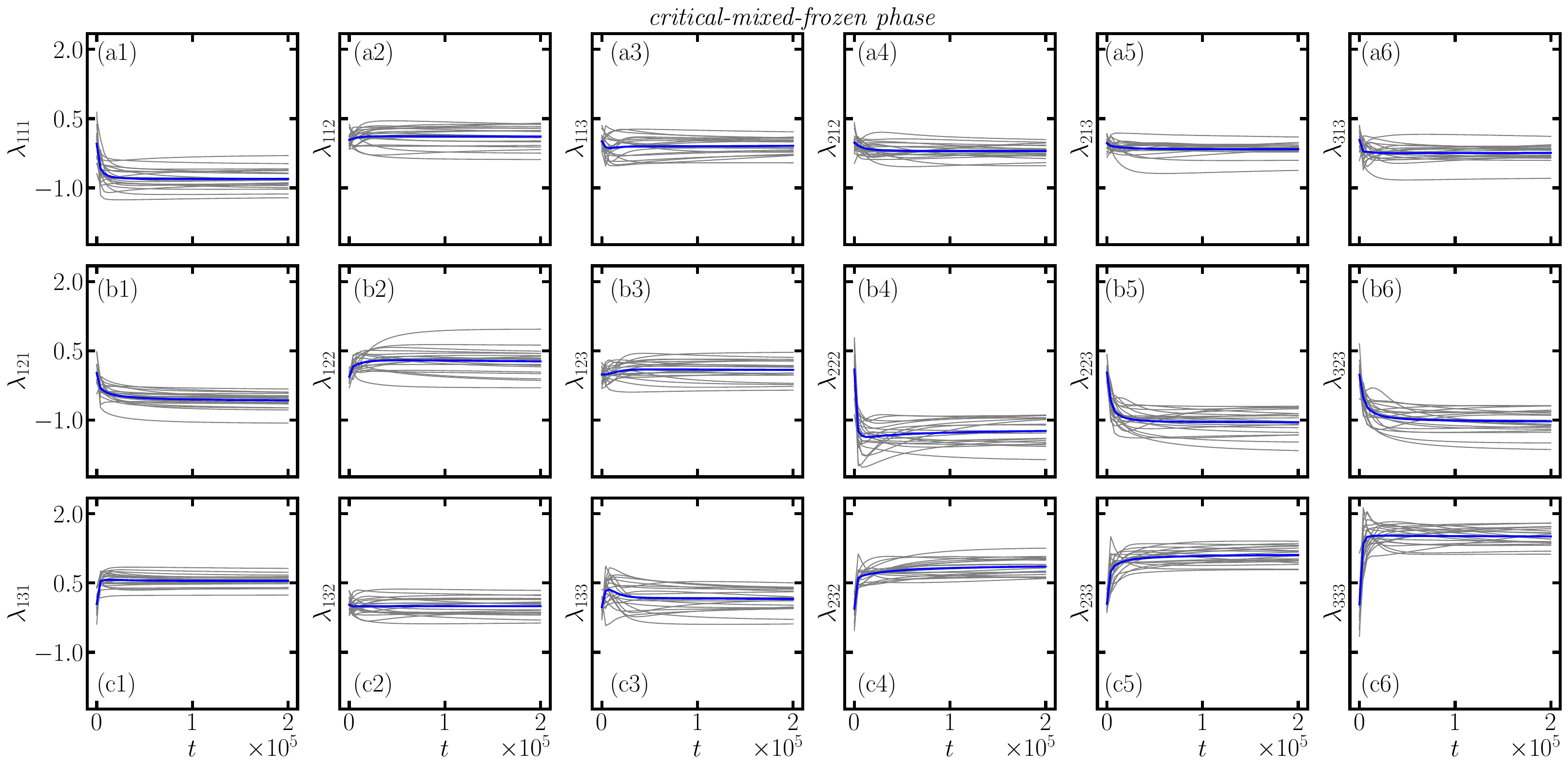}
    \caption{Dynamics of $\lambda_{\gamma \alpha \beta}$ for {\it critical-frozen-kernel dynamics} in Fig.~\ref{fig:crit_mix_fro} of the main text. Grey lines represent $\lambda_{\gamma \alpha \beta}$ of each random sample, and the blue lines represent the corresponding average. }
    \label{fig:lda_critical_mix_frozen}
\end{figure}

In Fig.~\ref{fig:B_exp} and Fig.~\ref{fig:B_poly}, we show the convergence of geometric quantity $\angle_{\alpha \beta}(t)$ towards a constant for dynamics in exponential and polynomial convergence class, which supports Lemma.~\ref{lemma_B} in the main text. Indeed, the converged constant \BZ{for each} dynamics \BZ{lies} within the range $[-1, 1]$, indicating the geometric interpretation discussed in the main text.

\begin{figure}[t]
    \centering
    \includegraphics[width=0.6\textwidth]{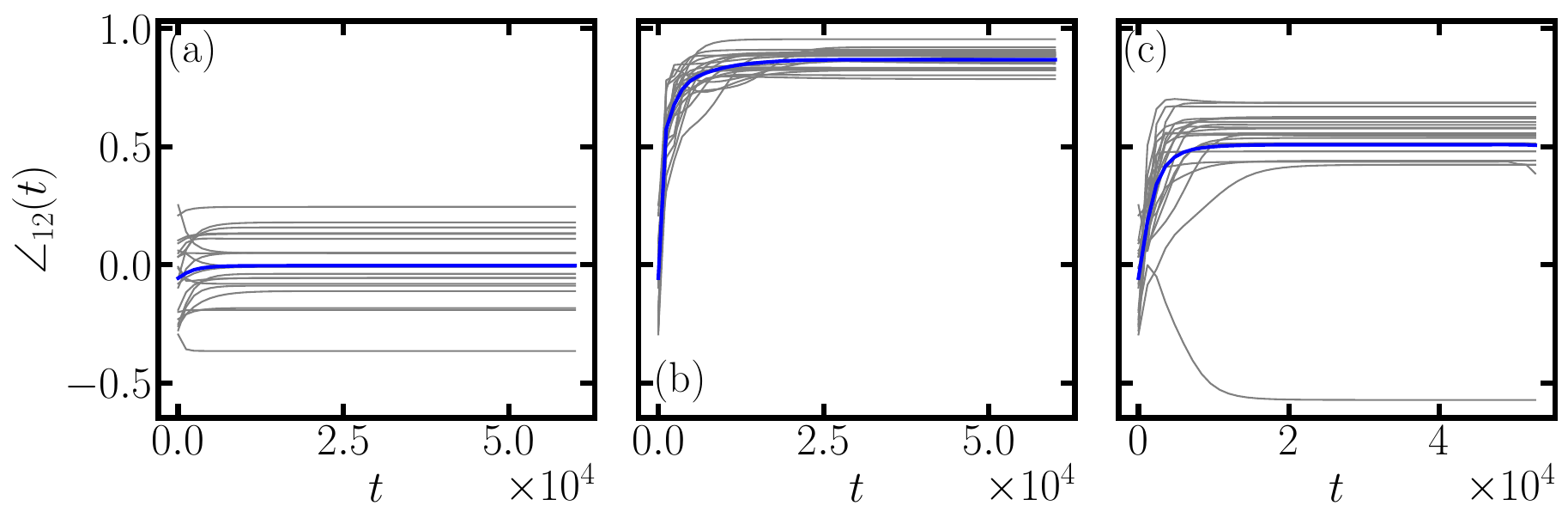}
    \caption{Dynamics of $\angle_{12}(t)$ for exponential convergence class. From left to right we show $\angle_{12}(t)$ for {\it frozen-kernel dynamics}, {\it frozen-error dynamics} and {\it mixed-frozen dynamics}. Grey lines represent $\angle_{12}$ of each random sample, and the blue lines represent the corresponding average. The settings follow Fig.~\ref{fig:exp_class_N2} of the main text.}
    \label{fig:B_exp}
\end{figure}

\begin{figure}[t]
    \centering
    \includegraphics[width=0.6\textwidth]{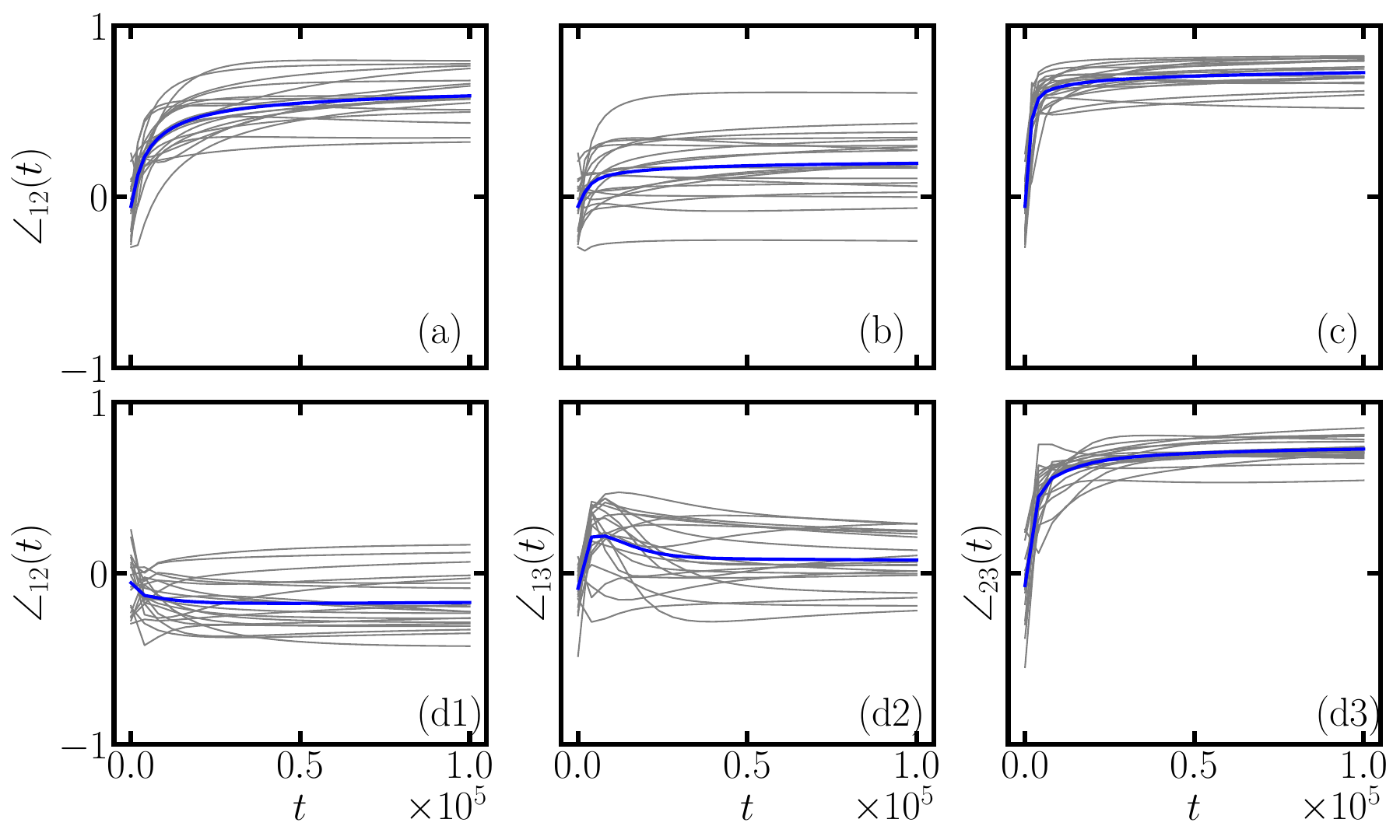}
    \caption{Dynamics of $\angle_{\alpha \beta}(t)$ for polynomial convergence class. 
    We show $\angle_{12}(t)$ for (a) {\it critical point}, (b) {\it critical-frozen-kernel dynamics} and (c) {\it critical-frozen-error dynamics}. We plot $\angle_{12}(t), \angle_{13}(t), \angle_{23}(t)$ in (d1)-(d3) for {\it critical-mixed-frozen dynamics}.
    Grey lines represent $\angle_{12}$ of each random sample, and the blue lines represent the corresponding average. The settings of top and bottom panels follow Fig.~\ref{fig:poly_class_N2} and Fig.~\ref{fig:crit_mix_fro} of the main text separately.}
    \label{fig:B_poly}
\end{figure}

% \newpage 

\section{Additional calculations on ensemble average results}
\label{app:ensemble_average_calculation}

In this section, we present calculations for ensemble average of QNTK and dQNTK. As they are defined in terms of first and second-order derivatives, we first show the expression for gradients.
From parameter-shift rule, the \BZ{derivative} of $\epsilon_\alpha = \braket{\psi_\alpha|U^\dagger O_\alpha U|\psi_\alpha}$ with $O_\alpha = \ketbra{\Phi_\alpha}{\Phi_\alpha}$ is 
\begin{align}
    \frac{\partial \epsilon_\alpha}{\partial \theta_\ell} = \frac{i}{2}\braket{\psi_\alpha|U_{\ell^-}^\dagger \left[X_\ell, O_{\alpha;\ell^+}\right]U_{\ell^-}|\psi_\alpha},
    \label{eq:supp_J1}
\end{align}
where we define the notation
\begin{align}
    U_{\ell^-} = \prod_{k=1}^{\ell-1} W_k V_k (\theta_k), U_{\ell^+} = \prod_{k=\ell}^L W_k V_k (\theta_k),
    \label{eq:U_ell_def}
\end{align}
and $O_{\alpha;\ell^+} = U_{\ell^+}^\dagger O_{\alpha}U_{\ell^+}$. Thus the unitary for whole circuit becomes $U = U_{\ell^+} U_{\ell^-}$. \BZ{One can show that Eq.~\eqref{eq:supp_J1} is equivalent to the parameter-shift rule~\cite{mitarai2018quantum, crooks2019gradients}, and the expression here is also utilized in previous related works~\cite{zhang2023dynamical, liu2023analytic}.} 

The second order gradient assuming $\ell_1<\ell_2$ and $\ell_1 = \ell_2 = \ell$ can be written in a similar way as
\begin{align}
    \frac{\partial^2 \epsilon_\alpha}{\partial \theta_{\ell_1}\partial\theta_{\ell_2}}  &= -\frac{1}{4}\braket{\psi_\alpha|U_{\ell_1^-}^\dagger[X_{\ell_1},U_{\ell_1 \shortto \ell_2}^\dagger[X_{\ell_2}, U_{\ell_2^+}^\dagger O_\alpha U_{\ell_2^+}]U_{\ell_1\shortto \ell_2}]U_{\ell_1^-}|\psi_\alpha} = -\frac{1}{4}\braket{\psi_\alpha|U_{\ell_1^-}^\dagger[X_{\ell_1},U_{\ell_1 \shortto \ell_2}^\dagger[X_{\ell_2}, O_{\alpha; \ell_2^+}] U_{\ell_1\shortto \ell_2}]U_{\ell_1^-}|\psi_\alpha}\\
    \frac{\partial^2 \epsilon_\alpha}{\partial\theta_\ell^2} &= -\frac{1}{4}\braket{\psi_\alpha|U_{\ell^-}^\dagger [X_\ell, [X_\ell, O_{\alpha; \ell^+}]]U_{\ell^-}|\psi_\alpha},
\end{align}
where
\begin{align}
    U_{\ell_1\shortto \ell_2} = \prod_{k = \ell_1}^{\ell_2 - 1} W_k V_k (\theta_k).
    \label{eq:U_1to2_def}
\end{align}

The ensemble average over Haar random unitaries are performed via symbolic calculation tools \texttt{RTNI}~\cite{fukuda2019rtni}.

\subsection{Average QNTK under restricted Haar ensemble}

For the QNTK $K_{\alpha \beta} = \sum_\ell \frac{\partial \epsilon_\alpha}{\partial \theta_\ell} \frac{\partial \epsilon_\beta}{\partial \theta_\ell}$, the restricted Haar ensemble average of product of derivatives become
\begin{align}
    &\mathbb{E}_{\calU_{\rm RH}}\left[\frac{\partial\epsilon_\alpha}{\partial \theta_\ell} \frac{\partial\epsilon_\beta}{\partial \theta_\ell}\right]
    = -\frac{1}{4}\int \diff U_{\ell^-}\diff U_{\ell^+} \tr\left(P_{\beta \alpha} U_{\ell^-}^\dagger\left[X_\ell, O_{\alpha;\ell^+}\right] U_{\ell^-} P_{\alpha \beta} U_{\ell^-}^\dagger \left[X_\ell, O_{\beta;\ell^+}\right]U_{\ell^-} \right)\\
    &= -\frac{1}{4} \int_{\calU_{\rm RH}}\diff U\int_{\calU_{\rm Haar}} \diff U_{\ell^-} \left[\tr\left(P_{\beta \alpha} U_{\ell^-}^\dagger X_\ell U_{\ell^-} O_{\alpha; U} P_{\alpha \beta} U_{\ell^-}^\dagger X_\ell U_{\ell^-} O_{\beta;U}\right) + \tr\left( P_{\beta \alpha} O_{\alpha;U} U_{\ell^-}^\dagger X_\ell U_{\ell^-}  P_{\alpha \beta} O_{\beta; U} U_{\ell^-}^\dagger X_\ell U_{\ell^-} \right)\right.\nonumber\\
    &\left. \qquad  \qquad \qquad \qquad \qquad \quad - \tr\left(P_{\beta \alpha} U_{\ell^-}^\dagger X_\ell U_{\ell^-} O_{\alpha;U} P_{\alpha \beta} O_{\beta;U} U_{\ell^-}^\dagger X_\ell U_{\ell^-} \right) - \tr\left(P_{\beta \alpha} O_{\alpha;U} U_{\ell^-}^\dagger X_\ell U_{\ell^-} P_{\alpha \beta} U_{\ell^-}^\dagger X_\ell U_{\ell^-} O_{\beta;U} \right)\right]\\
    &= -\frac{1}{4}\int_{\calU_{\rm RH}} \diff U \left[\frac{d\tr(O_{\alpha;U}P_{\alpha \beta})\tr(O_{\beta;U}P_{\beta \alpha}) - \tr(P_{\alpha \beta} O_{\beta;U} P_{\beta \alpha} O_{\alpha;U})}{d^2 - 1} + \frac{d\tr(P_{\alpha \beta} O_{\beta;U})\tr(P_{\beta \alpha} O_{\alpha;U}) - \tr(P_{\beta \alpha} O_{\alpha;U} P_{\alpha \beta} O_{\beta;U})}{d^2 - 1}\right.\nonumber\\
    & \left. \qquad \qquad\quad 
    - \frac{d\tr(P_{\beta \alpha})\tr(O_{\alpha;U} P_{\alpha \beta} O_{\beta; U})  - \tr(P_{\beta \alpha}O_{\alpha;U} P_{\alpha \beta}O_{\beta;U})}{d^2 - 1} - \frac{d\tr(P_{\alpha \beta})\tr(O_{\beta;U} P_{\beta\alpha } O_{\alpha; U})  - \tr(O_{\beta;U} P_{\beta \alpha} O_{\alpha;U} P_{\alpha \beta})}{d^2 - 1}\right]\\
    &= -\frac{d}{4} \int_{\calU_{\rm RH}} \diff U \frac{\tr(O_{\alpha;U}P_{\alpha \beta})\tr(O_{\beta;U}P_{\beta \alpha}) + \tr(P_{\alpha \beta} O_{\beta;U})\tr(P_{\beta \alpha} O_{\alpha;U}) - \tr(P_{\beta \alpha})\tr(O_{\alpha;U} P_{\alpha \beta} O_{\beta; U}) - \tr(P_{\alpha \beta})\tr(O_{\beta;U} P_{\beta\alpha } O_{\alpha; U})}{d^2 - 1}\\
    &= -\frac{d}{4} \int_{\calU_{\rm RH}} \diff U \frac{\tr(O_{\alpha;U}P_{\alpha \beta})\tr(O_{\beta;U}P_{\beta \alpha}) + \tr(P_{\alpha \beta} O_{\beta;U})\tr(P_{\beta \alpha} O_{\alpha;U}) - \braket{\psi_\alpha|\psi_\beta} \tr(O_{\alpha;U} P_{\alpha \beta} O_{\beta; U}) - \braket{\psi_\beta|\psi_\alpha} \tr(O_{\beta;U} P_{\beta\alpha } O_{\alpha; U})}{d^2 - 1}\\
    &= -\frac{d}{4(d^2-1)}\mathbb{E}_{\calU_{\rm RH}}\left[ T_{\alpha \beta}^* T_{\alpha \alpha} T_{\beta\alpha}^* T_{\beta \beta} + T_{\beta \beta}^* T_{\beta \alpha} T_{\alpha \alpha}^* T_{\alpha \beta} - \braket{\psi_\alpha|\psi_\beta} \braket{\Phi_\beta|\Phi_{\alpha}} T_{\alpha \alpha} T_{\beta \beta}^* - \braket{\psi_\beta|\psi_\alpha} \braket{\Phi_{\alpha}|\Phi_{\beta}}T_{\beta \beta} T_{\alpha \alpha}^*\right]\\
    &= -\frac{d}{4(d^2-1)} \mathbb{E}_{\calU_{\rm RH}} \left[T_{\alpha \beta}^* T_{\alpha \alpha} T_{\beta\alpha}^* T_{\beta \beta} - \delta_{\alpha \beta} T_{\alpha \alpha} T_{\beta \beta}^* + c.c.\right],
    \label{eq:gagb_rh}
\end{align}
\BZ{where $P_{\alpha \beta} = \ketbra{\alpha}{\beta}$ is an operator introduced for the convenience of unitary integral calculation utilizing \texttt{RTNI}} and $T_{\alpha \beta} \equiv \braket{\Phi_{\alpha}|U|\psi_\beta}$. Here $c.c.$ stands for complex conjugate.

For $\alpha = \beta$, we have
\begin{align}
    &\mathbb{E}_{\calU_{\rm RH}}\left[\frac{\partial\epsilon_\alpha}{\partial \theta_\ell} \frac{\partial\epsilon_\alpha}{\partial \theta_\ell}\right] =-\frac{d}{2(d^2-1)} \mathbb{E}_{\calU_{\rm RH}} \left[|T_{\alpha \alpha}|^4 - |T_{\alpha \alpha}|^2\right] = \frac{d}{2(d^2 - 1)}o_\alpha (1-o_\alpha),
\end{align}
where we utilize $|T_{\alpha\alpha}|^2 = |\braket{\Phi_\alpha|U|\psi_\alpha}|^2 = o_\alpha$. On the other hand, for $\alpha \neq \beta$, it becomes
\begin{align}
    \mathbb{E}_{\calU_{\rm RH}}\left[\frac{\partial\epsilon_\alpha}{\partial \theta_\ell} \frac{\partial\epsilon_\beta}{\partial \theta_\ell}\right] &= -\frac{d}{4(d^2-1)} \mathbb{E}_{\calU_{\rm RH}}\left[T_{\alpha \beta}^* T_{\alpha \alpha} T_{\beta\alpha}^* T_{\beta \beta} + c.c.\right]\\
    & = -\frac{d}{4(d^2 - 1)}\mathbb{E}_{\calU_{\rm RH}}\left[|T_{\alpha \beta}|e^{-i\phi_\beta} |T_{\alpha \alpha}| e^{i\phi_\alpha} |T_{\beta \alpha}|e^{-i\phi_\alpha} |T_{\beta \beta}|e^{i\phi_\beta} + c.c.\right]\\
    &= -\frac{d}{2(d^2 - 1)}\mathbb{E}_{\calU_{\rm RH}}\left[|T_{\alpha \beta}||T_{\alpha \alpha}|  |T_{\beta \alpha}| |T_{\beta \beta}|\right],
    % &\simeq -\frac{d}{2(d^2 - 1)(d - 1)} \sqrt{o_\alpha o_\beta (1-o_\alpha) (1-o_\beta)},
\end{align}
where in the second line, we utilize the definition of restricted Haar ensemble in Eq.~\eqref{eq:rh_def}. We see that the off-diagonal terms require extra information. 
% and in the last line, we utilize the mean-field approach to approximate $\mathbb{E}_{\calU_{\rm RH}}[|T_{\alpha \beta}|] \simeq \sqrt{(1-o_\beta)/(d-1)}$.

The average QNTK under restricted Haar ensemble becomes
\begin{align}
    \overline{K_{\alpha \alpha}(\infty)} &= L \mathbb{E}_{\calU_{\rm RH}}\left[\frac{\partial\epsilon_\alpha}{\partial \theta_\ell} \frac{\partial\epsilon_\alpha}{\partial \theta_\ell}\right] = \frac{Ld}{2(d^2 - 1)}o_\alpha (1-o_\alpha) \simeq \frac{L}{2d}o_\alpha(1-o_\alpha), \label{eq:K_diag_rh}\\
    \overline{K_{\alpha \beta}(\infty)} &= L \mathbb{E}_{\calU_{\rm RH}}\left[\frac{\partial\epsilon_\alpha}{\partial \theta_\ell} \frac{\partial\epsilon_\beta}{\partial \theta_\ell}\right] 
    % \simeq -\frac{Ld}{2(d^2-1)(d-1)} \sqrt{o_\alpha o_\beta (1-o_\alpha) (1-o_\beta)}
    % &\simeq -\frac{L}{2d^2} \sqrt{o_\alpha o_\beta (1-o_\alpha) (1-o_\beta)}, 
    = -\frac{L d}{2(d^2 - 1)}\mathbb{E}_{\calU_{\rm RH}}\left[|T_{\alpha \beta}||T_{\alpha \alpha}|  |T_{\beta \alpha}| |T_{\beta \beta}|\right] \simeq -\frac{L}{2d} \mathbb{E}_{\calU_{\rm RH}}\left[|T_{\alpha \beta}||T_{\alpha \alpha}|  |T_{\beta \alpha}| |T_{\beta \beta}|\right],
    \label{eq:K_offdiag_rh}
\end{align}
where we approximate them with $d \gg 1$ at the end. 
% A direct result can be concluded is that
% \begin{align}
%     \overline{\angle_{\alpha \beta}(\bm \theta)} \equiv \frac{\overline{K_{\alpha \beta} (\infty)}}{\sqrt{\overline{K_{\alpha \alpha}(\infty)} \: \overline{K_{\beta \beta}(\infty)}}} \simeq \frac{-L \sqrt{o_\alpha o_\beta (1-o_\alpha) (1-o_\beta)}/(2d^2)}{L\sqrt{o_\alpha(1-o_\alpha) o_\beta (1-o_\beta)}/(2d)} = -\frac{1}{d}.
% \end{align}

\subsection{Average relative dQNTK under restricted Haar ensemble}

Ahead of presenting the calculation details of relative QNTK, we summarize the results here.
\begin{align}
    \overline{\lambda_{\alpha \alpha \alpha}(\infty)} &= \frac{\overline{\mu_{\alpha \alpha \alpha}(\infty)}}{\overline{K_{\alpha \alpha}(\infty)}}  \simeq -\frac{1}{4d}\left[2(d o_\alpha - 2) + L(2o_\alpha - 1)\right].
    % , \label{eq:lda_diag_rh}\\
    % \overline{\lambda_{\alpha \alpha \beta}(\infty)} &= \frac{\overline{\mu_{\alpha \alpha \beta}(\infty)}}{\sqrt{\overline{K_{\alpha \alpha}(\infty)}\: \overline{K_{\beta \beta}(\infty)}}} \simeq \frac{1}{4d(d-1)}\left[\frac{2(d+2)}{(d+3)}\left((d+2) o_\alpha - 2\right) + \frac{(L-1)d^2 (2o_\alpha - 1)}{d^2 - 1}\right] \nonumber
    % \\
    % &\simeq \frac{1}{4d^2}\left[2(d o_\alpha - 2) + L(2o_\alpha - 1)\right]\label{eq:lda_offdiag_rh},
\end{align}

In this section, we evaluate the relative dQNTK $\overline{\lambda_{\gamma \alpha \beta}(\infty)} = \overline{\mu_{\gamma \alpha \beta}(\infty)}/\sqrt{\overline{K_{\alpha \alpha}(\infty) K_{\beta \beta}(\infty)}}$. We first calculate $\overline{\mu_{\gamma \alpha \beta}(\infty)}$. Recall that $\mu_{\gamma \alpha \beta} = \sum_{\ell, \ell'} \frac{\partial \epsilon_\gamma}{\partial \theta_{\ell}} \frac{\partial^2 \epsilon_\alpha}{\partial \theta_{\ell} \partial \theta_{\ell'}} \frac{\partial \epsilon_\beta}{\partial \theta_{\ell'}} = \sum_{\ell} \frac{\partial \epsilon_\gamma}{\partial \theta_{\ell}} \frac{\partial^2 \epsilon_\alpha}{\partial \theta_{\ell}^2} \frac{\partial \epsilon_\beta}{\partial \theta_\ell} + \sum_{\ell \neq \ell'} \frac{\partial \epsilon_\gamma}{\partial \theta_{\ell}} \frac{\partial^2 \epsilon_\alpha}{\partial \theta_{\ell} \partial \theta_{\ell'}} \frac{\partial \epsilon_\beta}{\partial \theta_{\ell'}}$, we then calculate the ensemble average of the two terms separately. As only $\lambda_{\alpha \alpha \beta}$ is utilized in the dynamical equations (see Eq.~\eqref{eq:eqs_full}), then we only consider ensemble average of $\mu_{\alpha \alpha \beta}$ in the following.

\subsubsection{$\mathbb{E}_{\calU_{\rm RH}}\left[\frac{\partial \epsilon_\alpha}{\partial \theta_\ell} \frac{\partial^2 \epsilon_\alpha}{\partial\theta_\ell^2}\frac{\partial \epsilon_\beta}{\partial \theta_\ell}\right]$ under restricted Haar ensemble \\}

We can expand it following the parameter-shift rule as
\small
\begin{align}
    & \mathbb{E}_{\calU_{\rm RH}}\left[\frac{\partial \epsilon_\alpha}{\partial \theta_\ell} \frac{\partial^2 \epsilon_\alpha}{\partial\theta_\ell^2} \frac{\partial \epsilon_\beta}{\partial \theta_\ell}\right] = \frac{1}{16} \int \diff U_{\ell^-}\diff U_{\ell^+} \tr( P_{\alpha \alpha} U_{\ell^-}^\dagger [X_\ell, [X_\ell, O_{\alpha; \ell^+}]] U_{\ell^-} P_{\alpha \beta} U_{\ell^-}^\dagger \left[X_\ell, O_{\beta;\ell^+}\right]U_{\ell^-} P_{\beta \alpha} U_{\ell^-}^\dagger \left[X_\ell, O_{\alpha;\ell^+}\right]U_{\ell^-}) \\
    &= \frac{2}{16}\int_{\calU_{\rm RH}} \diff U \int_{\calU_{\rm Haar}} \diff U_{\ell^-} \left[\tr( P_{\alpha \alpha} O_{\alpha; U} P_{\alpha \beta} U_{\ell^-}^\dagger X_\ell U_{\ell^-} O_{\beta;U}  P_{\beta \alpha} U_{\ell^-}^\dagger X_\ell  U_{\ell^-} O_{\alpha;U}) +  \tr( P_{\alpha \alpha} O_{\alpha; U} P_{\alpha \beta} O_{\beta; U} U_{\ell^-}^\dagger X_\ell U_{\ell^-} P_{\beta \alpha}  O_{\alpha;U} U_{\ell^-}^\dagger X_\ell U_{\ell^-}) \right.\nonumber\\
    & \qquad \qquad \qquad  \qquad \qquad -  \tr( P_{\alpha \alpha} O_{\alpha; U} P_{\alpha \beta} U_{\ell^-}^\dagger X_\ell U_{\ell^-} O_{\beta; U} P_{\beta \alpha}  O_{\alpha; U} U_{\ell^-}^\dagger X_\ell U_{\ell^-}) - \tr( P_{\alpha \alpha}  O_{\alpha; U} P_{\alpha \beta}   O_{\beta;U} U_{\ell^-}^\dagger X_\ell U_{\ell^-} P_{\beta \alpha} U_{\ell^-}^\dagger X_\ell U_{\ell^-} O_{\alpha;U} ) \nonumber\\
    & \qquad \qquad \qquad  \qquad \qquad + \tr( P_{\alpha \alpha} U_{\ell^-}^\dagger X_\ell U_{\ell^-} O_{\alpha; U} U_{\ell^-}^\dagger X_\ell U_{\ell^-} P_{\alpha \beta} U_{\ell^-}^\dagger X_\ell U_{\ell^-} O_{\beta; U}  P_{\beta \alpha} O_{\alpha; U} U_{\ell^-}^\dagger X_\ell U_{\ell^-}) \nonumber\\
    & \qquad \qquad \qquad  \qquad \qquad +  \tr( P_{\alpha \alpha} U_{\ell^-}^\dagger X_\ell U_{\ell^-} O_{\alpha; U} U_{\ell^-}^\dagger X_\ell U_{\ell^-} P_{\alpha \beta}  O_{\beta; U} U_{\ell^-}^\dagger X_\ell U_{\ell^-} P_{\beta \alpha} U_{\ell^-}^\dagger X_\ell U_{\ell^-} O_{\alpha; U} )\nonumber\\
    & \qquad \qquad \qquad  \qquad \qquad - \tr( P_{\alpha \alpha} U_{\ell^-}^\dagger X_\ell U_{\ell^-} O_{\alpha; U} U_{\ell^-}^\dagger X_\ell U_{\ell^-} P_{\alpha \beta} U_{\ell^-}^\dagger X_\ell U_{\ell^-} O_{\beta; U}  P_{\beta \alpha} U_{\ell^-}^\dagger X_\ell U_{\ell^-} O_{\alpha; U} )\nonumber\\
    & \qquad \qquad \qquad  \qquad \qquad \left. - \tr( P_{\alpha \alpha} U_{\ell^-}^\dagger X_\ell U_{\ell^-} O_{\alpha; U} U_{\ell^-}^\dagger X_\ell U_{\ell^-} P_{\alpha \beta}  O_{\beta; U} U_{\ell^-}^\dagger X_\ell U_{\ell^-} P_{\beta \alpha}  O_{\alpha; U}U_{\ell^-}^\dagger X_\ell U_{\ell^-} ) \right].
    \label{eq:ghg_l1}
\end{align}
\normalsize
The first term is
\begin{align}
     I_1 & \equiv \int_{\calU_{\rm RH}} \diff U \int_{\calU_{\rm Haar}} \diff U_{\ell^-} \tr( P_{\alpha \alpha} O_{\alpha; U} P_{\alpha \beta} U_{\ell^-}^\dagger X_\ell U_{\ell^-} O_{\beta;U}  P_{\beta \alpha} U_{\ell^-}^\dagger X_\ell  U_{\ell^-} O_{\alpha;U})\nonumber\\
     &= \int_{\calU_{\rm RH}} \diff U \frac{d \tr(O_{\beta;U} P_{\beta \alpha}) \tr(O_{\alpha;U}P_{\alpha \alpha} O_{\alpha;U} P_{\alpha \beta}) - \tr(O_{\alpha;U} P_{\alpha \alpha} O_{\alpha;U} P_{\alpha \beta} O_{\beta; U} P_{\beta \alpha})}{d^2 - 1} \\
     &= \frac{1}{d^2-1}\mathbb{E}_{\calU_{\rm RH}}\left[dT_{\beta \beta} T_{\beta \alpha}^*  T_{\alpha \alpha} T_{\alpha \alpha}^* T_{\alpha \alpha} T_{\alpha \beta}^* - T_{\alpha \alpha} T_{\alpha \alpha}^* T_{\alpha \alpha} T_{\beta \beta}^* T_{\beta \beta} T_{\alpha \alpha}^*\right]\\
     &= \frac{1}{d^2-1}\mathbb{E}_{\calU_{\rm RH}}\left[dT_{\beta \beta} T_{\beta \alpha}^*  T_{\alpha \alpha} |T_{\alpha \alpha}|^2 T_{\alpha \beta}^* - |T_{\alpha \alpha}|^4 |T_{\beta \beta}|^2\right].
\end{align}
The second term is
\begin{align}
    I_2 &\equiv \int_{\calU_{\rm RH}} \diff U \int_{\calU_{\rm Haar}} \diff U_{\ell^-} \tr( P_{\alpha \alpha} O_{\alpha; U} P_{\alpha \beta} O_{\beta; U} U_{\ell^-}^\dagger X_\ell U_{\ell^-} P_{\beta \alpha}  O_{\alpha;U} U_{\ell^-}^\dagger X_\ell U_{\ell^-}) \nonumber\\
    &= \int_{\calU_{\rm RH}} \diff U \frac{d \tr(P_{\beta \alpha} O_{\alpha; U}) \tr(P_{\alpha \alpha} O_{\alpha; U} P_{\alpha \beta} O_{\beta; U}) - \tr(P_{\alpha \alpha} O_{\alpha;U} P_{\alpha \beta} O_{\beta; U} P_{\beta \alpha} O_{\alpha; U})}{d^2 - 1}\\
    &= \frac{1}{d^2-1} \mathbb{E}_{\calU_{\rm RH}} \left[d T_{\alpha \beta} T_{\alpha \alpha}^* |T_{\alpha \alpha}|^2 T_{\beta \beta}^* T_{\beta \alpha} - |T_{\alpha \alpha}|^4 |T_{\beta \beta}|^2\right] = I_1^*.
\end{align}
The third term is
\begin{align}
     I_3 &\equiv \int_{\calU_{\rm RH}} \diff U \int_{\calU_{\rm Haar}} \diff U_{\ell^-} \tr( P_{\alpha \alpha} O_{\alpha; U} P_{\alpha \beta} U_{\ell^-}^\dagger X_\ell U_{\ell^-} O_{\beta; U} P_{\beta \alpha}  O_{\alpha; U} U_{\ell^-}^\dagger X_\ell U_{\ell^-}) \nonumber\\
     &= \int_{\calU_{\rm RH}} \diff U \frac{d \tr(O_{\beta;U} P_{\beta \alpha} O_{\alpha; U})\tr(P_{\alpha \alpha} O_{\alpha; U}P_{\alpha \beta}) - \tr(P_{\alpha \alpha}O_{\alpha; U} P_{\alpha \beta} O_{\beta; U} P_{\beta \alpha} O_{\alpha; U})}{d^2-1}\\
     &= \frac{1}{d^2-1} \mathbb{E}_{\calU_{\rm RH}} \left[ d T_{\beta \beta} T_{\alpha \alpha}^* \braket{\Phi_{\alpha}|\Phi_\beta} T_{\alpha \alpha}^* T_{\alpha \alpha} \braket{\psi_\beta| \psi_\alpha} -  |T_{\alpha \alpha}|^2 |T_{\alpha \alpha}|^2 |T_{\beta \beta}|^2
    \right]\\
    &= \frac{1}{d^2-1} \mathbb{E}_{\calU_{\rm RH}} \left[ d \delta_{\alpha \beta} T_{\beta \beta} T_{\alpha \alpha}^*  |T_{\alpha \alpha}|^2  -  |T_{\alpha \alpha}|^4 |T_{\beta \beta}|^2
    \right].
\end{align}
The forth term is
\begin{align}
    I_4 &\equiv \int_{\calU_{\rm RH}} \diff U \int_{\calU_{\rm Haar}} \diff U_{\ell^-} \tr( P_{\alpha \alpha}  O_{\alpha; U} P_{\alpha \beta}   O_{\beta;U} U_{\ell^-}^\dagger X_\ell U_{\ell^-} P_{\beta \alpha} U_{\ell^-}^\dagger X_\ell U_{\ell^-} O_{\alpha;U})\nonumber\\
    &=  \int_{\calU_{\rm RH}} \diff U \frac{d\tr(P_{\beta \alpha}) \tr(O_{\alpha;U}P_{\alpha \alpha} O_{\alpha;U}P_{\alpha \beta}O_{\beta; U}) - \tr(O_{\alpha;U} P_{\alpha \alpha} O_{\alpha;U} P_{\alpha \beta} O_{\beta;U} P_{\beta \alpha})}{d^2 - 1}\\
    &=  \frac{1}{d^2-1} \mathbb{E}_{\calU_{\rm RH}} \left[d \braket{\psi_\alpha|\psi_\beta} T_{\alpha \alpha} T_{\alpha \alpha}^* T_{\alpha \alpha} T_{\beta \beta}^* \braket{\Phi_\beta|\Phi_{\alpha}} - |T_{\alpha \alpha}|^2 |T_{\alpha \alpha}|^2 |T_{\beta \beta}|^2\right]\\
    &= \frac{1}{d^2-1} \mathbb{E}_{\calU_{\rm RH}} \left[d \delta_{\alpha \beta} T_{\alpha \alpha} |T_{\alpha \alpha}|^2 T_{\beta \beta}^*  - |T_{\alpha \alpha}|^4 |T_{\beta \beta}|^2 \right] = I_3^*.
\end{align}
The fifth term is
\begin{align}
    I_5 &\equiv \int_{\calU_{\rm RH}} \diff U \int_{\calU_{\rm Haar}} \diff U_{\ell^-} \tr( P_{\alpha \alpha} U_{\ell^-}^\dagger X_\ell U_{\ell^-} O_{\alpha; U} U_{\ell^-}^\dagger X_\ell U_{\ell^-} P_{\alpha \beta} U_{\ell^-}^\dagger X_\ell U_{\ell^-} O_{\beta; U}  P_{\beta \alpha} O_{\alpha; U} U_{\ell^-}^\dagger X_\ell U_{\ell^-}) \nonumber\\
    &= \delta_{\alpha \beta} \mathbb{E}_{\calU_{\rm RH}}\left[\frac{2 (d+2) T_{\beta \beta} T_{\alpha \alpha}^* - 2T_{\beta \beta} | T_{\alpha \alpha}| ^2 \left( T_{\alpha \beta}^*  +  T_{\beta \alpha}^* + T_{\alpha \alpha}^*\right)}{d^3+3 d^2-d-3}\right]\nonumber\\
    &\quad + \mathbb{E}_{\calU_{\rm RH}}\left[\frac{(d+2) T_{\alpha \alpha} T_{\beta \beta} | T_{\alpha \alpha}| ^2 T_{\alpha \beta}^* T_{\beta \alpha}^* - | T_{\alpha \alpha}| ^4 | T_{\beta \beta}| ^2 - 2 | T_{\alpha \alpha}| ^2 | T_{\beta \beta}| ^2}{d^3+3 d^2-d-3}\right].
\end{align}
The sixth term is
\begin{align}
    I_6 &\equiv \int_{\calU_{\rm RH}} \diff U \int_{\calU_{\rm Haar}} \diff U_{\ell^-} \tr( P_{\alpha \alpha} U_{\ell^-}^\dagger X_\ell U_{\ell^-} O_{\alpha; U} U_{\ell^-}^\dagger X_\ell U_{\ell^-} P_{\alpha \beta}  O_{\beta; U} U_{\ell^-}^\dagger X_\ell U_{\ell^-} P_{\beta \alpha} U_{\ell^-}^\dagger X_\ell U_{\ell^-} O_{\alpha; U} )\nonumber\\
    &= \delta_{\alpha \beta} \mathbb{E}_{\calU_{\rm RH}}\left[\frac{ 2 (d+2) T_{\alpha \alpha} T_{\beta \beta}^* -2 | T_{\alpha \alpha}| ^2 T_{\beta \beta}^* (T_{\alpha \alpha}+T_{\alpha \beta}+T_{\beta \alpha})}{d^3+3 d^2-d-3}\right] \nonumber\\
    &\quad + \mathbb{E}_{\calU_{\rm RH}}\left[\frac{(d+2) T_{\alpha \beta} T_{\beta \alpha} | T_{\alpha \alpha}| ^2 T_{\alpha \alpha}^* T_{\beta \beta}^* - | T_{\alpha \alpha}| ^4 | T_{\beta \beta}| ^2 - 2 | T_{\alpha \alpha}| ^2 | T_{\beta \beta}| ^2}{d^3+3 d^2-d-3} \right] = I_5^*.
\end{align}
The seventh term is
\begin{align}
    I_7 &\equiv \int_{\calU_{\rm RH}} \diff U \int_{\calU_{\rm Haar}} \diff U_{\ell^-} \tr( P_{\alpha \alpha} U_{\ell^-}^\dagger X_\ell U_{\ell^-} O_{\alpha; U} U_{\ell^-}^\dagger X_\ell U_{\ell^-} P_{\alpha \beta} U_{\ell^-}^\dagger X_\ell U_{\ell^-} O_{\beta; U}  P_{\beta \alpha} U_{\ell^-}^\dagger X_\ell U_{\ell^-} O_{\alpha; U})\nonumber\\
    &= \delta_{\alpha \beta} \mathbb{E}_{\calU_{\rm RH}}\left[\frac{T_{\beta \beta} | T_{\alpha \alpha}| ^2 \left( (d+2) T_{\alpha \alpha}^*-2  T_{\alpha \beta}^* - 2  T_{\beta \alpha }^*\right)}{d^3+3 d^2-d-3}\right] \nonumber\\
    &\quad + \mathbb{E}_{\calU_{\rm RH}}\left[\frac{2 (d+2) T_{\alpha \alpha} T_{\beta \beta} T_{\alpha \beta}^* T_{\beta \alpha }^*-2 T_{\alpha \alpha} T_{\beta \beta} | T_{\alpha \alpha}| ^2 T_{\alpha \beta}^* T_{\beta \alpha }^* - | T_{\alpha \alpha}| ^4 | T_{\beta \beta}| ^2-2 | T_{\alpha \alpha}| ^2 | T_{\beta \beta}| ^2}{d^3+3 d^2-d-3}\right].
\end{align}
The eighth (last) term is
\begin{align}
    I_8 &\equiv \int_{\calU_{\rm RH}} \diff U \int_{\calU_{\rm Haar}} \diff U_{\ell^-} \tr( P_{\alpha \alpha} U_{\ell^-}^\dagger X_\ell U_{\ell^-} O_{\alpha; U} U_{\ell^-}^\dagger X_\ell U_{\ell^-} P_{\alpha \beta}  O_{\beta; U} U_{\ell^-}^\dagger X_\ell U_{\ell^-} P_{\beta \alpha}  O_{\alpha; U}U_{\ell^-}^\dagger X_\ell U_{\ell^-} ) \nonumber\\
    &= \delta_{\alpha \beta} \mathbb{E}_{\calU_{\rm RH}}\left[\frac{T_{\beta \beta}^* | T_{\alpha \alpha}| ^2 \left((d+2) T_{\alpha \alpha} -2 T_{\alpha \beta} - 2T_{\beta \alpha}\right)}{d^3 + 3d^2 - d - 3}\right] \nonumber\\
    &\quad + \mathbb{E}_{\calU_{\rm RH}}\left[\frac{2 (d+2) T_{\alpha \beta} T_{\beta \alpha} T_{\alpha \alpha}^* T_{\beta \beta}^* - 2 T_{\alpha \beta} T_{\beta \alpha} | T_{\alpha \alpha}| ^2 T_{\alpha \alpha}^* T_{\beta \beta}^* - | T_{\alpha \alpha}| ^4 |T_{\beta \beta}|^2 - 2 | T_{\alpha \alpha}| ^2 |T_{\beta \beta}|^2}{d^3 +3d^2 - d - 3}\right] = I_7^*.
\end{align}

Then we have
\small
\begin{align}
    &\mathbb{E}_{\calU_{\rm RH}}\left[\frac{\partial \epsilon_\alpha}{\partial \theta_\ell} \frac{\partial^2 \epsilon_\alpha}{\partial\theta_\ell^2} \frac{\partial \epsilon_\beta}{\partial \theta_\ell}\right] = \frac{2}{16} \left(I_1 + I_2 - I_3 - I_4 + I_5 + I_6 - I_7 - I_8 \right)\nonumber\\
    &= \frac{1}{8} \left(I_1 - I_3 + I_5 - I_7 + c.c.\right)\\
    &= \frac{1}{8} \left(\frac{1}{d^2-1}\mathbb{E}_{\calU_{\rm RH}}\left[dT_{\beta \beta} T_{\beta \alpha}^*  T_{\alpha \alpha} |T_{\alpha \alpha}|^2 T_{\alpha \beta}^* - |T_{\alpha \alpha}|^4 |T_{\beta \beta}|^2\right] - \frac{1}{d^2-1} \mathbb{E}_{\calU_{\rm RH}} \left[ d \delta_{\alpha \beta} T_{\beta \beta} T_{\alpha \alpha}^*  |T_{\alpha \alpha}|^2  -  |T_{\alpha \alpha}|^4 |T_{\beta \beta}|^2
    \right] \right.\nonumber\\
    & + \delta_{\alpha \beta} \mathbb{E}_{\calU_{\rm RH}}\left[\frac{2 (d+2) T_{\beta \beta} T_{\alpha \alpha}^* - 2T_{\beta \beta} | T_{\alpha \alpha}| ^2 \left( T_{\alpha \beta}^*  +  T_{\beta \alpha}^* + T_{\alpha \alpha}^*\right)}{d^3+3 d^2-d-3}\right] + \mathbb{E}_{\calU_{\rm RH}}\left[\frac{(d+2) T_{\alpha \alpha} T_{\beta \beta} | T_{\alpha \alpha}| ^2 T_{\alpha \beta}^* T_{\beta \alpha}^* - | T_{\alpha \alpha}| ^2 | T_{\beta \beta}| ^2 \left(|T_{\alpha \alpha}|^2 + 2 \right)}{d^3+3 d^2-d-3}\right]\nonumber\\
    &\left. - \delta_{\alpha \beta} \mathbb{E}_{\calU_{\rm RH}}\left[\frac{T_{\beta \beta} | T_{\alpha \alpha}| ^2 \left( (d+2) T_{\alpha \alpha}^*-2  T_{\alpha \beta}^* - 2  T_{\beta \alpha }^*\right)}{d^3+3 d^2-d-3}\right]  - \mathbb{E}_{\calU_{\rm RH}}\left[\frac{2T_{\alpha \alpha} T_{\beta \beta}  T_{\alpha \beta}^* T_{\beta \alpha }^* \left( d+2 - | T_{\alpha \alpha}| ^2\right) - | T_{\alpha \alpha}| ^2 | T_{\beta \beta}| ^2 \left(| T_{\alpha \alpha}| ^2 +2\right)}{d^3+3 d^2-d-3}\right] + c.c.\right)\\
    &= \mathbb{E}_{\calU_{\rm RH}}\left[ \frac{1}{8(d^2 - 1)} \left(\frac{(d+2)^2}{d+3} |T_{\alpha \alpha}|^2 - \frac{2(d+2)}{d+3}\right) \left( T_{\alpha \alpha} T_{\beta \beta} T_{\alpha\beta}^* T_{\beta \alpha}^* - \delta_{\alpha \beta} T_{\alpha \alpha} T_{\beta \beta}^* \right) + c.c.\right].
    \label{eq:ghg_l2}
\end{align}
\normalsize
For $\alpha = \beta$, it is reduced to
\begin{align}
    \mathbb{E}_{\calU_{\rm RH}}\left[\frac{\partial \epsilon_\alpha}{\partial \theta_\ell} \frac{\partial^2 \epsilon_\alpha}{\partial\theta_\ell^2} \frac{\partial \epsilon_\alpha}{\partial \theta_\ell}\right] 
    &=\mathbb{E}_{\calU_{\rm RH}}\left[ \frac{1}{4(d^2 - 1)} \left(\frac{(d+2)^2}{d+3} |T_{\alpha \alpha}|^2 - \frac{2(d+2)}{d+3}\right) \left( |T_{\alpha \alpha}|^4  - |T_{\alpha \alpha}|^2  \right)\right]\\
    &= \frac{(d+2)(o_\alpha-1) o_\alpha ((d+2) o_\alpha-2)}{4(d^2-1)(d+3)},
\end{align}
where we denote $o_\alpha = \epsilon_\alpha(\infty) + y_\alpha$ for simplicity. On the other hand, for $\alpha \neq \beta$, it is reduced to
\begin{align}
    \mathbb{E}_{\calU_{\rm RH}}\left[\frac{\partial \epsilon_\alpha}{\partial \theta_\ell} \frac{\partial^2 \epsilon_\beta}{\partial\theta_\ell^2} \frac{\partial \epsilon_\alpha}{\partial \theta_\ell}\right] &= \mathbb{E}_{\calU_{\rm RH}}\left[ \frac{1}{8(d^2 - 1)} \left(\frac{(d+2)^2}{d+3} |T_{\alpha \alpha}|^2 - \frac{2(d+2)}{d+3}\right) T_{\alpha \alpha} T_{\beta \beta} T_{\alpha\beta}^* T_{\beta \alpha}^* + c.c.\right]\\
    &=  \mathbb{E}_{\calU_{\rm RH}}\left[ \frac{1}{4(d^2 - 1)} \left(\frac{(d+2)^2}{d+3} |T_{\alpha \alpha}|^2 - \frac{2(d+2)}{d+3}\right) |T_{\alpha \alpha}| |T_{\beta \beta}| |T_{\alpha\beta}| T_{\beta \alpha}|\right].
    % &= \frac{(d+2) ((d+2)o_\alpha - 2)}{4(d^2 - 1)(d+3)(d-1)} \sqrt{o_\alpha o_\beta (1-o_\beta) (1-o_\alpha) }.
\end{align}

\subsubsection{$\mathbb{E}_{\calU_{\rm RH}}[\frac{\partial \epsilon_\alpha}{\partial \theta_{\ell_1}} \frac{\partial^2 \epsilon_\alpha}{\partial \theta_{\ell_1} \partial\theta_{\ell_2}}  \frac{\partial \epsilon_\beta}{\partial \theta_{\ell_2}}]$ under restricted Haar ensemble}

The other part $\sum_{l_1\neq l_2}\mathbb{E}_{\calU_{\rm RH}}[\frac{\partial \epsilon_\alpha}{\partial \theta_{\ell_1}} \frac{\partial^2 \epsilon_\alpha}{\partial \theta_{\ell_1} \partial\theta_{\ell_2}}  \frac{\partial \epsilon_\beta}{\partial \theta_{\ell_2}}] = \sum_{\ell_1 < \ell_2}  \mathbb{E}_{\calU_{\rm RH}}\left[\frac{\partial^2 \epsilon_\alpha}{\partial \theta_{\ell_1} \partial\theta_{\ell_2}}  \left(\frac{\partial \epsilon_\alpha}{\partial \theta_{\ell_1}} \frac{\partial \epsilon_\beta}{\partial \theta_{\ell_2}} + \frac{\partial \epsilon_\alpha}{\partial \theta_{\ell_2}} \frac{\partial \epsilon_\beta}{\partial \theta_{\ell_1}} \right) \right]$,
and specifically for $\alpha = \beta$, it can be simplified to $2\sum_{{\ell_1} < {\ell_2}}\mathbb{E}_{\calU_{\rm RH}}[\frac{\partial^2 \epsilon_\alpha}{\partial \theta_{\ell_1} \partial\theta_{\ell_2}}  \frac{\partial \epsilon_\alpha}{\partial \theta_{\ell_1}} 
 \frac{\partial \epsilon_\alpha}{\partial \theta_{\ell_2}}]$. $\mathbb{E}_{\calU_{\rm RH}}\left[\frac{\partial^2 \epsilon_\alpha}{\partial \theta_{\ell_1} \partial\theta_{\ell_2}} \left(\frac{\partial \epsilon_\alpha}{\partial \theta_{\ell_1}} \frac{\partial \epsilon_\beta}{\partial \theta_{\ell_2}} + \frac{\partial \epsilon_\alpha}{\partial \theta_{\ell_2}} \frac{\partial \epsilon_\beta}{\partial \theta_{\ell_1}} \right) \right]$ becomes
\begin{align}
     &\mathbb{E}_{\calU_{\rm RH}}\left[\frac{\partial^2 \epsilon_\alpha}{\partial \theta_{\ell_1} \partial\theta_{\ell_2}} \left(\frac{\partial \epsilon_\alpha}{\partial \theta_{\ell_1}} \frac{\partial \epsilon_\beta}{\partial \theta_{\ell_2}} + \frac{\partial \epsilon_\alpha}{\partial \theta_{\ell_2}} \frac{\partial \epsilon_\beta}{\partial \theta_{\ell_1}} \right)\right] \nonumber \\
     &= \frac{1}{16}\int \diff U_{\ell_1^-} \diff U_{\ell_1 \shortto \ell_2} \diff U_{\ell_2^+} \tr(P_{\beta \alpha} U_{\ell_1^-}^\dagger \left[X_{\ell_1}, U_{\ell_1 \shortto \ell_2}^\dagger \left[X_{\ell_2}, O_{\alpha; \ell_2^+}\right] U_{\ell_1 \shortto \ell_2}\right] U_{\ell_1^-}P_{\alpha \alpha} U_{\ell_1^-}^\dagger \left[X_{\ell_1}, O_{\alpha; \ell_1^+}\right]U_{\ell_1^-} P_{\alpha \beta} U_{\ell_2^-}^\dagger \left[X_{\ell_2}, O_{\beta; \ell_2^+}\right]U_{\ell_2^-}) \nonumber\\
     & + \frac{1}{16}\int \diff U_{\ell_1^-} \diff U_{\ell_1 \shortto \ell_2} \diff U_{\ell_2^+} \tr(P_{\beta \alpha} U_{\ell_1^-}^\dagger \left[X_{\ell_1}, U_{\ell_1 \shortto \ell_2}^\dagger \left[X_{\ell_2}, O_{\alpha; \ell_2^+}\right] U_{\ell_1 \shortto \ell_2}\right] U_{\ell_1^-}P_{\alpha \alpha} U_{\ell_2^-}^\dagger \left[X_{\ell_2}, O_{\alpha; \ell_2^+}\right]U_{\ell_2^-} P_{\alpha \beta} U_{\ell_1^-}^\dagger \left[X_{\ell_1}, O_{\beta; \ell_1^+}\right]U_{\ell_1^-}).
\end{align}
We calculate the above two separately.
The first one becomes
\begin{align}
    &\mathbb{E}_{\calU_{\rm RH}}\left[\frac{\partial^2 \epsilon_\alpha}{\partial \theta_{\ell_1} \partial\theta_{\ell_2}} \frac{\partial \epsilon_\alpha}{\partial \theta_{\ell_1}} \frac{\partial \epsilon_\beta}{\partial \theta_{\ell_2}} \right] \nonumber\\
    &=\frac{1}{16}\int \diff U_{\ell_1^-} \diff U_{\ell_1 \shortto \ell_2} \diff U_{\ell_2^+} \tr(P_{\beta \alpha} U_{\ell_1^-}^\dagger \left[X_{\ell_1}, U_{\ell_1 \shortto \ell_2}^\dagger \left[X_{\ell_2}, O_{\alpha; \ell_2^+}\right] U_{\ell_1 \shortto \ell_2}\right] U_{\ell_1^-}P_{\alpha \alpha} U_{\ell_1^-}^\dagger \left[X_{\ell_1}, O_{\alpha; \ell_1^+}\right]U_{\ell_1^-} P_{\alpha \beta} U_{\ell_2^-}^\dagger \left[X_{\ell_2}, O_{\beta; \ell_2^+}\right]U_{\ell_2^-}) \nonumber\\
    &= \frac{1}{16} \int_{\calU_{\rm RH}} \diff U_{\rm RH} \int_{\calU_{\rm Haar}} \diff U_{\ell_1^-} \diff U_{\ell_1 \shortto \ell_2} \left[ \tr(P_{\beta \alpha} U_{\ell_1^-}^\dagger X_{\ell_1} X_{\ell_2, \ell_1 \shortto \ell_2} U_{\ell_1^-} O_{\alpha;U} P_{\alpha \alpha} U_{\ell_1^-}^\dagger X_{\ell_1} U_{\ell_1^-} O_{\alpha;U} P_{\alpha \beta} U_{\ell_1^-}^\dagger X_{\ell_2, \ell_1 \shortto \ell_2} U_{\ell_1^-} O_{\beta;U}) \right. \nonumber\\
    &\quad + \tr(P_{\beta \alpha} U_{\ell_1^-}^\dagger X_{\ell_1} X_{\ell_2, \ell_1 \shortto \ell_2} U_{\ell_1^-} O_{\alpha;U} P_{\alpha \alpha} O_{\alpha;U} U_{\ell_1^-}^\dagger X_{\ell_1} U_{\ell_1^-}  P_{\alpha \beta} O_{\beta;U} U_{\ell_1^-}^\dagger X_{\ell_2, \ell_1 \shortto \ell_2} U_{\ell_1^-}) \nonumber\\
    &\quad - \tr(P_{\beta \alpha} U_{\ell_1^-}^\dagger X_{\ell_1} X_{\ell_2, \ell_1 \shortto \ell_2} U_{\ell_1^-} O_{\alpha;U} P_{\alpha \alpha} O_{\alpha;U} U_{\ell_1^-}^\dagger X_{\ell_1} U_{\ell_1^-}  P_{\alpha \beta} U_{\ell_1^-}^\dagger X_{\ell_2, \ell_1 \shortto \ell_2} U_{\ell_1^-} O_{\beta;U}) \nonumber\\
    &\quad - \tr(P_{\beta \alpha} U_{\ell_1^-}^\dagger X_{\ell_1} X_{\ell_2, \ell_1 \shortto \ell_2} U_{\ell_1^-} O_{\alpha;U} P_{\alpha \alpha} U_{\ell_1^-}^\dagger X_{\ell_1} U_{\ell_1^-} O_{\alpha;U} P_{\alpha \beta}  O_{\beta;U} U_{\ell_1^-}^\dagger X_{\ell_2, \ell_1 \shortto \ell_2} U_{\ell_1^-}) \nonumber\\
    & \quad + \tr(P_{\beta \alpha} O_{\alpha;U} U_{\ell_1^-}^\dagger X_{\ell_2, \ell_1 \shortto \ell_2}  X_{\ell_1} U_{\ell_1^-}   P_{\alpha \alpha} U_{\ell_1^-}^\dagger X_{\ell_1} U_{\ell_1^-} O_{\alpha;U} P_{\alpha \beta} U_{\ell_1^-}^\dagger X_{\ell_2, \ell_1 \shortto \ell_2} U_{\ell_1^-} O_{\beta;U}) \nonumber\\
    &\quad + \tr(P_{\beta \alpha} O_{\alpha;U} U_{\ell_1^-}^\dagger X_{\ell_2, \ell_1 \shortto \ell_2}  X_{\ell_1} U_{\ell_1^-}   P_{\alpha \alpha} O_{\alpha;U} U_{\ell_1^-}^\dagger X_{\ell_1} U_{\ell_1^-} P_{\alpha \beta}  O_{\beta;U} U_{\ell_1^-}^\dagger X_{\ell_2, \ell_1 \shortto \ell_2} U_{\ell_1^-}) \nonumber\\
    &\quad - \tr(P_{\beta \alpha} O_{\alpha;U} U_{\ell_1^-}^\dagger X_{\ell_2, \ell_1 \shortto \ell_2}  X_{\ell_1} U_{\ell_1^-}   P_{\alpha \alpha}  O_{\alpha;U} U_{\ell_1^-}^\dagger X_{\ell_1} U_{\ell_1^-} P_{\alpha \beta} U_{\ell_1^-}^\dagger X_{\ell_2, \ell_1 \shortto \ell_2} U_{\ell_1^-} O_{\beta;U}) \nonumber\\
    &\quad - \tr(P_{\beta \alpha} O_{\alpha;U} U_{\ell_1^-}^\dagger X_{\ell_2, \ell_1 \shortto \ell_2}  X_{\ell_1} U_{\ell_1^-}   P_{\alpha \alpha} U_{\ell_1^-}^\dagger X_{\ell_1} U_{\ell_1^-} O_{\alpha;U} P_{\alpha \beta} O_{\beta;U} U_{\ell_1^-}^\dagger X_{\ell_2, \ell_1 \shortto \ell_2} U_{\ell_1^-} ) \nonumber\\
    &\quad + \tr(P_{\beta \alpha} U_{\ell_1^-}^\dagger X_{\ell_1} U_{\ell_1^-} O_{\alpha;U} U_{\ell_1^-}^\dagger X_{\ell_2, \ell_1 \shortto \ell_2} U_{\ell_1^-}  P_{\alpha \alpha} O_{\alpha;U} U_{\ell_1^-}^\dagger X_{\ell_1} U_{\ell_1^-}  P_{\alpha \beta} U_{\ell_1^-}^\dagger X_{\ell_2, \ell_1 \shortto \ell_2} U_{\ell_1^-} O_{\beta;U}) \nonumber\\
    &\quad + \tr(P_{\beta \alpha} U_{\ell_1^-}^\dagger X_{\ell_1} U_{\ell_1^-} O_{\alpha;U} U_{\ell_1^-}^\dagger X_{\ell_2, \ell_1 \shortto \ell_2} U_{\ell_1^-}  P_{\alpha \alpha} U_{\ell_1^-}^\dagger X_{\ell_1} U_{\ell_1^-} O_{\alpha;U} P_{\alpha \beta} O_{\beta;U} U_{\ell_1^-}^\dagger X_{\ell_2, \ell_1 \shortto \ell_2} U_{\ell_1^-} ) \nonumber\\
    &\quad - \tr(P_{\beta \alpha} U_{\ell_1^-}^\dagger X_{\ell_1} U_{\ell_1^-} O_{\alpha;U} U_{\ell_1^-}^\dagger X_{\ell_2, \ell_1 \shortto \ell_2} U_{\ell_1^-}  P_{\alpha \alpha} U_{\ell_1^-}^\dagger X_{\ell_1} U_{\ell_1^-} O_{\alpha;U} P_{\alpha \beta} U_{\ell_1^-}^\dagger X_{\ell_2, \ell_1 \shortto \ell_2} U_{\ell_1^-} O_{\beta;U}) \nonumber\\
    &\quad - \tr(P_{\beta \alpha} U_{\ell_1^-}^\dagger X_{\ell_1} U_{\ell_1^-} O_{\alpha;U} U_{\ell_1^-}^\dagger X_{\ell_2, \ell_1 \shortto \ell_2} U_{\ell_1^-}  P_{\alpha \alpha} O_{\alpha;U} U_{\ell_1^-}^\dagger X_{\ell_1} U_{\ell_1^-}  P_{\alpha \beta} O_{\beta;U} U_{\ell_1^-}^\dagger X_{\ell_2, \ell_1 \shortto \ell_2} U_{\ell_1^-} ) \nonumber\\
    & \quad + \tr(P_{\beta \alpha} U_{\ell_1^-}^\dagger X_{\ell_2, \ell_1 \shortto \ell_2} U_{\ell_1^-} O_{\alpha;U} U_{\ell_1^-}^\dagger X_{\ell_1} U_{\ell_1^-}     P_{\alpha \alpha} O_{\alpha;U} U_{\ell_1^-}^\dagger X_{\ell_1} U_{\ell_1^-}  P_{\alpha \beta} U_{\ell_1^-}^\dagger X_{\ell_2, \ell_1 \shortto \ell_2} U_{\ell_1^-} O_{\beta;U}) \nonumber\\
    & \quad + \tr(P_{\beta \alpha} U_{\ell_1^-}^\dagger X_{\ell_2, \ell_1 \shortto \ell_2} U_{\ell_1^-} O_{\alpha;U} U_{\ell_1^-}^\dagger X_{\ell_1} U_{\ell_1^-}     P_{\alpha \alpha} U_{\ell_1^-}^\dagger X_{\ell_1} U_{\ell_1^-} O_{\alpha;U} P_{\alpha \beta} O_{\beta;U} U_{\ell_1^-}^\dagger X_{\ell_2, \ell_1 \shortto \ell_2} U_{\ell_1^-} )  \nonumber\\
    & \quad - \tr(P_{\beta \alpha} U_{\ell_1^-}^\dagger X_{\ell_2, \ell_1 \shortto \ell_2} U_{\ell_1^-} O_{\alpha;U} U_{\ell_1^-}^\dagger X_{\ell_1} U_{\ell_1^-}     P_{\alpha \alpha} U_{\ell_1^-}^\dagger X_{\ell_1} U_{\ell_1^-} O_{\alpha;U} P_{\alpha \beta} U_{\ell_1^-}^\dagger X_{\ell_2, \ell_1 \shortto \ell_2} U_{\ell_1^-} O_{\beta;U}) \nonumber\\
    &\quad - \tr(P_{\beta \alpha} U_{\ell_1^-}^\dagger X_{\ell_2, \ell_1 \shortto \ell_2} U_{\ell_1^-} O_{\alpha;U} U_{\ell_1^-}^\dagger X_{\ell_1} U_{\ell_1^-}     P_{\alpha \alpha} O_{\alpha;U} U_{\ell_1^-}^\dagger X_{\ell_1} U_{\ell_1^-}  P_{\alpha \beta} O_{\beta;U} U_{\ell_1^-}^\dagger X_{\ell_2, \ell_1 \shortto \ell_2} U_{\ell_1^-} ). 
    \label{eq:ha12ga1gb2_l1}
\end{align}
The first term is
\begin{align}
    I_1 &\equiv \int \diff U_{\ell_1^-} \diff U_{\ell_1 \shortto \ell_2} \diff U_{\ell_2^+} \tr(P_{\beta \alpha} U_{\ell_1^-}^\dagger X_{\ell_1} X_{\ell_2, \ell_1 \shortto \ell_2} U_{\ell_1^-} O_{\alpha;U} P_{\alpha \alpha} U_{\ell_1^-}^\dagger X_{\ell_1} U_{\ell_1^-} O_{\alpha;U} P_{\alpha \beta} U_{\ell_1^-}^\dagger X_{\ell_2, \ell_1 \shortto \ell_2} U_{\ell_1^-} O_{\beta;U})\nonumber\\
    &= \mathbb{E}_{\calU_{\rm RH}}\left[\frac{dT_{\alpha \alpha}T_{\beta \beta} |T_{\alpha \alpha}| ^2T_{\alpha \beta}^*T_{\beta \alpha}^*-|T_{\alpha \alpha}| ^4 |T_{\beta \beta}| ^2}{(d-1) (d+1)^2} \right].
\end{align}
The second term is
\begin{align}
    I_2 &\equiv  \int \diff U_{\ell_1^-} \diff U_{\ell_1 \shortto \ell_2} \diff U_{\ell_2^+} \tr(P_{\beta \alpha} U_{\ell_1^-}^\dagger X_{\ell_1} X_{\ell_2, \ell_1 \shortto \ell_2} U_{\ell_1^-} O_{\alpha;U} P_{\alpha \alpha} O_{\alpha;U} U_{\ell_1^-}^\dagger X_{\ell_1} U_{\ell_1^-}  P_{\alpha \beta} O_{\beta;U} U_{\ell_1^-}^\dagger X_{\ell_2, \ell_1 \shortto \ell_2} U_{\ell_1^-}) \nonumber\\
    &= \mathbb{E}_{\calU_{\rm RH}}\left[\delta_{\alpha \beta} \frac{d  | T_{\alpha \alpha}| ^2 T_{\beta \beta}^* \left(d T_{\alpha \beta}-T_{\alpha \alpha}\right)}{\left(d^2-1\right)^2}+\frac{| T_{\alpha \alpha}| ^2 | T_{\beta \beta}| ^2 \left(| T_{\alpha \alpha}| ^2-d\right)}{\left(d^2-1\right)^2} \right].
\end{align}
The third term is
\begin{align}
    I_3 &\equiv \int \diff U_{\ell_1^-} \diff U_{\ell_1 \shortto \ell_2} \diff U_{\ell_2^+} \tr(P_{\beta \alpha} U_{\ell_1^-}^\dagger X_{\ell_1} X_{\ell_2, \ell_1 \shortto \ell_2} U_{\ell_1^-} O_{\alpha;U} P_{\alpha \alpha} O_{\alpha;U} U_{\ell_1^-}^\dagger X_{\ell_1} U_{\ell_1^-}  P_{\alpha \beta} U_{\ell_1^-}^\dagger X_{\ell_2, \ell_1 \shortto \ell_2} U_{\ell_1^-} O_{\beta;U}) \nonumber\\
    &= \mathbb{E}_{\calU_{\rm RH}} \left[\frac{| T_{\alpha \alpha}| ^2 \left(| T_{\beta \beta}| ^2 \left(| T_{\alpha \alpha}| ^2-d\right)-d T_{\alpha \alpha} T_{\beta \beta} T_{\alpha \beta}^* T_{\beta \alpha}^*\right)}{\left(d^2-1\right)^2} + \delta_{\alpha \beta}\frac{d^2  T_{\beta \beta} | T_{\alpha \alpha}| ^2 T_{\alpha \beta}^*}{\left(d^2-1\right)^2} \right].
\end{align}
The forth term is
\begin{align}
    I_4 &\equiv \int \diff U_{\ell_1^-} \diff U_{\ell_1 \shortto \ell_2} \diff U_{\ell_2^+} \tr(P_{\beta \alpha} U_{\ell_1^-}^\dagger X_{\ell_1} X_{\ell_2, \ell_1 \shortto \ell_2} U_{\ell_1^-} O_{\alpha;U} P_{\alpha \alpha} U_{\ell_1^-}^\dagger X_{\ell_1} U_{\ell_1^-} O_{\alpha;U} P_{\alpha \beta}  O_{\beta;U} U_{\ell_1^-}^\dagger X_{\ell_2, \ell_1 \shortto \ell_2} U_{\ell_1^-}) \nonumber\\
    &= \mathbb{E}_{\calU_{\rm RH}} \left[\delta_{\alpha \beta} \frac{d T_{\alpha \alpha} |T_{\alpha \alpha}| ^2T_{\beta \beta}^*}{(d-1) (d+1)^2}-\frac{|T_{\alpha \alpha}| ^4 |T_{\beta \beta}| ^2}{(d-1) (d+1)^2} \right].
\end{align}
The fifth term is
\begin{align}
    I_5 &\equiv \int \diff U_{\ell_1^-} \diff U_{\ell_1 \shortto \ell_2} \diff U_{\ell_2^+} \tr(P_{\beta \alpha} O_{\alpha;U} U_{\ell_1^-}^\dagger X_{\ell_2, \ell_1 \shortto \ell_2}  X_{\ell_1} U_{\ell_1^-}   P_{\alpha \alpha} U_{\ell_1^-}^\dagger X_{\ell_1} U_{\ell_1^-} O_{\alpha;U} P_{\alpha \beta} U_{\ell_1^-}^\dagger X_{\ell_2, \ell_1 \shortto \ell_2} U_{\ell_1^-} O_{\beta;U})\nonumber\\
    &= \mathbb{E}_{\calU_{\rm RH}} \left[\delta_{\alpha \beta}  \frac{d T_{\beta \beta} | T_{\alpha \alpha}| ^2 \left(d T_{\alpha \beta}^*-T_{\alpha \alpha}^*\right)}{\left(d^2-1\right)^2}+\frac{| T_{\alpha \alpha}| ^2 | T_{\beta \beta}| ^2 \left(| T_{\alpha \alpha}| ^2-d\right)}{\left(d^2-1\right)^2} \right] = I_2^*.
\end{align}
The sixth term is
\begin{align}
    I_6 &\equiv \int \diff U_{\ell_1^-} \diff U_{\ell_1 \shortto \ell_2} \diff U_{\ell_2^+} \tr(P_{\beta \alpha} O_{\alpha;U} U_{\ell_1^-}^\dagger X_{\ell_2, \ell_1 \shortto \ell_2}  X_{\ell_1} U_{\ell_1^-}   P_{\alpha \alpha} O_{\alpha;U} U_{\ell_1^-}^\dagger X_{\ell_1} U_{\ell_1^-} P_{\alpha \beta}  O_{\beta;U} U_{\ell_1^-}^\dagger X_{\ell_2, \ell_1 \shortto \ell_2} U_{\ell_1^-}) \nonumber\\
    &= \mathbb{E}_{\calU_{\rm RH}} \left[ \frac{d T_{\alpha \beta} T_{\beta \alpha} | T_{\alpha \alpha}| ^2 T_{\alpha \alpha}^* T_{\beta \beta}^*-| T_{\alpha \alpha}| ^4 | T_{\beta \beta}| ^2}{(d-1) (d+1)^2} \right] = I_1^*.
\end{align}
The seventh term is
\begin{align}
    I_7 &\equiv \int \diff U_{\ell_1^-} \diff U_{\ell_1 \shortto \ell_2} \diff U_{\ell_2^+} \tr(P_{\beta \alpha} O_{\alpha;U} U_{\ell_1^-}^\dagger X_{\ell_2, \ell_1 \shortto \ell_2}  X_{\ell_1} U_{\ell_1^-}   P_{\alpha \alpha}  O_{\alpha;U} U_{\ell_1^-}^\dagger X_{\ell_1} U_{\ell_1^-} P_{\alpha \beta} U_{\ell_1^-}^\dagger X_{\ell_2, \ell_1 \shortto \ell_2} U_{\ell_1^-} O_{\beta;U}) \nonumber\\
    &= \mathbb{E}_{\calU_{\rm RH}} \left[\delta_{\alpha \beta} \frac{d  T_{\beta \beta} | T_{\alpha \alpha}| ^2 T_{\alpha \alpha}^*}{(d-1) (d+1)^2}-\frac{| T_{\alpha \alpha}| ^4 | T_{\beta \beta}| ^2}{(d-1) (d+1)^2} \right] = I_4^*.
\end{align}
The eighth term is
\begin{align}
    I_8 &\equiv  \int \diff U_{\ell_1^-} \diff U_{\ell_1 \shortto \ell_2} \diff U_{\ell_2^+} \tr(P_{\beta \alpha} O_{\alpha;U} U_{\ell_1^-}^\dagger X_{\ell_2, \ell_1 \shortto \ell_2}  X_{\ell_1} U_{\ell_1^-}   P_{\alpha \alpha} U_{\ell_1^-}^\dagger X_{\ell_1} U_{\ell_1^-} O_{\alpha;U} P_{\alpha \beta} O_{\beta;U} U_{\ell_1^-}^\dagger X_{\ell_2, \ell_1 \shortto \ell_2} U_{\ell_1^-} ) \nonumber\\
    &= \mathbb{E}_{\calU_{\rm RH}} \left[ \frac{| T_{\alpha \alpha}| ^2 \left(| T_{\beta \beta}| ^2 \left(| T_{\alpha \alpha}| ^2-d\right)-d T_{\alpha \beta} T_{\beta \alpha} T_{\alpha \alpha}^* T_{\beta \beta}^*\right)}{\left(d^2-1\right)^2} + \delta_{\alpha \beta} \frac{d^2 T_{\alpha \beta} | T_{\alpha \alpha}| ^2 T_{\beta \beta}^*}{\left(d^2-1\right)^2} \right] = I_3^*.
\end{align}
% The first eighth terms in Eq.~\eqref{eq:ha12ga1gb2_l1} can be summarized as
% \begin{align}
%     S_1 & \equiv I_1 + I_2 - I_3 - I_4 + I_5 + I_6 - I_7 - I_8 = I_1 + I_2 - I_3 - I_4 + c.c.\\
%     &= \mathbb{E}_{\calU_{\rm RH}}\left[\frac{d^2| T_{\alpha \alpha}| ^2 T_{\alpha \alpha} T_{\beta \beta} T_{\alpha \beta}^* T_{\beta \alpha}^*}{\left(d^2-1\right)^2} + c.c. \right] + \delta_{\alpha \beta} \mathbb{E}_{\calU_{\rm RH}}\left[ \frac{d^2  | T_{\alpha \alpha}| ^2 \left(T_{\alpha \beta} T_{\beta \beta}^* - T_{\alpha \beta}^* T_{\beta \beta} - T_{\alpha \alpha} T_{\beta \beta}^*\right)}{\left(d^2-1\right)^2} + c.c.\right]
% \end{align}
The ninth term is
\begin{align}
    I_9 &\equiv \int \diff U_{\ell_1^-} \diff U_{\ell_1 \shortto \ell_2} \diff U_{\ell_2^+} \tr(P_{\beta \alpha} U_{\ell_1^-}^\dagger X_{\ell_1} U_{\ell_1^-} O_{\alpha;U} U_{\ell_1^-}^\dagger X_{\ell_2, \ell_1 \shortto \ell_2} U_{\ell_1^-}  P_{\alpha \alpha} O_{\alpha;U} U_{\ell_1^-}^\dagger X_{\ell_1} U_{\ell_1^-}  P_{\alpha \beta} U_{\ell_1^-}^\dagger X_{\ell_2, \ell_1 \shortto \ell_2} U_{\ell_1^-} O_{\beta;U}) \nonumber\\
    &= \mathbb{E}_{\calU_{\rm RH}}\left[\delta_{\alpha \beta} \frac{d  T_{\beta \beta} T_{\alpha \alpha}^* \left(d-| T_{\alpha \alpha}| ^2\right)}{\left(d^2-1\right)^2}-\frac{| T_{\alpha \alpha}| ^2 | T_{\beta \beta}| ^2 \left(d-| T_{\alpha \alpha}| ^2\right)}{\left(d^2-1\right)^2} \right].
\end{align}
The tenth term is
\begin{align}
    I_{10} &\equiv \int \diff U_{\ell_1^-} \diff U_{\ell_1 \shortto \ell_2} \diff U_{\ell_2^+} \tr(P_{\beta \alpha} U_{\ell_1^-}^\dagger X_{\ell_1} U_{\ell_1^-} O_{\alpha;U} U_{\ell_1^-}^\dagger X_{\ell_2, \ell_1 \shortto \ell_2} U_{\ell_1^-}  P_{\alpha \alpha} U_{\ell_1^-}^\dagger X_{\ell_1} U_{\ell_1^-} O_{\alpha;U} P_{\alpha \beta} O_{\beta;U} U_{\ell_1^-}^\dagger X_{\ell_2, \ell_1 \shortto \ell_2} U_{\ell_1^-} ) \nonumber \\
    &= \mathbb{E}_{\calU_{\rm RH}}\left[ \frac{d T_{\alpha \beta} T_{\beta \alpha} | T_{\alpha \alpha}| ^2 T_{\alpha \alpha}^* T_{\beta \beta}^*-| T_{\alpha \alpha}| ^4 | T_{\beta \beta}| ^2}{(d-1) (d+1)^2} \right] = I_1^*.
\end{align}
The eleventh term is
\begin{align}
    I_{11} &\equiv \int \diff U_{\ell_1^-} \diff U_{\ell_1 \shortto \ell_2} \diff U_{\ell_2^+} \tr(P_{\beta \alpha} U_{\ell_1^-}^\dagger X_{\ell_1} U_{\ell_1^-} O_{\alpha;U} U_{\ell_1^-}^\dagger X_{\ell_2, \ell_1 \shortto \ell_2} U_{\ell_1^-}  P_{\alpha \alpha} U_{\ell_1^-}^\dagger X_{\ell_1} U_{\ell_1^-} O_{\alpha;U} P_{\alpha \beta} U_{\ell_1^-}^\dagger X_{\ell_2, \ell_1 \shortto \ell_2} U_{\ell_1^-} O_{\beta;U}) \nonumber\\
    &= \mathbb{E}_{\calU_{\rm RH}}\left[ \delta_{\alpha \beta}\frac{d  T_{\beta \beta} | T_{\alpha \alpha}| ^2 T_{\alpha \alpha}^*}{(d-1) (d+1)^2}-\frac{| T_{\alpha \alpha}| ^4 | T_{\beta \beta}| ^2}{(d-1) (d+1)^2} \right] = I_4^*.
\end{align}
The twelfth term is
\begin{align}
    I_{12} &\equiv \int \diff U_{\ell_1^-} \diff U_{\ell_1 \shortto \ell_2} \diff U_{\ell_2^+} \tr(P_{\beta \alpha} U_{\ell_1^-}^\dagger X_{\ell_1} U_{\ell_1^-} O_{\alpha;U} U_{\ell_1^-}^\dagger X_{\ell_2, \ell_1 \shortto \ell_2} U_{\ell_1^-}  P_{\alpha \alpha} O_{\alpha;U} U_{\ell_1^-}^\dagger X_{\ell_1} U_{\ell_1^-}  P_{\alpha \beta} O_{\beta;U} U_{\ell_1^-}^\dagger X_{\ell_2, \ell_1 \shortto \ell_2} U_{\ell_1^-} ) \nonumber\\
    &= \mathbb{E}_{\calU_{\rm RH}}\left[ \frac{\left(d-| T_{\alpha \alpha}| ^2\right) \left(d T_{\alpha \beta} T_{\beta \alpha} T_{\alpha \alpha}^* T_{\beta \beta}^*-| T_{\alpha \alpha}| ^2 | T_{\beta \beta}| ^2\right)}{\left(d^2-1\right)^2} \right].
\end{align}
The thirteenth term is
\begin{align}
    I_{13} &\equiv \int \diff U_{\ell_1^-} \diff U_{\ell_1 \shortto \ell_2} \diff U_{\ell_2^+} \tr(P_{\beta \alpha} U_{\ell_1^-}^\dagger X_{\ell_2, \ell_1 \shortto \ell_2} U_{\ell_1^-} O_{\alpha;U} U_{\ell_1^-}^\dagger X_{\ell_1} U_{\ell_1^-}     P_{\alpha \alpha} O_{\alpha;U} U_{\ell_1^-}^\dagger X_{\ell_1} U_{\ell_1^-}  P_{\alpha \beta} U_{\ell_1^-}^\dagger X_{\ell_2, \ell_1 \shortto \ell_2} U_{\ell_1^-} O_{\beta;U}) \nonumber\\
    &= \mathbb{E}_{\calU_{\rm RH}}\left[ \frac{d T_{\alpha \alpha} T_{\beta \beta}| T_{\alpha \alpha}| ^2 T_{\alpha \beta}^* T_{\beta \alpha}^*-| T_{\alpha \alpha}| ^4 | T_{\beta \beta}| ^2}{(d-1) (d+1)^2} \right] = I_1.
\end{align}
The fourteenth term is
\begin{align}
    I_{14} &\equiv \int \diff U_{\ell_1^-} \diff U_{\ell_1 \shortto \ell_2} \diff U_{\ell_2^+} \tr(P_{\beta \alpha} U_{\ell_1^-}^\dagger X_{\ell_2, \ell_1 \shortto \ell_2} U_{\ell_1^-} O_{\alpha;U} U_{\ell_1^-}^\dagger X_{\ell_1} U_{\ell_1^-}     P_{\alpha \alpha} U_{\ell_1^-}^\dagger X_{\ell_1} U_{\ell_1^-} O_{\alpha;U} P_{\alpha \beta} O_{\beta;U} U_{\ell_1^-}^\dagger X_{\ell_2, \ell_1 \shortto \ell_2} U_{\ell_1^-} )  \nonumber\\
    &= \mathbb{E}_{\calU_{\rm RH}} \left[ \delta_{\alpha \beta} \frac{d T_{\alpha \alpha} T_{\beta \beta}^* \left(d-| T_{\alpha \alpha}| ^2\right)}{\left(d^2-1\right)^2}-\frac{| T_{\alpha \alpha}| ^2 | T_{\beta \beta}| ^2 \left(d-| T_{\alpha \alpha}| ^2\right)}{\left(d^2-1\right)^2}\right] = I_9^*.
\end{align}
The fifteenth term is
\begin{align}
    I_{15} &\equiv \int \diff U_{\ell_1^-} \diff U_{\ell_1 \shortto \ell_2} \diff U_{\ell_2^+} \tr(P_{\beta \alpha} U_{\ell_1^-}^\dagger X_{\ell_2, \ell_1 \shortto \ell_2} U_{\ell_1^-} O_{\alpha;U} U_{\ell_1^-}^\dagger X_{\ell_1} U_{\ell_1^-}     P_{\alpha \alpha} U_{\ell_1^-}^\dagger X_{\ell_1} U_{\ell_1^-} O_{\alpha;U} P_{\alpha \beta} U_{\ell_1^-}^\dagger X_{\ell_2, \ell_1 \shortto \ell_2} U_{\ell_1^-} O_{\beta;U}) \nonumber\\
    & = \mathbb{E}_{\calU_{\rm RH}} \left[ \frac{\left(d-| T_{\alpha \alpha}| ^2\right) \left(d T_{\alpha \alpha} T_{\beta \beta} T_{\alpha \beta}^* T_{\beta \alpha}^*-| T_{\alpha \alpha}| ^2 | T_{\beta \beta}| ^2\right)}{\left(d^2-1\right)^2} \right] = I_{12}^*.
\end{align}
The sixteenth term is
\begin{align}
    I_{16} &\equiv \int \diff U_{\ell_1^-} \diff U_{\ell_1 \shortto \ell_2} \diff U_{\ell_2^+} \tr(P_{\beta \alpha} U_{\ell_1^-}^\dagger X_{\ell_2, \ell_1 \shortto \ell_2} U_{\ell_1^-} O_{\alpha;U} U_{\ell_1^-}^\dagger X_{\ell_1} U_{\ell_1^-}     P_{\alpha \alpha} O_{\alpha;U} U_{\ell_1^-}^\dagger X_{\ell_1} U_{\ell_1^-}  P_{\alpha \beta} O_{\beta;U} U_{\ell_1^-}^\dagger X_{\ell_2, \ell_1 \shortto \ell_2} U_{\ell_1^-} ) \nonumber\\
    &= \mathbb{E}_{\calU_{\rm RH}} \left[ \delta_{\alpha \beta} \frac{d  T_{\alpha \alpha} | T_{\alpha \alpha}| ^2 T_{\beta \beta}^*}{(d-1) (d+1)^2}-\frac{| T_{\alpha \alpha}| ^4 | T_{\beta \beta}| ^2}{(d-1) (d+1)^2} \right] = I_4.
\end{align}
% Then the rest eighth terms in Eq.~\eqref{eq:ha12ga1gb2_l1} can be combined to
% \begin{align}
%     S_2 &= I_9 + I_{10} - I_{11} - I_{12} + I_{13} + I_{14} - I_{15} - I_{16} = I_9 + I_1^* - I_4^* - I_{12} + I_1 + I_9^* - I_{12}^* - I_4\\
%     &= I_9 + I_1 - I_4 - I_{12} + c.c.\\
%     &= \mathbb{E}_{\calU_{\rm RH}}\left[\frac{d^2 (|T_{\alpha \alpha}|^2-1) T_{\alpha \alpha} T_{\beta \beta} T_{\alpha \beta}^* T_{\beta \alpha}^* }{(d^2-1)^2} + c.c.\right] + \delta_{\alpha \beta}\mathbb{E}_{\calU_{\rm RH}}\left[\frac{d^2 T_{\alpha \alpha} T_{\beta \beta}^* (1-|T_{\alpha \alpha}|^2)}{(d^2-1)^2} + c.c.\right]
% \end{align}
Finally we have
\begin{align}
    &\mathbb{E}_{\calU_{\rm RH}}\left[\frac{\partial^2 \epsilon_\alpha}{\partial \theta_{\ell_1} \partial\theta_{\ell_2}} \frac{\partial \epsilon_\alpha}{\partial \theta_{\ell_1}} \frac{\partial \epsilon_\beta}{\partial \theta_{\ell_2}}\right] = \frac{1}{16}\sum_{i=1}^4 \left(I_{4i+1} + I_{4i+2} - I_{4i+3} - I_{4i+4}\right)\nonumber\\
    &= \frac{1}{16} \left(2I_1 + I_2 - I_3 - 2I_4 + I_9 - I_{12} + c.c.\right)\\
    &\left. \qquad + \mathbb{E}_{\calU_{\rm RH}}\left[\frac{d^2 (|T_{\alpha \alpha}|^2-1) T_{\alpha \alpha} T_{\beta \beta} T_{\alpha \beta}^* T_{\beta \alpha}^* }{(d^2-1)^2} + c.c.\right] + \delta_{\alpha \beta}\mathbb{E}_{\calU_{\rm RH}}\left[\frac{d^2 T_{\alpha \alpha} T_{\beta \beta}^* (1-|T_{\alpha \alpha}|^2)}{(d^2-1)^2} + c.c.\right]\right)\\
    &= \frac{1}{16} \mathbb{E}_{\calU_{\rm RH}}\left[ \frac{d^2 (2|T_{\alpha \alpha}|^2-1) \left(T_{\alpha \alpha} T_{\beta \beta} T_{\alpha \beta}^* T_{\beta \alpha}^* - \delta_{\alpha \beta} T_{\alpha \alpha} T_{\beta \beta}^*\right) }{(d^2 - 1)^2} + c.c.\right].
    \label{eq:ha12ga1gb2_l2}
\end{align}
Specifically, for $\alpha = \beta$, we have
\begin{align}
    \mathbb{E}_{\calU_{\rm RH}}\left[\frac{\partial^2 \epsilon_\alpha}{\partial \theta_{\ell_1} \partial\theta_{\ell_2}} \frac{\partial \epsilon_\alpha}{\partial \theta_{\ell_1}} \frac{\partial \epsilon_\alpha}{\partial \theta_{\ell_2}}\right] = \frac{d^2 o_\alpha (o_\alpha-1)(2o_\alpha-1)}{8(d^2-1)^2}.
\end{align}
On the other hand, for $\alpha \neq \beta$, we have
\begin{align}
    \mathbb{E}_{\calU_{\rm RH}}\left[\frac{\partial^2 \epsilon_\alpha}{\partial \theta_{\ell_1} \partial\theta_{\ell_2}} \frac{\partial \epsilon_\alpha}{\partial \theta_{\ell_1}} \frac{\partial \epsilon_\beta}{\partial \theta_{\ell_2}}\right] &=  \frac{1}{16} \mathbb{E}_{\calU_{\rm RH}}\left[ \frac{d^2 (2|T_{\alpha \alpha}|^2-1) T_{\alpha \alpha} T_{\beta \beta} T_{\alpha \beta}^* T_{\beta \alpha}^*}{(d^2 - 1)^2} + c.c.\right] \\
    &= \mathbb{E}_{\calU_{\rm RH}}\left[\frac{d^2 (2|T_{\alpha \alpha}|^2-1) |T_{\alpha \alpha}| |T_{\beta \beta}| |T_{\alpha \beta}| |T_{\beta \alpha}|}{8(d^2 - 1)^2} \right].
    % &= \frac{d^2 (2o_\alpha - 1)\sqrt{o_\alpha o_\beta (1-o_\alpha) (1-o_\beta)}}{8(d^2 - 1)^2 (d-1)}.
\end{align}

Similarly, we next work on
\begin{align}
    &\mathbb{E}_{\calU_{\rm RH}}\left[\frac{\partial^2 \epsilon_\alpha}{\partial \theta_{\ell_1} \partial\theta_{\ell_2}} \frac{\partial \epsilon_\alpha}{\partial \theta_{\ell_2}} \frac{\partial \epsilon_\beta}{\partial \theta_{\ell_1}} \right] \nonumber\\
    &= \frac{1}{16}\int \diff U_{\ell_1^-} \diff U_{\ell_1 \shortto \ell_2} \diff U_{\ell_2^+} \tr(P_{\beta \alpha} U_{\ell_1^-}^\dagger \left[X_{\ell_1}, U_{\ell_1 \shortto \ell_2}^\dagger \left[X_{\ell_2}, O_{\alpha; \ell_2^+}\right] U_{\ell_1 \shortto \ell_2}\right] U_{\ell_1^-}P_{\alpha \alpha} U_{\ell_2^-}^\dagger \left[X_{\ell_2}, O_{\alpha; \ell_2^+}\right]U_{\ell_2^-} P_{\alpha \beta} U_{\ell_1^-}^\dagger \left[X_{\ell_1}, O_{\beta; \ell_1^+}\right]U_{\ell_1^-}) \nonumber\\
    &= \frac{1}{16} \int_{\calU_{\rm RH}} \diff U_{\rm RH} \int_{\calU_{\rm Haar}} \diff U_{\ell_1^-} \diff U_{\ell_1 \shortto \ell_2} \left[ \tr(P_{\beta \alpha} U_{\ell_1^-}^\dagger X_{\ell_1} X_{\ell_2, \ell_1 \shortto \ell_2} U_{\ell_1^-} O_{\alpha;U} P_{\alpha \alpha} U_{\ell_1^-}^\dagger X_{\ell_2,\ell_1 \shortto \ell_2} U_{\ell_1^-} O_{\alpha;U} P_{\alpha \beta} U_{\ell_1^-}^\dagger X_{\ell_1} U_{\ell_1^-} O_{\beta;U}) \right. \nonumber\\
    &\quad + \tr(P_{\beta \alpha} U_{\ell_1^-}^\dagger X_{\ell_1} X_{\ell_2, \ell_1 \shortto \ell_2} U_{\ell_1^-} O_{\alpha;U} P_{\alpha \alpha} O_{\alpha;U} U_{\ell_1^-}^\dagger X_{\ell_2, \ell_1 \shortto \ell_2} U_{\ell_1^-}  P_{\alpha \beta} O_{\beta;U} U_{\ell_1^-}^\dagger X_{\ell_1} U_{\ell_1^-}) \nonumber\\
    &\quad - \tr(P_{\beta \alpha} U_{\ell_1^-}^\dagger X_{\ell_1} X_{\ell_2, \ell_1 \shortto \ell_2} U_{\ell_1^-} O_{\alpha;U} P_{\alpha \alpha} O_{\alpha;U} U_{\ell_1^-}^\dagger X_{\ell_2, \ell_1 \shortto \ell_2} U_{\ell_1^-}  P_{\alpha \beta} U_{\ell_1^-}^\dagger X_{\ell_1} U_{\ell_1^-} O_{\beta;U}) \nonumber\\
    &\quad - \tr(P_{\beta \alpha} U_{\ell_1^-}^\dagger X_{\ell_1} X_{\ell_2, \ell_1 \shortto \ell_2} U_{\ell_1^-} O_{\alpha;U} P_{\alpha \alpha} U_{\ell_1^-}^\dagger X_{\ell_2, \ell_1 \shortto \ell_2} U_{\ell_1^-} O_{\alpha;U} P_{\alpha \beta}  O_{\beta;U} U_{\ell_1^-}^\dagger  X_{\ell_1} U_{\ell_1^-} ) \nonumber\\
    & \quad + \tr(P_{\beta \alpha} O_{\alpha;U} U_{\ell_1^-}^\dagger X_{\ell_2, \ell_1 \shortto \ell_2}  X_{\ell_1} U_{\ell_1^-}   P_{\alpha \alpha} U_{\ell_1^-}^\dagger X_{\ell_2,\ell_1 \shortto \ell_2} U_{\ell_1^-} O_{\alpha;U} P_{\alpha \beta} U_{\ell_1^-}^\dagger X_{\ell_1} U_{\ell_1^-} O_{\beta;U} ) \nonumber\\
    &\quad + \tr(P_{\beta \alpha} O_{\alpha;U} U_{\ell_1^-}^\dagger X_{\ell_2, \ell_1 \shortto \ell_2}  X_{\ell_1} U_{\ell_1^-}   P_{\alpha \alpha} O_{\alpha;U} U_{\ell_1^-}^\dagger X_{\ell_2, \ell_1 \shortto \ell_2} U_{\ell_1^-}  P_{\alpha \beta} O_{\beta;U} U_{\ell_1^-}^\dagger X_{\ell_1} U_{\ell_1^-} ) \nonumber\\
    &\quad - \tr(P_{\beta \alpha} O_{\alpha;U} U_{\ell_1^-}^\dagger X_{\ell_2, \ell_1 \shortto \ell_2}  X_{\ell_1} U_{\ell_1^-}   P_{\alpha \alpha}  O_{\alpha;U} U_{\ell_1^-}^\dagger X_{\ell_2, \ell_1 \shortto \ell_2} U_{\ell_1^-}  P_{\alpha \beta} U_{\ell_1^-}^\dagger X_{\ell_1} U_{\ell_1^-} O_{\beta;U} ) \nonumber\\
    &\quad - \tr(P_{\beta \alpha} O_{\alpha;U} U_{\ell_1^-}^\dagger X_{\ell_2, \ell_1 \shortto \ell_2}  X_{\ell_1} U_{\ell_1^-}   P_{\alpha \alpha} U_{\ell_1^-}^\dagger X_{\ell_2, \ell_1 \shortto \ell_2} U_{\ell_1^-} O_{\alpha;U} P_{\alpha \beta}  O_{\beta;U} U_{\ell_1^-}^\dagger  X_{\ell_1} U_{\ell_1^-} ) \nonumber\\
    &\quad + \tr(P_{\beta \alpha} U_{\ell_1^-}^\dagger X_{\ell_1} U_{\ell_1^-} O_{\alpha;U} U_{\ell_1^-}^\dagger X_{\ell_2, \ell_1 \shortto \ell_2} U_{\ell_1^-}  P_{\alpha \alpha} O_{\alpha;U} U_{\ell_1^-}^\dagger X_{\ell_2, \ell_1 \shortto \ell_2} U_{\ell_1^-}  P_{\alpha \beta} U_{\ell_1^-}^\dagger X_{\ell_1} U_{\ell_1^-} O_{\beta;U} ) \nonumber\\
    &\quad + \tr(P_{\beta \alpha} U_{\ell_1^-}^\dagger X_{\ell_1} U_{\ell_1^-} O_{\alpha;U} U_{\ell_1^-}^\dagger X_{\ell_2, \ell_1 \shortto \ell_2} U_{\ell_1^-}  P_{\alpha \alpha} U_{\ell_1^-}^\dagger X_{\ell_2, \ell_1 \shortto \ell_2} U_{\ell_1^-} O_{\alpha;U} P_{\alpha \beta}  O_{\beta;U} U_{\ell_1^-}^\dagger  X_{\ell_1} U_{\ell_1^-} ) \nonumber\\
    &\quad - \tr(P_{\beta \alpha} U_{\ell_1^-}^\dagger X_{\ell_1} U_{\ell_1^-} O_{\alpha;U} U_{\ell_1^-}^\dagger X_{\ell_2, \ell_1 \shortto \ell_2} U_{\ell_1^-}  P_{\alpha \alpha} U_{\ell_1^-}^\dagger X_{\ell_2,\ell_1 \shortto \ell_2} U_{\ell_1^-} O_{\alpha;U} P_{\alpha \beta} U_{\ell_1^-}^\dagger X_{\ell_1} U_{\ell_1^-} O_{\beta;U} ) \nonumber\\
    &\quad - \tr(P_{\beta \alpha} U_{\ell_1^-}^\dagger X_{\ell_1} U_{\ell_1^-} O_{\alpha;U} U_{\ell_1^-}^\dagger X_{\ell_2, \ell_1 \shortto \ell_2} U_{\ell_1^-}  P_{\alpha \alpha} O_{\alpha;U} U_{\ell_1^-}^\dagger X_{\ell_2, \ell_1 \shortto \ell_2} U_{\ell_1^-}  P_{\alpha \beta} O_{\beta;U} U_{\ell_1^-}^\dagger X_{\ell_1} U_{\ell_1^-} ) \nonumber\\
    & \quad + \tr(P_{\beta \alpha} U_{\ell_1^-}^\dagger X_{\ell_2, \ell_1 \shortto \ell_2} U_{\ell_1^-} O_{\alpha;U} U_{\ell_1^-}^\dagger X_{\ell_1} U_{\ell_1^-}     P_{\alpha \alpha} O_{\alpha;U} U_{\ell_1^-}^\dagger X_{\ell_2, \ell_1 \shortto \ell_2} U_{\ell_1^-}  P_{\alpha \beta} U_{\ell_1^-}^\dagger X_{\ell_1} U_{\ell_1^-} O_{\beta;U} ) \nonumber\\
    &\quad + \tr(P_{\beta \alpha} U_{\ell_1^-}^\dagger X_{\ell_2, \ell_1 \shortto \ell_2} U_{\ell_1^-} O_{\alpha;U} U_{\ell_1^-}^\dagger X_{\ell_1} U_{\ell_1^-}     P_{\alpha \alpha} U_{\ell_1^-}^\dagger X_{\ell_2, \ell_1 \shortto \ell_2} U_{\ell_1^-} O_{\alpha;U} P_{\alpha \beta}  O_{\beta;U} U_{\ell_1^-}^\dagger  X_{\ell_1} U_{\ell_1^-} )\nonumber\\
    & \quad - \tr(P_{\beta \alpha} U_{\ell_1^-}^\dagger X_{\ell_2, \ell_1 \shortto \ell_2} U_{\ell_1^-} O_{\alpha;U} U_{\ell_1^-}^\dagger X_{\ell_1} U_{\ell_1^-}     P_{\alpha \alpha} U_{\ell_1^-}^\dagger X_{\ell_2,\ell_1 \shortto \ell_2} U_{\ell_1^-} O_{\alpha;U} P_{\alpha \beta} U_{\ell_1^-}^\dagger X_{\ell_1} U_{\ell_1^-} O_{\beta;U} ) \nonumber\\
    & \quad - \tr(P_{\beta \alpha} U_{\ell_1^-}^\dagger X_{\ell_2, \ell_1 \shortto \ell_2} U_{\ell_1^-} O_{\alpha;U} U_{\ell_1^-}^\dagger X_{\ell_1} U_{\ell_1^-}     P_{\alpha \alpha} O_{\alpha;U} U_{\ell_1^-}^\dagger X_{\ell_2, \ell_1 \shortto \ell_2} U_{\ell_1^-}  P_{\alpha \beta} O_{\beta;U} U_{\ell_1^-}^\dagger X_{\ell_1} U_{\ell_1^-} ).
    \label{eq:ha12ga2gb1_l1}
\end{align}
The first is 
\begin{align}
    I_1 &\equiv \tr(P_{\beta \alpha} U_{\ell_1^-}^\dagger X_{\ell_1} X_{\ell_2, \ell_1 \shortto \ell_2} U_{\ell_1^-} O_{\alpha;U} P_{\alpha \alpha} U_{\ell_1^-}^\dagger X_{\ell_2,\ell_1 \shortto \ell_2} U_{\ell_1^-} O_{\alpha;U} P_{\alpha \beta} U_{\ell_1^-}^\dagger X_{\ell_1} U_{\ell_1^-} O_{\beta;U}) \nonumber\\
    &= \mathbb{E}_{\calU_{\rm RH}}\left[ \frac{d T_{\alpha \alpha} T_{\beta \beta} | T_{\alpha \alpha}| ^2 T_{\alpha \beta}^* T_{\beta \alpha}^*-| T_{\alpha \alpha}| ^4 | T_{\beta \beta}| ^2}{(d-1) (d+1)^2} \right].
\end{align}
The second is 
\begin{align}
    I_2 &\equiv \tr(P_{\beta \alpha} U_{\ell_1^-}^\dagger X_{\ell_1} X_{\ell_2, \ell_1 \shortto \ell_2} U_{\ell_1^-} O_{\alpha;U} P_{\alpha \alpha} O_{\alpha;U} U_{\ell_1^-}^\dagger X_{\ell_2, \ell_1 \shortto \ell_2} U_{\ell_1^-}  P_{\alpha \beta} O_{\beta;U} U_{\ell_1^-}^\dagger X_{\ell_1} U_{\ell_1^-}) \nonumber\\
    &= \mathbb{E}_{\calU_{\rm RH}}\left[ \delta_{\alpha \beta} \frac{d  | T_{\alpha \alpha}| ^2 T_{\beta \beta}^* (d T_{\beta \alpha}-T_{\alpha \alpha})}{\left(d^2-1\right)^2}+\frac{| T_{\alpha \alpha}| ^2 | T_{\beta \beta}| ^2 \left(| T_{\alpha \alpha}| ^2-d\right)}{\left(d^2-1\right)^2} \right].
\end{align}
The third is
\begin{align}
    I_3 &\equiv \tr(P_{\beta \alpha} U_{\ell_1^-}^\dagger X_{\ell_1} X_{\ell_2, \ell_1 \shortto \ell_2} U_{\ell_1^-} O_{\alpha;U} P_{\alpha \alpha} O_{\alpha;U} U_{\ell_1^-}^\dagger X_{\ell_2, \ell_1 \shortto \ell_2} U_{\ell_1^-}  P_{\alpha \beta} U_{\ell_1^-}^\dagger X_{\ell_1} U_{\ell_1^-} O_{\beta;U}) \nonumber\\
    &= \mathbb{E}_{\calU_{\rm RH}}\left[ \frac{| T_{\alpha \alpha}| ^2 \left(| T_{\beta \beta}| ^2 \left(| T_{\alpha \alpha}| ^2-d\right)-d T_{\alpha \alpha} T_{\beta \beta} T_{\alpha \beta}^* T_{\beta \alpha}^*\right)}{\left(d^2-1\right)^2} + \delta_{\alpha \beta}\frac{d^2  T_{\beta \beta} | T_{\alpha \alpha}| ^2 T_{\beta \alpha}^*}{\left(d^2-1\right)^2} \right].
\end{align}
The forth is
\begin{align}
    I_4 &\equiv \tr(P_{\beta \alpha} U_{\ell_1^-}^\dagger X_{\ell_1} X_{\ell_2, \ell_1 \shortto \ell_2} U_{\ell_1^-} O_{\alpha;U} P_{\alpha \alpha} U_{\ell_1^-}^\dagger X_{\ell_2, \ell_1 \shortto \ell_2} U_{\ell_1^-} O_{\alpha;U} P_{\alpha \beta}  O_{\beta;U} U_{\ell_1^-}^\dagger  X_{\ell_1} U_{\ell_1^-} ) \nonumber\\
    &= \mathbb{E}_{\calU_{\rm RH}}\left[\delta_{\alpha \beta}\frac{d  T_{\alpha \alpha} | T_{\alpha \alpha}| ^2 T_{\beta \beta}^*}{(d-1) (d+1)^2} - \frac{| T_{\alpha \alpha}| ^4 | T_{\beta \beta}| ^2}{(d-1) (d+1)^2} \right].
\end{align}
The fifth is
\begin{align}
    I_5 &\equiv \tr(P_{\beta \alpha} O_{\alpha;U} U_{\ell_1^-}^\dagger X_{\ell_2, \ell_1 \shortto \ell_2}  X_{\ell_1} U_{\ell_1^-}   P_{\alpha \alpha} U_{\ell_1^-}^\dagger X_{\ell_2,\ell_1 \shortto \ell_2} U_{\ell_1^-} O_{\alpha;U} P_{\alpha \beta} U_{\ell_1^-}^\dagger X_{\ell_1} U_{\ell_1^-} O_{\beta;U} ) \nonumber\\
    &= \mathbb{E}_{\calU_{\rm RH}}\left[ \delta_{\alpha \beta}\frac{d  T_{\beta \beta} | T_{\alpha \alpha}| ^2 \left(d T_{\beta \alpha}^*-T_{\alpha \alpha}^*\right)}{\left(d^2-1\right)^2}+\frac{| T_{\alpha \alpha}| ^2 | T_{\beta \beta}| ^2 \left(| T_{\alpha \alpha}| ^2-d\right)}{\left(d^2-1\right)^2} \right] = I_2^*.
\end{align}
The sixth is
\begin{align}
    I_6 &\equiv \tr(P_{\beta \alpha} O_{\alpha;U} U_{\ell_1^-}^\dagger X_{\ell_2, \ell_1 \shortto \ell_2}  X_{\ell_1} U_{\ell_1^-}   P_{\alpha \alpha} O_{\alpha;U} U_{\ell_1^-}^\dagger X_{\ell_2, \ell_1 \shortto \ell_2} U_{\ell_1^-}  P_{\alpha \beta} O_{\beta;U} U_{\ell_1^-}^\dagger X_{\ell_1} U_{\ell_1^-} ) \nonumber\\
    &= \mathbb{E}_{\calU_{\rm RH}}\left[ \frac{d T_{\alpha \beta} T_{\beta \alpha} | T_{\alpha \alpha}| ^2 T_{\alpha \alpha}^* T_{\beta \beta}^*-| T_{\alpha \alpha}| ^4 | T_{\beta \beta}| ^2}{(d-1) (d+1)^2} \right] = I_1^*.
\end{align}
The seventh is
\begin{align}
    I_7 &\equiv \tr(P_{\beta \alpha} O_{\alpha;U} U_{\ell_1^-}^\dagger X_{\ell_2, \ell_1 \shortto \ell_2}  X_{\ell_1} U_{\ell_1^-}   P_{\alpha \alpha}  O_{\alpha;U} U_{\ell_1^-}^\dagger X_{\ell_2, \ell_1 \shortto \ell_2} U_{\ell_1^-}  P_{\alpha \beta} U_{\ell_1^-}^\dagger X_{\ell_1} U_{\ell_1^-} O_{\beta;U} ) \nonumber\\
    &= \mathbb{E}_{\calU_{\rm RH}}\left[ \delta_{\alpha \beta}\frac{d  T_{\beta \beta} | T_{\alpha \alpha}| ^2 T_{\alpha \alpha}^*}{(d-1) (d+1)^2}-\frac{| T_{\alpha \alpha}| ^4 | T_{\beta \beta}| ^2}{(d-1) (d+1)^2} \right] = I_4^*.
\end{align}
The eighth is
\begin{align}
    I_8 &\equiv \tr(P_{\beta \alpha} O_{\alpha;U} U_{\ell_1^-}^\dagger X_{\ell_2, \ell_1 \shortto \ell_2}  X_{\ell_1} U_{\ell_1^-}   P_{\alpha \alpha} U_{\ell_1^-}^\dagger X_{\ell_2, \ell_1 \shortto \ell_2} U_{\ell_1^-} O_{\alpha;U} P_{\alpha \beta}  O_{\beta;U} U_{\ell_1^-}^\dagger  X_{\ell_1} U_{\ell_1^-} ) \nonumber\\
    &= \mathbb{E}_{\calU_{\rm RH}}\left[ \frac{| T_{\alpha \alpha}| ^2 \left(| T_{\beta \beta}| ^2 \left(| T_{\alpha \alpha}| ^2-d\right)-d T_{\alpha \beta} T_{\beta \alpha} T_{\alpha \alpha}^* T_{\beta \beta}^*\right)}{\left(d^2-1\right)^2} + \delta_{\alpha \beta}\frac{d^2  T_{\beta \alpha} | T_{\alpha \alpha}| ^2 T_{\beta \beta}^*}{\left(d^2-1\right)^2} \right] = I_3^*.
\end{align}
The ninth is
\begin{align}
    I_{9} &\equiv \tr(P_{\beta \alpha} U_{\ell_1^-}^\dagger X_{\ell_1} U_{\ell_1^-} O_{\alpha;U} U_{\ell_1^-}^\dagger X_{\ell_2, \ell_1 \shortto \ell_2} U_{\ell_1^-}  P_{\alpha \alpha} O_{\alpha;U} U_{\ell_1^-}^\dagger X_{\ell_2, \ell_1 \shortto \ell_2} U_{\ell_1^-}  P_{\alpha \beta} U_{\ell_1^-}^\dagger X_{\ell_1} U_{\ell_1^-} O_{\beta;U} ) \nonumber\\
    &= \mathbb{E}_{\calU_{\rm RH}}\left[ \frac{d T_{\alpha \alpha} T_{\beta \beta} | T_{\alpha \alpha}| ^2 T_{\alpha \beta}^* T_{\beta \alpha}^*-| T_{\alpha \alpha}| ^4 | T_{\beta \beta}| ^2}{(d-1) (d+1)^2} \right] = I_1.
\end{align}
The tenth is
\begin{align}
    I_{10} &\equiv \tr(P_{\beta \alpha} U_{\ell_1^-}^\dagger X_{\ell_1} U_{\ell_1^-} O_{\alpha;U} U_{\ell_1^-}^\dagger X_{\ell_2, \ell_1 \shortto \ell_2} U_{\ell_1^-}  P_{\alpha \alpha} U_{\ell_1^-}^\dagger X_{\ell_2, \ell_1 \shortto \ell_2} U_{\ell_1^-} O_{\alpha;U} P_{\alpha \beta}  O_{\beta;U} U_{\ell_1^-}^\dagger  X_{\ell_1} U_{\ell_1^-} ) \nonumber\\
    &= \mathbb{E}_{\calU_{\rm RH}}\left[ \delta_{\alpha \beta}\frac{d  T_{\alpha \alpha} T_{\beta \beta}^* \left(d-| T_{\alpha \alpha}| ^2\right)}{\left(d^2-1\right)^2}-\frac{| T_{\alpha \alpha}| ^2 | T_{\beta \beta}| ^2 \left(d-| T_{\alpha \alpha}| ^2\right)}{\left(d^2-1\right)^2} \right].
\end{align}
The eleventh is
\begin{align}
    I_{11} &\equiv \tr(P_{\beta \alpha} U_{\ell_1^-}^\dagger X_{\ell_1} U_{\ell_1^-} O_{\alpha;U} U_{\ell_1^-}^\dagger X_{\ell_2, \ell_1 \shortto \ell_2} U_{\ell_1^-}  P_{\alpha \alpha} U_{\ell_1^-}^\dagger X_{\ell_2,\ell_1 \shortto \ell_2} U_{\ell_1^-} O_{\alpha;U} P_{\alpha \beta} U_{\ell_1^-}^\dagger X_{\ell_1} U_{\ell_1^-} O_{\beta;U} ) \nonumber\\
    &= \mathbb{E}_{\calU_{\rm RH}}\left[ \frac{\left(d-| T_{\alpha \alpha}| ^2\right) \left(d T_{\alpha \alpha} T_{\beta \beta} T_{\alpha \beta}^* T_{\beta \alpha}^*-| T_{\alpha \alpha}| ^2 | T_{\beta \beta}| ^2\right)}{\left(d^2-1\right)^2} \right].
\end{align}
The twelfth is
\begin{align}
    I_{12} &\equiv \tr(P_{\beta \alpha} U_{\ell_1^-}^\dagger X_{\ell_1} U_{\ell_1^-} O_{\alpha;U} U_{\ell_1^-}^\dagger X_{\ell_2, \ell_1 \shortto \ell_2} U_{\ell_1^-}  P_{\alpha \alpha} O_{\alpha;U} U_{\ell_1^-}^\dagger X_{\ell_2, \ell_1 \shortto \ell_2} U_{\ell_1^-}  P_{\alpha \beta} O_{\beta;U} U_{\ell_1^-}^\dagger X_{\ell_1} U_{\ell_1^-} ) \nonumber\\
    &= \mathbb{E}_{\calU_{\rm RH}}\left[ \delta_{\alpha \beta}\frac{d  T_{\alpha \alpha} | T_{\alpha \alpha}| ^2 T_{\beta \beta}^*}{(d-1) (d+1)^2}-\frac{| T_{\alpha \alpha}| ^4 | T_{\beta \beta}| ^2}{(d-1) (d+1)^2} \right] = I_4.
\end{align}
The thirteenth is
\begin{align}
    I_{13} &\equiv \tr(P_{\beta \alpha} U_{\ell_1^-}^\dagger X_{\ell_2, \ell_1 \shortto \ell_2} U_{\ell_1^-} O_{\alpha;U} U_{\ell_1^-}^\dagger X_{\ell_1} U_{\ell_1^-}     P_{\alpha \alpha} O_{\alpha;U} U_{\ell_1^-}^\dagger X_{\ell_2, \ell_1 \shortto \ell_2} U_{\ell_1^-}  P_{\alpha \beta} U_{\ell_1^-}^\dagger X_{\ell_1} U_{\ell_1^-} O_{\beta;U} ) \nonumber\\
    &= \mathbb{E}_{\calU_{\rm RH}}\left[ \delta_{\alpha \beta}\frac{d  T_{\beta \beta} T_{\alpha \alpha}^* \left(d-| T_{\alpha \alpha}| ^2\right)}{\left(d^2-1\right)^2}-\frac{| T_{\alpha \alpha}| ^2 | T_{\beta \beta}| ^2 \left(d-| T_{\alpha \alpha}| ^2\right)}{\left(d^2-1\right)^2} \right] = I_{10}^*.
\end{align}
The fourteenth is
\begin{align}
    I_{14} &\equiv \tr(P_{\beta \alpha} U_{\ell_1^-}^\dagger X_{\ell_2, \ell_1 \shortto \ell_2} U_{\ell_1^-} O_{\alpha;U} U_{\ell_1^-}^\dagger X_{\ell_1} U_{\ell_1^-}     P_{\alpha \alpha} U_{\ell_1^-}^\dagger X_{\ell_2, \ell_1 \shortto \ell_2} U_{\ell_1^-} O_{\alpha;U} P_{\alpha \beta}  O_{\beta;U} U_{\ell_1^-}^\dagger  X_{\ell_1} U_{\ell_1^-} ) \nonumber\\
    &= \mathbb{E}_{\calU_{\rm RH}}\left[ \frac{d T_{\alpha \beta} T_{\beta \alpha} | T_{\alpha \alpha}| ^2 T_{\alpha \alpha}^* T_{\beta \beta}^*-| T_{\alpha \alpha}| ^4 | T_{\beta \beta}| ^2}{(d-1) (d+1)^2} \right] = I_1^*.
\end{align}
The fifteenth is
\begin{align}
    I_{15} &\equiv  \tr(P_{\beta \alpha} U_{\ell_1^-}^\dagger X_{\ell_2, \ell_1 \shortto \ell_2} U_{\ell_1^-} O_{\alpha;U} U_{\ell_1^-}^\dagger X_{\ell_1} U_{\ell_1^-}     P_{\alpha \alpha} U_{\ell_1^-}^\dagger X_{\ell_2,\ell_1 \shortto \ell_2} U_{\ell_1^-} O_{\alpha;U} P_{\alpha \beta} U_{\ell_1^-}^\dagger X_{\ell_1} U_{\ell_1^-} O_{\beta;U} ) \nonumber\\
    &= \mathbb{E}_{\calU_{\rm RH}}\left[ \delta_{\alpha \beta} \frac{d T_{\beta \beta} | T_{\alpha \alpha}| ^2 T_{\alpha \alpha}^*}{(d-1) (d+1)^2}-\frac{| T_{\alpha \alpha}| ^4 | T_{\beta \beta}| ^2}{(d-1) (d+1)^2} \right] = I_4^*.
\end{align}
The sixteenth (last) is
\begin{align}
    I_{16} &\equiv \tr(P_{\beta \alpha} U_{\ell_1^-}^\dagger X_{\ell_2, \ell_1 \shortto \ell_2} U_{\ell_1^-} O_{\alpha;U} U_{\ell_1^-}^\dagger X_{\ell_1} U_{\ell_1^-}     P_{\alpha \alpha} O_{\alpha;U} U_{\ell_1^-}^\dagger X_{\ell_2, \ell_1 \shortto \ell_2} U_{\ell_1^-}  P_{\alpha \beta} O_{\beta;U} U_{\ell_1^-}^\dagger X_{\ell_1} U_{\ell_1^-} ) \nonumber\\
    &= \mathbb{E}_{\calU_{\rm RH}}\left[ \frac{\left(d-| T_{\alpha \alpha}| ^2\right) \left(d T_{\alpha \beta} T_{\beta \alpha} T_{\alpha \alpha}^* T_{\beta \beta}^*-| T_{\alpha \alpha}| ^2 | T_{\beta \beta}| ^2\right)}{\left(d^2-1\right)^2} \right] = I_{11}^*.
\end{align}
Therefore, we have
\begin{align}
    &\mathbb{E}_{\calU_{\rm RH}}\left[\frac{\partial^2 \epsilon_\alpha}{\partial \theta_{\ell_1} \partial\theta_{\ell_2}} \frac{\partial \epsilon_\alpha}{\partial \theta_{\ell_2}} \frac{\partial \epsilon_\beta}{\partial \theta_{\ell_1}} \right] = \frac{1}{16} \sum_{i=1}^4\left(I_{4i+1} + I_{4i+2} - I_{4i+3} - I_{4i+4} \right) \nonumber\\
    &= \frac{1}{16}\left(2I_1 + I_2 - I_3 - 2I_4  + I_{10} - I_{11} + c.c. \right)\\
    &= \frac{1}{16} \mathbb{E}_{\calU_{\rm RH}}\left[ \frac{d^2 (2|T_{\alpha \alpha}|^2 - 1) \left( T_{\alpha \alpha} T_{\beta \beta} T_{\alpha \beta}^* T_{\beta \alpha}^* - \delta_{\alpha \beta} T_{\alpha \alpha} T_{\beta \beta}^*\right) }{\left(d^2-1\right)^2} + c.c.\right],
    \label{eq:ha12ga2gb1_l2}
\end{align}
which \BZ{matches} Eq.~\eqref{eq:ha12ga1gb2_l2}. 

\subsubsection{Summary}

Combining Eq.~\eqref{eq:ghg_l2}, Eq.~\eqref{eq:ha12ga1gb2_l2} and~\eqref{eq:ha12ga2gb1_l2}, we finally have
\begin{align}
    \overline{\mu_{\alpha \alpha \beta}(\infty)} &=  L \mathbb{E}_{\calU_{\rm RH}}\left[\frac{\partial \epsilon_\gamma}{\partial \theta_{\ell}} \frac{\partial^2 \epsilon_\alpha}{\partial \theta_{\ell}^2}\right] + L(L - 1)  \mathbb{E}_{\calU_{\rm RH}} \left[\frac{\partial \epsilon_\gamma}{\partial \theta_{\ell_1}} \frac{\partial^2 \epsilon_\alpha}{\partial \theta_{\ell_1} \partial \theta_{\ell_2}} \frac{\partial \epsilon_\beta}{\partial \theta_{\ell_2}}\right]\\
    &= L \mathbb{E}_{\calU_{\rm RH}}\left[ \frac{d+2}{8(d^2 - 1)(d+3)} \left((d+2) |T_{\alpha \alpha}|^2 - 2\right) \left( T_{\alpha \alpha} T_{\beta \beta} T_{\alpha\beta}^* T_{\beta \alpha}^* - \delta_{\alpha \beta} T_{\alpha \alpha} T_{\beta \beta}^* \right) + c.c.\right] \nonumber\\
    & \quad + L(L-1) \mathbb{E}_{\calU_{\rm RH}}\left[ \frac{d^2 (2|T_{\alpha \alpha}|^2 - 1) \left( T_{\alpha \alpha} T_{\beta \beta} T_{\alpha \beta}^* T_{\beta \alpha}^* - \delta_{\alpha \beta} T_{\alpha \alpha} T_{\beta \beta}^*\right) }{16\left(d^2-1\right)^2} + c.c.\right].
\end{align}
For $\alpha = \beta$, it can be simplified to
\begin{align}
    &\overline{\mu_{\alpha \alpha \alpha}(\infty)} \nonumber\\
    &= \mathbb{E}_{\calU_{\rm RH}}\left[ \frac{L (d+2)}{4(d^2 - 1)(d+3)} \left((d+2) |T_{\alpha \alpha}|^2 - 2\right) \left( |T_{\alpha \alpha}|^4 - |T_{\alpha \alpha}|^2 \right)
    + L(L-1) \frac{d^2 (2|T_{\alpha \alpha}|^2 - 1) \left( |T_{\alpha \alpha}|^4 - |T_{\alpha \alpha}|^2 \right) }{8\left(d^2-1\right)^2}\right]\\
    &= \frac{L o_\alpha (o_\alpha-1)}{8(d^2-1)}\left[\frac{2 (d+2)}{d+3} \left((d+2) o_\alpha - 2\right) + \frac{(L-1) d^2 (2 o_\alpha - 1)}{d^2 - 1}\right],
\end{align}
and for $\alpha \neq \beta$, it becomes
\begin{align}
    &\overline{\mu_{\alpha \alpha \beta}(\infty)} \nonumber\\
    &= \mathbb{E}_{\calU_{\rm RH}}\left[ \frac{L(d+2)}{8(d^2 - 1)(d+3)} \left((d+2) |T_{\alpha \alpha}|^2 - 2\right)  T_{\alpha \alpha} T_{\beta \beta} T_{\alpha\beta}^* T_{\beta \alpha}^* +  \frac{L(L-1) d^2 (2|T_{\alpha \alpha}|^2 - 1)  T_{\alpha \alpha} T_{\beta \beta} T_{\alpha \beta}^* T_{\beta \alpha}^*  }{16\left(d^2-1\right)^2} + c.c.\right]\\
    &= \mathbb{E}_{\calU_{\rm RH}}\left[ \frac{L(d+2)}{4(d^2 - 1)(d+3)} \left((d+2) |T_{\alpha \alpha}|^2 - 2\right)  |T_{\alpha \alpha}| |T_{\beta \beta}| |T_{\alpha\beta}| |T_{\beta \alpha}| +  \frac{L(L-1) d^2 (2|T_{\alpha \alpha}|^2 - 1)  |T_{\alpha \alpha}| |T_{\beta \beta}| |T_{\alpha \beta}| |T_{\beta \alpha}|  }{8\left(d^2-1\right)^2} \right] \label{eq:mu_off_diag}
    % &\simeq \frac{L}{8(d^2-1)(d-1)}\left[\frac{2(d+2)}{(d+3)}\left((d+2) o_\alpha - 2\right) + \frac{(L-1)d^2 (2o_\alpha - 1)}{d^2 - 1}\right]\sqrt{o_\alpha o_\beta (1-o_\beta) (1-o_\alpha)}.
\end{align}
With one more step, we can obtain the ensemble average relative dQNTK as
\begin{align}
    \overline{\lambda_{\alpha \alpha \alpha}(\infty)} &= \frac{\overline{\mu_{\alpha \alpha \alpha}(\infty)}}{\overline{K_{\alpha \alpha}(\infty)}} = -\frac{1}{4d} \left[\frac{2 (d+2)}{d+3} \left((d+2) o_\alpha - 2\right) + \frac{(L-1) d^2 (2 o_\alpha - 1)}{d^2 - 1}\right] \simeq -\frac{1}{4d}\left[2(d o_\alpha - 2) + L(2o_\alpha - 1)\right], \label{eq:lda_diag_rh}\\
    % \overline{\lambda_{\alpha \alpha \beta}(\infty)} &= \frac{\overline{\mu_{\alpha \alpha \beta}(\infty)}}{\sqrt{\overline{K_{\alpha \alpha}(\infty)}\: \overline{K_{\beta \beta}(\infty)}}}
    % \simeq \frac{1}{4d(d-1)}\left[\frac{2(d+2)}{(d+3)}\left((d+2) o_\alpha - 2\right) + \frac{(L-1)d^2 (2o_\alpha - 1)}{d^2 - 1}\right] \nonumber
    % \\
    % &\simeq \frac{1}{4d^2}\left[2(d o_\alpha - 2) + L(2o_\alpha - 1)\right]\label{eq:lda_offdiag_rh},
\end{align}
where we approximate it with $L, d \gg 1$ at the end. The off-diagonal part $\overline{\lambda_{\alpha \alpha \beta}(\infty)} = \frac{\overline{\mu_{\alpha \alpha \beta}(\infty)}}{\sqrt{\overline{K_{\alpha \alpha}(\infty)}\: \overline{K_{\beta \beta}(\infty)}}}$ can be found from Eq.~\eqref{eq:mu_off_diag} and~\eqref{eq:K_diag_rh}.

\end{widetext}


\begin{thebibliography}{69}%
\makeatletter
\providecommand \@ifxundefined [1]{%
 \@ifx{#1\undefined}
}%
\providecommand \@ifnum [1]{%
 \ifnum #1\expandafter \@firstoftwo
 \else \expandafter \@secondoftwo
 \fi
}%
\providecommand \@ifx [1]{%
 \ifx #1\expandafter \@firstoftwo
 \else \expandafter \@secondoftwo
 \fi
}%
\providecommand \natexlab [1]{#1}%
\providecommand \enquote  [1]{``#1''}%
\providecommand \bibnamefont  [1]{#1}%
\providecommand \bibfnamefont [1]{#1}%
\providecommand \citenamefont [1]{#1}%
\providecommand \href@noop [0]{\@secondoftwo}%
\providecommand \href [0]{\begingroup \@sanitize@url \@href}%
\providecommand \@href[1]{\@@startlink{#1}\@@href}%
\providecommand \@@href[1]{\endgroup#1\@@endlink}%
\providecommand \@sanitize@url [0]{\catcode `\\12\catcode `\$12\catcode
  `\&12\catcode `\#12\catcode `\^12\catcode `\_12\catcode `\%12\relax}%
\providecommand \@@startlink[1]{}%
\providecommand \@@endlink[0]{}%
\providecommand \url  [0]{\begingroup\@sanitize@url \@url }%
\providecommand \@url [1]{\endgroup\@href {#1}{\urlprefix }}%
\providecommand \urlprefix  [0]{URL }%
\providecommand \Eprint [0]{\href }%
\providecommand \doibase [0]{https://doi.org/}%
\providecommand \selectlanguage [0]{\@gobble}%
\providecommand \bibinfo  [0]{\@secondoftwo}%
\providecommand \bibfield  [0]{\@secondoftwo}%
\providecommand \translation [1]{[#1]}%
\providecommand \BibitemOpen [0]{}%
\providecommand \bibitemStop [0]{}%
\providecommand \bibitemNoStop [0]{.\EOS\space}%
\providecommand \EOS [0]{\spacefactor3000\relax}%
\providecommand \BibitemShut  [1]{\csname bibitem#1\endcsname}%
\let\auto@bib@innerbib\@empty
%</preamble>
\bibitem [{\citenamefont {Peruzzo}\ \emph {et~al.}(2014)\citenamefont
  {Peruzzo}, \citenamefont {McClean}, \citenamefont {Shadbolt}, \citenamefont
  {Yung}, \citenamefont {Zhou}, \citenamefont {Love}, \citenamefont
  {Aspuru-Guzik},\ and\ \citenamefont {Obrien}}]{peruzzo2014variational}%
  \BibitemOpen
  \bibfield  {author} {\bibinfo {author} {\bibfnamefont {A.}~\bibnamefont
  {Peruzzo}}, \bibinfo {author} {\bibfnamefont {J.}~\bibnamefont {McClean}},
  \bibinfo {author} {\bibfnamefont {P.}~\bibnamefont {Shadbolt}}, \bibinfo
  {author} {\bibfnamefont {M.-H.}\ \bibnamefont {Yung}}, \bibinfo {author}
  {\bibfnamefont {X.-Q.}\ \bibnamefont {Zhou}}, \bibinfo {author}
  {\bibfnamefont {P.~J.}\ \bibnamefont {Love}}, \bibinfo {author}
  {\bibfnamefont {A.}~\bibnamefont {Aspuru-Guzik}},\ and\ \bibinfo {author}
  {\bibfnamefont {J.~L.}\ \bibnamefont {Obrien}},\ }\bibfield  {title}
  {\bibinfo {title} {A variational eigenvalue solver on a photonic quantum
  processor},\ }\href@noop {} {\bibfield  {journal} {\bibinfo  {journal} {Nat.
  Commun.}\ }\textbf {\bibinfo {volume} {5}},\ \bibinfo {pages} {4213}
  (\bibinfo {year} {2014})}\BibitemShut {NoStop}%
\bibitem [{\citenamefont {Farhi}\ \emph {et~al.}(2014)\citenamefont {Farhi},
  \citenamefont {Goldstone},\ and\ \citenamefont {Gutmann}}]{farhi2014quantum}%
  \BibitemOpen
  \bibfield  {author} {\bibinfo {author} {\bibfnamefont {E.}~\bibnamefont
  {Farhi}}, \bibinfo {author} {\bibfnamefont {J.}~\bibnamefont {Goldstone}},\
  and\ \bibinfo {author} {\bibfnamefont {S.}~\bibnamefont {Gutmann}},\
  }\bibfield  {title} {\bibinfo {title} {A quantum approximate optimization
  algorithm},\ }\href@noop {} {\bibfield  {journal} {\bibinfo  {journal}
  {arXiv:1411.4028}\ } (\bibinfo {year} {2014})}\BibitemShut {NoStop}%
\bibitem [{\citenamefont {McClean}\ \emph {et~al.}(2016)\citenamefont
  {McClean}, \citenamefont {Romero}, \citenamefont {Babbush},\ and\
  \citenamefont {Aspuru-Guzik}}]{mcclean2016theory}%
  \BibitemOpen
  \bibfield  {author} {\bibinfo {author} {\bibfnamefont {J.~R.}\ \bibnamefont
  {McClean}}, \bibinfo {author} {\bibfnamefont {J.}~\bibnamefont {Romero}},
  \bibinfo {author} {\bibfnamefont {R.}~\bibnamefont {Babbush}},\ and\ \bibinfo
  {author} {\bibfnamefont {A.}~\bibnamefont {Aspuru-Guzik}},\ }\bibfield
  {title} {\bibinfo {title} {The theory of variational hybrid quantum-classical
  algorithms},\ }\href@noop {} {\bibfield  {journal} {\bibinfo  {journal} {New
  J. Phys.}\ }\textbf {\bibinfo {volume} {18}},\ \bibinfo {pages} {023023}
  (\bibinfo {year} {2016})}\BibitemShut {NoStop}%
\bibitem [{\citenamefont {McClean}\ \emph {et~al.}(2018)\citenamefont
  {McClean}, \citenamefont {Boixo}, \citenamefont {Smelyanskiy}, \citenamefont
  {Babbush},\ and\ \citenamefont {Neven}}]{mcclean2018barren}%
  \BibitemOpen
  \bibfield  {author} {\bibinfo {author} {\bibfnamefont {J.~R.}\ \bibnamefont
  {McClean}}, \bibinfo {author} {\bibfnamefont {S.}~\bibnamefont {Boixo}},
  \bibinfo {author} {\bibfnamefont {V.~N.}\ \bibnamefont {Smelyanskiy}},
  \bibinfo {author} {\bibfnamefont {R.}~\bibnamefont {Babbush}},\ and\ \bibinfo
  {author} {\bibfnamefont {H.}~\bibnamefont {Neven}},\ }\bibfield  {title}
  {\bibinfo {title} {Barren plateaus in quantum neural network training
  landscapes},\ }\href@noop {} {\bibfield  {journal} {\bibinfo  {journal} {Nat.
  Commun.}\ }\textbf {\bibinfo {volume} {9}},\ \bibinfo {pages} {4812}
  (\bibinfo {year} {2018})}\BibitemShut {NoStop}%
\bibitem [{\citenamefont {McArdle}\ \emph {et~al.}(2020)\citenamefont
  {McArdle}, \citenamefont {Endo}, \citenamefont {Aspuru-Guzik}, \citenamefont
  {Benjamin},\ and\ \citenamefont {Yuan}}]{mcardle2020quantum}%
  \BibitemOpen
  \bibfield  {author} {\bibinfo {author} {\bibfnamefont {S.}~\bibnamefont
  {McArdle}}, \bibinfo {author} {\bibfnamefont {S.}~\bibnamefont {Endo}},
  \bibinfo {author} {\bibfnamefont {A.}~\bibnamefont {Aspuru-Guzik}}, \bibinfo
  {author} {\bibfnamefont {S.~C.}\ \bibnamefont {Benjamin}},\ and\ \bibinfo
  {author} {\bibfnamefont {X.}~\bibnamefont {Yuan}},\ }\bibfield  {title}
  {\bibinfo {title} {Quantum computational chemistry},\ }\href@noop {}
  {\bibfield  {journal} {\bibinfo  {journal} {Rev. Mod. Phys.}\ }\textbf
  {\bibinfo {volume} {92}},\ \bibinfo {pages} {015003} (\bibinfo {year}
  {2020})}\BibitemShut {NoStop}%
\bibitem [{\citenamefont {Cerezo}\ \emph
  {et~al.}(2021{\natexlab{a}})\citenamefont {Cerezo}, \citenamefont
  {Arrasmith}, \citenamefont {Babbush}, \citenamefont {Benjamin}, \citenamefont
  {Endo}, \citenamefont {Fujii}, \citenamefont {McClean}, \citenamefont
  {Mitarai}, \citenamefont {Yuan}, \citenamefont {Cincio} \emph
  {et~al.}}]{cerezo2021variational}%
  \BibitemOpen
  \bibfield  {author} {\bibinfo {author} {\bibfnamefont {M.}~\bibnamefont
  {Cerezo}}, \bibinfo {author} {\bibfnamefont {A.}~\bibnamefont {Arrasmith}},
  \bibinfo {author} {\bibfnamefont {R.}~\bibnamefont {Babbush}}, \bibinfo
  {author} {\bibfnamefont {S.~C.}\ \bibnamefont {Benjamin}}, \bibinfo {author}
  {\bibfnamefont {S.}~\bibnamefont {Endo}}, \bibinfo {author} {\bibfnamefont
  {K.}~\bibnamefont {Fujii}}, \bibinfo {author} {\bibfnamefont {J.~R.}\
  \bibnamefont {McClean}}, \bibinfo {author} {\bibfnamefont {K.}~\bibnamefont
  {Mitarai}}, \bibinfo {author} {\bibfnamefont {X.}~\bibnamefont {Yuan}},
  \bibinfo {author} {\bibfnamefont {L.}~\bibnamefont {Cincio}}, \emph
  {et~al.},\ }\bibfield  {title} {\bibinfo {title} {Variational quantum
  algorithms},\ }\href@noop {} {\bibfield  {journal} {\bibinfo  {journal} {Nat.
  Rev. Phys.}\ }\textbf {\bibinfo {volume} {3}},\ \bibinfo {pages} {625}
  (\bibinfo {year} {2021}{\natexlab{a}})}\BibitemShut {NoStop}%
\bibitem [{\citenamefont {Killoran}\ \emph {et~al.}(2019)\citenamefont
  {Killoran}, \citenamefont {Bromley}, \citenamefont {Arrazola}, \citenamefont
  {Schuld}, \citenamefont {Quesada},\ and\ \citenamefont
  {Lloyd}}]{killoran2019continuous}%
  \BibitemOpen
  \bibfield  {author} {\bibinfo {author} {\bibfnamefont {N.}~\bibnamefont
  {Killoran}}, \bibinfo {author} {\bibfnamefont {T.~R.}\ \bibnamefont
  {Bromley}}, \bibinfo {author} {\bibfnamefont {J.~M.}\ \bibnamefont
  {Arrazola}}, \bibinfo {author} {\bibfnamefont {M.}~\bibnamefont {Schuld}},
  \bibinfo {author} {\bibfnamefont {N.}~\bibnamefont {Quesada}},\ and\ \bibinfo
  {author} {\bibfnamefont {S.}~\bibnamefont {Lloyd}},\ }\bibfield  {title}
  {\bibinfo {title} {Continuous-variable quantum neural networks},\ }\href
  {https://doi.org/10.1103/PhysRevResearch.1.033063} {\bibfield  {journal}
  {\bibinfo  {journal} {Phys. Rev. Res.}\ }\textbf {\bibinfo {volume} {1}},\
  \bibinfo {pages} {033063} (\bibinfo {year} {2019})}\BibitemShut {NoStop}%
\bibitem [{\citenamefont {Niu}\ \emph {et~al.}(2022)\citenamefont {Niu},
  \citenamefont {Zlokapa}, \citenamefont {Broughton}, \citenamefont {Boixo},
  \citenamefont {Mohseni}, \citenamefont {Smelyanskyi},\ and\ \citenamefont
  {Neven}}]{niu2022}%
  \BibitemOpen
  \bibfield  {author} {\bibinfo {author} {\bibfnamefont {M.~Y.}\ \bibnamefont
  {Niu}}, \bibinfo {author} {\bibfnamefont {A.}~\bibnamefont {Zlokapa}},
  \bibinfo {author} {\bibfnamefont {M.}~\bibnamefont {Broughton}}, \bibinfo
  {author} {\bibfnamefont {S.}~\bibnamefont {Boixo}}, \bibinfo {author}
  {\bibfnamefont {M.}~\bibnamefont {Mohseni}}, \bibinfo {author} {\bibfnamefont
  {V.}~\bibnamefont {Smelyanskyi}},\ and\ \bibinfo {author} {\bibfnamefont
  {H.}~\bibnamefont {Neven}},\ }\bibfield  {title} {\bibinfo {title}
  {Entangling quantum generative adversarial networks},\ }\href
  {https://doi.org/10.1103/PhysRevLett.128.220505} {\bibfield  {journal}
  {\bibinfo  {journal} {Phys. Rev. Lett.}\ }\textbf {\bibinfo {volume} {128}},\
  \bibinfo {pages} {220505} (\bibinfo {year} {2022})}\BibitemShut {NoStop}%
\bibitem [{\citenamefont {Kandala}\ \emph {et~al.}(2017)\citenamefont
  {Kandala}, \citenamefont {Mezzacapo}, \citenamefont {Temme}, \citenamefont
  {Takita}, \citenamefont {Brink}, \citenamefont {Chow},\ and\ \citenamefont
  {Gambetta}}]{kandala2017hardware}%
  \BibitemOpen
  \bibfield  {author} {\bibinfo {author} {\bibfnamefont {A.}~\bibnamefont
  {Kandala}}, \bibinfo {author} {\bibfnamefont {A.}~\bibnamefont {Mezzacapo}},
  \bibinfo {author} {\bibfnamefont {K.}~\bibnamefont {Temme}}, \bibinfo
  {author} {\bibfnamefont {M.}~\bibnamefont {Takita}}, \bibinfo {author}
  {\bibfnamefont {M.}~\bibnamefont {Brink}}, \bibinfo {author} {\bibfnamefont
  {J.~M.}\ \bibnamefont {Chow}},\ and\ \bibinfo {author} {\bibfnamefont
  {J.~M.}\ \bibnamefont {Gambetta}},\ }\bibfield  {title} {\bibinfo {title}
  {Hardware-efficient variational quantum eigensolver for small molecules and
  quantum magnets},\ }\href@noop {} {\bibfield  {journal} {\bibinfo  {journal}
  {Nature}\ }\textbf {\bibinfo {volume} {549}},\ \bibinfo {pages} {242}
  (\bibinfo {year} {2017})}\BibitemShut {NoStop}%
\bibitem [{\citenamefont {Ebadi}\ \emph {et~al.}(2022)\citenamefont {Ebadi},
  \citenamefont {Keesling}, \citenamefont {Cain}, \citenamefont {Wang},
  \citenamefont {Levine}, \citenamefont {Bluvstein}, \citenamefont {Semeghini},
  \citenamefont {Omran}, \citenamefont {Liu}, \citenamefont {Samajdar} \emph
  {et~al.}}]{ebadi2022quantum}%
  \BibitemOpen
  \bibfield  {author} {\bibinfo {author} {\bibfnamefont {S.}~\bibnamefont
  {Ebadi}}, \bibinfo {author} {\bibfnamefont {A.}~\bibnamefont {Keesling}},
  \bibinfo {author} {\bibfnamefont {M.}~\bibnamefont {Cain}}, \bibinfo {author}
  {\bibfnamefont {T.~T.}\ \bibnamefont {Wang}}, \bibinfo {author}
  {\bibfnamefont {H.}~\bibnamefont {Levine}}, \bibinfo {author} {\bibfnamefont
  {D.}~\bibnamefont {Bluvstein}}, \bibinfo {author} {\bibfnamefont
  {G.}~\bibnamefont {Semeghini}}, \bibinfo {author} {\bibfnamefont
  {A.}~\bibnamefont {Omran}}, \bibinfo {author} {\bibfnamefont {J.-G.}\
  \bibnamefont {Liu}}, \bibinfo {author} {\bibfnamefont {R.}~\bibnamefont
  {Samajdar}}, \emph {et~al.},\ }\bibfield  {title} {\bibinfo {title} {Quantum
  optimization of maximum independent set using rydberg atom arrays},\
  }\href@noop {} {\bibfield  {journal} {\bibinfo  {journal} {Science}\ }\textbf
  {\bibinfo {volume} {376}},\ \bibinfo {pages} {1209} (\bibinfo {year}
  {2022})}\BibitemShut {NoStop}%
\bibitem [{\citenamefont {Cong}\ \emph {et~al.}(2019)\citenamefont {Cong},
  \citenamefont {Choi},\ and\ \citenamefont {Lukin}}]{cong2019quantum}%
  \BibitemOpen
  \bibfield  {author} {\bibinfo {author} {\bibfnamefont {I.}~\bibnamefont
  {Cong}}, \bibinfo {author} {\bibfnamefont {S.}~\bibnamefont {Choi}},\ and\
  \bibinfo {author} {\bibfnamefont {M.~D.}\ \bibnamefont {Lukin}},\ }\bibfield
  {title} {\bibinfo {title} {Quantum convolutional neural networks},\
  }\href@noop {} {\bibfield  {journal} {\bibinfo  {journal} {Nature Physics}\
  }\textbf {\bibinfo {volume} {15}},\ \bibinfo {pages} {1273} (\bibinfo {year}
  {2019})}\BibitemShut {NoStop}%
\bibitem [{\citenamefont {Chen}\ \emph {et~al.}(2021)\citenamefont {Chen},
  \citenamefont {Wossnig}, \citenamefont {Severini}, \citenamefont {Neven},\
  and\ \citenamefont {Mohseni}}]{chen2021universal}%
  \BibitemOpen
  \bibfield  {author} {\bibinfo {author} {\bibfnamefont {H.}~\bibnamefont
  {Chen}}, \bibinfo {author} {\bibfnamefont {L.}~\bibnamefont {Wossnig}},
  \bibinfo {author} {\bibfnamefont {S.}~\bibnamefont {Severini}}, \bibinfo
  {author} {\bibfnamefont {H.}~\bibnamefont {Neven}},\ and\ \bibinfo {author}
  {\bibfnamefont {M.}~\bibnamefont {Mohseni}},\ }\bibfield  {title} {\bibinfo
  {title} {Universal discriminative quantum neural networks},\ }\href@noop {}
  {\bibfield  {journal} {\bibinfo  {journal} {Quantum Machine Intelligence}\
  }\textbf {\bibinfo {volume} {3}},\ \bibinfo {pages} {1} (\bibinfo {year}
  {2021})}\BibitemShut {NoStop}%
\bibitem [{\citenamefont {Zhang}\ and\ \citenamefont
  {Zhuang}(2022)}]{zhang2022fast}%
  \BibitemOpen
  \bibfield  {author} {\bibinfo {author} {\bibfnamefont {B.}~\bibnamefont
  {Zhang}}\ and\ \bibinfo {author} {\bibfnamefont {Q.}~\bibnamefont {Zhuang}},\
  }\bibfield  {title} {\bibinfo {title} {Fast decay of classification error in
  variational quantum circuits},\ }\href@noop {} {\bibfield  {journal}
  {\bibinfo  {journal} {Quantum Science and Technology}\ }\textbf {\bibinfo
  {volume} {7}},\ \bibinfo {pages} {035017} (\bibinfo {year}
  {2022})}\BibitemShut {NoStop}%
\bibitem [{\citenamefont {Zhuang}\ and\ \citenamefont
  {Zhang}(2019)}]{zhuang2019}%
  \BibitemOpen
  \bibfield  {author} {\bibinfo {author} {\bibfnamefont {Q.}~\bibnamefont
  {Zhuang}}\ and\ \bibinfo {author} {\bibfnamefont {Z.}~\bibnamefont {Zhang}},\
  }\bibfield  {title} {\bibinfo {title} {Physical-layer supervised learning
  assisted by an entangled sensor network},\ }\href
  {https://doi.org/10.1103/PhysRevX.9.041023} {\bibfield  {journal} {\bibinfo
  {journal} {Phys. Rev. X}\ }\textbf {\bibinfo {volume} {9}},\ \bibinfo {pages}
  {041023} (\bibinfo {year} {2019})}\BibitemShut {NoStop}%
\bibitem [{\citenamefont {Xia}\ \emph {et~al.}(2021)\citenamefont {Xia},
  \citenamefont {Li}, \citenamefont {Zhuang},\ and\ \citenamefont
  {Zhang}}]{xia2021}%
  \BibitemOpen
  \bibfield  {author} {\bibinfo {author} {\bibfnamefont {Y.}~\bibnamefont
  {Xia}}, \bibinfo {author} {\bibfnamefont {W.}~\bibnamefont {Li}}, \bibinfo
  {author} {\bibfnamefont {Q.}~\bibnamefont {Zhuang}},\ and\ \bibinfo {author}
  {\bibfnamefont {Z.}~\bibnamefont {Zhang}},\ }\bibfield  {title} {\bibinfo
  {title} {Quantum-enhanced data classification with a variational entangled
  sensor network},\ }\href {https://doi.org/10.1103/PhysRevX.11.021047}
  {\bibfield  {journal} {\bibinfo  {journal} {Phys. Rev. X}\ }\textbf {\bibinfo
  {volume} {11}},\ \bibinfo {pages} {021047} (\bibinfo {year}
  {2021})}\BibitemShut {NoStop}%
\bibitem [{\citenamefont {Farhi}\ and\ \citenamefont
  {Neven}(2018)}]{farhi2018classification}%
  \BibitemOpen
  \bibfield  {author} {\bibinfo {author} {\bibfnamefont {E.}~\bibnamefont
  {Farhi}}\ and\ \bibinfo {author} {\bibfnamefont {H.}~\bibnamefont {Neven}},\
  }\bibfield  {title} {\bibinfo {title} {Classification with quantum neural
  networks on near term processors},\ }\href@noop {} {\bibfield  {journal}
  {\bibinfo  {journal} {arXiv:1802.06002}\ } (\bibinfo {year}
  {2018})}\BibitemShut {NoStop}%
\bibitem [{\citenamefont {Li}\ \emph {et~al.}(2022)\citenamefont {Li},
  \citenamefont {Lu},\ and\ \citenamefont {Deng}}]{li2022quantum}%
  \BibitemOpen
  \bibfield  {author} {\bibinfo {author} {\bibfnamefont {W.}~\bibnamefont
  {Li}}, \bibinfo {author} {\bibfnamefont {Z.-d.}\ \bibnamefont {Lu}},\ and\
  \bibinfo {author} {\bibfnamefont {D.-L.}\ \bibnamefont {Deng}},\ }\bibfield
  {title} {\bibinfo {title} {Quantum neural network classifiers: A tutorial},\
  }\href@noop {} {\bibfield  {journal} {\bibinfo  {journal} {SciPost Physics
  Lecture Notes}\ ,\ \bibinfo {pages} {061}} (\bibinfo {year}
  {2022})}\BibitemShut {NoStop}%
\bibitem [{\citenamefont {Grant}\ \emph {et~al.}(2018)\citenamefont {Grant},
  \citenamefont {Benedetti}, \citenamefont {Cao}, \citenamefont {Hallam},
  \citenamefont {Lockhart}, \citenamefont {Stojevic}, \citenamefont {Green},\
  and\ \citenamefont {Severini}}]{grant2018hierarchical}%
  \BibitemOpen
  \bibfield  {author} {\bibinfo {author} {\bibfnamefont {E.}~\bibnamefont
  {Grant}}, \bibinfo {author} {\bibfnamefont {M.}~\bibnamefont {Benedetti}},
  \bibinfo {author} {\bibfnamefont {S.}~\bibnamefont {Cao}}, \bibinfo {author}
  {\bibfnamefont {A.}~\bibnamefont {Hallam}}, \bibinfo {author} {\bibfnamefont
  {J.}~\bibnamefont {Lockhart}}, \bibinfo {author} {\bibfnamefont
  {V.}~\bibnamefont {Stojevic}}, \bibinfo {author} {\bibfnamefont {A.~G.}\
  \bibnamefont {Green}},\ and\ \bibinfo {author} {\bibfnamefont
  {S.}~\bibnamefont {Severini}},\ }\bibfield  {title} {\bibinfo {title}
  {Hierarchical quantum classifiers},\ }\href@noop {} {\bibfield  {journal}
  {\bibinfo  {journal} {npj Quantum Information}\ }\textbf {\bibinfo {volume}
  {4}},\ \bibinfo {pages} {65} (\bibinfo {year} {2018})}\BibitemShut {NoStop}%
\bibitem [{\citenamefont {Li}\ \emph {et~al.}(2015)\citenamefont {Li},
  \citenamefont {Liu}, \citenamefont {Xu},\ and\ \citenamefont
  {Du}}]{li2015experimental}%
  \BibitemOpen
  \bibfield  {author} {\bibinfo {author} {\bibfnamefont {Z.}~\bibnamefont
  {Li}}, \bibinfo {author} {\bibfnamefont {X.}~\bibnamefont {Liu}}, \bibinfo
  {author} {\bibfnamefont {N.}~\bibnamefont {Xu}},\ and\ \bibinfo {author}
  {\bibfnamefont {J.}~\bibnamefont {Du}},\ }\bibfield  {title} {\bibinfo
  {title} {Experimental realization of a quantum support vector machine},\
  }\href@noop {} {\bibfield  {journal} {\bibinfo  {journal} {Physical review
  letters}\ }\textbf {\bibinfo {volume} {114}},\ \bibinfo {pages} {140504}
  (\bibinfo {year} {2015})}\BibitemShut {NoStop}%
\bibitem [{\citenamefont {Havl{\'\i}{\v{c}}ek}\ \emph
  {et~al.}(2019)\citenamefont {Havl{\'\i}{\v{c}}ek}, \citenamefont
  {C{\'o}rcoles}, \citenamefont {Temme}, \citenamefont {Harrow}, \citenamefont
  {Kandala}, \citenamefont {Chow},\ and\ \citenamefont
  {Gambetta}}]{havlivcek2019supervised}%
  \BibitemOpen
  \bibfield  {author} {\bibinfo {author} {\bibfnamefont {V.}~\bibnamefont
  {Havl{\'\i}{\v{c}}ek}}, \bibinfo {author} {\bibfnamefont {A.~D.}\
  \bibnamefont {C{\'o}rcoles}}, \bibinfo {author} {\bibfnamefont
  {K.}~\bibnamefont {Temme}}, \bibinfo {author} {\bibfnamefont {A.~W.}\
  \bibnamefont {Harrow}}, \bibinfo {author} {\bibfnamefont {A.}~\bibnamefont
  {Kandala}}, \bibinfo {author} {\bibfnamefont {J.~M.}\ \bibnamefont {Chow}},\
  and\ \bibinfo {author} {\bibfnamefont {J.~M.}\ \bibnamefont {Gambetta}},\
  }\bibfield  {title} {\bibinfo {title} {Supervised learning with
  quantum-enhanced feature spaces},\ }\href@noop {} {\bibfield  {journal}
  {\bibinfo  {journal} {Nature}\ }\textbf {\bibinfo {volume} {567}},\ \bibinfo
  {pages} {209} (\bibinfo {year} {2019})}\BibitemShut {NoStop}%
\bibitem [{\citenamefont {Larocca}\ \emph {et~al.}(2023)\citenamefont
  {Larocca}, \citenamefont {Ju}, \citenamefont {Garc{\'\i}a-Mart{\'\i}n},
  \citenamefont {Coles},\ and\ \citenamefont {Cerezo}}]{larocca2023theory}%
  \BibitemOpen
  \bibfield  {author} {\bibinfo {author} {\bibfnamefont {M.}~\bibnamefont
  {Larocca}}, \bibinfo {author} {\bibfnamefont {N.}~\bibnamefont {Ju}},
  \bibinfo {author} {\bibfnamefont {D.}~\bibnamefont
  {Garc{\'\i}a-Mart{\'\i}n}}, \bibinfo {author} {\bibfnamefont {P.~J.}\
  \bibnamefont {Coles}},\ and\ \bibinfo {author} {\bibfnamefont
  {M.}~\bibnamefont {Cerezo}},\ }\bibfield  {title} {\bibinfo {title} {Theory
  of overparametrization in quantum neural networks},\ }\href@noop {}
  {\bibfield  {journal} {\bibinfo  {journal} {Nat. Comput. Sci.}\ }\textbf
  {\bibinfo {volume} {3}},\ \bibinfo {pages} {542} (\bibinfo {year}
  {2023})}\BibitemShut {NoStop}%
\bibitem [{\citenamefont {Liu}\ \emph {et~al.}(2022)\citenamefont {Liu},
  \citenamefont {Tacchino}, \citenamefont {Glick}, \citenamefont {Jiang},\ and\
  \citenamefont {Mezzacapo}}]{liu2022representation}%
  \BibitemOpen
  \bibfield  {author} {\bibinfo {author} {\bibfnamefont {J.}~\bibnamefont
  {Liu}}, \bibinfo {author} {\bibfnamefont {F.}~\bibnamefont {Tacchino}},
  \bibinfo {author} {\bibfnamefont {J.~R.}\ \bibnamefont {Glick}}, \bibinfo
  {author} {\bibfnamefont {L.}~\bibnamefont {Jiang}},\ and\ \bibinfo {author}
  {\bibfnamefont {A.}~\bibnamefont {Mezzacapo}},\ }\bibfield  {title} {\bibinfo
  {title} {Representation learning via quantum neural tangent kernels},\
  }\href@noop {} {\bibfield  {journal} {\bibinfo  {journal} {PRX Quantum}\
  }\textbf {\bibinfo {volume} {3}},\ \bibinfo {pages} {030323} (\bibinfo {year}
  {2022})}\BibitemShut {NoStop}%
\bibitem [{\citenamefont {Liu}\ \emph {et~al.}(2023)\citenamefont {Liu},
  \citenamefont {Najafi}, \citenamefont {Sharma}, \citenamefont {Tacchino},
  \citenamefont {Jiang},\ and\ \citenamefont {Mezzacapo}}]{liu2023analytic}%
  \BibitemOpen
  \bibfield  {author} {\bibinfo {author} {\bibfnamefont {J.}~\bibnamefont
  {Liu}}, \bibinfo {author} {\bibfnamefont {K.}~\bibnamefont {Najafi}},
  \bibinfo {author} {\bibfnamefont {K.}~\bibnamefont {Sharma}}, \bibinfo
  {author} {\bibfnamefont {F.}~\bibnamefont {Tacchino}}, \bibinfo {author}
  {\bibfnamefont {L.}~\bibnamefont {Jiang}},\ and\ \bibinfo {author}
  {\bibfnamefont {A.}~\bibnamefont {Mezzacapo}},\ }\bibfield  {title} {\bibinfo
  {title} {Analytic theory for the dynamics of wide quantum neural networks},\
  }\href@noop {} {\bibfield  {journal} {\bibinfo  {journal} {Phys. Rev. Lett.}\
  }\textbf {\bibinfo {volume} {130}},\ \bibinfo {pages} {150601} (\bibinfo
  {year} {2023})}\BibitemShut {NoStop}%
\bibitem [{\citenamefont {Liu}\ \emph {et~al.}(2024)\citenamefont {Liu},
  \citenamefont {Lin},\ and\ \citenamefont {Jiang}}]{liu2022laziness}%
  \BibitemOpen
  \bibfield  {author} {\bibinfo {author} {\bibfnamefont {J.}~\bibnamefont
  {Liu}}, \bibinfo {author} {\bibfnamefont {Z.}~\bibnamefont {Lin}},\ and\
  \bibinfo {author} {\bibfnamefont {L.}~\bibnamefont {Jiang}},\ }\bibfield
  {title} {\bibinfo {title} {Laziness, barren plateau, and noises in machine
  learning},\ }\href@noop {} {\bibfield  {journal} {\bibinfo  {journal} {Mach.
  Learn.: Sci. Technol.arXiv:2206.09313}\ }\textbf {\bibinfo {volume} {5}},\
  \bibinfo {pages} {015058} (\bibinfo {year} {2024})}\BibitemShut {NoStop}%
\bibitem [{\citenamefont {Wang}\ \emph {et~al.}(2022)\citenamefont {Wang},
  \citenamefont {Liu}, \citenamefont {Liu}, \citenamefont {Luo}, \citenamefont
  {Du},\ and\ \citenamefont {Tao}}]{wang2022symmetric}%
  \BibitemOpen
  \bibfield  {author} {\bibinfo {author} {\bibfnamefont {X.}~\bibnamefont
  {Wang}}, \bibinfo {author} {\bibfnamefont {J.}~\bibnamefont {Liu}}, \bibinfo
  {author} {\bibfnamefont {T.}~\bibnamefont {Liu}}, \bibinfo {author}
  {\bibfnamefont {Y.}~\bibnamefont {Luo}}, \bibinfo {author} {\bibfnamefont
  {Y.}~\bibnamefont {Du}},\ and\ \bibinfo {author} {\bibfnamefont
  {D.}~\bibnamefont {Tao}},\ }\bibfield  {title} {\bibinfo {title} {Symmetric
  pruning in quantum neural networks},\ }\href@noop {} {\bibfield  {journal}
  {\bibinfo  {journal} {arXiv:2208.14057}\ } (\bibinfo {year}
  {2022})}\BibitemShut {NoStop}%
\bibitem [{\citenamefont {Yu}\ \emph {et~al.}(2023)\citenamefont {Yu},
  \citenamefont {Li}, \citenamefont {Ye}, \citenamefont {Lu}, \citenamefont
  {Han},\ and\ \citenamefont {Deng}}]{yu2023expressibility}%
  \BibitemOpen
  \bibfield  {author} {\bibinfo {author} {\bibfnamefont {L.-W.}\ \bibnamefont
  {Yu}}, \bibinfo {author} {\bibfnamefont {W.}~\bibnamefont {Li}}, \bibinfo
  {author} {\bibfnamefont {Q.}~\bibnamefont {Ye}}, \bibinfo {author}
  {\bibfnamefont {Z.}~\bibnamefont {Lu}}, \bibinfo {author} {\bibfnamefont
  {Z.}~\bibnamefont {Han}},\ and\ \bibinfo {author} {\bibfnamefont {D.-L.}\
  \bibnamefont {Deng}},\ }\bibfield  {title} {\bibinfo {title}
  {Expressibility-induced concentration of quantum neural tangent kernels},\
  }\href@noop {} {\bibfield  {journal} {\bibinfo  {journal} {arXiv:2311.04965}\
  } (\bibinfo {year} {2023})}\BibitemShut {NoStop}%
\bibitem [{\citenamefont {Zhang}\ \emph
  {et~al.}(2023{\natexlab{a}})\citenamefont {Zhang}, \citenamefont {Liu},
  \citenamefont {Wu}, \citenamefont {Jiang},\ and\ \citenamefont
  {Zhuang}}]{zhang2023dynamical}%
  \BibitemOpen
  \bibfield  {author} {\bibinfo {author} {\bibfnamefont {B.}~\bibnamefont
  {Zhang}}, \bibinfo {author} {\bibfnamefont {J.}~\bibnamefont {Liu}}, \bibinfo
  {author} {\bibfnamefont {X.-C.}\ \bibnamefont {Wu}}, \bibinfo {author}
  {\bibfnamefont {L.}~\bibnamefont {Jiang}},\ and\ \bibinfo {author}
  {\bibfnamefont {Q.}~\bibnamefont {Zhuang}},\ }\bibfield  {title} {\bibinfo
  {title} {Dynamical phase transition in quantum neural networks with large
  depth},\ }\href@noop {} {\bibfield  {journal} {\bibinfo  {journal}
  {arXiv:2311.18144}\ } (\bibinfo {year} {2023}{\natexlab{a}})}\BibitemShut
  {NoStop}%
\bibitem [{\citenamefont {Helstrom}(1967)}]{helstrom1967minimum}%
  \BibitemOpen
\bibfield  {journal} {  }\bibfield  {author} {\bibinfo {author} {\bibfnamefont
  {C.~W.}\ \bibnamefont {Helstrom}},\ }\bibfield  {title} {\bibinfo {title}
  {Minimum mean-squared error of estimates in quantum statistics},\ }\href@noop
  {} {\bibfield  {journal} {\bibinfo  {journal} {Physics letters A}\ }\textbf
  {\bibinfo {volume} {25}},\ \bibinfo {pages} {101} (\bibinfo {year}
  {1967})}\BibitemShut {NoStop}%
\bibitem [{\citenamefont {Helstrom}(1969)}]{helstrom1969quantum}%
  \BibitemOpen
  \bibfield  {author} {\bibinfo {author} {\bibfnamefont {C.~W.}\ \bibnamefont
  {Helstrom}},\ }\bibfield  {title} {\bibinfo {title} {Quantum detection and
  estimation theory},\ }\href@noop {} {\bibfield  {journal} {\bibinfo
  {journal} {Journal of Statistical Physics}\ }\textbf {\bibinfo {volume}
  {1}},\ \bibinfo {pages} {231} (\bibinfo {year} {1969})}\BibitemShut {NoStop}%
\bibitem [{\citenamefont {Zhang}\ \emph
  {et~al.}(2023{\natexlab{b}})\citenamefont {Zhang}, \citenamefont {Allcock},
  \citenamefont {Wan}, \citenamefont {Liu}, \citenamefont {Sun}, \citenamefont
  {Yu}, \citenamefont {Yang}, \citenamefont {Qiu}, \citenamefont {Ye},
  \citenamefont {Chen} \emph {et~al.}}]{zhang2023tensorcircuit}%
  \BibitemOpen
  \bibfield  {author} {\bibinfo {author} {\bibfnamefont {S.-X.}\ \bibnamefont
  {Zhang}}, \bibinfo {author} {\bibfnamefont {J.}~\bibnamefont {Allcock}},
  \bibinfo {author} {\bibfnamefont {Z.-Q.}\ \bibnamefont {Wan}}, \bibinfo
  {author} {\bibfnamefont {S.}~\bibnamefont {Liu}}, \bibinfo {author}
  {\bibfnamefont {J.}~\bibnamefont {Sun}}, \bibinfo {author} {\bibfnamefont
  {H.}~\bibnamefont {Yu}}, \bibinfo {author} {\bibfnamefont {X.-H.}\
  \bibnamefont {Yang}}, \bibinfo {author} {\bibfnamefont {J.}~\bibnamefont
  {Qiu}}, \bibinfo {author} {\bibfnamefont {Z.}~\bibnamefont {Ye}}, \bibinfo
  {author} {\bibfnamefont {Y.-Q.}\ \bibnamefont {Chen}}, \emph {et~al.},\
  }\bibfield  {title} {\bibinfo {title} {Tensorcircuit: a quantum software
  framework for the nisq era},\ }\href@noop {} {\bibfield  {journal} {\bibinfo
  {journal} {Quantum}\ }\textbf {\bibinfo {volume} {7}},\ \bibinfo {pages}
  {912} (\bibinfo {year} {2023}{\natexlab{b}})}\BibitemShut {NoStop}%
\bibitem [{\citenamefont {Cerezo}\ \emph
  {et~al.}(2021{\natexlab{b}})\citenamefont {Cerezo}, \citenamefont {Sone},
  \citenamefont {Volkoff}, \citenamefont {Cincio},\ and\ \citenamefont
  {Coles}}]{cerezo2021cost}%
  \BibitemOpen
  \bibfield  {author} {\bibinfo {author} {\bibfnamefont {M.}~\bibnamefont
  {Cerezo}}, \bibinfo {author} {\bibfnamefont {A.}~\bibnamefont {Sone}},
  \bibinfo {author} {\bibfnamefont {T.}~\bibnamefont {Volkoff}}, \bibinfo
  {author} {\bibfnamefont {L.}~\bibnamefont {Cincio}},\ and\ \bibinfo {author}
  {\bibfnamefont {P.~J.}\ \bibnamefont {Coles}},\ }\bibfield  {title} {\bibinfo
  {title} {Cost function dependent barren plateaus in shallow parametrized
  quantum circuits},\ }\href@noop {} {\bibfield  {journal} {\bibinfo  {journal}
  {Nat. Commun.}\ }\textbf {\bibinfo {volume} {12}},\ \bibinfo {pages} {1791}
  (\bibinfo {year} {2021}{\natexlab{b}})}\BibitemShut {NoStop}%
\bibitem [{\citenamefont {Nakata}\ and\ \citenamefont
  {Murao}(2013)}]{nakata2013diagonal}%
  \BibitemOpen
  \bibfield  {author} {\bibinfo {author} {\bibfnamefont {Y.}~\bibnamefont
  {Nakata}}\ and\ \bibinfo {author} {\bibfnamefont {M.}~\bibnamefont {Murao}},\
  }\bibfield  {title} {\bibinfo {title} {Diagonal-unitary 2-design and their
  implementations by quantum circuits},\ }\href@noop {} {\bibfield  {journal}
  {\bibinfo  {journal} {International Journal of Quantum Information}\ }\textbf
  {\bibinfo {volume} {11}},\ \bibinfo {pages} {1350062} (\bibinfo {year}
  {2013})}\BibitemShut {NoStop}%
\bibitem [{\citenamefont {Roberts}\ and\ \citenamefont
  {Yoshida}(2017)}]{roberts2017chaos}%
  \BibitemOpen
  \bibfield  {author} {\bibinfo {author} {\bibfnamefont {D.~A.}\ \bibnamefont
  {Roberts}}\ and\ \bibinfo {author} {\bibfnamefont {B.}~\bibnamefont
  {Yoshida}},\ }\bibfield  {title} {\bibinfo {title} {Chaos and complexity by
  design},\ }\href@noop {} {\bibfield  {journal} {\bibinfo  {journal} {Journal
  of High Energy Physics}\ }\textbf {\bibinfo {volume} {2017}},\ \bibinfo
  {pages} {1} (\bibinfo {year} {2017})}\BibitemShut {NoStop}%
\bibitem [{\citenamefont {Bergholm}\ \emph {et~al.}(2018)\citenamefont
  {Bergholm}, \citenamefont {Izaac}, \citenamefont {Schuld}, \citenamefont
  {Gogolin}, \citenamefont {Ahmed}, \citenamefont {Ajith}, \citenamefont
  {Alam}, \citenamefont {Alonso-Linaje}, \citenamefont {AkashNarayanan},
  \citenamefont {Asadi} \emph {et~al.}}]{bergholm2018pennylane}%
  \BibitemOpen
  \bibfield  {author} {\bibinfo {author} {\bibfnamefont {V.}~\bibnamefont
  {Bergholm}}, \bibinfo {author} {\bibfnamefont {J.}~\bibnamefont {Izaac}},
  \bibinfo {author} {\bibfnamefont {M.}~\bibnamefont {Schuld}}, \bibinfo
  {author} {\bibfnamefont {C.}~\bibnamefont {Gogolin}}, \bibinfo {author}
  {\bibfnamefont {S.}~\bibnamefont {Ahmed}}, \bibinfo {author} {\bibfnamefont
  {V.}~\bibnamefont {Ajith}}, \bibinfo {author} {\bibfnamefont {M.~S.}\
  \bibnamefont {Alam}}, \bibinfo {author} {\bibfnamefont {G.}~\bibnamefont
  {Alonso-Linaje}}, \bibinfo {author} {\bibfnamefont {B.}~\bibnamefont
  {AkashNarayanan}}, \bibinfo {author} {\bibfnamefont {A.}~\bibnamefont
  {Asadi}}, \emph {et~al.},\ }\bibfield  {title} {\bibinfo {title} {Pennylane:
  Automatic differentiation of hybrid quantum-classical computations},\
  }\href@noop {} {\bibfield  {journal} {\bibinfo  {journal} {arXiv preprint
  arXiv:1811.04968}\ } (\bibinfo {year} {2018})}\BibitemShut {NoStop}%
\bibitem [{\citenamefont {{Qiskit contributors}}(2023)}]{Qiskit}%
  \BibitemOpen
  \bibfield  {author} {\bibinfo {author} {\bibnamefont {{Qiskit
  contributors}}},\ }\href {https://doi.org/10.5281/zenodo.2573505} {\bibinfo
  {title} {Qiskit: An open-source framework for quantum computing}} (\bibinfo
  {year} {2023})\BibitemShut {NoStop}%
\bibitem [{\citenamefont {Thanasilp}\ \emph {et~al.}(2023)\citenamefont
  {Thanasilp}, \citenamefont {Wang}, \citenamefont {Nghiem}, \citenamefont
  {Coles},\ and\ \citenamefont {Cerezo}}]{thanasilp2023subtleties}%
  \BibitemOpen
  \bibfield  {author} {\bibinfo {author} {\bibfnamefont {S.}~\bibnamefont
  {Thanasilp}}, \bibinfo {author} {\bibfnamefont {S.}~\bibnamefont {Wang}},
  \bibinfo {author} {\bibfnamefont {N.~A.}\ \bibnamefont {Nghiem}}, \bibinfo
  {author} {\bibfnamefont {P.}~\bibnamefont {Coles}},\ and\ \bibinfo {author}
  {\bibfnamefont {M.}~\bibnamefont {Cerezo}},\ }\bibfield  {title} {\bibinfo
  {title} {Subtleties in the trainability of quantum machine learning models},\
  }\href@noop {} {\bibfield  {journal} {\bibinfo  {journal} {Quantum Machine
  Intelligence}\ }\textbf {\bibinfo {volume} {5}},\ \bibinfo {pages} {21}
  (\bibinfo {year} {2023})}\BibitemShut {NoStop}%
\bibitem [{\citenamefont {Ragone}\ \emph {et~al.}(2024)\citenamefont {Ragone},
  \citenamefont {Bakalov}, \citenamefont {Sauvage}, \citenamefont {Kemper},
  \citenamefont {Ortiz~Marrero}, \citenamefont {Larocca},\ and\ \citenamefont
  {Cerezo}}]{ragone2024lie}%
  \BibitemOpen
  \bibfield  {author} {\bibinfo {author} {\bibfnamefont {M.}~\bibnamefont
  {Ragone}}, \bibinfo {author} {\bibfnamefont {B.~N.}\ \bibnamefont {Bakalov}},
  \bibinfo {author} {\bibfnamefont {F.}~\bibnamefont {Sauvage}}, \bibinfo
  {author} {\bibfnamefont {A.~F.}\ \bibnamefont {Kemper}}, \bibinfo {author}
  {\bibfnamefont {C.}~\bibnamefont {Ortiz~Marrero}}, \bibinfo {author}
  {\bibfnamefont {M.}~\bibnamefont {Larocca}},\ and\ \bibinfo {author}
  {\bibfnamefont {M.}~\bibnamefont {Cerezo}},\ }\bibfield  {title} {\bibinfo
  {title} {A lie algebraic theory of barren plateaus for deep parameterized
  quantum circuits},\ }\href@noop {} {\bibfield  {journal} {\bibinfo  {journal}
  {Nature Communications}\ }\textbf {\bibinfo {volume} {15}},\ \bibinfo {pages}
  {7172} (\bibinfo {year} {2024})}\BibitemShut {NoStop}%
\bibitem [{\citenamefont {You}\ \emph {et~al.}(2023)\citenamefont {You},
  \citenamefont {Chakrabarti}, \citenamefont {Chen},\ and\ \citenamefont
  {Wu}}]{you2023analyzing}%
  \BibitemOpen
  \bibfield  {author} {\bibinfo {author} {\bibfnamefont {X.}~\bibnamefont
  {You}}, \bibinfo {author} {\bibfnamefont {S.}~\bibnamefont {Chakrabarti}},
  \bibinfo {author} {\bibfnamefont {B.}~\bibnamefont {Chen}},\ and\ \bibinfo
  {author} {\bibfnamefont {X.}~\bibnamefont {Wu}},\ }\bibfield  {title}
  {\bibinfo {title} {Analyzing convergence in quantum neural networks:
  Deviations from neural tangent kernels},\ }\href@noop {} {\bibfield
  {journal} {\bibinfo  {journal} {arXiv preprint arXiv:2303.14844}\ } (\bibinfo
  {year} {2023})}\BibitemShut {NoStop}%
\bibitem [{\citenamefont {Crooks}(2019)}]{crooks2019gradients}%
  \BibitemOpen
  \bibfield  {author} {\bibinfo {author} {\bibfnamefont {G.}~\bibnamefont
  {Crooks}},\ }\bibfield  {title} {\bibinfo {title} {Gradients of parameterized
  quantum gates using the parameter-shift rule and gate decomposition},\
  }\href@noop {} {\bibfield  {journal} {\bibinfo  {journal} {arXiv preprint
  arXiv:1905.13311}\ } (\bibinfo {year} {2019})}\BibitemShut {NoStop}%
\bibitem [{\citenamefont {Mitarai}\ \emph {et~al.}(2018)\citenamefont {Mitarai},
  \citenamefont {Negoro}, \citenamefont {Kitagawa},\ and\ \citenamefont
  {Fujii}}]{mitarai2018quantum}%
  \BibitemOpen
  \bibfield  {author} {\bibinfo {author} {\bibfnamefont {K.}~\bibnamefont
  {Mitarai}}, \bibinfo {author} {\bibfnamefont {M.}~\bibnamefont {Negoro}},
  \bibinfo {author} {\bibfnamefont {M.}~\bibnamefont {Kitagawa}},\ and\ \bibinfo
  {author} {\bibfnamefont {K.}~\bibnamefont {Fujii}},\ }\bibfield  {title}
  {\bibinfo {title} {Quantum circuit learning},\ }\href@noop {} {\bibfield
  {journal} {\bibinfo  {journal} {Physical Review A}\ }\textbf {\bibinfo
  {volume} {98}},\ \bibinfo {pages} {032309} (\bibinfo {year} {2018})}%
  \BibitemShut {NoStop}%
\bibitem [{\citenamefont {Fukuda}\ \emph {et~al.}(2019)\citenamefont {Fukuda},
  \citenamefont {K\"onig},\ and\ \citenamefont {Nechita}}]{fukuda2019rtni}%
  \BibitemOpen
  \bibfield  {author} {\bibinfo {author} {\bibfnamefont {M.}~\bibnamefont
  {Fukuda}}, \bibinfo {author} {\bibfnamefont {R.}~\bibnamefont {K\"onig}},\
  and\ \bibinfo {author} {\bibfnamefont {I.}~\bibnamefont {Nechita}},\
  }\bibfield  {title} {\bibinfo {title} {{RTNI}---A symbolic integrator for
  Haar-random tensor networks},\ }\href@noop {} {\bibfield  {journal} {\bibinfo
  {journal} {Journal of Physics A: Mathematical and Theoretical}\ }\textbf
  {\bibinfo {volume} {52}},\ \bibinfo {pages} {425303} (\bibinfo {year}
  {2019})}\BibitemShut {NoStop}%
\end{thebibliography}
\end{document}